\documentclass[3p,preprint,times]{elsarticle}

\usepackage[bookmarks,pdfhighlight=/O,pdfstartview=FitH,colorlinks=false]{hyperref}
\usepackage[utf8]{inputenc}
\usepackage{amsmath,amstext,amsthm,amssymb,amsfonts,slashed,qmech}
\usepackage{mathtools}
\usepackage{dsfont}
\usepackage{graphicx,color}
\usepackage{multirow}
\usepackage{booktabs}
\usepackage{tabularx}
\usepackage{array}
\usepackage[most]{tcolorbox}
\usepackage{blindtext}
\usepackage{lipsum}
\newtcolorbox[auto counter, number within=chapter]{NumBox}[1][]{
    enhanced,
    breakable,
    fonttitle=\scshape,
    title={Numerical methods: \thetcbcounter},
    #1
}
\usepackage{rotating}
\usepackage{hyperref}
\hypersetup{
    colorlinks = false,
    linkbordercolor = {white}
}

\usepackage[font=footnotesize, justification=RaggedRight]{subfig}

\newcommand{\fmc}{\mathrm{fm}/c}
\newcommand{\ee}{\ensuremath{\mathrm{e}}}
\newcommand{\fbF}{\ensuremath{f_{\bar{\text{F}}}}}
\newcommand{\muB}{\mu_{\text{B}}}

\begin{document}

\begin{frontmatter}

\title{Kinetics of the chiral phase transition in a quark-meson $\sigma$-model}

\author[1]{Alex Meistrenko}
\ead{alex.meistrenko@googlemail.com}

\author[1]{Hendrik van Hees}
\ead{hees@itp.uni-frankfurt.de}

\author[1]{Carsten Greiner}
\ead{Carsten.Greiner@th.physik.uni-frankfurt.de}

\address[1]{Institut f{\"u}r Theoretische Physik, Goethe-Universit{\"a}t
  Frankfurt am Main, Max-von-Laue-Straße 1, D-60438 Frankfurt am Main, Germany
}

\begin{abstract}
  In this study an effective description in the 2PI effective-action
  formalism for systems of quarks and mesons in and out of equilibrium
  within a numerical approach is developed, allowing to approximate the
  complexity of QCD by taking only the lightest and most relevant
  degrees of freedom into account. In particular the temporarily
  building up of fluctuations of the net-baryon number encoded by the
  fourth-order cumulant (or the rescaled curtosis) for lower momenta is
  being demonstrated when the phase transition occurs near the critical
  point, or even stronger when the phase transition is of first order,
  although the initial system is prepared with purely Gaussian
  fluctuations in the net baryon number. This is the result of the
  evolving slow and critical order parameter, i.e., the
  $\sigma$-field. On the other hand, depending on the speed of the
  (Hubble-)expansion scale, the final dissipative evolution due to the
  collisions among the mesons, the quarks and anti-quarks and the order
  field weakens the final fluctuations considerably.
\end{abstract}


\end{frontmatter}

\section{Introduction}
\label{chap:introduction}

The experimental approach for studying fundamental properties of quantum
chromodynamics (QCD) is based on heavy ion physics at relativistic
energies, requiring large accelerating facilities like the Conseil
Europ{\'e}en pour la Recherche Nucl{\'e}aire (CERN) and the Relativistic
Heavy Ion Collider (RHIC) as well as much smaller but more specialized
facilities like the Gesellschaft für Schwerionenforschung (GSI), the
Facility for Antiproton and Ion Research (FAIR) as well as the
Nuclotron-based Ion Collider fAcility (NICA), where highly ionized atoms
are accelerated to almost the speed of light. Nowadays, it is
well-established, that there exists a numerous number of different phases
within QCD, which are characterized by different effective degrees of
freedom. Besides the hadronic phase, there is a second highly prominent
phase, known as the quark gluon plasma (QGP) \cite{ARSENE20051},
denoting a quasi-free state of quarks and gluons at high energy and/or
particle densities several times higher compared to the usual values of
nuclear matter. The existence of a QGP can be expected in the inner core
of neutron stars as well as in the early Universe during the first few
$\mu \text{s}$ after the Big Bang.  However, those conditions are not
accessible for experimental physics. The only possible way of studying
QGP properties on Earth is encoded into the dynamical evolution of an
expanding fireball created in a heavy-ion collision as described before,
suggesting that the created medium of quarks and gluons exists only for
a very short period of time
$\Delta t\simeq 10\,\fmc\sim 10^{-23}\,\mathrm{s}$.

For small values of the chemical potential $\muB$ it has been found
within lattice QCD, that there is a smooth crossover between the
quark-gluon plasma and the hadronic phase at a relatively high
temperature of $T_{\text{c}} \simeq 155\,\MeV$
\cite{FODOR200287,Aoki:2006we}, resulting in the limit of vanishing
light-quark masses in a second-order phase transition with
$\mathrm{O}(4)$ as the underlying universality class.
Depending on the quark masses one expects different orders of phase
transitions for strongly interacting matter (see e.g.
Ref.~\cite{doi:10.1142/9789814663717_0001}). However, the connection
between the most interesting symmetry groups, being responsible for
deconfinement and chiral symmetry restoration, remains a subject of
study. Based on effective models like the Nambu-Jona-Lasinio (NJL),
quark-meson (QM) model with constituent quarks
\cite{PhysRev.122.345,PhysRev.124.246,Jungnickel:1995fp,doi:10.1142/S0217751X03014034}
and their Polyakov-loop extended versions (PNJL), (PQM)
\cite{PhysRevD.73.014019,PhysRevD.75.034007,PhysRevD.76.074023,HERBST201158}
as well as universality arguments, for high values of the baryon
chemical potential one expects a first-order chiral phase transition,
ending in a critical point of second order
\cite{MASAYUKI1989668,PhysRevD.58.096007,doi:10.1142/S0217751X92001757,PhysRevD.75.085015}.
The exploration of the QCD phase diagram is one of the most important
goals for relativistic heavy-ion experiments.

In the vicinity of a critical point the relaxation time diverges, due to
growing correlation lengths, leading probably to a phase transition out
of equilibrium. Assuming then a static medium in a stationary condition
of equilibrium could be a crude and insufficient approximation,
resulting in overestimated or significantly false predictions in case of
most promising observables \cite{Asakawa:2019kek}. Even a first-order
boundary with a coexistence of two different phases could require a
non-equilibrium description, when the system is driven out of the
equilibrium state during the evolution process of a heavy-ion
collision. Furthermore, a heavy-ion collision forms a finite volume of
the medium, expanding rapidly in time and thus leading to a limited
formation of long-range correlations. In detail, the size of correlated
domains cannot significantly exceed a scale of $\xi\sim 5\,\fm$,
comparable with the initial radius of a heavy-ion collision.
 
Consequently, those effects could drastically modify predicted
observables for the phase transition, which have been derived from
thermodynamic quantities. Such observables of interest are mainly the
cumulants and cumulant ratios of conserved quantities
\cite{PhysRevD.60.114028,PhysRevLett.85.2072}, which are directly
related to thermodynamic susceptibilities via the correlation length of
the system
\cite{PhysRevLett.102.032301,PhysRevLett.103.262301,PhysRevD.82.074008}. The
net-baryon number as well as the net-charge number are the most
prominent and promising examples of conserved quantities. A recent
calculation shows the behaviour of fluctuations encoded in cumulant
ratios for the net-quark (respectively net-baryon) number within a
functional renormalization group (FRG) approach (see
Refs.~\cite{WETTERICH199390,doi:10.1142/S0217751X94000972,BERGES2002223})
of a Polyakov-loop extended quark-meson model \cite{PhysRevC.83.054904}.
In particular, the results show a critical behavior of higher-order
cumulant ratios\footnote{With increasing order of cumulants the
  dependence on the correlation length becomes more prominent.} for
several choices of the quark chemical potential to temperature
$\mu_{\text{q}}/T$, standing for linear trajectories in the
$T$-$\mu_{\text{q}}$ plane. Analogously, different choices of the ratio
between the entropy and the net-quark density refer to non-linear
trajectories in the $T$-$\mu_{\text{q}}$ plane, being more related to
the expansion process in a heavy-ion collision. The ratio
$R_{4,2}=c_4/c_2\equiv\kappa\sigma^2_{\text{q,net}}$ for generalized
susceptibilities (see Ref.~\cite{PhysRevC.83.054904} for scaling
behavior at finite chemical potential),
\begin{equation*}
\begin{split}
  c_n\left (T\right )&=\frac{\partial^n\left [p\left (T,\mu_{\text{q}}\right )/T^4\right ]}{\partial\left (\mu_{\text{q}}/T\right )^n}\,,\\
  c_1&=\frac{N_{\text{q,net}}}{VT^3}\,,\quad c_2=\frac{1}{VT^3}\left
    <\left (N_{\text{q,net}}-\left <N_{\text{q,net}}\right >\right
    )^2\right
  >\equiv\frac{1}{VT^3}\sigma_{\text{q,net}}^2 \, ,\\
  c_3&=\frac{1}{VT^3}\left <\left (N_{\text{q,net}}-\left
        <N_{\text{q,net}}\right >\right )^3\right >\,,\quad
  c_4=\frac{1}{VT^3}\left [\left <\left (N_{\text{q,net}}-\left
          <N_{\text{q,net}}\right >\right )^4\right
    >-3\sigma_{\text{q,net}}^4\right ]\,,
\end{split}
\end{equation*} 
shows an oscillatory behavior and changes its sign below a certain value
of the temperature, denoted with $T_{\text{c}}$ in the
following. Thereby, different choices for $\mu_{\text{q}}/T$ refer to
different and experimentally accessible center-of-mass energies,
motivating the energy-beam scan at RHIC. Similar studies of cumulants
are also known from lattice-QCD calculations, where due to the sign
problem of fermions in MC-based path-integral formulations a Taylor
expansion in powers of the thermodynamic potential\footnote{Note, that a
  Taylor expansion with a finite number of terms would break down in
  case of a critical point, requiring a careful analysis of the
  convergence radius.} is required to compute cumulants of conserved
quantities for non-vanishing chemical potentials. Thereby, one
introduces thermodynamic bulk quantities (i.e., pressure, energy and
entropy densities), which depend on the chemical potentials of baryon
number, strangeness and electric charge\footnote{Often simple referred
  to as equation of state (EoS) with non-vanishing chemical potentials.}
(see \cite{FODOR200287,PhysRevD.66.074507,PhysRevD.95.054504} and
references therein).

An experimental confirmation of the critical point would be crucial to
understand fundamental properties of QCD in form of the present
symmetries. Furthermore, it would significantly help to confirm and
improve effective models. Following that, from first experimental data
it seems to be unlikely that the critical point can be found below
$\muB \equiv 3 \mu_q \sim 200\,\MeV$ \cite{PhysRevLett.105.022302},
being in full agreement with lattice-QCD calculations. More interesting
are the net-proton number fluctuations\footnote{Due to high statistics
  net-proton number fluctuations are used for estimating the overall
  net-baryon number fluctuations.} in terms of cumulants, skewness and
kurtosis as presented by the STAR Collaboration at RHIC and summarized
in \cite{LUO201675}. The preliminary results on the fourth order
cumulant ratio, encoded in terms of the rescaled kurtosis
$\kappa\sigma^2$, show a non-trivial dependence on the center-of-mass
energy in central Au+Au collisions, which to date cannot be completely
explained by established transport models (see \cite{HE2016296} and
references therein), even though some of the observed features are
present in the quark-meson model of Ref.~\cite{PhysRevC.83.054904}. The
remarkable drop of the rescaled kurtosis from $\sim 1$ to even negative
values within the error bars with a subsequent increase of this
observable for decreasing center of mass energies could indicate a
critical point near a center of mass energy of $\sqrt{s_{NN}}=20\,\GeV$.

Such a behavior is expected in a similar way for the order parameter
$\sigma$ in an effective description of QCD
\cite{PhysRevLett.107.052301}. For the fourth-order cumulant
$\kappa_4\left (\sigma_V\right ):=\left <\left (\delta\sigma_V\right
  )^4\right >-3\left <\left (\delta\sigma_V\right )^2\right >^2$, where
$\sigma_V$ denotes the volume-averaged order parameter for the zero mode
$\sigma_V=\int\dd^3\,\vec x\,\sigma\left (x\right )$ and
$\left (\delta\sigma_V\right )^n$ stands for the centralized moment of
order $n$. Here, the cumulant of fourth order is calculated as a
function of the Ising parameters
$t=\left (T-T_{\text{c}}\right )/T_{\text{c}}$ (reduced temperature) and
$H$ (magnetic field), which can be analytically mapped to the relevant
parameters $T$ and $\muB$ of the QCD phase diagram. This calculation
requires a specification of the universal equation of state around the
critical point ~\cite{PhysRevLett.107.052301}. The analysis shows that
the fourth-order cumulant is negative along the crossover line with
$t>0$ and strongly positive in the region of the first-order phase
transition $t<0$ for all values of $H\neq 0$. Qualitatively, this can be
understood by considering the probability distribution of the
volume-averaged order parameter $\sigma_V$, having a Gaussian
distribution in the crossover region (negative kurtosis), which is a
direct consequence of the central limit theorem. With decreasing $t$ the
Gaussian distribution function develops to a broader probability
distribution alongside with a further decrease of the kurtosis. Below
$t=0$ the probability distribution of $\sigma_V$ has a two-peak
structure in the mixed phase of the first order, resulting for $t<0$ in
a positive kurtosis around the dominant peak for all values of
$H\neq 0$.

The density plot for the cumulant $\kappa_4$ of the volume averaged
order paramater $\sigma_V$ can be transformed in a similar behavior of
the cumulants of conserved quantities, which can be motivated by
considering for instance the fluctuation of the proton distribution
function due to fluctuations of the order parameter and pure statistical
fluctuations around the equilibrium distribution function
$f_{\text{p}}^{\text{eq}}$:
\begin{equation*} \delta f_p=\delta f_{\text{p}}^{\text{stat}}+\frac{\partial
f_{\text{p}}^{\text{eq}}}{\partial m}g\delta\sigma\,,
\end{equation*} 
where the second term follows from the mass relation $m=g\sigma$ with
$g$ denoting the coupling constant of the linear $\sigma$
model. Integrating over the phase
space results in
\begin{equation*} \delta N_p=\delta N_{\text{p}}^{\text{stat}}+\delta\sigma_V
gd\int\frac{\dd^3\vec p}{\left (2\pi\right )^3}\frac{\partial f_{\text{p}}^{\text{eq}}}{\partial
m}
\end{equation*} 
with $d$ denoting the degeneracy factor. Assuming that
$\delta N_{\text{p}}^{\text{stat}}$ and $\delta\sigma_V$ are stochastically independent
variables and focusing on most singular terms near the critical point,
leads to the following expression:
\begin{equation*}
\begin{split}
  \kappa_4\left (N_{\text{p}}\right )&=\kappa_4\left
    (N_{\text{p}}^{\text{stat}}\right )+\kappa_4\left (\sigma_V\right
  )\left (gd \int\frac{\dd^3\vec p}{\left (2\pi\right )^3}\frac{\partial
      f_{\text{p}}^{\text{eq}}}{\partial
      m}\right )^4+\cdots \\
  &=\left <N_{\text{p}}\right >+\kappa_4\left (\sigma_V\right )\left
    (\frac{gd}{T} \int\frac{\dd^3\vec p}{\left (2\pi\right
      )^3}\frac{\partial E_{\text{p}}}{\partial
        m}f_{\text{p}}^{\text{eq}}\right )^4+...\,,
\end{split}
\end{equation*} 
where $E_{\text{p}}$ denotes the dispersion relation and the last
equality follows from the reasonable assumption of Poisson statistics
for $\delta N_{\text{p}}^{\text{stat}}$. Following that, a negative
value of $\kappa_4\left (\sigma_V\right )$ leads to a reduction of the
corresponding fourth-order cumulant of the proton distribution function
in comparison to the Poisson value $\left <N_{\text{p}} \right >$
(analogously for net protons), leading to an effective mechanism for the
experimentally observed behavior with a
significant deviation from the cumulants of the expected Skellam
distribution function for net-protons.

For a discovery of the QCD critical point it is crucial to develop
dynamical models for fluctuations of the net-baryon number that can be
embedded in simulations of heavy-ion collisions: Recently it has been
emphasized that the relation of the higher-order cumulants being
proportional to some higher power in the correlation length being valid
at ideal thermodynamical equilibrium close to the critical point can not
truely be established in a non-equilibrium situation envisaged in
relativistic heavy ion collisions. The diffusion of the conserved
net-baryon number is typically understood by elastic collisions of the
quarks or protons, whereas the order parameter and thus the potential
onset of correlated domains obeys a generalized wave equation
\cite{Asakawa:2019kek}. The
crucial step is thus to develop a nonequilibrium description of
transport equations of the quarks and the coupled chiral fields in order
to learn and understand the possible formation of (critical)
fluctuations in the net-baryon number and thus the build up of higher
order cumulants.

In addition, it requires improved theoretical, dynamical predictions and
experiments with highly increased statistics to measure the overall
net-baryon fluctuations and to exclude also possible technical issues,
concerning, for instance, a limited detector acceptance. Last but not
least, also physical effects due to finite-volume fluctuations have to
be taken into account carefully.

With all this said, the central point of this paper is an effective
description of the chiral phase transition by means of the quark-meson
model, resulting from an approximate realization of the chiral
symmetry. Thereby, the chiral phase transition is studied for systems of
quarks and mesons in and out of equilibrium within a numerical approach
of the coupled transport and wave equations. We are interested in
investigating critical phenomena at the phase transition and the
critical point, where time-dependent long-range correlations can
arise. In particular, this work focuses on the restoration of the chiral
symmetry within the linear $\sigma$ model with constituent quarks. Its
first-order phase transition ends in a critical point of second order,
and non-equilibrium effects near the phase transition can significantly
modify the critical behavior, such as the fluctuation of the net-baryon
number (density), being the most important observable for the phase
transition in this model.

The paper is organized as follows: We will start our discussion in
Sect.~\ref{sec:IntroQCD} with some fundamental properties of QCD,
focusing on the restoration of the chiral symmetry in a quark-meson
model at high temperatures, based on the motivation described in
Ref.~\cite{Meistrenko:2013yya}. In Sect.~\ref{chap:EvolEq} we derive a
set of coupled evolution equations by considering a particular
truncation and approximate scheme of the underlying 2PI effective
quantum action\footnote{A more phenomenological study, motivated by the
  Langevin equation, is discussed in Ref.~\cite{Wesp2018}, where a novel
  statistical approach for the scattering effects between quarks as
  quasi particles and mean fields is introduced. Very recently a similar
  representation of the 2PI of the quark-meson model is given in
  \cite{Shen:2020jya}. Here the full Kadonoff-Baym evolution equations
  are derived and solved for a static, non-equilibrium situation in the
  crossover at net baryon density being zero and to demonstrate the
  appearance of full thermal equilibrium and spectral properties of the
  mesons and quarks.}. Therefore, we review main aspects of the
so-called real-time Schwinger-Keldysh formalism and focus on the
importance of the interaction between soft and hard modes, encoded in a
dissipation kernel. After applying reasonable approximations to the
effective quantum action and discussing thermodynamic properties of the
effective thermodynamic potential as well as effective mass terms,
neglecting divergent vacuum contributions, we end up with a set of
coupled kinetic Boltzmann-type and dissipative mean-field (wave-type)
equations for the interactions among the quarks, the mesons and the
chiral fields.

Finally, within this full approach with mean-field and transport
equations we show the dynamical study of the chiral phase transition in
Sect.~\ref{chap:DynamicalEvol}, where the equations of motion are
modified to include an expanding three-dimensional Hubble-geometry for
homogeneous and isotropic systems. Here, the cumulants and cumulant
ratios of the net-quark number are considered with respect to different
momentum bins and different initial truely Gaussian configurations of
the net-quark number in large sets of independent runs, allowing to
study a time-dependent evolution of cumulants in expanding systems.

To keep the main text readable, more detailed calculations are performed
in the appendix.

\section{Chiral effective model of QCD}
\label{sec:IntroQCD}

Within the Standard Model of elementary particle physics the strong
interaction is described by Quantum Chromodynamics (QCD), which is a
gauge theory based on the ``color gauge group''
$\text{SU}(3)_{\text{c}}$. It describes the strong interaction between
quarks, anti-quarks, and gluons. One important feature of the
perturbative aspects of the theory is that the running gauge coupling of
the theory is small at high energy scales (``asymptotic freedom'') and
large at low energy scales and thus perturbative methods become invalid
in the low-energy range.

Phenomenologically the observed asymptotic free states of the theory are
not quarks and gluons but rather hadrons, i.e., bound states of three
quarks (baryons) or quark-antiquark pairs (mesons). Correspondingly the
absence of observable states carrying color is known as the
confinement of quarks and gluons within color-neutral bound
states. Thus at low energies it is desirable to develop effective field
theories describing hadrons.

A guiding principle for the formulation of such effective models are the
symmetries of the QCD Lagrangian \cite{dgh92},
\begin{equation}
\mathcal{L}=-\frac{1}{4} F_{\mu \nu}^a F^{a \mu \nu} +\bar{\psi} (\ii
\slashed{D}-\hat{M}) \psi.
\end{equation}
Here the fields $\psi =\equiv \psi_{ia}$ are Dirac fields for the 6
quark flavors $i$ (up, down, charm, strange, top, and bottom) and 3
colors $a$. The mass matrix $M=\mathrm{diag}(m_u,m_d,m_c,m_s,m_t,m_b)$
acts in flavor space.

The $\text{SU}_{\text{c}}(3)$ gluon field $A^{\mu}=T^a A^{a
  \mu}$ with the hermitean matrices $T^a=\lambda^a/2$ ($\lambda^a$:
Gell-Mann matrices) acting in color space define the covariant
derivative,
\begin{equation}
\D_{\mu} = \partial_{\mu} - \ii g A_{\mu}. 
\end{equation}
The field-strength tensor $F_{\mu \nu}$ is defined by
\begin{equation}
F_{\mu \nu}^a=\partial_{\mu} A_{\nu}^a - \partial_{\nu} A_{\nu}^a -
g f^{abc} A_{\mu}^a A_{\nu}^b.
\end{equation}
In the light-quark sector (up and down quarks) the most important
symmetry in the context of effective hadronic theories is the chiral
symmetry in the limit of massless light quarks. The assumption of this
approximate symmetry is justified by the observation that the ``current
masses'' of the light up- and down-quarks are around a few MeV, which is
small compared to a typical hadronic mass scale of around $1\;\GeV$ and
can thus be treated as a perturbation. Since from now on we are
interested only in the lightest quarks, we understand that the flavor
index runs only over $i \in \{\text{up},\text{down} \}$.

As it turns out, besides the weak explicit breaking of the chiral
symmetry the symmetry is also explicitly broken by the formation of a
non-vanishing quark condensate, i.e., the vacuum-expectation value
$\langle 0| \bar{\psi} \psi |0 \rangle \neq 0$. This leads to the
spontaneous breaking of the
$\text{SU}(2)_{\text{L}} \times \text{SU}(2)_{\text{R}}$ chiral symmetry
(acting on the left- and right-handed parts,
$\psi_{\text{R}/\text{L}} = (1 \pm \gamma_5)/2 \psi$ of the doublet
formed by the up- and down-quark fields) to the approximate isospin
symmetry $\text{SU}(2)_{\text{V}}$ with the up- and down-quark fields
transforming as a doublet under this transformation. At the hadronic
level the spontaneous breaking of the chiral symmetry manifests itself
in the mass splitting between chiral-partner hadrons as well as the
presence of the very light pions, which are considered the to be
(pseudo-)Goldstone bosons of the spontaneously broken symmetry.

As detailed in the Introduction in the following we are interested in
the understanding of the phase transition from the deconfined phase of
strongly interacting matter, where at high temperatures and/or
net-baryon densities quark- and gluon-like quasiparticles rather than
hadrons become the relevant degrees of freedom. As lattice-QCD
calculations show, at vanishing net-baryon density, i.e., vanishing
baryo-chemical potential $\mu_{\text{B}}=0$, this transition is a
cross-over transition occuring at a ``pseudo-critical temperture''
$T_{\text{c}} \simeq 155 \; \MeV$. The deconfinement transition is
characterized by a strong increase in the thermodynamical variables
(energy density, entropy, pressure, etc.). In the same temperature
region also the quark condensate drops quite rapidly to 0. As the order
parameter of the approximate chiral symmetry this indicates the
restoration of this symmetry.

In the following we thus use an effective chiral model, a
linear-$\sigma$ model with (constituent) quarks, $\sigma$-mesons and
pions as a model to investigate the phase structure of strongly
interacting matter. This model is then used, applying the real-time
formalism of off-equilibrium quantum field theory, to derive a transport
equation with the goal to study the dynamics of the chiral transition at
various net-baryon densities for an expanding fireball of strongly
interacting matter. Of particular interest in the context of heavy-ion
collisions are the fluctuations of conserved charges like the net-baryon
number or the electric charge as possible observables related to
different types of phase transitions (cross-over, first-order, and
second-order in the region of a critical endpoint of the first-order
transition line in the phase diagram).

\subsection{Chiral symmetry}
\label{sec:ChiralSymmetryQCD}

In the following we denote by $\psi=(\psi_{\text{u}},\psi_\text{d})$ the iso-spinor for
the up- and down-quarks with usual Dirac-spinor fields $\psi_{\text{u}}$
and $\psi_{\text{d}}$, which we decompose into left- and right-handed
parts,
\begin{equation}
\label{eq:LeftRightQuark}
\psi_{\text{L}} = \frac{1-\gamma_5}{2} \psi, \quad
\psi_{\text{R}}=\frac{1+\gamma_5}{2} \psi.
\end{equation}
In the limit of massless quarks, the QCD-Lagrangian is invariant under
the chiral $\text{SU}(2)_{\text{L}} \times \text{SU}(2)_{\text{R}}$
symmetry, or equivalently under the vector and axialvector iso-spin
transformations with Pauli matrices $\tau^k$ being the generators of
isospin rotations:
\begin{equation}
\begin{split}
&\Lambda_V:=\exp\left (-\ii \,\frac{\vec\tau\cdot\vec\theta}{2}\right
)\,,\quad\psi \longrightarrow\left (1-\ii \,\frac{\vec\tau\cdot\vec\theta}{2}\right )\psi
\,,\quad\bar\psi \longrightarrow\bar\psi\left (1+\ii\,\frac{\vec\tau\cdot\vec\theta}{2}\right )\,,\\ 
&\Lambda_A:=\exp\left
  (-\ii\,\gamma_5\frac{\vec\tau\cdot\vec\theta}{2}\right )\,,\quad\psi
\longrightarrow\left (1-\ii \,\gamma_5\frac{\vec\tau\cdot\vec\theta}{2}\right )\psi
\,,\quad\bar\psi \longrightarrow\bar\psi\left (1-\ii\,\gamma_5\frac{\vec\tau\cdot\vec\theta}{2}\right )\,.
\label{eq:AxialVectorTrafo}
\end{split}
\end{equation}
According to Nother's theorem these symmetries 
lead to the following conserved vector and axial-vector currents,
\begin{equation}
V_\mu^k=\bar\chi \gamma_\mu\frac{\tau^k}{2}\chi\,,\qquad\qquad A_\mu^k=\bar\chi\gamma_\mu\gamma_5\frac{\tau^k}{2}\chi\,.
\label{eq:ConservedCurrents}
\end{equation}
The invariance with respect to vector and axial-vector transformations
$\mathrm{SU}_V\left (2\right )\times\mathrm{SU}_A\left (2\right )$ is
called chiral symmetry.

Introducing massive quarks breaks the axial-vector symmetry explicitly
and therefore also the chiral symmetry, meaning that the axial-vector
current $A_\mu^k$ is only partially conserved as can be directly seen by
transforming the mass term of the Lagrangian
$\mathcal{L}_{\text{m}}=-\bar{\psi} \hat{M} \psi$. With the quark-mass
matrix in isospin space,
$\hat{M}=\mathrm{diag}(m_{\text{u}},m_{\text{d}})$. However, as
mentioned before the masses of light quarks are small compared to
$\Lambda_{\mathrm{QCD}}$, therefore it is reasonable to assume that the
chiral symmetry should be still a good candidate for an approximate
symmetry with observable realization in nature.

Applying vector and axial-vector transformations to pseudo-scalar
iso-vector pion-like states $\pi^k=i\bar\psi\tau^k\gamma_5\psi$, forming
the lightest mesons of QCD and the scalar iso-scalar
$\sigma = \ii \bar{\psi} \psi$-like state, leads to the following
rotations in isospin-space\footnote{The $\sigma$-meson is usually
  identified with the $f^0\left (500\right )$-state.}:
\begin{equation}
\begin{split}
&\Lambda_V:\quad\vec\pi=\ii
\bar\psi\vec\tau\gamma_5\psi\,\longrightarrow\, \ii \bar\psi\vec\tau\gamma_5\psi+
\vec\theta\times\left (\ii\bar\psi\vec\tau\gamma_5\psi\right
)=\vec\pi+\vec\theta\times\vec\pi\,,\\
&\Lambda_V: \quad \sigma=\ii \bar{\psi} \, \psi \longrightarrow \, \ii
\bar{\psi} \psi = \sigma\,, \\
&\Lambda_A:\quad\vec\pi=\ii\bar\psi\vec\tau\gamma_5\psi\,\longrightarrow\, \ii\bar\psi\vec\tau\gamma_5\psi+
\vec\theta\left (\ii\bar\psi\psi\right )=\vec\pi+\vec\theta\sigma\,, \\
&\Lambda_A: \quad \sigma = \ii \bar{\psi} \psi \, \longrightarrow \, \ii
\bar{\psi \psi}-\ii \vec{\Theta} \cdot \bar{\psi} \gamma_5 \vec{\tau}
\psi = \psi -\vec{\theta} \cdot \vec{\pi}\,.
\label{eq:PionSigmaTrafo}
\end{split}
\end{equation}
Thus the vector and axial-vector transformations are realized as an
$\text{SO}(4)$ symmetry on the fields $(\sigma,\vec{\pi})$. Analogous
relations hold also for $\rho$- and $a_1$-like states of QCD (see
Ref.~\cite{doi:10.1142/S0218301397000147}). The rotations in
isospin-space \eqref{eq:PionSigmaTrafo} explain the existence of the
isospin triplet of pions with almost equal masses $m_\pi\sim 140\,\MeV$.
However, it is not true for the $\sigma$-like state, which is believed
to have a mass of about $m_\sigma\sim 500\,\MeV$ based on experimental
data \cite{Patrignani:2016xqp}.\footnote{The vector mesons $\rho$ and
  $a_1$ show also a huge mass gap of about $500\,\MeV$.  Naively, one
  would just expect that those vector mesons are degenerate states with
  equal masses.}  Consequently, the symmetry
$\mathrm{SU}_{\text{L}} \left (2\right )\times\mathrm{SU}_{\text{R}}
\left (2\right )$ is not realized in the real world and it seems
unrealistic to explain the absolute values and the mass splitting of
$m_\sigma$ and $m_{\pi^k}$ by taking only the explicit breaking of the
chrial symmetry into account, since the quark masses $m_u\approx m_d$
are by two orders of magnitude smaller compared with the lightest
mesons. The solution of this problem is the spontaneous breaking of
chiral symmetry at low temperatures. For quantum chromodynamics the
origin of the broken chiral symmetry is encoded in a finite value for
the quark condensate in vacuum, which can be summarized in the so-called
Gell-Mann-Oakes-Renner relation \cite{PhysRev.175.2195}:
\begin{equation}
-\left <0|m\bar\psi\psi|0\right >=-\frac{m_u+m_d}{2}\left <0|\bar\psi_u\psi_u+\bar\psi_d\psi_d|0\right >=m_\pi^2f_\pi^2\neq 0
\label{eq:GellMannOakes}
\end{equation}
with $f_\pi=93\,\MeV$ denoting the pion decay constant. This condensate
dissolves with increasing temperature and the chiral symmetry becomes
restored for $T \gtrsim 150\,\MeV$ up to the explicit breaking of the
chiral symmetry. Due to a $\sigma$-like structure of the quark
condensate \eqref{eq:GellMannOakes}, it is reasonable to construct a
mesonic model with $\sigma$ being the order parameter of the chiral
phase transition, which mimics the most important properties of the
chiral symmetry in QCD as discussed in more detail in the following
sections.\footnote{Note that extending the system of light quarks by the
  inclusion of the strange quark breaks the chiral symmetry
  significantly, explaining large mass splits for multiplets of more
  massive mesons in the real world. Consequently, beyond the low-energy
  limit the chiral symmetry looses its importance.} Since $m_{\text{u}}$
and $m_{\text{d}}$ are both small one can assume
$m_{\text{u}} \simeq m_{\text{d}}$, which explains that the
$\text{SU}_{\text{V}}(2)$ transformations is an approximate symmetry of
the ground state. This explains the low mass of the pions, which in the
chiral limit of massless quarks are the Nambu-Goldstone modes of the
spontaneous chiral-symmetry breaking due to the formation of the quark
condensate. In the mesonic description this implies that the
$\sigma$-meson field provides also the order parameter of this
spontaneous chiral-symmetry breaking, i.e., at low temperatures and
densities $\langle 0 |\sigma|0 \rangle \neq 0$.

\subsection{Effective description of spontaneously and explicitly broken chiral symmetry}
\label{sec:EffectiveDescriptionChiralSymmetry}

We consider the $\mathrm{SO}\left (N\right )$-Lagrangian of a field
theory with a field
$\Phi=\left (\Phi_1,...,\Phi_N\right ) \in \R^N$, including quadratic and quartic
self-interactions,
\begin{equation}
  \mathcal{L}=\frac{1}{2} \left (\partial_\mu\Phi\right ) \left (\partial^{\mu}
  \Phi\right ) +\frac{1}{2}\mu^2\Phi^2-\frac{\lambda}{4}\left
  (\Phi^2\right )^2\,.
\label{eq:lsm101}
\end{equation}
From defining the last two terms as an effective potential
\begin{equation}
V:=-\frac{1}{2}\mu^2\Phi^2+\frac{\lambda}{4}\left (\Phi^2\right )^2\,,
\label{eq:lsm102}
\end{equation}
it becomes obvious, that the symmetry of the Lagrangian with the origin
as global minimum can be broken for $\mu^2>0$, meaning that $V$ has a
minimum at $\Phi^2 \neq 0$,
\begin{equation}
\frac{\partial V}{\partial\Phi^2}=-\frac{1}{2}\mu^2+\frac{\lambda}{2}\Phi^2\stackrel{!}{=}0
\quad\Rightarrow\quad\Phi_0^2=\frac{\mu^2}{\lambda}=:\nu_0^2\,.
\label{eq:lsm103}
\end{equation}
More precisely, there is an infinite number of possible minima with
respect to different configurations of the components for the field
$\Phi$. The quantity $\nu_0^2$ is related to the length of the vector
$\Phi_0$, which defines the global minimum of the potential $V$ and
denotes the degenerate ground states of the interacting system
\eqref{eq:lsm101}.  By expanding the Lagrangian \eqref{eq:lsm101} around
the global minimum it becomes\footnote{Note that constant terms with
  respect to the chiral field do not change the physics, described by
  the Lagrangian.},
\begin{equation}
\mathcal{L}=\frac{1}{2}\left (\partial_\mu\Phi\right )^2-\frac{\lambda}{4}\left (\Phi^2-\nu_0^2\right )^2\,.
\label{eq:lsm104}
\end{equation}
A certain realization for the global minimum is called spontaneous
symmetry breaking in favor of a specified direction for the chiral
field.

Switching to the $\mathrm{SO}\left (4\right )$ linear sigma model the
chiral field becomes $\Phi=\left (\sigma,\vec\pi\right )$. Historically
this model has been widely used to describe the chiral symmetry in the
low-energy limit of hadron physics. Here, large values of nucleon mass
require a mechanism for breaking the chiral symmetry without destroying
the underlying chiral invariance of the Lagrangian. In case of the
linear sigma model the spontaneous symmetry breaking concerns directly
the $\sigma$-direction and can be achieved by setting the global minimum
\eqref{eq:lsm103} in $\sigma$-direction. Consequently, the
$\sigma$-component obtains a finite vacuum expectation value
$\left <\sigma\right >_0$, whereas the vacuum expectation values of
pions vanish
\begin{equation}
\left <\sigma\right >_0=\nu_0:=f_\pi\quad \text{and} \quad\left <\vec\pi\right >_0=0\,.
\label{eq:lsm105a}
\end{equation}
The $\vec\pi$-direction is given by a rotation in the potential minimum
around the origin. Due to a vanishing curvature in every partial
direction of $\vec\pi$, one obtains well-known relations for the
effective mass of sigma and pions
\begin{equation}
m_\sigma^2:=\left .\frac{\partial^2 V}{\partial\sigma^2}\right |_{\pi=\left <\vec\pi\right >_0}^{\sigma=\left <\sigma\right >_0}=2\lambda f_\pi^2\,,\qquad 
m_\pi^2:=\left .\frac{\partial^2 V}{\partial\pi^2}\right |_{\pi=\left <\pi\right >_0}^{\sigma=\left <\sigma\right >_0}=0\,,
\label{eq:lsm105b}
\end{equation}
with pions becoming massless modes as described by the Goldstone theorem
\cite{PhysRev.127.965}.

By introducing a term linear in the $\sigma$-field component an explicit
symmetry breaking can be taken into account. With identical vacuum
expectation values from the previous calculation, one obtains the
following relation for the ground state of the system:
\begin{equation}
\begin{split}
U&:=\frac{\lambda}{4}\left (\sigma^2+\vec\pi-\nu^ 2\right )^2-h_{\text{q}}\sigma\quad\Rightarrow\quad\nu^2=-\frac{h_{\text{q}}}{\lambda f_\pi}+f_\pi^2\,,\\
U_0&:= U\Big|_{\pi=\left <\vec\pi\right >_0}^{\sigma=\left <\sigma\right >_0}=\frac{1}{4}\frac{h_{\text{q}}^2}{\lambda f_\pi^2}-h_{\text{q}}f_\pi\,.
\end{split}
\label{es:lsm201}
\end{equation}
Obviously, the parameters of the theory have to be specified by matching
to observables, given here by the pion mass $m_\pi$ and the pion decay
constant $f_\pi$ (see Tab.~\ref{tab:LinSigmaModel}). The parameter $h_{\text{q}}$
follows directly from the modified mass terms:
\begin{equation}
m_\sigma^2:=\left .\frac{\partial^2 U}{\partial\sigma^2}\right |_{\pi=\left <\vec\pi\right >_0}^{\sigma=\left <\sigma\right >_0}=2\lambda f_\pi^2+m_\pi^2\,,\qquad 
m_\pi^2:=\left .\frac{\partial^2 U}{\partial\pi^2}\right |_{\pi=\left <\pi\right >_0}^{\sigma=\left <\sigma\right >_0}=\frac{h_{\text{q}}}{f_\pi}\,.
\label{eq:lsm105c}
\end{equation}
Additionally, the linear sigma model can be extended to include
constituent quarks for an effective description of the low-energy limit
of QCD. This full version of the $\mathrm{SO}\left(4\right)$-model
\cite{PhysRevC.64.045202} is suited for studying the chiral phase
transition. Because of the spontaneously and explicitly broken chiral
symmetry the mesonic part of this theory consists of a massive scalar
$\sigma$ field and three isoscalar pion fields (for a more detailed
discussion see \cite{doi:10.1142/S0218301397000147}).  Now, the $\sigma$
field represents the order parameter for the chiral phase transition and
mimics the properties of the quark condensate in QCD from
Sec.~\ref{sec:ChiralSymmetryQCD}, since both transform equally under
chiral transformation. Without explicit symmetry breaking an
$\mathrm{SU_L}\left(2\right)\times \mathrm{SU_R}\left(2\right)$ symmetry
transformation would let the full Lagrangian invariant:
\begin{equation}
\mathcal L=\sum_{i=1} \bar\psi_i \Big[\ii \partial\!\!\!/-g\left(\sigma+\ii
  \gamma_5\vec\pi\cdot\vec\tau\right)\Big]\psi_i 
+\frac{1}{2}\left(\partial_\mu\sigma\partial^\mu\sigma
+\partial_\mu\vec\pi
\partial^\mu\vec\pi\right)-\frac{\lambda}{4}\left(\sigma^2+\vec\pi^2-\nu^2\right)^2+f_{\pi}m_\pi^2\sigma+U_0\, ,
\label{eq:LinSigmaModell}
\end{equation}
where $i \in \{\text{u},\text{d} \}$. The field-shift term as well as
the zero potential constant follow from \eqref{es:lsm201} and
\eqref{eq:lsm105c}, resulting in $\nu^2=f_\pi^2-m_\pi^2/\lambda$,
$U_0=m^4_\pi/\left(4\lambda\right)-f_\pi^2m_\pi^2$. Thereby, all
parameters (see also Tab.~\ref{tab:LinSigmaModel}) are adjusted to match
the vacuum values of the pion decay constant $f_\pi=93\,\MeV$ and the
pion mass $m_\pi=138\,\MeV$, leading to an estimated value for the sigma
mass of $m_\sigma\approx604\,\MeV$.

\begin{figure*}
 \includegraphics[width=10cm]{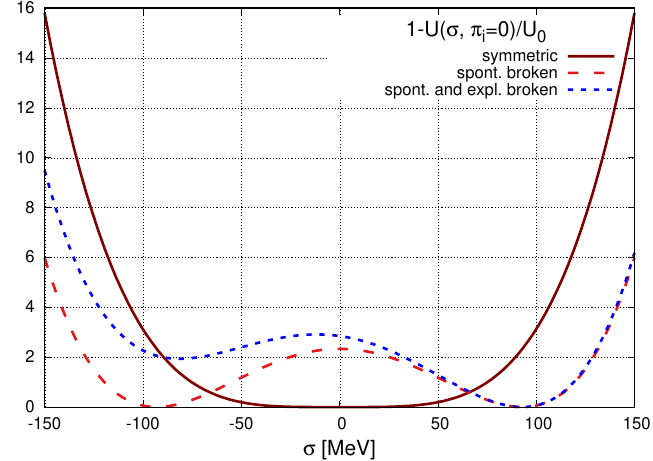} \centering
 \caption{Effective potential of the linear sigma model for symmetric,
   spontaneously as well as explicitly broken chiral symmetry with
   respect to $\sigma$ and $\pi_i=0$.}
 \label{fig:EffectivePotential1d}
\end{figure*}

\begin{figure*}
\subfloat{\includegraphics[width=7cm]{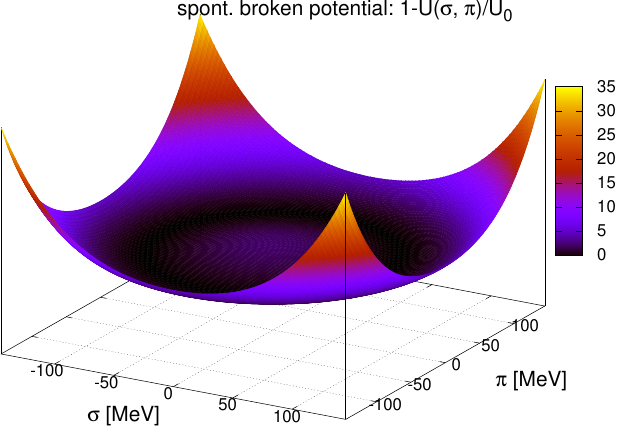}}
\subfloat{\includegraphics[width=7cm]{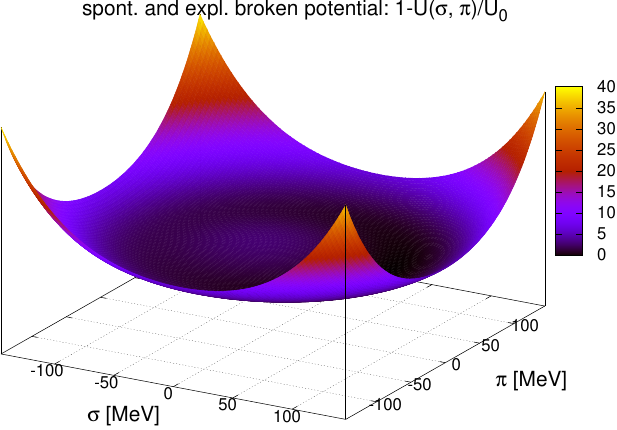}}
\centering
\caption{Effective potential of the linear sigma model for spontaneously
  as well as explicitly broken chiral symmetry with respect to $\sigma$
  and $\pi=\left|\vec\pi\right |$.}
\label{fig:EffectivePotential2d}
\end{figure*}

All discussed features of the model are shown schematically in
Fig.~\ref{fig:EffectivePotential1d} and
Fig.~\ref{fig:EffectivePotential2d} for the projections of the
effective potential on one- as well as two-dimensional
subspaces. Thereby, the symmetric potential denotes a configuration with
a vanishing vacuum expectation value.

\begin{table}
\centering
\begin{tabular}{|p{2.0cm}|p{3.5cm}|p{7.5cm}|}
\hline
parameter & value & description \\
\hline
\hline
$\lambda$ & 20 & coupling constant for $\sigma$ and $\vec\pi$\\
\hline
$g$ & $2-5$ & coupling constant between $\sigma,\vec\pi$ and $\psi_i$\\
\hline
$f_\pi$ & $93\,\MeV$ & pion decay constant \\
\hline
$m_\pi$ & $138\,\MeV$ & pion mass \\
\hline
$\nu^2$ & $f_\pi^2-m_\pi^2/\lambda$ & field shift term \\
\hline
$U_0$ & $m_\pi^4/\left (4\lambda\right )-f_\pi^2m_\pi^2$ & ground state \\
\hline
\end{tabular}
\caption{Parameters of the qaurk-meson model with the Lagrangian \eqref{eq:LinSigmaModell}.} 
\label{tab:LinSigmaModel}
\end{table}

\subsection{Semiclassical mean-field dynamics}
\label{sec:MeanFieldDynamics}

The classical action is formulated with respect to the Lagrangian of the
linear sigma model \eqref{eq:LinSigmaModell}:
\begin{equation}
\begin{split}
S\left [\Phi,\bar\psi,\psi\right ]&=\int\dd^4x\Bigg[\sum_i\bar\psi_i \left [\ii \partial\!\!\!/-g\left(\sigma+\ii \gamma_5\vec\pi\cdot\vec\tau\right)\right ]\psi_i\\
&\,\qquad\qquad+\frac{1}{2}\left(\partial_\mu\sigma\partial^\mu\sigma+\partial_\mu\vec\pi\partial^\mu\vec\pi\right )-\frac{\lambda}{4}\left(\sigma^2+\vec\pi^2-\nu^2\right)^2+f_{\pi}m_\pi^2\sigma+U_0\Bigg]\\
&:= S_0\left [\bar\psi,\psi\right ]+S_{I}\left [\Phi,\bar\psi,\psi\right ]+S_0[\Phi]+S_{I}\left [\Phi\right ]\,,
\label{eq:ClassicalAction}
\end{split}
\end{equation}
where the last line defines different contributions, given by the free
fermionic and bosonic parts:
\begin{equation}
S_0\left [\bar\psi,\psi\right ]=\int\dd^4x\left [\sum_i\bar\psi_i \ii \partial\!\!\!/\psi_i\right ]\,,\quad S_0\left [\Phi\right ]=\int\dd^4x\left [\frac{1}{2}\left(\partial_\mu\sigma\partial^\mu\sigma+\partial_\mu\vec\pi\partial^\mu\vec\pi\right )\right ]
\label{eq:LagrangianFree}
\end{equation}
as well as the corresponding interaction parts:
\begin{equation}
\begin{split}
S_I\left [\Phi,\bar\psi,\psi\right ]&=\int\dd^4x\left [-g\sum_i\bar\psi_i\left(\sigma+\ii \gamma_5\vec\pi\cdot\vec\tau\right)\psi_i\right ]\\
S_I\left [\Phi\right ]&=\int\dd^4x\left [-\frac{\lambda}{4}\left(\sigma^2+\vec\pi^2-\nu^2\right)^2+f_{\pi}m_\pi^2\sigma+U_0\right ]\,.
\end{split}
\end{equation}
The set of inhomogeneous Klein-Gordon equations of motion follow directly from variation of $S\left [\Phi,\bar\psi,\psi\right ]$ with 
respect to the components $\sigma$ and $\vec\pi$ of the chiral field and treating quarks at one-loop 
level:
\begin{equation}
\begin{aligned}
\partial_\mu\partial^\mu\sigma+\lambda\left(\sigma^2+\vec\pi^2-\nu^2\right)\sigma-f_\pi m_\pi^2+g\left<\bar\psi\psi\right \rangle&=0\,,\\
\partial_\mu\partial^\mu\vec\pi+\lambda\left(\sigma^2+\vec\pi^2-\nu^2\right)\vec\pi+g\left<\bar\psi \ii \gamma_5\vec\tau\psi\right \rangle&=0\,.
\end{aligned}
\label{eqn:KleinGordonInhom}
\end{equation}
Thereby, the scalar and pseudoscalar densities are given by\footnote{The
  exact derivation of scalar and pseudoscalar densities from fermionic
  one-loop integral can be found in the next section, when the model is
  extended to a functional approach.}:
\begin{equation}
\begin{aligned}
\left<\bar\psi\psi\right \rangle\left(t,\vec x\right)&=gd_\psi\sigma\left(t,\vec x\right)\int \frac{\dd^3 p}{\left(2\pi\right)^3}\,\frac{f_\psi\left(t,\vec x,\vec p\right)+f_{\bar\psi}\left(t,\vec x,\vec p\right)}{E\left(t,\vec x,\vec p\right)}\,,\\
\left<\bar\psi \ii \gamma_5\vec\tau\psi\right \rangle\left(t,\vec x\right)&=gd_\psi\vec\pi\left(t,\vec x\right)\int \frac{\dd^3 p}{\left(2\pi\right)^3}\,\frac{f_\psi\left(t,\vec x,\vec p\right)+f_{\bar\psi}\left(t,\vec x,\vec p\right)}{E\left(t,\vec x,\vec p\right)}\,,
\end{aligned}
\label{eqn:ScalarPseudoScalarDensity}
\end{equation}
with $f_\psi$ and $f_{\bar\psi}$ denoting the phase-space distribution
functions of quarks and antiquarks with their degeneracy factor
$d_\psi=N_cN_sN_f=12$ for $N_c=3$ colors, $N_s=2$ spins and $N_f=2$
flavors.  In a dynamical simulation, one of the most simple descriptions
of quarks would be to propagate them according to the Vlasov equation:
\begin{equation}
\left[\partial_t+\frac{\vec p}{E\left(t,\vec x,\vec p\right)}\cdot\nabla_{\vec x}-\nabla_{\vec x} E\left(t,\vec x,\vec p\right)\nabla_{\vec p}\right]f_{\psi,\bar\psi}\left(t,\vec x,\vec p\right)=0\,,
\label{eqn:a3b}
\end{equation}
where the force term is dynamically generated through the effective mass
of quarks, depending on the mean-field values,
\begin{equation}
E\left(t,\vec x,\vec p\right)=\sqrt{\vec{p}^2\left(t\right)+M_\psi^2\left(t,\vec x\right)}\,,\quad 
M_\psi^2\left(t,\vec x\right)=g^2\left[\sigma^2\left(t,\vec x\right)+\vec\pi^2\left(t,\vec x\right)\right]\,.
\label{eqn:a3c}
\end{equation}
With the inverse temperature $\beta =1/T$ and quark chemical potential
$\mu_\psi$ the effective thermodynamic potential of the semiclassical
approximation is given by:
\begin{equation}
\begin{split}
  \Omega_{\text{eff}}^{\mathrm{MF}}\left (T,\mu_\psi\right
  )=&\,\frac{\lambda}{4}\left (\sigma^2+\vec\pi^2-\nu ^2\right )^2-h_{\text{q}}\sigma-U_0\\
  &-d_\psi\frac{1}{\beta}\int\frac{\dd^3\vec p}{\left (2\pi\right
    )^3}\left [\ln\left (1+\ee^{-\beta\left (E_\psi-\mu_\psi\right
        )}\right
    )+\ln \left (1+\ee^{-\beta\left (E_{\bar\psi}+\mu_\psi\right )}\right )\right ]\,,\\
\label{eq:EffMeanFieldPotential}
\end{split}
\end{equation}
where we introduced the logarithmic terms, since their derivatives with
respect to the components of the chiral field reproduces the scalar and
pseudoscalar contributions with the densities of
Eq.~\ref{eqn:ScalarPseudoScalarDensity} as required:
\begin{equation}
\begin{split}
\frac{\partial\Omega_{\text{eff}}^{\mathrm{MF}}}{\partial\sigma}&=\lambda\left(\sigma^2+\vec\pi^2-\nu^2\right)\sigma-f_\pi m_\pi^2+g\left<\bar\psi\psi\right \rangle\stackrel{!}{=}0\\
\frac{\partial\Omega_{\text{eff}}^{\mathrm{MF}}}{\partial\vec\pi}&=\lambda\left(\sigma^2+\vec\pi^2-\nu^2\right)\vec\pi+g\left<\bar\psi
  \ii \gamma_5\vec\tau\psi\right \rangle \stackrel{!}{=}0
\label{eq:DerivativeEffPot}
\end{split}
\end{equation}
The chiral phase diagram is obtained by numericallly solving
self-consistently the first equation in \eqref{eq:DerivativeEffPot} for
arbitrary combinations of $T\geqq 0$ and $\mu_\psi$, since the
expectation value $\left <\vec\pi\right >$ vanishes in equilibrium. For
this purpose, one applies the bisection or Newton's method.  In the
second step, one computes the numerical derivatives of
Eqs.~\eqref{eq:SemiclassicalEffMasses} by applying a finite difference
method of higher order.

Here, the distribution functions in scalar and pseudoscalar densities
are given by the Fermi distribution and the last conditions account for
a stationary state. Now, the effective masses follow from the curvature
of the effective potential with respect to the chiral components, which
are evaluated at equilibrium values of the chiral field:
\begin{equation}
m_\sigma^2=\left .\frac{\partial^2\Omega_{\text{eff}}^{\mathrm{MF}}}{\partial\sigma^2}\right |_{\Phi=\left <\Phi\right >_{eq}}\,,
\qquad m_\pi^2=\left .\frac{\partial^2\Omega_{\text{eff}}^{\mathrm{MF}}}{\partial\vec\pi^2}\right |_{\Phi=\left <\Phi\right >_{eq}}\,.
\label{eq:SemiclassicalEffMasses}
\end{equation}
\begin{figure*}
\subfloat{\includegraphics[width=7cm]{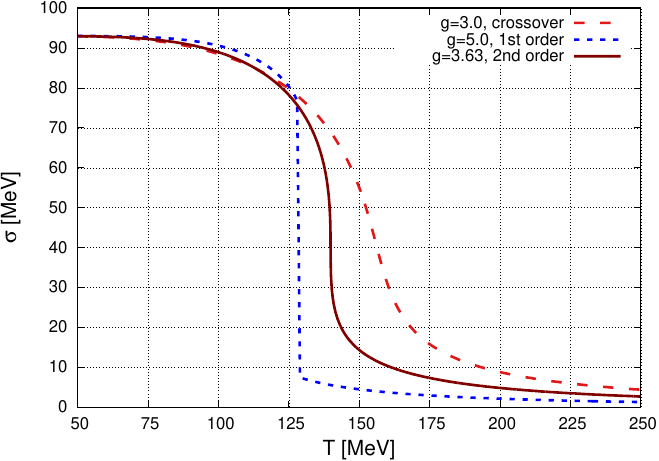}}
\subfloat{\includegraphics[width=7cm]{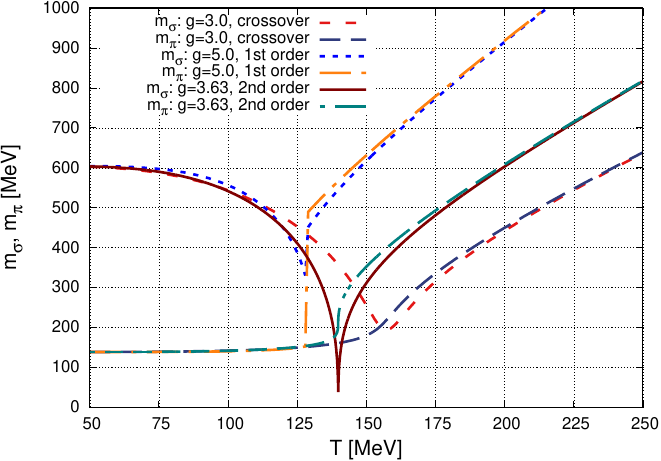}}\\
\centering
\caption{Order parameter, $\sigma$ and pion mass for the semiclassical
  approximation as given by Eqs.~\eqref{eqn:KleinGordonInhom}.  The
  coupling constant $g$ is choosen to reproduce different kinds of the
  chiral phase transition at vanishing chemical potential of quarks.}
\label{fig:OrderParameterAndMass}
\end{figure*}

In case of this semiclassical approach the most crucial and problematic
approximation is the fact of neglecting fluctuations around the mean
field, being justified only at very low or high temperatures, since
these thermal fluctuations can be interpreted as mesonic interactions
(see Sec.~\ref{chap:EvolEq}). In the vicinity of the critical point the
mass of $\sigma$ mesons becomes small as seen from the
Fig.~\ref{fig:OrderParameterAndMass}, so that thermal fluctuations due
to mesons can significantlly contribute to the total rate of
interactions between the different components of the chiral field and
quarks.

The resulting equilibrium properties of such a semiclassical description
are shown in Fig.~\ref{fig:OrderParameterAndMass} for the order
parameter as well as effective $\sigma$ and pion masses as a function of
the temperature $T$.  Depending on the Yukawa coupling constant $g$
different orders of the chiral-phase transition occur.  More results,
also with a constant binary cross section for quarks, can be found in
\cite{vanHees:2013qla,Wesp:2014xpa,Greiner:2015tra,Wesp2015}, where the
semiclassical model has been extended to include mean-field fluctuations
by introducing a new stochastic formalism for an effective interaction
between the $\sigma$ field and quarks via intermediate mesonic
states. In the next section we shall derive a consistent set of
equations for the chiral order parameter, $\erw{\sigma}$, as well as the
fluctuations in terms of a coupled set of (non-Markovian) kinetic
equations.

\section{Transport equations}
\label{chap:EvolEq}

The first part of this Sect.\ briefly introduces fundamental techniques
in non-equilibrium quantum field theory by starting from the general
concept of the Green's function in the real-time formalism, which has
been developed by Schwin\-ger and Keldysh
\cite{doi:10.1063/1.1703727,Keldysh:1964ud}. It is worth to mention,
that $n$-point Green's functions do not fulfill the properties of a
physical observable, since they cannot be directly measured. However,
they are connected to physical transition amplitudes (see for instance
\cite{book:16449}) and contain all accessible information about the
system, encoded in the underlying quantum field theory.  Based on the
real-time technique, we introduce the well-established concept of a
generating functional for connected diagrams and discuss the so-called
two-particle-irreducible (2PI) quantum effective action. This functional
approach acts as a starting point for deriving selfconsistent and exact
evolution equations on the level of one- and two-point functions in
quantum field theories out of equilibrium, known as Schwinger-Dyson and
Kadanoff-Baym equations.

After applying the gradient expansion to the exact evolution equations
for one- and two-point functions of the quark-meson model from
Sec.~\ref{sec:MeanFieldDynamics}, one obtains generalized transport
equations, allowing to interpret the full dynamics of two-point
functions in terms of mesonic interactions. Finally, a truncated version
of the 2PI effective action is considered, which still contains the most
relevant diagrams, describing also the dissipation with and without
memory effects, arising from the interaction between the slowly changing
chiral mean-field and mesonic excitations.

\subsection{Imaginary and real-time formalisms}
\label{sec:RealTime}

The physics of many-body systems in thermal equilibrium can be studied
by means of Matsubara's widely used Euclidean-time formalism
\cite{doi:10.1143/PTP.14.351}.  This approach is based on the similarity
between the partition function and transition amplitudes in the
path-integral formalism, allowing to connect both quantities by formally
identifying the imaginary time $t=-\ii \tau,\,\tau \in [0,\beta ]$ with
the inverse temperature $\beta=1/T$.  For an arbitrary function $f$ we define
space-time and energy-momentum integrations by
\begin{equation}
\begin{split}
  \int_xf\left (x\right )&=\int_0^{-i\beta}\dd t\int\dd^3\vec x\,f\left
    (t,\vec x\right )\,,\qquad \int_kf\left (k\right
  )=\frac{1}{-i\beta}\sum_{n=-\infty}^\infty\int\frac{\dd^3\vec k}{\left
      (2\pi\right )^3}f\left (i\omega_n ,\vec k\right )\,,
\label{eq:ImaginarySpaceTimeIntegration}
\end{split}
\end{equation}
where the so-called bosonic and fermionic Matsubara frequencies are
given by $\omega_n=2\pi n/\beta$, respectively
$\omega_n=2\pi\left (n+1\right )/\beta$ with $n \in \mathbb{Z}$. These discrete frequencies
result from the finite time interval and periodic/antiperiodic boundary
conditions (see Ref.~\cite{PhysRevD.9.3320} for more details).

In general, for non-equilibrium systems the initial density matrix
$\rho_D$ deviates significantly from the corresponding thermal
equilibrium density matrix (for instance
$\rho_D^{\text{eq}}\sim e^{-\beta H}$) of the same system.
Consequently, the correct treatment of a system out of equilibrium
requires to introduce a non-equilibrium effective action and to take the
initial correlations via $\rho_D\left (t_i\right )$ into account
\cite{doi:10.1063/1.1843591}.  The exact form of such an effective
action and corresponding generating functionals will be discussed in the
following (see Sec.~\ref{sec:1PIAction} and Sec.~\ref{sec:2PIAction}).

\begin{figure}
 \includegraphics[width=12cm]{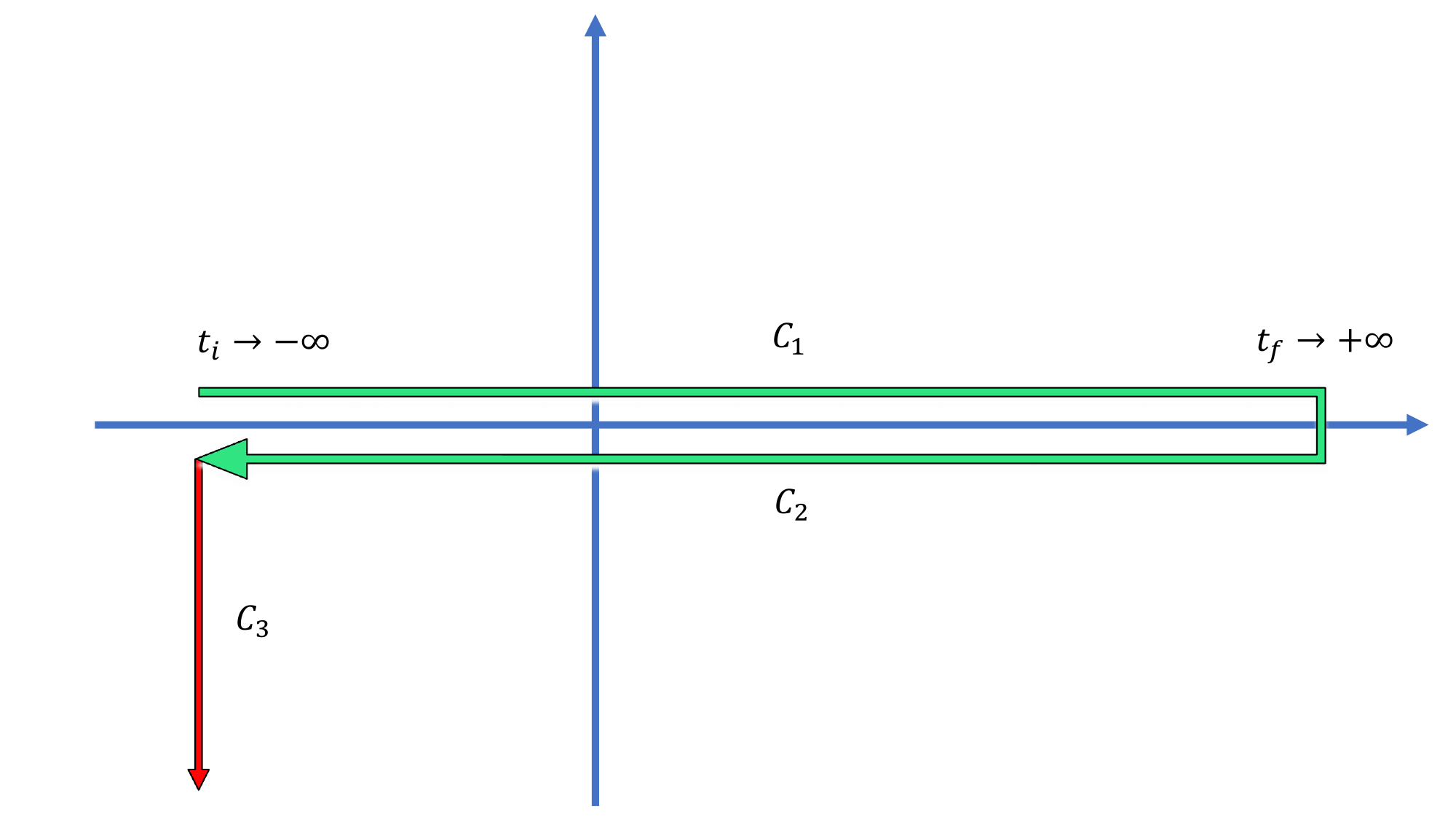} \centering
 \caption{Extended real-time contour of the CPT formalism by Schwinger and Keldysh.}
 \label{fig:CTP}
\end{figure}

In the Heisenberg picture, the time dependence of a non-equilibrium
system is encoded in the field operators, whereas the initial density
matrix $\rho_D\left (t_i\right )$, for evaluating expectation values,
stays constant. Therefore, the full time evolution requires to start
from the initial time $t_i\in\left (-\infty,\infty\right )$ and
propagate the system chronologically on the upper branch $\mathcal{C}_1$
to some maximum value of time $t_{f}> t_i$ as shown in
Fig.\ \ref{fig:CTP}. From the final time the operator evaluation runs in
antichronological order on the lower branch $\mathcal{C}_2$ back to $t_i$.
Finally, the vertical line to $t_i-i\beta$ allows to treat also
equilibrium systems within the same framework (compare with
Eq.~\eqref{eq:ImaginarySpaceTimeIntegration}).  The full contour line
represents the extended real-time contour of the closed-time-path (CTP)
method by Schwinger and Keldysh.  Consequently, time integrals along the
extended real-time contour
$\mathcal{C}:=\mathcal{C}_1\cup\mathcal{C}_2\cup\mathcal{C}_3$ can be
separated with respect to the upper, lower and vertical branches:
\begin{equation}
\int_\mathcal{C}\dd t=\int_{\mathcal{C}_1}\dd t+\int_{\mathcal{C}_2}\dd t+\int_{\mathcal{C}_3}\dd t=\int_{t_i}^{t_f}\dd t-\int_{t_i}^{t_f}\dd t+\int_{t_i}^{t_i-i\beta}\dd t\,.
\label{eq:IntCTP}
\end{equation}
A consistent description within the real-time formalism requires only
the contour $\mathcal{C}=\mathcal{C}_1\cup\mathcal{C}_2$, whereas the
imaginary-time formalism is based on the contour
$\mathcal{C}=\mathcal{C}_3$ (see
Refs.~\cite{Gelis:1994dp,Gelis:1999nx}). The functional approach
developed in the following sections stays true for equilibrium and
non-equilibrium systems and one has only to keep track of the right
formalism with its corresponding integration contour. Following that a
contour ordered two-point Green's function of a general bosonic or
fermionic field $\varphi\left (x\right )$ reads
\begin{equation}
  G_{\mathcal{C}}\left (x,y\right ):=\left <\mathcal{T}_{\mathcal
      C}\left (\varphi\left (x\right )\varphi^\dagger\left (y\right
      )\right )\right > = \,\Theta_{\mathcal{C}}\left (x^0,y^0\right
  )\left <\varphi\left (x\right )\varphi^\dagger\left (y\right )\right
  >\pm\Theta_{\mathcal{C}}\left (y^0,x^0\right )\left
    <\varphi^\dagger\left (y\right )\varphi\left (x\right )\right >\,,
\label{eq:CTPGreenFunc}
\end{equation}
where an odd number of permutations leads to a sign change for the
expectation value of fermionic fields.
The time-ordering operator $\mathcal{T}_{\mathcal C}$ along the contour
can be replaced by using the contour ordered Heaviside function,
\begin{equation}
\Theta_{\mathcal{C}}\left (x^0,y^0\right )\equiv\Theta_{\mathcal{C}}\left (x^0-y^0\right ):=
\begin{cases}
\Theta\left (x^0-y^0\right )\qquad\forall\,x^0,y^0\in\mathcal{C}_1\\
0\qquad\qquad\qquad\,\,\,\forall\,x^0\in\mathcal{C}_1,y^0\in\mathcal{C}_2\,,\\
1\qquad\qquad\qquad\,\,\,\forall\,y^0\in\mathcal{C}_1,x^0\in\mathcal{C}_2\,,\\
\Theta\left (y^0-x^0\right )\qquad\forall\,x^0,y^0\in\mathcal{C}_2\,.
\end{cases}
\label{eq:HeavysideFunc}
\end{equation}
Equivalently, one can introduce a matrix with four components along the contour,
\begin{equation}
\begin{split}
  G^{11}\left (x,y\right )&\equiv \ii G^c\left (x,y\right ):=\left <\mathcal{T}_c\left (\varphi\left (x\right )\varphi^\dagger\left (y\right )\right )\right >\\
  &=\Theta\left (x^0-y^0\right )\left <\varphi\left (x\right )\varphi^\dagger\left (y\right )\right >\pm\Theta\left (y^0-x^0\right )\left <\varphi^\dagger\left (y\right )\varphi\left (x\right )\right >\,\,\quad\qquad\forall\,x^0,y^0\in\mathcal{C}_1\,,\\
  G^{22}\left (x,y\right )&\equiv \ii G^a\left (x,y\right ):=\left <\mathcal{T}_a\left (\varphi\left (x\right )\varphi^\dagger\left (y\right )\right )\right >\\
  &=\Theta\left (y^0-x^0\right )\left <\varphi\left (x\right )\varphi^\dagger\left (y\right )\right >\pm\Theta\left (x^0-y^0\right )\left <\varphi^\dagger\left (y\right )\varphi\left (x\right )\right >\,\,\quad\qquad\forall\,x^0,y^0\in\mathcal{C}_2\,,\\
  G^{12}\left (x,y\right )&\equiv \ii G^<\left (x,y\right ):=\pm\left <\varphi^\dagger\left (y\right )\varphi\left (x\right )\right >\quad\qquad\qquad\qquad\qquad\qquad\,\,\,\,\,\forall\,x^0\in\mathcal{C}_1\,,y^0\in\mathcal{C}_2\,,\\
  G^{21}\left (x,y\right )&\equiv \ii G^>\left (x,y\right ):=\left
    <\varphi\left (x\right )\varphi^\dagger\left (y\right )\right
  >\qquad\qquad\qquad\qquad\qquad\qquad\,\,\,\,\,\forall\,y^0\in\mathcal{C}_1,x^0\in\mathcal{C}_2\,.
\label{eq:GreenFunc}
\end{split}
\end{equation}
Obviously, one of the four Green's functions can be eliminated and
instead of \eqref{eq:GreenFunc} one often finds physical expressions in
the literature, which are formulated in terms of retarded and advanced
Green's functions, 
\begin{equation}
\begin{split}
\ii G^{\mathrm{ret}}\left (x,y\right )&=G^{11}\left (x,y\right )-G^{12}\left (x,y\right )=G^{21}\left (x,y\right )-G^{22}\left (x,y\right )\\
&=\Theta\left (x^0-y^0\right )\left (\ii G^>\left (x,y\right )- \ii
  G^<\left (x,y\right )\right ) \, ,\\
\ii G^{\mathrm{adv}}\left (x,y\right )&=G^{11}\left (x,y\right )-G^{21}\left (x,y\right )=G^{12}\left (x,y\right )-G^{22}\left (x,y\right )\\
&=-\Theta\left (y^0-x^0\right )\left (\ii G^>\left (x,y\right )- \ii G^<\left (x,y\right )\right )\,.\\
\label{eq:PhysGreenFunc}
\end{split}
\end{equation}
Because of the form \eqref{eq:PhysGreenFunc}, one often introduces two additional quantities,
\begin{equation}
\begin{split}
A_{B/F}:=&\left <\left [\varphi\left (x\right ),\varphi^\dagger\left
      (y\right )\right ]_\mp\right >=\ii \left (G^>\left (x,y\right )- G^<\left (x,y\right )\right )\,,\\
F_{B/F}:=&\left <\left [\varphi\left (x\right ),\varphi^\dagger\left
      (y\right )\right ]_\pm\right >=\ii \left (G^>\left (x,y\right )+ G^<\left (x,y\right )\right )\,,
\label{eq:SpectStatFunc}
\end{split}
\end{equation}
which contain spectral and statistical information about the system.

In equilibrium the real-time propagators depend effectively only on the
difference with respect to space-time coordinates and it becomes
convenient to represent them in momentum space. Especially, one obtains
a dispersion relation between the retarded Green's function and the
spectral function\footnote{For non-equilibrium system the Fourier
  transformation is replaced by the Wigner transformation (see
  Sec.~\ref{sec:GenBoltzmannEquation}).},
\begin{equation}
G^{\mathrm{ret}}\left (k\right
)=\int\frac{\dd\omega}{2\pi}\frac{A_{B/F}(\omega,\vec k)}{k^0-\omega+\ii
  \epsilon}\,,
\label{eq:DispRel}
\end{equation}
which follows directly from the transformation properties for a product
of two functions in combination with the regularized form of a
$\theta$-function:
\begin{equation}
\begin{split}
\theta_\epsilon\left (x^0\right ):=\,\theta\left (x^0\right )\exp\left (-\epsilon x^0\right )
\quad \Rightarrow \quad\theta_\epsilon\left (k^0\right )=\int_0^\infty\dd
x^0\exp\left (\ii k^0x^0-\epsilon x^0\right )=\frac{\ii}{k^0+\ii \epsilon}\,.
\label{eq:RegularizedThetaFunc}
\end{split}
\end{equation}
For noninteracting bosonic or fermionic fields with mass $m$, the
Green's functions are solutions of the corresponding homogeneous
Klein-Gordon equations. Their explicit form can be derived from
analyzing the Kubo-Martin-Schwinger (KMS) boundary
conditions\footnote{Periodic for bosons and antiperiodic for fermions.}
of the imaginary-time contour $\mathcal C_3$ and by considering the
Fourier transform of the Wightman propagators $\ii G^\gtrless$ as shown
in Ref.~\cite{PhysRevD.9.3320}.  By introducing the Bose distribution
function $\fB \left (k^0\right )$ and making use of the general relations
\eqref{eq:GreenFunc}, \eqref{eq:PhysGreenFunc},
\eqref{eq:SpectStatFunc}, one obtains the following relations for bosons:
\begin{equation}
\begin{split}
A_{\text{B}}\left (k\right )&=2\pi\sign\left (k^0\right )\delta\left (k^2-m^2\right )\,,\quad \ii G^{\mathrm{ret}}\left (k\right )=\frac{i}{k^2-m^2+\ii \epsilon\sign\left (k^0\right )}\,,\\
\ii G^{\text{c}}\left (k\right )&=\frac{\ii}{k^2-m^2+\ii \epsilon}+2\pi\delta\left (k^2-m^2\right )\fB\left (|k^0|\right )\,,\\
\ii G^{\text{a}}\left (k\right )&=\frac{-\ii}{k^2-m^2-\ii \epsilon}+2\pi\delta\left (k^2-m^2\right )\fB\left (|k^0|\right )\,,\\
\ii G^{<}\left (k\right )&=2\pi\delta\left (k^2-m^2\right )\left (\theta\left (-k^0\right )+\fB\left (|k^0|\right )\right )\,,\\
\ii G^{>}\left (k\right )&=2\pi\delta\left (k^2-m^2\right )\left (\theta\left (k^0\right )+\fB\left (|k^0|\right )\right )\,.\\
\label{eq:ThermalPropagatorsBosons}
\end{split}
\end{equation}
An analogous calculation for fermions, with $\fF$ standing for the Fermi
distribution function and $\fbF$ referring to the corresponding
antiparticles, leads to:
\begin{equation}
\begin{split}
  A_{\text{F}}\left (k\right )&=2\pi\sign\left (k^0\right )\delta\left
    (k^2-m^2\right )\left (k\!\!\!/+m\right )\,,\quad \ii
  D^{\text{ret}}\left (k\right )=\frac{\ii \left (k\!\!\!/+m\right
    )}{k^2-m^2+\ii \epsilon\sign\left (k^0\right )}\\
  \ii D^{\text{c}}\left (k\right )&=\left [\frac{\ii}{k^2-m^2+\ii \epsilon}-2\pi\delta\left (k^2-m^2\right )\left (\theta\left (k^0\right )f_{F}\left (k^0\right )+\theta\left (-k^0\right )f_{\bar F}\left (k^0\right )\right )\right ]\left (k\!\!\!/+m\right )\,,\\
  \ii D^{\text{a}}\left (k\right )&=\left [\frac{-i}{k^2-m^2-\ii \epsilon}-2\pi\delta\left (k^2-m^2\right )\left (\theta\left (k^0\right )f_{F}\left (k^0\right )+\theta\left (-k^0\right )f_{\bar F}\left (k^0\right )\right )\right ]\left (k\!\!\!/+m\right )\,,\\
  \ii D^{<}\left (k\right )&=2\pi\delta\left (k^2-m^2\right )\left (\theta\left (-k^0\right )\left (1-f_{\bar F}\left (k^0\right )\right )-\theta\left (k^0\right )f_F\left (k^0\right )\right )\left (k\!\!\!/+m\right )\,,\\
  \ii D^{>}\left (k\right )&=2\pi\delta\left (k^2-m^2\right )\left
    (\theta\left (k^0\right )\left (1-f_F\left (k^0\right )\right
    )-\theta\left (-k^0\right )f_{\bar F}\left (k^0\right )\right )\left
    (k\!\!\!/+m\right )\,.
\label{eq:ThermalPropagatorsFermions}
\end{split}
\end{equation}
The expressions for the self-energies\footnote{The connection between
  self-energies and Green's functions is discussed in
  Sec.~\ref{sec:2PIAction}.} along the contour are similar to the
definition of Green's functions \eqref{eq:GreenFunc}.  However, in
general self-energies contain a singular part for diagonal elements (see
also Sec.~\ref{sec:2PIAction}) :
\begin{equation}
\begin{split}
\Pi^{11}\left (x,y\right )&=\delta^4\left (x-y\right
)\Pi^{\text{s}}\left (x\right )+\Theta\left (x^0-y^0\right )\Pi^>\left
  (x,y\right )+\Theta\left (y^0-x^0\right )\Pi^<\left (x,y\right )\,,\\
\Pi^{22}\left (x,y\right )&=\delta^4\left (x-y\right
)\Pi^{\text{s}}\left (x\right )+\Theta\left (x^0-y^0\right )\Pi^<\left
  (x,y\right )+\Theta\left (y^0-x^0\right )\Pi^>\left (x,y\right )\,.\\ 
\label{eq:SelfEnergyContour}
\end{split}
\end{equation}
In analogy to \eqref{eq:PhysGreenFunc} one obtains the following
relations for the retarded and advanced self-energies:
\begin{equation}
\begin{split}
\ii \Pi^{\mathrm{ret}}\left (x,y\right )&=\Theta\left (x^0-y^0\right
)\left (\ii \Pi^>\left (x,y\right )-\ii\Pi^<\left (x,y\right )\right )\,,\\
\ii \Pi^{\mathrm{adv}}\left (x,y\right )&=-\Theta\left (y^0-x^0\right
)\left (\ii \Pi^>\left (x,y\right )-\ii \Pi^<\left (x,y\right )\right )\,.\\
\label{eq:SelfEnergyRetAdv}
\end{split}
\end{equation}
\subsection{1PI quantum effective action}
\label{sec:1PIAction}

As discussed before, we prefer not to be restricted to mean-field
dynamics, but are interested in deriving transport equations for quantum
fluctuations around the mean field, which can be partially interpreted
as mesonic interactions. Therefore, we start directly with the commonly
used concept of a generating functional for connected diagrams
$W[J,\theta,\bar\theta]$, formulated here with respect to one-point
sources,
$J:=\left (J_0,...,J_3\right ),\,\theta:=\left (\theta_1,\theta_2\right
),\,\bar\theta:=\left (\bar\theta_1,\bar\theta_2\right )$. The source
terms act as auxiliary functions to derive connected diagrams and have
to vanish for the physical case, leading to generalized equations of
motion. To avoid an overloaded notation, one should just keep in mind,
that for systems out of equilibrium all source and field terms carry an
additional but hidden index with respect to the upper or lower branch of
the real-time contour $\mathcal{C}$. By introducing the shorthand
notation
\begin{equation}
\int_x:=\int_{C}\dd x^0\int\dd^3x\,,\qquad x\equiv\left (x^0,\vec x\right )
\label{eq:ShortInt}
\end{equation}
and summing up over identical indices for involved fields, the generating functional becomes
\begin{equation}
\begin{split}
  \ii W[J,\theta,\bar\theta]\equiv&\,\ln \left (Z\left [J,\theta,\bar\theta\right ]\right ),\\
  Z[J,\theta,\bar\theta]=&\,\mathcal{N}\int\mathcal{D}\Phi\mathcal{D}\bar\psi\mathcal{D}\psi\exp
  \Bigg( \ii \Bigg[S\left [\Phi,\bar\psi,\psi\right ] +\int_x J_a\left
    (x\right )\Phi_a\left (x\right ) +\int_x\bar\psi_i\left (x\right
  )\theta_i\left (x\right )+\int_x\bar\theta_i\left (x\right
  )\psi_i\left (x\right )\Bigg]\Bigg)\,,
\label{eq:1PIGenFunctional}
\end{split}
\end{equation}
with $\mathcal{N}$ denoting a normalization constant, which does not
depend on the source terms and will be skipped in the following without
loss of generality. The relevant expectation values of the quantum field
theory follow from the generating functional by taking functional
derivatives with respect to the corresponding source terms,
\begin{equation}
\begin{split}
  \left <\Phi_a\left (x\right )\right >&=\frac{\delta W[J,\theta,\bar\theta]}{\delta J_a\left (x\right )}=\phi_a\left (x\right )\,,\\
  \left <\bar\psi_i\left (x\right )\right >&=-\frac{\delta W[J,\theta,\bar\theta]}{\delta \theta_i\left (x\right )}\stackrel{!}{=}0\,,\qquad \left <\psi_i\left (x\right )\right >=\frac{\delta W[J,\theta,\bar\theta]}{\delta\bar\theta_i\left (x\right )}\stackrel{!}{=}0\,,\\
  G_{\Phi_a\Phi_b}\left (x,y\right )&=\left <\mathcal{T}_{\mathcal
      C}\left
      (\Phi_a\left (x\right )\Phi_b\left (y\right )\right )\right >=-\ii \frac{\delta^2 W}{\delta J_a\left (x\right )\delta J_b\left (y\right )}\,,\\
  D_{\psi_i\psi_j}\left (x,y\right )&=\left <\mathcal{T}_{\mathcal
      C}\left (\psi_i\left (x\right )\bar\psi_j\left (y\right )\right
    )\right >=-\ii \frac{\delta^2 W}{\delta\theta_j\left (y\right
    )\delta\bar\theta_i\left (x\right )} \,,
\label{eq:QuantumMeanFields}
\end{split}
\end{equation}
where we exploit the physical condition of vanishing fermionic mean
fields due to Lorentz invariance\footnote{Physical observables have to
  be bilinear in fermionic fields.}. Obviously, the name ``generating
functional'' of \eqref{eq:1PIGenFunctional} refers to the possibility of
deriving arbitrary $n$-point Green's functions by subsequent functional
differentiation with respect to the external sources.

With the properties \eqref{eq:QuantumMeanFields} one can introduce an
effective action by performing a functional Legendre transformation with
respect to one-point sources of the chiral field,
\begin{equation}
\Gamma[\left <\Phi\right >,\theta,\bar\theta]:=W\left [J,\theta,\bar\theta\right ]-\int_xJ_a\left (x\right )\phi_a\left (x\right )\,,
\label{eq:EffAction1}
\end{equation}
allowing to specify the source terms via the functional derivative,
\begin{equation}
J_a\left (x\right )=-\frac{\delta\Gamma\left [\phi_a\right ]}{\delta\phi_a\left (x\right )}\,.
\label{eq:SourceTerms}
\end{equation}
Furthermore, this relation defines dynamical mean-field equations for
the physical case of $J_a\left (x\right )\stackrel{!}{=}0$.  Making use
of the path transformation $\Phi\rightarrow\left <\Phi\right >+\varphi$
in Eq.~\eqref{eq:1PIGenFunctional}, results in:
\begin{equation}
\begin{split}
  W\left [J,\theta,\bar\theta\right
  ]=&\,-\ii\ln\Bigg[\int\mathcal{D}\varphi\mathcal{D}\bar\psi\mathcal{D}\psi\exp\Bigg(\ii
  \Bigg[S\left
    [\left <\Phi\right >+\varphi,\bar\psi,\psi\right ]\\
  &\,+\int_x J_a\left (x\right )\left (\left <\Phi_a\left (x\right )\right >+\varphi_a\left (x\right )\right )+\int_x\bar\psi_i\left (x\right )\theta_i\left (x\right )+\int_x\bar\theta_i\left (x\right )\psi_i\left (x\right )\Bigg]\Bigg)\Bigg]\\
  =&\,-i\ln\Bigg[\int\mathcal{D}\varphi\mathcal{D}\bar\psi\mathcal{D}\psi\exp\Bigg(\ii
  \Bigg[S\left [\left <\Phi\right >+\varphi,\bar\psi,\psi\right ]-S\left [\left <\Phi\right >\right ]\\
  &\qquad\qquad\,+\int_x J_a\left (x\right )\varphi_a\left (x\right )+\int_x\bar\psi_i\left (x\right )\theta_i\left (x\right )+\int_x\bar\theta_i\left (x\right )\psi_i\left (x\right )\Bigg]\Bigg)\Bigg]\\
  &\,+S\left [\left <\Phi\right >\right ]+\int_xJ_a\left (x\right )\phi_a\left (x\right )\,.\\
\label{eq:TrafoGenFunctional}
\end{split}
\end{equation}
With the definition,
\begin{equation}
\Gamma_1\left [\left <\Phi\right >,\theta,\bar\theta\right ]:= W\left [J,\theta,\bar\theta\right ]-\int_xJ_a\left (x\right )\phi_a\left (x\right )-S\left [\left <\Phi\right >\right ]=\Gamma\left [\left <\Phi\right >,\theta,\bar\theta\right ]-S\left [\left <\Phi\right >\right ]\,,
\label{eq:Gamma1}
\end{equation}
we obtain a new relation for the effective action, where
$\Gamma_1\left [\left <\Phi\right >,\theta,\bar\theta\right ]$ contains
one- and multi-loop diagrams, as can be seen in the following.

Before proceeding with the further analysis of
Eq.~\eqref{eq:TrafoGenFunctional}, we rewrite the purely bosonic part
$S\left [\Phi\right ]$ of the classical action
\eqref{eq:ClassicalAction}, evaluated at $\left <\Phi\right >+\varphi$,
by expanding it around the mean-field $\left <\Phi\right >$ in the
functional sense,
\begin{equation}
\begin{split}
S\left [\left <\Phi\right >+\varphi\right ]=\,S\left [\left <\Phi\right
  >\right ]+S_{I}\left [\left <\Phi\right >,\varphi^3\right ] +\int_x\left .\varphi_a\left (x\right )\frac{\delta S\left [\varphi\right ]}{\delta\varphi_a\left (x\right )}\right |_{\varphi=\left <\Phi\right >}
+\frac{1}{2}\int_{x,y}\varphi_a\left (x\right )\left .\frac{\delta^2 S\left [\varphi\right ]}{\delta\varphi_a\left (x\right )\delta\varphi_b\left (y\right )}\right |_{\varphi=\left <\Phi\right >}\varphi_b\left (y\right )\,,
\label{eq:ActionExpansion}
\end{split}
\end{equation}
where $S_I\left [\left <\Phi\right >,\varphi^3\right ]$ includes only
terms with at least cubic interaction with respect to the fluctuation
$\varphi$.

For notational reasons, we introduce the so-called ``classical'' or
modified form of the free inverse propagator, defined by the second
functional derivative of the bosonic part of the classical action
\eqref{eq:ClassicalAction} as given in the second integral of
Eq.~\eqref{eq:ActionExpansion}:
\begin{equation}
\begin{split}
\ii G_{0,\varphi_a\varphi_b}^{-1}\left (x,y\right ):=&\,\left .\frac{\delta^2 S\left [\varphi\right ]}{\delta\varphi_a\left (x\right )\delta\varphi_b\left (y\right )}\right |_{\varphi=\left <\Phi\right >}=\frac{\delta^2 S\left [\left <\Phi\right >\right ]}{\delta\phi_a\left (x\right )\delta\phi_b\left (y\right )}\\
=&\,-\left [\partial_\mu\partial^\mu+\lambda\left (\phi_c\left (x\right )\phi_c\left (x\right )-\nu^2\right )\right ]\delta_{ab}\delta_{\mathcal{C}}\left (x-y\right )\\
&\,-2\lambda\phi_a\left (x\right )\phi_b\left (x\right )\delta_{\mathcal{C}}\left (x-y\right )\,.\\
\label{eq:FreePropagators1}
\end{split}
\end{equation}
This representation allows to profit from the Gaussian form of path
integrals as will be seen in the following.  In particular, we obtain
for the diagonal elements
\begin{equation}
\begin{split}
\ii G_{0,\sigma\sigma}^{-1}\left (x,y\right )=&\,-\left (\partial_\mu\partial^\mu+\lambda\left (3\sigma^2+\sum_i\pi_i^2-\nu^2\right )\right )\delta_{\mathcal{C}}\left (x-y\right )\,,\\
\ii G_{0,\pi_i\pi_i}^{-1}\left (x,y\right )=&\,-\left (\partial_\mu\partial^\mu+\lambda\left (\sigma^2+3\pi_i^2+\sum_{j\neq i}\pi_j^2-\nu^2\right )\right )\delta_{\mathcal{C}}\left (x-y\right )\,.
\label{eq:DiagFreePropagators}
\end{split}
\end{equation}
For massive fermions the usual definition of an inverse free propagator,
\begin{equation}
\begin{split}
\ii D_{0,\psi_i\psi_j}^{-1}\left (x,y\right ):=&\,\left (\ii \partial\!\!\!/-M_{\psi_i}\right )\delta_{ij}\delta_{\mathcal{C}}\left (x-y\right )\,,
\label{eq:FreePropagators2}
\end{split}
\end{equation}
is used, where $M_{\psi_i}=\sigma+\ii \gamma_5\vec\pi\cdot\vec\tau$ acts as
a mass matrix\footnote{Even though the Lagrangian has no explicit mass
  dependence, an effective mass term is generated via interactions with
  the chiral field.}.

Now, we are equipped to reformulate Eq.~\eqref{eq:TrafoGenFunctional} in
a more convenient form. By making use of relations
\eqref{eq:ActionExpansion}-\eqref{eq:FreePropagators2} and applying them
to the full expression of the classical action, we can introduce a
shifted action,
\begin{equation}
\begin{split}
\hat S\left [\left <\Phi\right >,\varphi,\bar\psi,\psi\right ]:=&\,S\left [\left <\Phi\right >+\varphi,\bar\psi,\psi\right ]-S\left [\left <\Phi\right >\right ]-\int_x\varphi_a\left (x\right )\left .\frac{\delta S\left [\varphi\right ]}{\delta\varphi_a\left (x\right )}\right |_{\varphi=\left <\Phi\right >}\\
=&\,\frac{1}{2}\int_{x,y}\varphi_a\left (x\right )\ii G_{0,\varphi_a\varphi_b}^{-1}\left (x,y\right )\varphi_b\left (y\right )+S_I\left [\left <\Phi\right >,\varphi^3\right ]\\
&+\int_{x,y}\bar\psi_i\left (x\right ) \ii D_{0,\psi_i\psi_j}^{-1}\left (x,y\right )\psi_j\left (y\right )+S_I\left [\left <\Phi\right >+\varphi,\bar\psi,\psi\right ]\,.
\label{eq:ShiftedAction}
\end{split}
\end{equation}
For \eqref{eq:Gamma1}
\begin{equation}
\begin{split}
\Gamma_1\left [\left <\Phi\right >,\theta,\bar\theta\right ]=&\,-\ii \ln\Bigg[\int\mathcal{D}\varphi\mathcal{D}\bar\psi\mathcal{D}\psi\exp\Bigg(i\Bigg[\hat S\left [\left <\Phi\right >,\varphi,\bar\psi,\psi\right ]\\
&\qquad\qquad\,+\int_x\left (\left .\frac{\delta S\left [\varphi\right ]}{\delta\varphi_a\left (x\right )}\right |_{\varphi=\left <\Phi\right >}+J_a\left (x\right )\right )\varphi_a\left (x\right )\\
&\qquad\qquad\,+\int_x\bar\psi_i\left (x\right )\theta_i\left (x\right )+\int_x\bar\theta_i\left (x\right )\psi_i\left (x\right )\Bigg]\Bigg)\Bigg]\,
\label{eq:ShiftedGamma1}
\end{split}
\end{equation}
follows Comparing \eqref{eq:ShiftedGamma1} with
\eqref{eq:1PIGenFunctional} leads to the statement that
$\Gamma_1[\left <\Phi\right >,\theta,\bar\theta]$ can be understood as a
generating functional of connected diagrams for a theory with the
classical action
$\hat S\left [\left <\Phi\right >,\varphi,\bar\psi,\psi\right ]$, where
$J_{1,a}\left (x\right ):=\left (\left .\frac{\delta S\left
        [\varphi\right ]}{\delta\varphi_a\left (x\right )}\right
  |_{\varphi=\left <\Phi\right >}+J_a\left (x\right )\right )$ act as
source terms for the components of the chiral field. In case of scalar
fields an explicit calculation for this property can be found in
Ref.~\cite{PhysRevD.9.1686}.

We can explicitly evaluate the effective action to one-loop order for bosons by exploiting Gaussian integration in \eqref{eq:ShiftedGamma1} over bosonic as well as fermionic fields with Grassmann algebra:
\begin{equation}
\begin{split}
\int\mathcal{D}\varphi\exp\left [-\frac{1}{2}\int_{x,y}\varphi_a\left (x\right )G_{0,\varphi_a\varphi_b}^{-1}\left (x,y\right )\varphi_b\left (y\right )\right ]=&\,\sqrt{2\pi}\det\left (G_{0,\varphi_a\varphi_b}^{-1}\right )^{-\frac{1}{2}}\\
\int\mathcal{D}\bar\psi\mathcal{D}\psi\exp\left [-\int_{x,y}\bar\psi_i\left (x\right ) D_{0,\psi_i\psi_j}^{-1}\left (x,y\right )\psi_j\left (y\right )\right ]=&\,\det\left (D_{0,\psi_i\psi_j}^{-1}\right )
\label{eq:GaussianIntegrals}
\end{split}
\end{equation}
and employing the property
$\Tr \ln\left (M\right )=\ln\left (\det\left (M\right )\right )$
for a diagonalizable matrix $M$, where traces are taken with respect to
field and Dirac (for fermions) indices\footnote{In this final form all
  constant contributions are skipped, since they do not change the
  evolution equations, resulting from the effective action.}. Finally,
we arrive at
\begin{equation}
\begin{split}
\Gamma\left [\left <\Phi\right >,\theta,\bar\theta\right ]=&\,S\left [\left <\Phi\right >\right ]+\frac{\ii}{2}\Tr \ln\left (G_{0,\varphi_a\varphi_b}^{-1}\right )-\ii \Tr \ln\left (D_{0,\psi_i\psi_j}^{-1}\right )+\Gamma_2\left [\left <\Phi\right >,\theta,\bar\theta\right ]\,,
\label{eq:GammaFinal}
\end{split}
\end{equation}
where all connected and diagrams with two or more purely bosonic loops
are absorbed in the last term.  The bosonic part
$S_I\left [\left <\Phi\right >,\varphi^3\right ]$ of the shifted action
\eqref{eq:ShiftedAction} has no vertices with respect to only one
fluctuating field $\varphi$, meaning that different parts of purely
bosonic diagrams in
$\Gamma_2\left [\left <\Phi\right >,\theta,\bar\theta\right ]$ are
connected by at least two propagators.  Consequently, those diagrams
cannot be separated by cutting one inner line and
$\Gamma\left [\left <\Phi\right >,\theta,\bar\theta\right ]$ becomes the
1PI effective action for the chiral field, since also contributions from
zero- and one-loop order are one particle-irreducible.

\subsection{2PI quantum effective action}
\label{sec:2PIAction}

One of the most succesful approaches for describing non-equilibrium
systems is based on Baym's $\Phi$-functionals (see
Ref.~\cite{phd:HendrikVanHees} for more details) and is known as the 2PI
effective action, which is a technique to sum large classes of
perturbative 1PI diagrams in and out of equilibrium, closely related to
the well-known Cornwall–Jackiw–Tomboulis (CJT) formalism at zero and
finite temperature \cite{PhysRevD.10.2428, PhysRevD.47.2356}. In more
detail, the 2PI effective is a generating functional $\Gamma$ for
connected diagrams with fully dressed propagators as internal lines,
allowing to derive self-consistent off-shell evolution equations with
respect to the mean fields and Green's functions \eqref{eq:GreenFunc} of
the theory. These evolution equations preserve global symmetries of the
original theory and can be renormalized with vacuum counter terms
\cite{PhysRevD.65.025010,PhysRevD.65.105005}. They also guarantee
thermodynamic consistency \cite{PhysRev.127.1391} and recover the
correct equilibrium limit. Furthermore, the functional approach allows
for a systematic inclusion of collisional memory effects
\cite{IVANOV1999413, IVANOV2000313, KNOLL2001126, Ivanov2003}. The
abbreviation 2PI stands for two-particle irreducible, which means that
diagrams from the corresponding functional do not become disconnected by
cutting $2$ inner lines.

With the knowledge from the previous section, we are prepared to extend
the concept of the 1PI effective action to the 2PI generating functional
for connected diagrams with respect to one- $(J_a)$ and two-point
($K_{a,b},L_{i,j}$) sources. Thereby, we directly exploit the property
of vanishing expectation values for the fermionic fields and start with
the general form,
\begin{equation}
\begin{split}
\ii W[J,K,L]\equiv&\,\ln\left (Z\left [J,K,L\right ]\right ),\\
Z[J,K,L]=&\,\mathcal{N}\int\mathcal{D}\Phi\mathcal{D}\bar\psi\mathcal{D}\psi\exp\Bigg(\ii\Bigg[S^{K,L}\left [\Phi,\bar\psi,\psi\right ]+\int_xJ_a\left (x\right )\Phi_a\left (x\right )\Bigg]\Bigg)\,,
\label{eq:GenFunctional}
\end{split}
\end{equation}
where the modified expression for the classical action is given by
\begin{equation}
\begin{split}
S^{K,L}\left [\Phi,\bar\psi,\psi\right ]:=\,S\left
  [\Phi,\bar\psi,\psi\right ]+\frac{1}{2}\int_{x,y}\Phi_a\left (x\right
)K_{a,b}\left (x,y\right )\Phi_b\left (y\right ) +\int_{x,y}\bar\psi_i\left (x\right ) L_{i,j}\left (x,y\right )\psi_j\left (y\right )\,.
\label{eq:ShiftedActionKL}
\end{split}
\end{equation}
After performing a functional Legendre transfomation of
\eqref{eq:GenFunctional} with respect to the source $J$,
\begin{equation}
\Gamma^{K,L}\left [\left <\Phi\right >\right ]=W\left [J,K,L\right ]-\int_xJ_a\left (x\right )\phi_a\left (x\right )\,,
\label{eq:1GammaJKL}
\end{equation}
the generating functional \eqref{eq:1GammaJKL} can be formally
identified as the 1PI effective action for a theory with the modified
classical action \eqref{eq:ShiftedActionKL}. Thereby, the quadratic
field contributions with two-point sources act as effective
time- and space-dependent mass terms (see also
Ref.~\cite{doi:10.1063/1.1843591} for scalar field theory).

In analogy to Eq.~\eqref{eq:Gamma1}, we write beyond the zero-loop part:
\begin{equation}
\begin{split}
\Gamma_1^{K,L}\left [\left <\Phi\right >\right ]=&\,\Gamma^{K,L}\left [\left <\Phi\right >\right ]-S^{K,L}\left [\left <\Phi\right >\right ]\,.
\label{eq:2GammaJKL}
\end{split}
\end{equation}
According to expressions \eqref{eq:ShiftedGamma1} and
\eqref{eq:GammaFinal}, one obtains:
\begin{equation}
\Gamma_1^{K,L}\left [\left <\Phi\right >\right ]:=\,-\ii\ln\Bigg[\int\mathcal{D}\varphi\mathcal{D}\bar\psi\mathcal{D}\psi\exp\Bigg(\ii\Bigg[\hat S^{K,L}\left [\left <\Phi\right >,\varphi,\bar\psi,\psi\right ]-\int_x\frac{\delta\Gamma_1^{K,L}\left [\left <\Phi\right >\right ]}{\delta\phi_a}\varphi_a\left (x\right )\Bigg]\Bigg)\Bigg]\,,
\label{eq:Gamma1JKL}
\end{equation}
where the modified and shifted classical action follows directly from
comparing with the relation \eqref{eq:ShiftedAction}:
\begin{equation}
\begin{split}
\hat S^{K,L}\left [\left <\Phi\right >,\varphi,\bar\psi,\psi\right ]=&\,S_I\left [\left <\Phi\right >,\varphi^3\right ]+S_I\left [\left <\Phi\right >+\varphi,\bar\psi,\psi\right ]\\
&+\int_{x,y}\bar\psi_i\left (x\right )i\left
  (D_{0,\psi_i\psi_j}^{-1}\left (x,y\right )-\ii L_{i,j}\left (x,y\right )\right )\psi_j\left (y\right )\\
&+\frac{1}{2}\int_{x,y}\varphi_a\left (x\right )i\left
  (G_{0,\varphi_a\varphi_b}^{-1}\left (x,y\right )-\ii K_{a,b}\left (x,y\right )\right )\varphi_b\left (y\right )\,.
\label{eq:ShiftedModifiedAction}
\end{split}
\end{equation}
Before perfoming Gaussian integrals over field configurations, by
replacing the inverse classical propagators by modified expressions of
\eqref{eq:ShiftedModifiedAction}, we apply a second functional Legendre
transformation with respect to the two-point sources
\begin{equation}
\Gamma\left [\left <\Phi\right >,G,D\right ]:=\Gamma^{K,L}\left [\left <\Phi\right >\right ]-\int_{x,y}K_{b,a}\left (y,x\right )\frac{\delta\Gamma^{K,L}\left [\left <\Phi\right >\right ]}{\delta K_{b,a}\left (y,x\right )}-\int_{x,y}L_{j,i}\left (y,x\right )\frac{\delta\Gamma^{K,L}\left [\left <\Phi\right >\right ]}{\delta L_{j,i}\left (y,x\right )}\,.
\label{eq:Gamma2PI}
\end{equation}
With the requirement of independent one- and two-point sources, the
functional derivatives with respect to two-point sources lead to the
following relations:
\begin{equation}
\begin{split}
\frac{\delta\Gamma^{K,L}\left [\left <\Phi\right >\right ]}{\delta K_{b,a}\left (y,x\right )}&=\frac{\delta W\left [J,K,L\right ]}{\delta K_{b,a}\left (y,x\right )}\\
&=\left <\mathcal{T}_\mathcal{C}\left (\Phi_b\left (y\right )\Phi_a\left (x\right )\right )\right >=\frac{1}{2}\left (\phi_a\left (x\right )\phi_b\left (y\right )+G_{\varphi_a\varphi_b}\left (x,y\right )\right )\,,\\
\frac{\delta\Gamma^{K,L}\left [\left <\Phi\right >\right ]}{\delta L_{j,i}\left (y,x\right )}&=\frac{\delta W\left [J,K,L\right ]}{\delta L_{j,i}\left (y,x\right )}\\
&=\left <\mathcal{T}_\mathcal{C}\left (\bar\psi_j\left (y\right )\psi_i\left (x\right )\right )\right >=-D_{\psi_i\psi_j}\left (x,y\right )\,.
\label{eq:DerivativeTwoPoint}
\end{split}
\end{equation}
Consequently, we conclude that derivatives with respect to two-point
sources generate the full propagators for the chiral field, which
include also products of the mean-fields.

Now, without loss of generality we introduce a parametrization for inverse propagators:
\begin{equation}
\begin{split}
G^{-1}_{\varphi_a\varphi_b}\left(x,y\right)&=G^{-1}_{0,\varphi_a\varphi_b}\left(x,y\right)-\ii
K_{a,b}\left (x,y\right )-\Pi_{\varphi_a\varphi_b}\left(x,y\right)\,,\\
D^{-1}_{\psi_i\psi_j}\left(x,y\right)&=D^{-1}_{0,\psi_i\psi_j}\left(x,y\right)-\ii
L_{i,j}\left (x,y\right )-\Sigma_{\psi_i\psi_j}\left(x,y\right)\,,
\label{eq:ParamPropagators}
\end{split}
\end{equation}
where the source contributions vanish in physical cases, leading to the
general form of dressed propagators with
$\Pi_{\varphi_a\varphi_b}\left(x,y\right)$ and
$\Sigma_{\psi_i\psi_j}\left(x,y\right)$ denoting the proper
self-energies.

Combining \eqref{eq:Gamma2PI}, \eqref{eq:DerivativeTwoPoint},
\eqref{eq:ParamPropagators} and exploiting Gaussian integration in
analogy to \eqref{eq:GaussianIntegrals}, \eqref{eq:GammaFinal}, results
in
\begin{equation}
\begin{split}
\Gamma\left [\left <\Phi\right >,G,D\right ]&=S^{K,L}\left [\left <\Phi\right >\right ]+\Gamma_1^{K,L}\left [\left <\Phi\right >\right ]\\
&\quad-\frac{1}{2}\int_{x,y}K_{b,a}\left (y,x\right )\left [\phi_a\left (x\right )\phi_b\left (y\right )+G_{\varphi_a\varphi_b}\left (x,y\right )\right ]\\
&\quad+\int_{x,y}L_{j,i}\left (y,x\right )D_{\psi_i\psi_j}\left (x,y\right )\\
&=S\left [\left <\Phi\right >\right ]-\frac{1}{2}\Tr \left [K_{b,a}G_{\varphi_a\varphi_b}\right ]+\Tr \left [L_{j,i}D_{\psi_i\psi_j}\right ]+\Gamma^{K,L}_1\left [\left <\Phi\right >\right ]\\
&=S\left [\left <\Phi\right >\right ]+\frac{\ii}{2}\Tr \ln\left (G^{-1}_{\varphi_a\varphi_b}\right )-\ii \Tr \ln\left (D^{-1}_{\psi_i\psi_j}\right )\\
&\quad +\frac{\ii}{2}\Tr \left
  [G^{-1}_{0,\varphi_b\varphi_a}G_{\varphi_a\varphi_b}\right ]-\ii \Tr \left [D_{0,\psi_j\psi_i}^{-1}D_{\psi_i\psi_j}\right ]\\
&\quad -\frac{\ii}{2}\Tr \left [\Pi_{\varphi_b\varphi_a}
  G_{\varphi_a\varphi_b}\right ]+\Tr \left [\Sigma_{\psi_j\psi_i}
  D_{\psi_i\psi_j}\right ] +\text{const.}\\
&\quad +\left (\Gamma^{K,L}_1\left [\left <\Phi\right >\right
  ]-\frac{\ii}{2}\Tr \ln\left (G^{-1}_{\varphi_a\varphi_b}\right
  )+\ii \Tr \ln\left (D^{-1}_{\psi_i\psi_j}\right )\right )\\
&=:S\left [\left <\Phi\right >\right ]+\frac{\ii}{2}\Tr \ln\left
  (G^{-1}_{\varphi_a\varphi_b}\right )-\ii \Tr \ln\left
  (D^{-1}_{\psi_i\psi_j}\right )\\ 
&\quad +\frac{\ii}{2}\Tr \left
  [G^{-1}_{0,\varphi_b\varphi_a}G_{\varphi_a\varphi_b}\right ]-\ii \Tr \left [D_{0,\psi_j\psi_i}^{-1}D_{\psi_i\psi_j}\right ]+\Gamma_2\left [\left <\Phi\right >,G,D\right ]+\text{const.}\\
\label{eq:2PIAction}
\end{split}
\end{equation}
For the final result, we adjust the expression for the 2PI effective
action \eqref{eq:2PIAction} to a simplified and more usual notation,
which will be regularly used in the following:
\begin{equation}
\Gamma\left [\sigma,\vec\pi,G,D\right ]= S\left [\sigma,\vec\pi\right
]+\frac{\ii}{2}\Tr \ln
G^{-1}+\frac{\ii}{2}\Tr \,G_0^{-1}G-\ii \Tr \ln
D^{-1}-\ii \Tr D_0^{-1}D+\Gamma_2\left [\sigma,\vec\pi,G,D\right ]\,. 
\label{eq:2PIfull}
\end{equation}
Here, the mean-fields and propators stand for
\begin{equation}
\begin{split}
\left (\sigma,\vec\pi\right )=\left <\Phi\left (x\right )\right >\,,\qquad G_0^{-1}&=G^{-1}_{0,\varphi_a\varphi_b}\left (x,y\right )\,,\qquad D_0^{-1}=D^{-1}_{0,\psi_i\psi_j}\left (x,y\right )\,,\\
G&=G_{\varphi_a\varphi_b}\left (x,y\right )\,,\quad\qquad\,\,\, D=D_{\psi_i\psi_j}\left (x,y\right )\,.
\label{eq:PropagatorNotation}
\end{split}
\end{equation}
Taking functional derivatives of \eqref{eq:2PIAction} with respect to
the full propagators,
\begin{equation}
\begin{split}
\frac{\delta\Gamma\left [\left <\Phi\right >,G,D\right ]}{\delta G_{\varphi_b\varphi_a}\left (y,x\right )}&=-\frac{\ii}{2}\left [G^{-1}_{\varphi_a\varphi_b}\left (x,y\right )-G^{-1}_{0,\varphi_a\varphi_b}\left (x,y\right )
+2\ii\frac{\delta\Gamma_2\left [\left <\Phi\right >,G,D\right ]}{\delta G_{\varphi_b\varphi_a}\left (y,x\right )}\right ]\\
&=-\frac{1}{2}K_{a,b}\left (x,y\right )\,,\\
\frac{\delta\Gamma\left [\left <\Phi\right >,G,D\right ]}{\delta D_{\psi_j\psi_i}\left (y,x\right )}&=\ii\left [D^{-1}_{\psi_i\psi_j}\left (x,y\right )-D^{-1}_{0,\psi_i\psi_j}\left (x,y\right )
-\ii\frac{\delta\Gamma_2\left [\left <\Phi\right >,G,D\right ]}{\delta D_{\psi_j\psi_i}\left (y,x\right )}\right ]\\
&=L_{i,j}\left (x,y\right )\,,
\label{eq:DefGamma2}
\end{split}
\end{equation}
where the last equalities follow from comparing with expression \eqref{eq:Gamma2PI}, 
allows to define the proper self-energies of Eqs.~\eqref{eq:ParamPropagators} within the functional approach:
\begin{equation}
\Pi_{\varphi_a\varphi_b}\left(x,y\right)=2 \ii\frac{\delta\Gamma_2\left [\left <\Phi\right >,G,D\right ]}{\delta G_{\varphi_b\varphi_a}\left(y,x\right)}\,,\quad \Sigma_{\psi_i\psi_j}\left(x,y\right)=-\ii\frac{\delta\Gamma_2\left [\left <\Phi\right >,G,D\right ]}{\delta D_{\psi_j\psi_i}\left(y,x\right)}\,.
\label{eq:ProperSelfEnergy}
\end{equation}
Since the self-energies consist of one-particle irreducible diagrams,
the functional $\Gamma_2$ cannot contain two independent parts in the
same diagram, which are only connected by two propagators of type $G$ or
$D$.  To be more specific, deriving of $\Gamma_2$ with respect to one of
the fully dressed propagators is equivalent to opening the corresponding
propagator line in the diagramatic representation.  Consequently, the
functional $\Gamma_2$ stands for the sum of all connected two-particle
irreducible diagrams, which can be constructed from vertices of the
shifted classical action \eqref{eq:ShiftedModifiedAction}, connected by
bosonic $G$ as well as fermionic $D$ propagators. Indeed, in case of the
2PI effective action $\Gamma_2$ denotes the two-loop and higher order
for bosonic as well as fermionic diagrams, when the local part of the
fermionic self-energy $\Sigma$ is absorbed in the free inverse
propagator as an effective mass contribution\footnote{The local part of
  the fermionic self-energy arises from the interaction between the
  fermions and the chiral mean field, resulting in an effective mass
  contribution for fermions.}.

So, the final form \eqref{eq:2PIAction} is a significant simplification of the original and rather formal expression \eqref{eq:Gamma2PI}.\\
In case of vanishing source terms, the relations in \eqref{eq:DefGamma2}
define evolution equations for the propagators $G$ and $D$, fullfilling
the corresponding Schwinger-Dyson equations\footnote{Here, a symbolic
  matrix notation is used.}:
\begin{equation}
\begin{split}
G_0^{-1}G-\Pi G&=\mathds{1}\quad\Rightarrow\quad G=G_0+G_0\Pi G\,,\\
D_0^{-1}D-\Sigma D&=\mathds{1}\quad\Rightarrow\quad D=D_0+D_0\Sigma D\,.
\label{eq:SchwingerDyson}
\end{split}
\end{equation}
Because of the recursive structure of these differential equations, all
internal lines $G$, $D$ are indeed fully dressed propagators, which can
be expressed as infinite series with respect to the classical
propagators $G_0$, $D_0$ as well as self-energies $\Pi$, $\Sigma$.
Consequently, every diagram from the 2PI effective action resums an
infinite number of 1PI diagrams from Sec.~\eqref{sec:1PIAction}.

\subsection{Exact evolution equations from 2PI effective action}
\label{sec:ExactEvolEq}

Based on the previous section, we are prepared to recover exact
evolution equations on the level of fully dressed propagators, which
fullfill the Schwinger-Dyson equations \eqref{eq:SchwingerDyson} as well
as self-consistent mean-field equations.  Therefore, we explicitly
derive the 2PI effective action \eqref{eq:2PIfull} with respect to the
relevant mean fields of the chiral components and propagators, where the
spatial coordinates are taken on the upper or lower branch, denoted with
latin indices $a,b,c\in\{1,2\}$.\footnote{Note, that off-diagonal branch
  elements do not contribute to mean-field equations.}  From this, we
obtain stationary conditions, leading then to equations of motion with
traces running over flavor and Dirac indices:
\begin{equation}
\begin{split}
&\frac{\delta\Gamma}{\delta\sigma^c}=-J_0^c\stackrel{!}{=}0\\
&\quad\Rightarrow\quad\left (\partial_\mu\partial^\mu+\lambda\left(\left (\sigma^{\text{c}}\right )^2+\left (\vec\pi^{\text{c}}\right )^2-\nu^2\right)+3\lambda G^{cc}_{\sigma\sigma}+\lambda\sum_\ii G^{cc}_{\pi_i\pi_i}\right )\sigma^c\\
&\qquad\qquad =f_\pi m_\pi^2+g\Tr \left (D^{cc}_{\psi_i\psi_i}\right )+\frac{\delta\Gamma_2}{\delta\sigma^c}\,,\\
&\frac{\delta\Gamma}{\delta\pi_i^c}=-J_i^c\stackrel{!}{=}0\\
&\quad\Rightarrow\quad\left (\partial_\mu\partial^\mu+\lambda\left(\left (\sigma^c\right )^2+\left (\vec\pi^c\right )^2-\nu^2\right)+3\lambda G^{cc}_{\pi_i\pi_i}+\lambda\sum_{j\neq i}G_{\pi_j\pi_j}^{cc}+\lambda G^{cc}_{\sigma\sigma}\right )\pi_i^c\\
&\qquad\qquad =g\Tr \left (\ii \gamma_5\tau_i D^{cc}_{\psi_i\psi_i}\right )+\frac{\delta\Gamma_2}{\delta\pi_i^c}\,.
\label{eq:MeanFieldEquations}
\end{split}
\end{equation}
In the following the two quark flavors will be often treated as
degenerate states with equal masses, allowing to skip the
explicit dependence on the flavor index,
\begin{equation}
\begin{split}
&\frac{\delta\Gamma}{\delta G^{a,b}_{\sigma\sigma}}=-\frac{1}{2}K^{a,b}_{\sigma\sigma}\stackrel{!}{=}0\\
&\quad\Rightarrow\quad-\left (\partial_\mu\partial^\mu+\lambda\left (3\sigma^2+\sum_i\pi_i^2-\nu^2\right )\right )G_{\sigma\sigma}^{a,b}\left (x,y\right )\\
&\qquad\quad\,\, =\ii \int_z\Pi_{\sigma\sigma}^{a,c}\left (x,z\right )G_{\sigma\sigma}^{c,b}\left (z,y\right )+\ii\delta^{a,b}_{\mathcal{C}}\left (x-y\right )\,,\\
&\frac{\delta\Gamma}{\delta G^{a,b}_{\pi_i\pi_i}}=-\frac{1}{2}K^{a,b}_{\pi_i\pi_i}\stackrel{!}{=}0\\
&\quad\Rightarrow\quad -\left (\partial_\mu\partial^\mu+\lambda\left (\sigma^2+3\pi_i^2+\sum_{j\neq i}\pi_j^2-\nu^2\right )\right )G_{\pi_i\pi_i}^{a,b}\left (x,y\right )\\
&\qquad\quad\,\, =\ii \int_z\Pi_{\pi_i\pi_i}^{a,c}\left (x,z\right )G^{c,b}_{\pi_i\pi_i}\left (z,y\right )+\ii \delta^{a,b}_{\mathcal{C}}\left (x-y\right )\,,\\
&\frac{\delta\Gamma}{\delta D^{a,b}_{\psi_i\psi_i}}=L^{a,b}_{\psi_i\psi_i}\stackrel{!}{=}0\,,\\
&\quad\Rightarrow\quad \left (\ii\partial\!\!\!/-M_{\psi_i}\right )D_{\psi\psi}^{a,b}\left (x,y\right )=\ii \int_z\Sigma_{\psi_i\psi_i}^{a,c}\left (x,z\right )D^{c,b}_{\psi_i\psi_i}\left (z,y\right )+\ii \delta^{a,b}_{\mathcal{C}}\left (x-y\right )\,.\\
\label{eq:KadanoffBaymEquations}
\end{split}
\end{equation}
Thereby, the integro-differential evolution equations for propagators
can be reformulated in a rather usual way by explicitly evaluating the
time arguments on the contour line, leading to the famous Kadanoff-Baym
equations. In case of the important Wightman functions, one obtains
following expressions:
\begin{equation}
\begin{split}
\left (\partial_\mu\partial^\mu+M^2_{\sigma}\left (x\right )\right )\ii G^{\lessgtr}_{\sigma\sigma}\left (x,y\right )=&-\int_{t_0}^{t_1}\dd z_0\int\dd^3z\left [\Pi_{\sigma\sigma}^{>}\left (x,z\right )-\Pi_{\sigma\sigma}^{<}\left (x,z\right )\right ]\ii G^{\lessgtr}_{\sigma\sigma}\left (z,y\right )\\
&+\int_{t_0}^{t_2}\dd z_0\int\dd^3z\,\Pi_{\sigma\sigma}^{\lessgtr}\left (x,z\right )\left [\ii G^{>}_{\sigma\sigma}\left (z,y\right )-\ii G^{<}_{\sigma\sigma}\left (z,y\right )\right ]\,,\\
\left (\partial_\mu\partial^\mu+M^2_{\pi_i}\left (x\right )\right )\ii G^{\lessgtr}_{\pi_i\pi_i}\left (x,y\right )=&-\int_{t_0}^{t_1}\dd z_0\int\dd^3z\left [\Pi_{\pi_i\pi_i}^{>}\left (x,z\right )-\Pi_{\pi_i\pi_i}^{<}\left (x,z\right )\right ]\ii G^{\lessgtr}_{\pi_i\pi_i}\left (z,y\right )\\
&+\int_{t_0}^{t_2}\dd z_0\int\dd^3z\,\Pi_{\pi_i\pi_i}^{\lessgtr}\left (x,z\right )\left [\ii G^{>}_{\pi_i\pi_i}\left (z,y\right )-\ii G^{<}_{\pi_i\pi_i}\left (z,y\right )\right ]\,,\\
\left (\ii \partial\!\!\!/-M_{\psi_i}\left (x\right )\right )\ii D^{\lessgtr}_{\psi_i\psi_i}\left (x,y\right )=&+\int_{t_0}^{t_1}\dd z_0\int\dd^3z\left [\Sigma_{\psi_i\psi_i}^{>}\left (x,z\right )-\Sigma_{\psi_i\psi_i}^{<}\left (x,z\right )\right ]\ii D^{\lessgtr}_{\pi_i\pi_i}\left (z,y\right )\\
&-\int_{t_0}^{t_2}\dd z_0\int\dd^3z\,\Sigma_{\psi_i\psi_i}^{\lessgtr}\left (x,z\right )\left [\ii D^{>}_{\psi_i\psi_i}\left (z,y\right )-\ii D^{<}_{\psi_i\psi_i}\left (z,y\right )\right ]\,,\\
\label{eq:KadanoffBaymWightmann}
\end{split}
\end{equation}
where we introduced a shorthand notation for the effective mass terms,
depending on the local part of the self-energy (compare with
\eqref{eq:SelfEnergyHartree}):
\begin{equation}
\begin{split}
M_{\sigma}^2\left (x\right ):=&\,\lambda\left (3\sigma^2+\sum_i\pi_i^2-\nu^2\right )+\Pi_{\sigma\sigma}^{\mathrm{loc.}}\left (x\right )\,,\qquad\quad\quad\Pi_{\sigma\sigma}^{\mathrm{loc.}}\left (x\right ):=\ii \Pi^{11}_{\sigma\sigma}\left (x,x\right )\,,\\
M_{\pi_i}^2\left (x\right ):=&\,\lambda\left (\sigma^2+3\pi_i^2+\sum_{j\neq i}\pi_j^2-\nu^2\right )+\Pi_{\pi_i\pi_i}^{\mathrm{loc.}}\left (x\right )\,,\quad\,\Pi_{\pi_i\pi_i}^{\mathrm{loc.}}\left (x\right ):=\ii \Pi^{11}_{\pi_i\pi_i}\left (x,x\right )\,,\\
M^2_{\psi_i}\left (x\right ):=&\,M_{\psi_i}^\dagger M_{\psi_i}=g^2\left (\sigma-\ii \gamma_5\vec\pi\cdot\vec\tau\right )\left (\sigma+\ii \gamma_5\vec\pi\cdot\vec\tau\right )\equiv g^2\left (\sigma^2+\sum_i\pi_i^2\right )\,.
\label{eq:EffMassGeneral}
\end{split}
\end{equation}
The evolution equations for retarded and advanced propagators can be
derived in full analogy to Eqs.~\eqref{eq:KadanoffBaymWightmann}.

\begin{figure}[h]
\captionsetup[subfloat]{labelformat=empty}
\qquad\mbox{$\Gamma_2\sim$}
\subfloat[$\int_{\mathcal{C}}G_{\sigma\sigma}^2$]{\includegraphics[width=2.4cm]{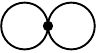}}\quad +
\subfloat[$\int_{\mathcal{C}}G_{\pi_i\pi_i}^2$]{\includegraphics[width=2.4cm]{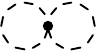}}\quad +
\subfloat[$\int_{\mathcal{C}}G_{\sigma\sigma}G_{\pi_i\pi_i}$]{\includegraphics[width=2.4cm]{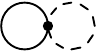}}\quad +
\subfloat[$\int_{\mathcal{C}}G_{\pi_i\pi_i}G_{\pi_j\pi_j}$]{\includegraphics[width=2.4cm]{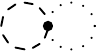}}\\
\hspace*{1.5cm}+\subfloat[$\int_{\mathcal{C}}G_{\sigma\sigma}^4$]{\includegraphics[width=2.4cm]{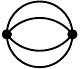}}\quad +
\subfloat[$\int_{\mathcal{C}}G_{\pi_i\pi_i}^4$]{\includegraphics[width=2.4cm]{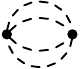}}\quad +
\subfloat[$\int_{\mathcal{C}}G_{\sigma\sigma}^2G_{\pi_i\pi_i}^2$]{\includegraphics[width=2.4cm]{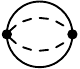}}\quad +
\subfloat[$\int_{\mathcal{C}}G_{\pi_i\pi_i}^2G_{\pi_j\pi_j}^2$]{\includegraphics[width=2.4cm]{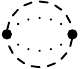}}\\
\hspace*{1.5cm}+\subfloat[$\int_{\mathcal{C}}\phi\,G_{\sigma\sigma}^3\phi$]{\includegraphics[width=2.4cm]{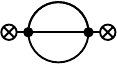}}\quad +
\subfloat[$\int_{\mathcal{C}}\phi\,G_{\sigma\sigma}G_{\pi_i\pi_i}^2\phi$]{\includegraphics[width=2.4cm]{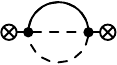}}\quad +
\subfloat[$\int_{\mathcal{C}}D^2G_{\sigma\sigma}$]{\includegraphics[width=2.4cm]{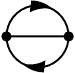}}\quad +
\subfloat[$\int_{\mathcal{C}}D^2G_{\pi_i\pi_i}$]{\includegraphics[width=2.4cm]{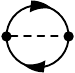}}\\
\caption{Included 2PI part of the effective action with 1st line:
  Hartree diagrams, 2nd line: basketball diagrams, 3rd line: sunset
  diagrams, where solid lines stand for the $\sigma$ propagator, dashed
  and pointed lines for the pion propagators and solid lines with arrows
  for the fermion propagator. The circle with a cross represents a
  mean-field. For all diagrams a schematic integral representation in
  terms of formal expressions for the propagators is shown.}
\label{fig:2PIpart}
\end{figure}

For the rest of this section, we will consider a truncated version of
the 2PI effective action to simplify the dynamics encoded in the exact
evolution equations \eqref{eq:MeanFieldEquations} and
\eqref{eq:KadanoffBaymEquations}.  The $\Gamma_2$-part of the truncated
version is given by its representation in Fig.~\ref{fig:2PIpart},
consisting only of one- and two-point diagrams.  For those diagrams a
first order gradient expansion of Kadanoff-Baym equations in Wigner
space reduces to a Markov-like collisional dynamics without memory
effects \cite{IVANOV1999413, IVANOV2000313, KNOLL2001126,
  Ivanov2003}. However, since we are interested in describing
dissipation phenomena from the mean-fields to mesons, it becomes
necessary to go beyond the simple gradient expansion by partially
including memory effects at least for the bosonic sunset diagram.
Thereby, one has to emphasize and keep in mind that all finite
truncations of the 2PI effective action cause serious difficulties,
concerning the Ward-Takahashi-identities of global and local symmetries,
which are violated in the first neglected order of the expansion
parameter. This violation of the symmetries is a direct consequence of
the resummation in the two-point function and it even follows, that also
the Goldstone theorem is violated
\cite{PhysRevD.15.2897,PhysRevD.66.025028}, leading to a non-vanishing
and temperature dependent mass of pions in the broken phase of the
linear sigma model, even when the chiral symmetry is not explicitly
broken. We refer to Ref.~\cite{PILAFTSIS2013594}, where the authors
discuss some possible modifications for a symmetry-improved 2PI
effective action.

\subsection{Generalized Boltzmann Equation}
\label{sec:GenBoltzmannEquation}

From a numerical point of view a direct solution of the Kadanoff-Baym
equations \eqref{eq:KadanoffBaymEquations} is a costly task, since all
relevant two-point functions have to be stored over the time, requiring
a rapidly increasing amount of memory even for a moderately growing
number of space-time grid points.  Furthermore, with every additional
time step the computational time for this step increases due to the
memory aspect of Kadanoff-Baym equations\footnote{In numerical
  computations one usually introduces a maximum value for the difference
  between two time points, which is chosen to be large compared to the
  relaxation time of memory effects.}.  It is also worth to mention,
that from a physical point of view a direct and exact solution of the
Kadanoff-Baym equations is difficult to interpret, since the dynamics is
encoded in abstract two-point functions. Of course, the propagators
allow to extract various physical observables, but they offer only a
limited insight to the physics in terms of elementary scattering
processes, which allow to understand the underlying transport effects as
long as the system is not too far away from the equilibrium state.
Because of these reasons, we will simplify the Kadanoff-Baym equations
by deriving transport evolution equations for quasi-particle
distribution functions. Thereby, the convolutions of propagators and
self-energies can be interpreted in terms of collisional integrals for
quasi-particle distribution functions.  For this approach, it is
suitable to switch to the Wigner transform with respect to the
difference of space-time positions $\Delta x:=x-y$ and the averaged
space-time variable $X:=\frac{x+y}{2}$.

Denote $B\left (x,y\right )$ an arbitrary two-point function, its Wigner
transform and the inverse are given by
\begin{equation}
\begin{split}
\mathcal{F}_W\left [B\left (x,y\right )\right ]:=&\,\int\dd^4\Delta x\,\ee^{\ii p^\mu\Delta x_\mu}B\left (X+\frac{\Delta x}{2},X-\frac{\Delta x}{2}\right )\\
\equiv&\,\int\dd^4\Delta x\,\ee^{\ii p^\mu\Delta x_\mu}B\left (X,\Delta x\right )=\tilde B\left (X,p\right )\,,\\
\mathcal{F}^{-1}_W\left [\tilde{B}\left (X,p\right )\right ]=&\,\int\frac{\dd^4 p}{\left (2\pi\right )^4}\,\ee^{-\ii p^\mu\Delta x_\mu}\tilde B\left (X,p\right )\,.
\label{eq:WignerTransform}
\end{split}
\end{equation}
For products of two-point functions, one obtains a convolution in
momentum space, as known from the usual Fourier transform
\begin{equation}
\begin{split}
\mathcal{F}_W\left [B\left (x,y\right )C\left (x,y\right )\right ]=\int\frac{\dd^4 k}{\left (2\pi\right )^4}\tilde B\left (X,p+k\right )\tilde C\left (X,k\right )=\left (\tilde A*\tilde B\right )\left (X,p\right )\,.
\label{eq:WignerConvolution}
\end{split}
\end{equation}
Transforming the Kadanoff-Baym equations
\eqref{eq:KadanoffBaymEquations}, \eqref{eq:KadanoffBaymWightmann}
requires also to evaluate spatial convolutions of the form:
\begin{equation}
\begin{split}
D\left (x,y\right ):=\int\dd^4 z\, B\left (x,z\right )C\left (z,y\right )\,.
\label{eq:WignerConvolutionSpace}
\end{split}
\end{equation}
This can be done by expanding the formal expressions (compare with
Eq.~\eqref{eq:WignerTransform} and see \cite{book:1046942,book:304428}
for further details),
\begin{equation}
B\left (x,z\right )\equiv B\left (X+\frac{z-y}{2},x-z\right )\,,\qquad C\left (z,y\right )\equiv C\left (X+\frac{z-x}{2},z-y\right )\,,
\label{eq:TaylorForm}
\end{equation}
in a Taylor series with respect to the first variable and making use of
the property, that $\partial_X$ is a generator of translations.  To be
more precise, for an arbitrary function $f\left (X\right )$
\begin{equation}
f\left (X+a\right )=\ee^{a\partial_X}f\left (X\right )\,.
\label{eq:TaylorGenerator}
\end{equation}
After applying this relation to Eq.~\eqref{eq:TaylorForm}, one obtains
the following expression for the convolution:
\begin{equation}
\begin{split}
D\left (x,y\right )=&\,\int\dd^4z\,\left [\ee^{\frac{z-y}{2}\partial^B_X}B\left (X,x-z\right )\right ]\left [\ee^{\frac{z-x}{2}\partial_X^C}C\left (X,z-y\right )\right ]=A'\ast B'\,,\\
B'\left (X,x-z\right ):=&\,B\left (X,x-z\right )\ee^{-\frac{x-z}{2}\partial_X^C}\,,\quad C'\left (X,z-y\right ):=C\left (X,z-y\right )\ee^{\frac{z-y}{2}\partial^B_X}\,,
\label{eq:ConvolutionTaylorGenerator}
\end{split}
\end{equation}
where the operators act to the left or right side, flagged by the
corresponding superscript.  A Fourier transform of this expression
results then in:
\begin{equation}
\begin{split}
\mathcal{F}\left [B'\ast C'\right ]&=\mathcal{F}\left [B'\right ]\mathcal{F}\left [C'\right ]\,,\\
\mathcal{F}\left [B'\right ]&=\int\dd^4u\,\ee^{\ii p_\mu u^\mu}B\left (X,u\right )\ee^{-\frac{u}{2}\partial_X^C}=B'\left (X,p\right )=\tilde B\left (X,p\right )\ee^{\frac{\ii}{2}\partial_X^C\partial_p^B}\,,\,\,\\
\mathcal{F}\left [C'\right ]&=\int\dd^4u\,\ee^{\ii p_\mu u^\mu}C\left (X,u\right )\ee^{\frac{u}{2}\partial_X^B}=C'\left (X,p\right )=\tilde C\left (X,p\right )\ee^{-\frac{\ii}{2}\partial_X^B\partial_p^C}\,,
\label{eq:FourierProduct}
\end{split}
\end{equation}
where the last relation follows from the general property:
\begin{equation}
\mathcal{F}\left [\ee^{au}f\left (u\right )\right ]=f\left (p-ia\right )=\ee^{-ia\partial_p}f\left (p\right )\,.
\end{equation}
Finally, after introducing the diamond operator,
\begin{equation}
\begin{split}
\Diamond\{\cdot\}\{\cdot\}:=&\,\frac{1}{2}\left (\partial^{\left (1\right )}_X\partial^{\left (2\right )}_p-\partial_p^{\left (1\right )}\partial^{\left (2\right )}_X\right )\{\cdot\}\{\cdot\}\,,
\label{eq:WignerConvSpace}
\end{split}
\end{equation}
the Wigner transform of a spatial convolution
\eqref{eq:WignerConvolutionSpace} reads
\begin{equation}
\tilde D\left (X,p\right ):=\int\dd^4\Delta x\,\ee^{\ii p^\mu\Delta x_\mu}\int\dd^4 z\, B\left (x,z\right )C\left (z,y\right )=\ee^{-\ii\Diamond}\{\tilde B\left (X,p\right )\}\{\tilde C\left (X,p\right )\}\,.
\label{eq:WignerConvFinal}
\end{equation}
In transport theory one usually assumes that all relevant two-point
functions change slowly with respect to the averaged coordinate,
allowing to skip higher-order derivatives,
\begin{equation}
\tilde D\left (X,p\right )=\tilde{B}\left (X,p\right )\tilde{C}\left (X,p\right )-\ii\Diamond\{\tilde B\left (X,p\right )\}\{\tilde C\left (X,p\right )\}+\mathcal{O}\left (\partial_X^2\right )\,.
\label{eq:FirstOrder}
\end{equation}
Now, we start with rewriting the left-hand side of the Kadanoff-Baym
equations\footnote{Here, the flavor indices are skipped, also in the
  following, to avoid an overloaded notation as well as formally equal
  calculations for sigma and pion propagators.} for Wightman functions
in terms of the Wigner transform and its coordinates $\Delta x$, $X$:
\begin{equation}
\begin{split}
\mathcal{F}_W\left [\partial_\mu\partial^\mu \ii G^{\lessgtr}\left (x,y\right )\right ]&=\int\dd^4\Delta x\,\ee^{\ii p\Delta x}\partial_\mu\partial^\mu \ii G^{\lessgtr}\left (x,y\right )\\
&=\int\dd^4\Delta x\,\ee^{\ii p\Delta x}\left (\frac{1}{4}\partial_X^2+\partial_X\partial_{\Delta x}+\partial_{\Delta x}^2\right )\ii G^{\lessgtr}\left (x,y\right )\\
&=\left (\frac{1}{4}\partial_X^2-\ii p\partial_X-p^2\right )\ii \tilde G^{\lessgtr}\left (X,p\right )\,,
\label{eq:WignerOperators1}
\end{split}
\end{equation}
\begin{equation}
\begin{split}
\mathcal{F}_W\left [M^2\left (x\right )\ii G^{\lessgtr}\left (x,y\right )\right ]&=\int\dd^4\Delta x\,\ee^{\ii p\Delta x}M^2\left (x\right )\ii G^{\lessgtr}\left (x,y\right )\\
&=\int\dd^4\Delta x\,\ee^{\ii p\Delta x}\int\dd^4z\,\delta\left (x-z\right )M^2\left (z\right )\ii G^{\lessgtr}\left (z,y\right )\\
&=\ee^{-\ii \Diamond}\{M^2\left (X\right )\}\{\ii \tilde G^{\lessgtr}\left (X,p\right )\}\,,
\label{eq:WignerMass}
\end{split}
\end{equation}
\begin{equation}
\begin{split}
\mathcal{F}_W\left [\left (\ii \partial\!\!\!/-M_\psi\left (x\right )\right ) \ii D^{\lessgtr}\left (x,y\right )\right ]&=\int\dd^4\Delta x\,\ee^{\ii p\Delta x}\left (\ii \frac{1}{2}\partial\!\!\!/_X+\ii \partial\!\!\!/_{\Delta x}-M_\psi\left (x\right )\right ) \ii D^{\lessgtr}\left (x,y\right )\\
&=\left (\ii \frac{1}{2}\partial\!\!\!/_X+p\!\!\!/\right )\ii \tilde D^{\lessgtr}\left (X,p\right )-\ee^{-\ii \Diamond}\{M_\psi\left (X\right )\}\{\ii \tilde D^{\lessgtr}\left (X,p\right )\}\,.
\label{eq:WignerOperators2}
\end{split}
\end{equation}
The Wigner transform of the right-hand side follows from a straightforward
calculation by extending the time contour from $-\infty$ to $+\infty$
and taking into account the final time by a theta function, leading to
retarded and advanced expressions inside the memory integrals:
\begin{equation}
\begin{split}
\left (p^2+\ii p\partial_X-\frac{1}{4}\partial_X^2\right )\ii \tilde G^{\lessgtr}\left (X,p\right )-\ee^{-\ii \Diamond}\{M^2\left (X\right )+\tilde\Pi^{\mathrm{ret}}\}\{\ii \tilde G^{\lessgtr}\}&=\ee^{-\ii \Diamond}\{\tilde\Pi^{\lessgtr}\}\{\ii \tilde G^{\mathrm{adv}}\}\,,\\
\left (p\!\!\!/+\ii \frac{1}{2}\partial\!\!\!/_X\right )\ii \tilde D^{\lessgtr}\left (X,p\right )-\ee^{-\ii \Diamond}\{M_\psi\left (X\right )+\tilde\Sigma^{\mathrm{ret}}\}\{\ii \tilde D^{\lessgtr}\left (X,p\right )\}&=\ee^{-\ii \Diamond}\{\tilde\Sigma^{\lessgtr}\}\{\ii \tilde D^{\mathrm{adv}}\}\,.
\label{eq:WignerKadanoffBaym}
\end{split}
\end{equation}
To first order in gradient expansion, it follows
\begin{equation}
\begin{split}
&\left (p^2+\ii p\partial_X-M^2-\tilde\Pi^{\mathrm{ret}}\right )\ii \tilde G^{\lessgtr}+\ii \Diamond\{M^2+\tilde\Pi^{\mathrm{ret}}\}\{\ii \tilde G^{\lessgtr}\}=\tilde\Pi^{\lessgtr}\ii \tilde G^{\mathrm{adv}}-\ii \Diamond\{\tilde\Pi^{\lessgtr}\}\{\ii \tilde G^{\mathrm{adv}}\}\,,\\
&\left (p\!\!\!/+\ii \frac{1}{2}\partial\!\!\!/_X-M_\psi-\tilde\Sigma^{\mathrm{ret}}\right )\ii \tilde D^{\lessgtr}+\ii \Diamond\{M_\psi+\tilde\Sigma^{\mathrm{ret}}\}\{\ii \tilde D^{\lessgtr}\}
=\tilde\Sigma^{\lessgtr}\ii \tilde D^{\mathrm{adv}}-\ii \Diamond\{\tilde\Sigma^{\lessgtr}\}\{\ii \tilde D^{\mathrm{adv}}\}
\label{eq:FirstOrderGradientExpansion}
\end{split}
\end{equation}
Since real and imaginary parts are independent, it is convenient to
separate them, by exploiting the property of
$\ii \tilde G^{\lessgtr},\,\ii \tilde\Pi^{\lessgtr}$ being real
functions in Wigner space. Furthermore, one should decompose the
retarded and advanced propagators as well as corresponding self-energies
with respect to the real and imaginary parts in Wigner space. For bosons
one obtains following relations\footnote{Corresponding expressions for
  fermions can be introduced analogously, keeping only in mind, that the
  real and imaginary parts are considered with respect to each Lorentz
  component.}:
\begin{equation}
\begin{split}
\tilde G^{\text{ret,adv}}&=\operatorname{Re}\{\tilde G^{\mathrm{ret}}\}\pm \ii\operatorname{Im}\{\tilde G^{\mathrm{ret}}\}=:\operatorname{Re}\{\tilde G^{\mathrm{ret}}\}\mp \ii \tilde A/2\,,\\
\tilde\Pi^{\text{ret,adv}}&=\operatorname{Re}\{\tilde\Pi^{\mathrm{ret}}\}\pm \ii\operatorname{Im}\{\tilde\Pi^{\mathrm{ret}}\}=:\operatorname{Re}\{\tilde\Pi^{\mathrm{ret}}\}\mp \ii \tilde\Gamma/2\,.
\label{eq:RetSelfEnegyDecompose}
\end{split}
\end{equation}
For instance, the Wigner transform of the retarded self-energy is given by:
\begin{equation}
\begin{split}
\tilde\Pi^{\mathrm{ret}}\left (X,k\right )&=\int\dd^4\Delta x\,\ee^{\ii k\Delta x}\Theta\left (x^0-y^0\right )\left (\Pi^>\left (x,y\right )-\Pi^<\left (x,y\right )\right )\\
&=\int\frac{\dd k^{0'}}{2\pi}\frac{\ii \tilde\Pi^>\left (X,k'\right )-\ii \tilde\Pi^<\left (X,k'\right )}{k^0-k^{0'}+\ii \epsilon}\,.\\
\label{eq:RetSelfEnergy}
\end{split}
\end{equation}
Decomposing the retarded self-energy with respect to the real and imaginary parts, leads to:
\begin{equation}
\begin{split}
\operatorname{Im}\left [\tilde\Pi^{\mathrm{ret}}\left (X,k\right )\right ]&=\int\frac{\dd k^{0'}}{2\pi}\frac{-\epsilon}{\left (k^0-k^{0'}\right )^2+\epsilon^2}\left (\ii \tilde\Pi^>\left (X,k'\right )-\ii \tilde\Pi^<\left (X,k'\right )\right )\\
&\quad\underset{\epsilon\rightarrow 0}{\longrightarrow}-\frac{1}{2}\left (\ii \tilde\Pi^>\left (X,k\right )-\ii \tilde\Pi^<\left (X,k\right )\right )\,,\\
\operatorname{Re}\left [\tilde\Pi^{\mathrm{ret}}\left (X,k\right )\right ]&=\int\frac{\dd k^{0'}}{2\pi}\frac{k^0-k^{0'}}{\left (k^0-k^{0'}\right )^2+\epsilon^2}\left (\ii \tilde\Pi^>\left (X,k'\right )-\ii \tilde\Pi^<\left (X,k'\right )\right )\\
&\quad\underset{\epsilon\rightarrow 0}{\longrightarrow}\mathcal{P}\int\frac{\dd k^{0'}}{2\pi}\frac{-2\operatorname{Im}\left [\tilde\Pi^{\mathrm{ret}}\left (X,k'\right )\right ]}{k^0-k^{0'}}
=\mathcal{P}\int\frac{\dd k^{0'}}{2\pi}\frac{\tilde\Gamma}{k^0-k^{0'}}\,,\\
\label{eq:ReImRetSelfEnergy}
\end{split}
\end{equation}
where the last relation is known as the Kramers–Kronig relation, which
defines an analytic connection between the real and imaginary part.

Inserting the decomposition \eqref{eq:RetSelfEnegyDecompose} into
\eqref{eq:FirstOrderGradientExpansion} leads then to a set of two
equations for every particle species. In detail, one obtains from the
imaginary parts the so-called generalized transport equations (see also
Ref.~\cite{Juchem:2004cs} for scalar field theory),

\begin{equation}
\begin{split}
2p_\mu\partial_X^\mu \ii \tilde G^{\lessgtr}+2\Diamond\{M^2+\operatorname{Re}\{\tilde\Pi^{\mathrm{ret}}\}\}\{\ii \tilde G^{\lessgtr}\}
+2\Diamond\{\ii \tilde\Pi^{\lessgtr}\}\{\operatorname{Re}\{\tilde G^{\mathrm{ret}}\}\}&=\ii \tilde\Pi^<\ii \tilde G^>-\ii \tilde\Pi^>\ii \tilde G^<\,,\\
\partial\!\!\!/_X \ii \tilde D^{\lessgtr}+2\Diamond\{M_\psi+\operatorname{Re}\{\tilde\Sigma^{\mathrm{ret}}\}\}\{\ii \tilde D^{\lessgtr}\}
+2\Diamond\{\ii \tilde\Sigma^{\lessgtr}\}\{\operatorname{Re}\{\tilde D^{\mathrm{ret}}\}\}&=\ii \tilde\Sigma^<\ii \tilde D^>-\ii \tilde\Sigma^>\ii \tilde D^<\,,\\
\label{eq:ImaginaryPartFirstOrderGradienExpansion}
\end{split}
\end{equation}
whereas the real parts lead to the generalized mass-shell equations,
\begin{equation}
\begin{split}
\left (p^2-M^2-\operatorname{Re}\{\tilde\Pi^{\mathrm{ret}}\}\right )\ii \tilde G^{\lessgtr}&=\ii \tilde\Pi^{\lessgtr}\operatorname{Re}\{\tilde G^{\mathrm{ret}}\}
+\frac{1}{2}\Diamond\{\ii \tilde\Pi^{<}\}\{\ii \tilde G^>\}-\frac{1}{2}\{\ii \tilde\Pi^>\}\{\ii \tilde G^<\}\,,\\
\left (p\!\!\!/-M_\psi-\operatorname{Re}\{\tilde\Sigma^{\mathrm{ret}}\}\right )\ii \tilde D^{\lessgtr}&=\ii \tilde\Sigma^{\lessgtr}\operatorname{Re}\{\tilde D^{\mathrm{ret}}\}
+\frac{1}{2}\Diamond\{\ii \tilde\Sigma^{<}\}\{\ii \tilde D^>\}-\frac{1}{2}\{\ii \tilde\Sigma^>\}\{\ii \tilde D^<\}\,.\\
\label{eq:RealPartFirstOrderGradientExpansion}
\end{split}
\end{equation}
In Wigner space the propagator structure becomes more obvious by writing
$\ii \tilde{G}^{\lessgtr}$ as a product of the spectral function
$\tilde A\left (X,p\right )$ and the generalized quasi-particle
distribution function $\tilde{N}^{\lessgtr}\left (X,p\right )$. This
separation method is known as the Kadanoff-Baym ansatz, which
generalizes the KMS condition for propagators in equilibrium to systems
out of equilibrium. With the identity
$\ii \tilde G^<\left (X,-p\right )=\ii \tilde G^>\left (X,p\right )$ and
in accordance with the definition of the spectral function, one obtains
the following expressions for bosonic propagators:
\begin{equation}
\begin{split}
\ii \tilde{G}^{>}\left (X,p\right )&=\tilde{A}\left (X,p\right )\tilde{N}^{>}\left (X,p\right )\,,\quad\,\,
\tilde{N}^{>}\left (X,p\right )=
\begin{cases}
 1+f\left (X,\vec p\right )\,,\qquad\,\,\,\,\, p^0=E_p\\
 -f\left (X,-\vec p\right )\,,\,\,\qquad\,\,\,\, p^0=-E_p\,,
\end{cases}\\
\ii \tilde{G}^{<}\left (X,p\right )&=\tilde{A}\left (X,p\right )\tilde{N}^{<}\left (X,p\right )\,,\quad\,\,
\tilde{N}^{<}\left (X,p\right )=
\begin{cases}
f\left (X,\vec p\right )\,,\qquad\qquad\,\,\,\, p^0=E_p\\
-\left (1+f\left (X,-\vec p\right )\right )\,,\,\,\, p^0=-E_p\,,
\end{cases}\\
\end{split}
\label{eq:WignerPropagatorBoson}
\end{equation}
where $f\left (X,\vec p\right )$ denotes the usual one-particle distribution function.
For fermionic Wightman propagators, it follows: 
\begin{equation}
\begin{split}
\ii \tilde{D}^{>}\left (X,p\right )&=\tilde{A}_\psi\left (X,p\right )\tilde{N}^{>}_\psi\left (X,p\right )\,,\quad\,
\tilde{N}_\psi^{>}\left (X,p\right )=
\begin{cases}
 1-f_\psi\left (X,\vec p\right )\,,\quad\,\,\,\,\, p^0=E_p,\\
 f_{\bar\psi}\left (X,-\vec p\right )\,,\qquad\,\,\,\,\, p^0=-E_p,
\end{cases}\\
\ii \tilde{D}^{<}\left (X,p\right )&=-\tilde{A}_\psi\left (X,p\right )\tilde{N}^{<}_\psi\left (X,p\right )\,,\,\,
\tilde{N}_\psi^{<}\left (X,p\right )=
\begin{cases}
f_\psi\left (X,\vec p\right )\,,\qquad\quad\,\,\,\, p^0=E_p,\\
\left (1-f_{\bar\psi}\left (X,-\vec p\right )\right )\,,\,\,\, p^0=-E_p.
\end{cases}\\
\end{split}
\label{eq:WignerPropagatorFermion}
\end{equation}
The spectral functions $\tilde A,\,\tilde A_\psi$ with corresponding
widths $\tilde\Gamma,\,\tilde\Gamma_\psi$ follow directly from the
first-order gradient expansion of the Kadanoff-Baym equations
\eqref{eq:KadanoffBaymEquations} for retarded and advanced propagators
(see also \cite{Juchem:2004cs,Juchem:2003bi}). In detail, one obtains
for bosons the well-known Breit-Wigner form (analogously for fermions):
\begin{equation}
\tilde{A}=\frac{\tilde{\Gamma}}{\tilde{\Omega}^2+\tilde\Gamma^2/4}\,,\qquad
\tilde\Omega^2\left (X,p\right )=p^2-\tilde{M}^2\left (X\right )-\text{Re}\left (\tilde\Pi^{\mathrm{ret}}\left (X,p\right )\right )\,.
\label{eq:eeq02}
\end{equation}
Since the limit of the Lorentz sequence results in a $\delta$-function,
\begin{equation}
\delta_\epsilon\left (x\right )=\frac{1}{\pi}\frac{\epsilon}{x^2+\epsilon^2}\xrightarrow[\epsilon\rightarrow 0]{}\delta\left (x\right ),
\label{eq:eeq03}
\end{equation}
vanishing widths $\tilde\Gamma,\,\tilde\Gamma_\psi$ of quasi-particles
lead to a simplified form for the spectral functions, known as on-shell
approximation,
\begin{equation}
\begin{split}
\tilde{A}&=2\pi\sign\left (p^0\right )\delta\left (\tilde\Omega^2\right )=\frac{\pi}{E\left (X,p\right )}\left (\delta\left (p^0-E\left (X,p\right )\right )-\delta\left (p^0+E\left (X,p\right )\right )\right )\,,\\
\tilde{A}_\psi&=2\pi\sign\left (p^0\right )\delta\left (\tilde\Omega^2_\psi\right )\left (p\!\!\!/+M_\psi\right )\\
&=\left (p\!\!\!/+M_\psi\right )\frac{\pi}{E\left (X,p\right )}\left (\delta\left (p^0-E\left (X,p\right )\right )-\delta\left (p^0+E\left (X,p\right )\right )\right )\,.
\label{eq:SpectralFunctions}
\end{split}
\end{equation}
Now, it is straightforward to derive generalized transport equations of
Boltzmann type by making use of the decompositions
\eqref{eq:WignerPropagatorBoson}, \eqref{eq:WignerPropagatorFermion} and
their on-shell interpretations with
\eqref{eq:SpectralFunctions}. Thereby, in the limit of vanishing
quasi-particle widths $\tilde\Gamma,\,\tilde\Gamma_\psi$ the
contribution from the real part of the retarded self-energy can be
neglected as can be seen from the Kramers-Kronig relation
\eqref{eq:ReImRetSelfEnergy}.  This approximation scheme will also be
discussed in the next section. Consequently, the dispersion relation
obtains its usual
form $E=\sqrt{\vec{p}^2+\tilde{M}^2\left (X\right )}$.

In comparison to the generalized mass-shell equations
\eqref{eq:RealPartFirstOrderGradientExpansion}, the generalized
transport equations \eqref{eq:ImaginaryPartFirstOrderGradienExpansion}
have a rather natural form for transport equations, with a drift term, a
Vlasov term from the local part of the self-energy and a mass shift due
to the real part of the retarded self-energy. Furthermore, the right
side of the equation contains a typical gain and loss structure. The
second diamond operator on the left-hand side does not explicitly depend on
the quasi-particle distribution functions and thus purely describes the
off-shell evolution of the system. This part is of no relevance for
Boltzmann like evolutions equations as pointed out in
\cite{Juchem:2003bi} and it is indeed sufficient to solve
\eqref{eq:ImaginaryPartFirstOrderGradienExpansion} instead of
\eqref{eq:RealPartFirstOrderGradientExpansion}, since the
solutions differ only by contributions of higher order in gradients.

\subsection{Thermodynamic properties}
\label{sec:EffMass}

Before proceeding with non-equilibrium studies, it is convenient to
discuss some thermodynamic properties of the effective potential, which
is directly related to the effective action:
\begin{equation}
\begin{split}
\Omega_{\mathrm{\text{eff}}}\left [\sigma,\vec\pi,G,D\right ]=-\frac{1}{\beta V}i\Gamma\left [\sigma,\vec\pi,G,D\right ]\,.
\label{eq:EffPotential2PI}
\end{split}
\end{equation}
In the following, we firstly focus on the effective bosonic mass,
arising from the Hartree approximation of the $\Gamma_2$-part and will
then discuss an extension to include quark contributions to the
effective mass of bosons. Diagrammatically, the Hartree approximation is
given by the first line in Fig.~\ref{fig:2PIpart} and includes the most
relevant contributions to the effective potential of bosons by taking
into account all one-point diagrams from the full truncation of the
effective action $\Gamma\left [\sigma,\vec\pi,G,D\right ]$. In terms of
1PI diagrams the Hartree approximation sums the so-called daisy diagrams
for tree level propagators, which form a large and important class of
thermal diagrams. The full 2PI effective action for this approximation
reads

\begin{equation}
\begin{split}
\Gamma^{\mathrm{h}}\left [\sigma,\vec\pi,G,D\right ]=& S\left [\sigma,\vec\pi\right ]+\frac{\ii}{2}\Tr \ln G^{-1}+\frac{\ii}{2}\Tr \,G_{0}^{-1}G
-\ii \Tr \ln D^{-1}-\ii \Tr D_0^{-1}D\\
&-\frac{3}{4}\lambda\int_{\mathcal C}\dd^4x\,G_{\sigma\sigma}^2\left (x,x\right )-\frac{1}{2}\lambda\sum_i\int_{\mathcal C}\dd^4x\,G_{\sigma\sigma}\left (x,x\right )G_{\pi_i\pi_i}\left (x,x\right )\\\
&-\frac{3}{4}\lambda\sum_i\int_{\mathcal C}\dd^4x\,G_{\pi_i\pi_i}^2\left (x,x\right )-\frac{1}{2}\lambda\sum_{i,j\neq i}\int_{\mathcal C}\dd^4x\,G_{\pi_i\pi_i}\left (x,x\right )G_{\pi_j\pi_j}\left (x,x\right )\,.
\label{eq:HartreeEffAction}
\end{split}
\end{equation}

The restriction to one-point diagrams allows to calculate the full
expression of the effective potential by simply using the imaginary time
formalism with on-shell propagators as will be discussed in the
following.

Within the Hartree approximation the general form for fully dressed
propagators of the Schwinger-Dyson equation \eqref{eq:SchwingerDyson}
for bosons and fermions reduce to the rather simple expressions
\begin{equation}
\begin{split}
\ii G_{\varphi_a\varphi_a}^{-1}\left (x,y\right )&=\ii G_{0,\varphi_a\varphi_a}^{-1}\left (x,y\right )-\ii \Pi_{\varphi_a\varphi_a}\left (x,y\right )
=\left (-\partial_\mu\partial^\mu-M_{\varphi_a}^2\left (x\right )\right )\delta_{\mathcal{C}}\left (x-y\right )\,,\\
\ii D_{\psi_i\psi_i}^{-1}\left (x,y\right )&=\ii D_{0,\psi_i\psi_i}^{-1}\left (x,y\right )-i\Sigma_{\psi_i\psi_i}\left (x,y\right )=\left (\ii \partial\!\!\!/-M_{\psi_i}\left (x\right )\right )\delta_{\mathcal{C}}\left (x-y\right )\,,
\label{eq:HartreePropagator}
\end{split}
\end{equation}
containing only local self-energy contributions and leading to effective
mass terms (compare also with Eq.~\eqref{eq:EffMassGeneral}).  As
already discussed in Sec.~\ref{sec:RealTime} the equilibrium propagators
as two-point functions depend only on the relative space-time
difference, allowing to perform all relevant calculations in momentum
space. Furthermore, within the imaginary time formalism with
$k=(\omega_n,\vec k)=:k_n$
\begin{equation}
\begin{split}
\ii G_{\varphi_a\varphi_a}^{-1}\left (k\right )&=k^2-M_{\varphi_a}^2\quad\Rightarrow\quad G_{\varphi_a\varphi_a}\left (k\right )=\frac{i}{k_n^2-M_{\varphi_a}^2}\,,\\
\ii D_{\psi_i\psi_i}^{-1}\left (k\right )&=k\!\!\!/-M_{\psi_i}\quad\,\,\,\Rightarrow\quad D_{\psi_i\psi_i}\left (k\right )=\frac{i}{k\!\!\!/_n-M_{\psi_i}}\,,
\label{eq:HartreePropagatorMomentum}
\end{split}
\end{equation}
where the last step becomes possible as long as the mass terms are
positive, since the norm of the Euclidean momentum vector with the
metric convention $g^{\mu\nu}=\left (-1,-1,-1,-1\right )$ is always
spacelike $k_n^2\leq 0$ .  This calculation would be incorrect for
real-time propagators, where the Minkowski metric leads to the existence
of possible poles in the denominator, which have to be taken into
account, leading to a form with an explicit vacuum and a thermal part as
already known from the thermal propagators of a noninteracting field
theory (see relations \eqref{eq:ThermalPropagatorsBosons}
and \eqref{eq:ThermalPropagatorsFermions}).

In \ref{sec:ThermoPotentialImaginary}, we calculate the logarithmic
terms as well as the loop integrals for the effective thermodynamic
potential \eqref{eq:HartreeEffAction}.  With the following definition of
a loop integral,
\begin{equation}
Q\left (M_{\varphi_a}\right ):=\int\frac{\dd^3\vec p}{\left (2\pi\right )^3}\frac{1}{E_{\varphi_a}}f_{\varphi_a}\left (t,\vec x,\vec p\right )=\int\frac{\dd^3\vec p}{\left (2\pi\right )^3}\frac{1}{E_{\varphi_a}}\frac{1}{\ee^{\beta E_{\varphi_a}}-1}\,,
\label{eq:AbbreviationLoopInt}
\end{equation}
the trace relations at one-loop order in equilibrium lead to
\begin{equation}
\begin{split}
\frac{\ii}{2}\Tr \,G_{0}^{-1}G=&\,-\frac{i\beta V}{2}\sum_a\left [M_{\varphi_a}^2-\lambda\left (3\varphi_a^2+\sum_{b\neq a}\varphi_b^2-\nu^2\right )\right ]Q\left (M_{\varphi_a}\right )\,,\\
\ii \Tr D_0^{-1}D=&\,\text{const.}
\label{eq:TraceRelations}
\end{split}
\end{equation}
We are then able to give an explicit and rather lengthy result for the effective potential,
\begin{equation}
\begin{split}
\Omega_{\text{eff}}
=&\,\Omega_{\text{eff}}^{\mathrm{MF}}+\frac{1}{\beta}\int\frac{\dd^3\vec p}{\left (2\pi\right )^3}\ln\left (1-\ee^{-\beta E_\sigma}\right )+\sum_i\frac{1}{\beta}\int\frac{\dd^3\vec p}{\left (2\pi\right )^3}\ln\left (1-\ee^{-\beta E_{\pi_i}}\right )\\
&\,\,\qquad -\frac{1}{2}\left [M_\sigma^2-\lambda\left (3\sigma^2+\sum_i\pi_i^2-\nu^2\right )\right ]Q\left (M_\sigma\right )\\
&\,\,\qquad -\frac{1}{2}\sum_i\left [M_{\pi_i}^2-\lambda\left (\sigma^2+3\pi_i^2+\sum_{j\neq i}\pi_j^2-\nu^2\right )\right ]Q\left (M_{\pi_i}\right )\\
&\,\,\qquad +\frac{3}{4}\lambda\left [Q\left (M_\sigma\right )\right ]^2+\frac{3}{4}\lambda\sum_i\left [Q\left (M_{\pi_i}\right )\right ]^2 \\
&\,\,\qquad +\frac{1}{2}\lambda\sum_i\left [Q\left (M_\sigma\right )Q\left (M_{\pi_i}\right )\right ]
+\frac{1}{2}\lambda\sum_{i,j\neq i}\left [Q\left (M_{\pi_i}\right )Q\left (M_{\pi_j}\right )\right ]+R_{\text{eff}}\\
&=:\,\Omega_{\text{eff}}^{\mathrm{MF}}+\Omega_{\text{eff}}^{\varphi_a}+R_{\text{eff}}\,,
\label{eq:EffPotential}
\end{split}
\end{equation}
where $\Omega_{\text{eff}}^{\varphi_a}$ contains all mesonic
contributions and $\Omega^{\mathrm{MF}}_{\text{eff}}$ is already known
from Sec.~\ref{sec:MeanFieldDynamics} (see
Eq.~\eqref{eq:EffMeanFieldPotential}). Finally, divergent integral
contributions are absorbed into the term $R_{\text{eff}}$, which require
a proper renormalization scheme\footnote{Note that the divergent terms
  are not purely vacuum contributions, since the effective mass terms
  are determined self-consistently and depend on the medium properties
  via the one-particle distribution functions.} as discussed in
Refs.~\cite{Rischke:1998qy,0954-3899-26-4-309} for a purely mesonic
theory. However, we will simply skip those terms, since the
renormalization of the mass leads qualitatively to a rather small
modification in the vicinity of the phase transition as already known
for the mesonic part from Ref.~\cite{0954-3899-26-4-309}.

We also note that a proper renormalization-scheme in this full
off-equilibrium situation is very challenging. For the equilibrium case
it has been shown in
\cite{PhysRevD.65.025010,PhysRevD.65.105005,PhysRevD.66.025028,0954-3899-26-4-309,Blaizot:2003br}
that self-consistent renormalization schemes based on the 2PI/CJT action
formalism can be renormalized with vacuum counter terms, but that also
``hidden divergencies'' have to be properly subtracted involving a
self-consistent solution of the corresponding subdiagrams involving
overlapping divergences. For that reason it is also not sufficient to
simply use a cutoff-regularization since the divergent contributions to
the non-perturbative sub-divergencies cannot be controlled. As already
stated, we skip the terms $R_{\text{eff}}$ containing the divergent
contributions to define a feasible set of kinetic equations to study the
fluctuations of conserved charges in the medium undergoing a cross-over
or phase transition.

In comparison to the effective potential of the semi-classical
mean-field approach, which is simply given by the term
$\Omega_{\text{eff}}^{\mathrm{MF}}$, the full Hartree approximation
leads to a significant modification, extending the model by a large
number of resummed mesonic diagrams.

\begin{figure*}
\centering
\fbox{\includegraphics[width=0.7\linewidth]{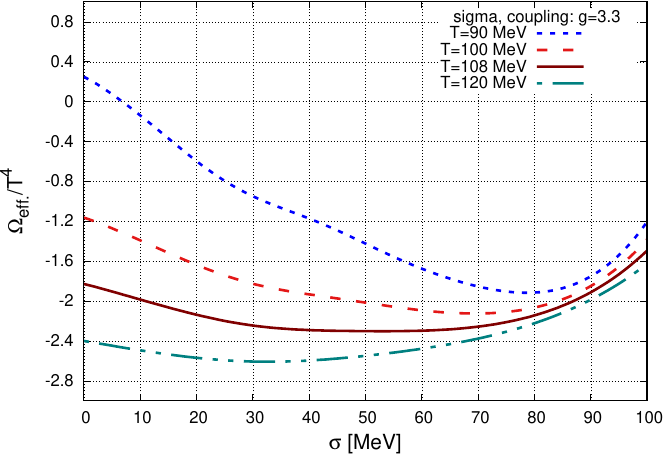}}\\
\fbox{\includegraphics[width=0.7\linewidth]{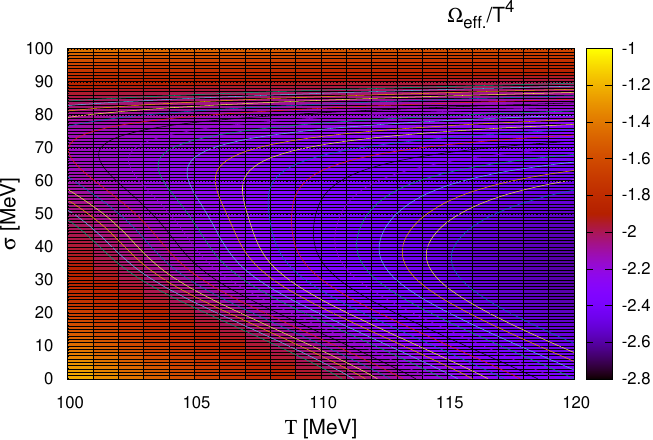}}
\caption{Normalized form of the effective potential from Hartree approximation \eqref{eq:EffPotential} for a fixed value of the quark chemical potential $\mu=157\,\MeV$.}
\label{fig:EffectivePotential}
\end{figure*}

The upper plot of Fig.~\ref{fig:EffectivePotential} shows a normalized
form of the effective potential $\Omega_{\text{eff}}/T^4$ from
Eq.~\eqref{eq:EffPotential} for several values of the temperature $T$
and a fixed quark chemical potential $\mu=157\,\MeV$. The effective
potential has a very flat form for $T=108\,\MeV$, being also supported
by the plot below, showing equipotential lines in the
$\sigma$-$T$-plane, which become almost parallel to the $\sigma$-axis in
the range of $T=108-110\,\MeV$. In the following we consider also the
derivatives of the effective potential to confirm more precisely the
position of the critical point.

Now, we derive explicit expression for the stationary value of the chiral field $\left <\Phi\right >=\left (\left <\sigma\right >,\left <\vec\pi\right >\right )$ as well as effective mass terms 
of sigma and pions, resulting from partial derivatives of the thermodynamic potential,
\begin{equation}
\begin{split}
\frac{\partial\Omega_{\text{eff}}}{\partial\sigma}&=\frac{\partial\Omega_{\text{eff}}^{\mathrm{MF}}}{\partial\sigma}+\frac{\partial\Omega_{\text{eff}}^{\varphi_a}}{\partial\sigma}\stackrel{!}{=}0\\
&=\lambda\left [\left (\sigma^2+\sum_i\pi_i^2-\nu^2\right )+3Q\left (M_\sigma\right )+\sum_iQ\left (M_{\pi_i}\right )\right ]\sigma-f_\pi m_\pi^2+g\left <\bar\psi\psi\right >\\
\frac{\partial\Omega_{\text{eff}}}{\partial\pi_i}&=\frac{\partial\Omega_{\text{eff}}^{MF}}{\partial\pi_i}+\frac{\partial\Omega_{\text{eff}}^{\varphi_a}}{\partial\pi_i}\stackrel{!}{=}0\\
&=\lambda\left [\left (\sigma^2+\sum_i\pi_i^2-\nu^2\right )+Q\left (M_\sigma\right )+3Q\left (M_{\pi_i}\right )+\sum_{j\neq i}Q\left (M_{\pi_j}\right )\right ]\pi_i+g\left <\bar\psi \ii \gamma_5\tau_i\psi\right >
\label{eq:ThermoPotenialDerivative}
\end{split}
\end{equation}

\begin{equation}
\begin{aligned}
\frac{\partial^2\Omega_{\text{eff}}}{\partial\sigma^2}=&\,\lambda\left [\left(3\sigma^2+\sum_i\pi_i^2-\nu^2\right)+3Q\left (M_\sigma\right )+\sum_i Q\left (M_{\pi_i}\right )\right ]\\
&\quad+\frac{\partial}{\partial\sigma}g\left <\bar\psi\psi\right >\Bigg|_{\sigma=\left <\sigma\right >}=: M_\sigma^2\,,\\
\frac{\partial^2\Omega_{\text{eff}}}{\partial\pi_i^2}=&\,\lambda\left [\left(\sigma^2+3\pi_i^2+\sum_{j\neq i}\pi_j^2-\nu^2\right)
+3Q\left (M_{\pi_i}\right )+\sum_{j\neq i}Q\left (M_{\pi_j}\right )+Q\left (M_{\sigma}\right )\right ]\\
&\quad+\frac{\partial}{\partial\pi_i}g\left<\bar\psi \ii \gamma_5\tau_i\psi\right \rangle\Bigg|_{\pi_i=\left <\pi_i\right >}=:M_{\pi_i}^2\,,
\end{aligned}
\label{eqn:ThermoPotentialMass}
\end{equation}
with mass terms being evaluated at the stationary value of the chiral
field
$\left <\Phi\right >=\left (\left <\sigma\right >,\left <\vec\pi\right
  >\right )$.  We note that with vanishing scalar- and pseudoscalar
densities the mass relations of Eq.~\eqref{eqn:ThermoPotentialMass}
reduce to purely bosonic expressions, being equal to the effective mass
terms from the propagator gap equation of Hartree approximation without
quarks (compare also with Ref.~\cite{0954-3899-26-4-309}). However,
additional contributions to the propagator from the fermionic sunset
diagrams are crucial to reproduce a phase diagram as expected from a
quark-meson model, with the order of the phase transition depending on
the chemical potential of quarks\footnote{Out of equilibrium the phase
  transition is governed by the effective quark number.}. An inclusion
of two-loop diagrams from Fig.~\ref{fig:2PIpart} makes it necessary to
consider the effective mass as a momentum dependent expression, since
also the real part of the retarded self-energy has to be taken into
account (compare with Eq.~\eqref{eq:eeq02}), leading to a modification
for the propagator gap equations.  In numerical calculations we checked
for the zero mode, that a contribution from the retarded part of the
fermionic sunset diagram to the effective mass in the sense of
Kramers-Kronig relation fails to reproduce a phase diagram with a second
order phase transition. Consequently, the first gradient expansion with
the on-shell ansatz for propagators is not sufficient for our studies
and we decided to use effectively derived mass terms of
Eq.~\eqref{eqn:ThermoPotentialMass}.  However, neglecting all two-loop
diagrams from $\Gamma_2$ in Fig.~\ref{fig:2PIpart} is a reasonable
approximation for the thermodynamic potential as long as the system's
effective particle density is small for higher order processes to
contribute significantly to the real part of the retarded
self-energy.\footnote{Such an argument does not hold for the vacuum part
  of the self-energy and one has to keep this limitation in mind.}

For systems out of equilibrium, we simply rewrite the effective mass
terms of Eq.~\eqref{eqn:ThermoPotentialMass} by replacing the thermal
loop integrals with the general expressions, which are directly
calculated in the real-time formalism (compare therefore
Eq.~\eqref{eq:ThermoPotenialDerivative} with exact evolution equations
for the chiral field \eqref{eq:MeanFieldEquations}).  Since the loop
integrals are local in space and time, the Wigner transform depends
effectively only on the variable $X$ and the momentum dependence can be
integrated out, leading to the following loop integrals for bosons,
\begin{equation}
\begin{split}
\ii G^{\text{c}}_{\varphi_a\varphi_a}\left (x,x\right )&=\frac{1}{2}\left (\ii G^>_{\varphi_a\varphi_a}\left(x,x\right)+\ii G^<_{\varphi_a\varphi_a}\left(x,x\right)\right )=
\frac{1}{2}\int\frac{\dd^4p}{\left (2\pi\right )^4}\tilde F_{\varphi_a\varphi_a}\left (X,p\right )\\ 
&=\frac{1}{2}\int\frac{\dd^4p}{\left (2\pi\right )^4}\tilde A_{\varphi_a}\left (X,p\right )\left (N^>_{\varphi_a}\left (X,p\right )+N_{\varphi_a}^<\left (X,p\right )\right )\\
&=\frac{1}{2}\int\frac{d^3p}{\left(2\pi\right)^3}\frac{1+2f^{\varphi_a}\left(t,\vec x,\vec p\right)}{E^{\varphi_a}_{\vec p}}
=:\int\frac{d^3p}{\left(2\pi\right)^3}\frac{f^{\varphi_a}\left(t,\vec x,\vec p\right)}{E^{\varphi_a}_{\vec p}}+R_{\varphi_a}\left (x\right )
\label{eq:LoopBosonic}
\end{split}
\end{equation}
and fermions,
\begin{equation}
\begin{split}
\Tr \left (\ii D^{\text{c}}\left (x,x\right )\right )&=\frac{1}{2}\Tr \left (\ii D^{>}\left (x,x\right )+\ii D^{<}\left (x,x\right )\right )=\frac{1}{2}\int\frac{\dd^4p}{\left (2\pi\right )^4}\Tr \left (\tilde F_{\psi\psi}\left (X,p\right )\right )\\
&=\frac{1}{2}\int\frac{\dd^4p}{\left (2\pi\right )^4}\Tr \left (\tilde{A}_\psi\left (X,p\right )\left (\tilde{N}^{>}_\psi\left (X,p\right )-\tilde{N}^{<}_\psi\left (X,p\right )\right )\right )\\
&=-d_\psi g\sigma\int\frac{\dd^3\vec p}{\left (2\pi\right )^3}\frac{1}{E_{\vec p}^\psi}\left [f^\psi\left (t,\vec x,\vec p\right )+f^{\bar\psi}\left (t,\vec x,\vec p\right )-1\right ]\\
&=:-\left <\bar\psi\psi\right >+R_{\sigma,\psi}\left (x\right )\,,\\
\Tr \left (\ii \gamma_5\tau_j \ii D^{\text{c}}\left (x,x\right )\right )&=:-\left <\bar\psi \ii \gamma_5\tau_j\psi\right >+R_{\pi_j,\psi}\left (x\right )
\label{eq:LoopFermionic}
\end{split}
\end{equation}
where the traces are evaluated with respect to Dirac and Lorentz indices
as well as Pauli indices in isospin-space,
$\Tr \left (p\!\!\!/+M_\psi^\dagger\right )=4g\sigma$ and
$\Tr \left (\ii \gamma_5\tau_j\left (p\!\!\!/+M_\psi^\dagger\right
  )\right )=4g\pi_j$.  The scalar and pseudoscalar densities
$\left <\bar\psi\psi\right >$,
$\left <\bar\psi \ii \gamma_5\tau_j\psi\right >$ are equal to the
definitions given in Eq.~\eqref{eqn:ScalarPseudoScalarDensity}.
Finally, $R_{\varphi_a,\psi}\left (x\right )$ absorbs divergent
contributions, requiring once more a proper renormalization due to the
mass dependence of the loop integral. With a similar argument as before
we will simply skip these terms in the following.  Note that a
calculation of logarithmic terms in Eq.~\eqref{eq:HartreeEffAction}
within the real-time formalism is more technical and not required for
the dynamical evolution equations, since they are not explicitly present
in the evolution equations as seen from \eqref{eq:MeanFieldEquations}
and \eqref{eq:KadanoffBaymEquations}.

The self-consistent set of equations \eqref{eq:ThermoPotenialDerivative}
and \eqref{eqn:ThermoPotentialMass} is solved for equilibrium
one-particle distribution functions by using for instance an improved
version of Newton's algorithm for finding minima (respectively roots) of
multidimensional functions. In case of a problematic initial input
vector
$\left (\sigma_0,\,\vec\pi_0,\,m_{\sigma,0}^2,\,m_{\pi,0}^2\right )$ for
Newton's algorithm an additional stochastic distortion to the initial
guess can be applied.

\begin{figure*}
\centering
\subfloat{\includegraphics[width=7cm]{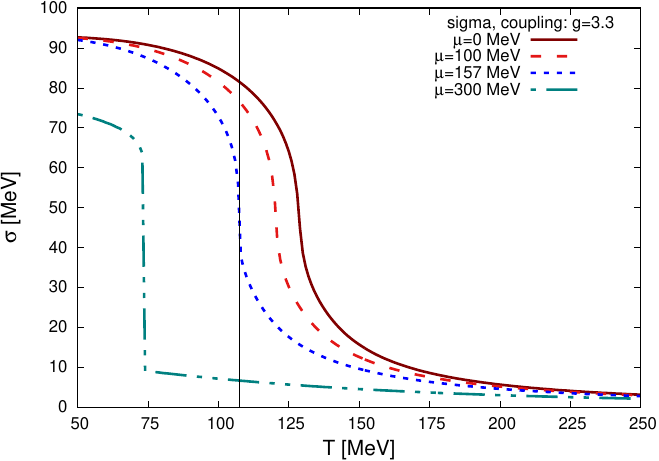}}
\subfloat{\includegraphics[width=7cm]{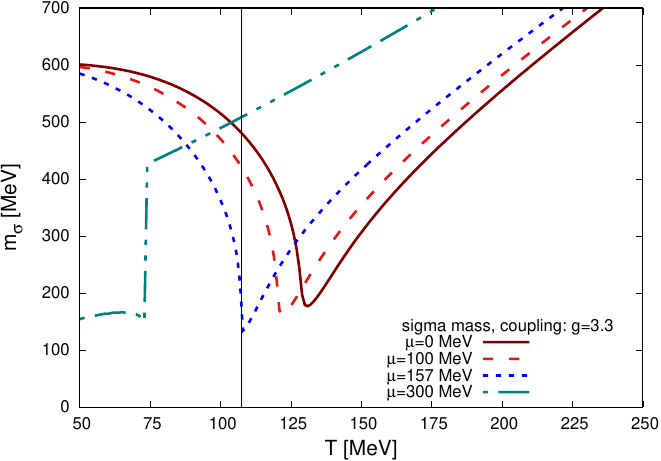}}\\
\caption{Equilibrium phase diagram of the linear sigma model from self-consistent solving of Eqs.~\eqref{eq:ThermoPotenialDerivative} and \eqref{eqn:ThermoPotentialMass} 
for the coupling constant $g=3.3$. Showing the order 
parameter $\sigma$ and the effective mass $m_\sigma$ as a function of the temperature $T$ and several values of the quark chemical potential $\mu$.}
\label{fig:PhaseDiagramFull}
\end{figure*}

\begin{figure*}
\centering
\fbox{\subfloat{\includegraphics[width=7cm]{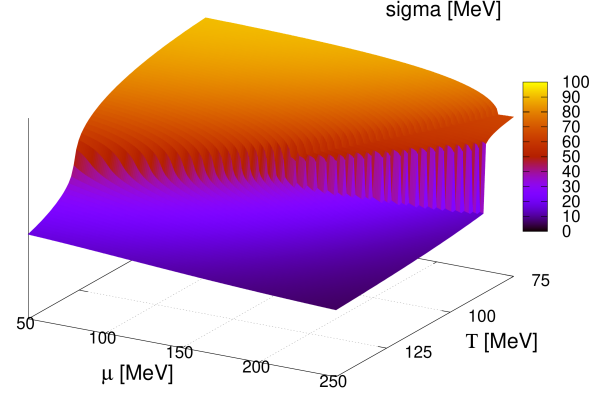}}
\subfloat{\includegraphics[width=7cm]{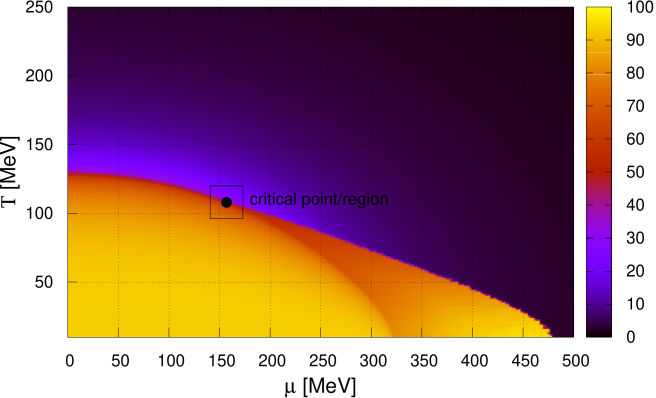}}}\\
\fbox{\subfloat{\includegraphics[width=7cm]{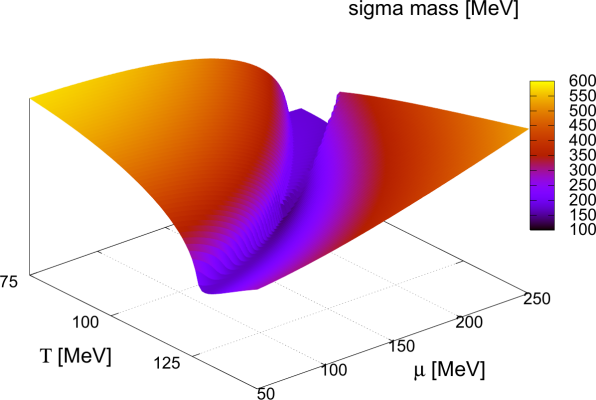}}
\subfloat{\includegraphics[width=7cm]{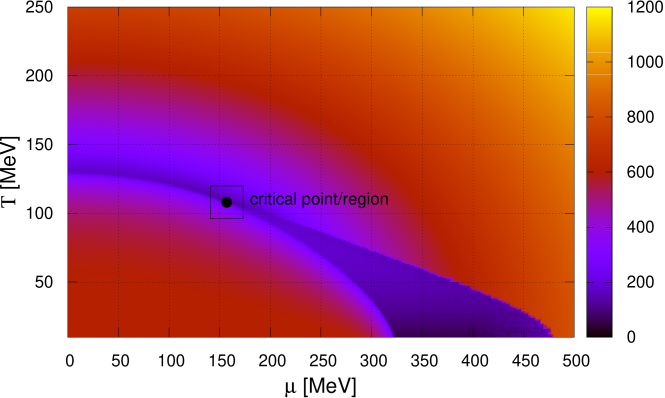}}}\\
\fbox{\subfloat{\includegraphics[width=7cm]{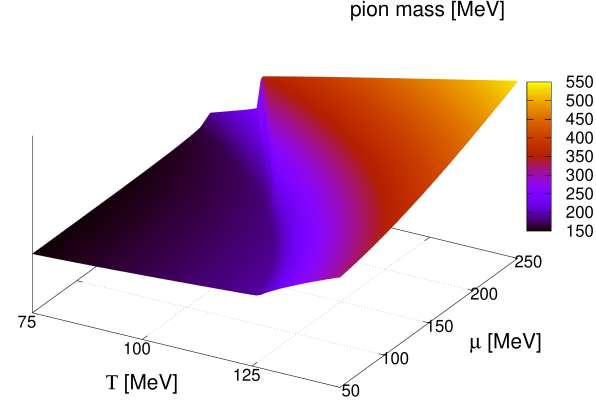}}
\subfloat{\includegraphics[width=7cm]{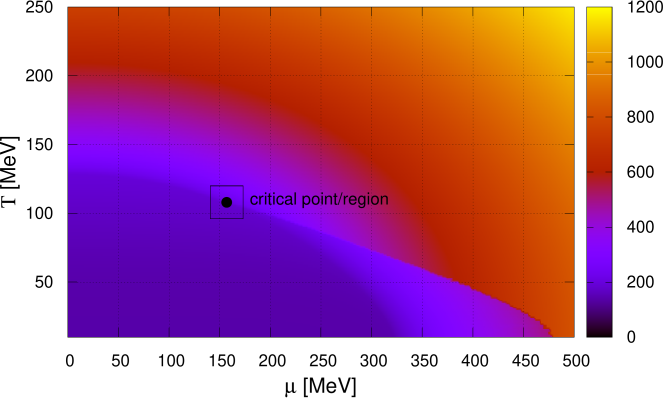}}}\\
\centering
\caption{Equilibrium phase diagram of the linear sigma model from
  self-consistent solving of Eqs.~\eqref{eq:ThermoPotenialDerivative}
  and \eqref{eqn:ThermoPotentialMass} for the coupling constant
  $g=3.3$. The first line shows the order parameter $\sigma$ as a
  function of the temperature $T$ and quark chemical potential $\mu$
  (3-dimensional and 2-dimensional) Analougsly, the second and third
  lines show similar information for the sigma and pion mass.}
\label{fig:PhaseDiagram3d}
\end{figure*}

Fig.~\ref{fig:PhaseDiagramFull} and Fig.~\ref{fig:PhaseDiagram3d} show
the numerical solution for the order parameter
\eqref{eq:ThermoPotenialDerivative} as well as effective mass terms
\eqref{eqn:ThermoPotentialMass} with respect to the temperature and
several values of the quark chemical potential $\mu$. Effectively, such
a behavior of $\sigma$ and $m_\sigma$ is already known from
Sec.~\ref{sec:MeanFieldDynamics}, where the variation of the coupling
constant $g$ between the chiral field and quarks led to different orders
of the phase transition at zero chemical potential (see
Fig.~\ref{fig:OrderParameterAndMass}).


In detail, below $\mu\approx 157\,\MeV$ we find the system in a
crossover like state with a smooth structure of the order parameter,
which changes for large values of $\mu$ to a first-order phase
transition with a gap at the correspondig value of the critical
temperature. Following that, the phase transition of the first-order has
to end up in a critical point of second order with decreasing $\mu$,
which is identified in the region of
$\left (T\approx 108\,\MeV, \; \mu\approx 157\,\MeV\right )$. Here, it has
to be pointed out that the required property of a vanishing sigma mass
is not perfectly fulfilled at the critical point, being a consequence of
explicit symmetry breaking, renormalization as well as truncation
effects of the 2PI action as discussed before, requiring a more detailed
study of those contributions in an upcoming work. Note, that the
position of the critical point is not a fixed and universal value of the
model, more precisely it depends strongly on the value of the coupling
constant $g$ to quarks. Therefore, we adjusted the coupling constant in
such a way, that the crossover temperature at zero quark chemical
potential is still comparable with QCD. Furthermore, the quark chemical
potential stays in a range, being still accessible to heavy-ion
experiments\footnote{Indeed, it cannot be excluded that the value of the
  chemical potential for the critical point in QCD, if it exists, is
  significantly below $T\simeq 100\,\MeV$. In this case it would be
  hardly possible to find a signal for the phase transition in
  experimental data.} and beyond the region, which is already excluded
by experiments and lattice QCD calculations (see
Sec.~\ref{chap:introduction}).

\subsection{Dissipation term}

Now, we are prepared to derive a dissipation term, which is consistent
with the on-shell approximation of Sec.~\ref{sec:GenBoltzmannEquation},
but will partially recover the memory properties of the mean-field and
Kadanoff-Baym equations. Doing so, we extend the usual formalism of the
inhomogeneous Klein-Gordon as well as Boltzmann-Uehling-Uhlenbeck equation to
non-local interactions in time.

As discussed before, for the study of the chiral phase transition we
focus on systems with a non-vanishing mean-field for the first component
$\phi:=\sigma$ of the chiral field $\Phi=\left (\sigma,\vec\pi\right )$
and $\vec\pi$ being zero.  Thereby, the leading
contribution\footnote{Since the coupling constant is large at the phase
  transition, the relevant parameter is given by the number of involved
  fields.} to the dissipative part of the mean-field equation follows
from the bosonic sunset diagrams:
\begin{equation}
\begin{split}
\Gamma_2^{\text{b.s.}}=\,&3 \ii \lambda^2\int_{\mathcal
  C}\dd^4x\int_{\mathcal C}\dd^4y\,\phi\left (x\right
)\,G_{\sigma\sigma}^3\left (x,y\right )\,\phi\left (y\right )\\
& +\ii \lambda^2\sum_{i}\int_{\mathcal C}\dd^4x\int_{\mathcal
  C}\dd^4y\,\phi\left (x\right )\,G_{\sigma\sigma}\left (x,y\right
)G_{\pi_i\pi_i}^2\left (x,y\right )\,\phi\left (y\right )\,. 
\label{eq:Damping}
\end{split}
\end{equation}
As shown in the following, this formal expression can be rewritten in
terms of a convolution, describing the past-dependent interaction
between the mean-field and a memory kernel, which is generated by
non-zero modes.

\subsection{On-shell approximation of the dissipation term}
\label{sec:DissipationTerm}

After taking the functional derivative of \eqref{eq:Damping} with
respect to the mean-field evaluated on the upper branch of the real-time
contour and splitting the contour integrals, we obtain
\begin{equation}
\begin{split}
\frac{\delta \Gamma_2^{\text{b.s.}}}{\delta\phi^1\left (z\right )}=&3\ii \lambda^2\int\dd^4y\,\left (G_{\sigma\sigma}^{11}\left (z,y\right )\right )^3\phi^1\left (y\right )+3\ii \lambda^2\int\dd^4x\,\phi^1\left (x\right )\left (G_{\sigma\sigma}^{11}\left (x,z\right )\right )^3\\
&-3\ii \lambda^2\int\dd^4y\,\left (G_{\sigma\sigma}^{12}\left (z,y\right )\right )^3\phi^2\left (y\right )-3\ii \lambda^2\int\dd^4x\,\phi^2\left (x\right )\left (G_{\sigma\sigma}^{21}\left (x,z\right )\right )^3\\
&+\ii \lambda^2\sum_i\int\dd^4y\,G_{\sigma\sigma}^{11}\left (z,y\right )\left (G_{\pi_i\pi_i}^{11}\left (z,y\right )\right )^2\phi^1\left (y\right )\\
&+\ii \lambda^2\sum_i\int\dd^4x\,\phi^1\left (x\right )G_{\sigma\sigma}^{11}\left (x,z\right )\left (G_{\pi_i\pi_i}^{11}\left (x,z\right )\right )^2\\
&-\ii \lambda^2\sum_i\int\dd^4y\,G_{\sigma\sigma}^{12}\left (z,y\right )\left (G_{\pi_i\pi_i}^{12}\left (z,y\right )\right )^2\phi^2\left (y\right )\\
&-\ii \lambda^2\sum_i\int\dd^4x\,\phi^2\left (x\right )G_{\sigma\sigma}^{21}\left (x,z\right )\left (G_{\pi_i\pi_i}^{21}\left (x,z\right )\right )^2\,.\\
\label{eq:DerivPhi}
\end{split}
\end{equation}
A similar expression follows also for the lower branch. However, it is
sufficient to consider only one of the two branches for the mean-field
eqution as seen from the general form \eqref{eq:MeanFieldEquations}.
Applying the relations for the contour Green's functions
\eqref{eq:GreenFunc} to \eqref{eq:DerivPhi} and taking the physical
solution $\phi^1\stackrel{!}{=}\phi^2\stackrel{!}{=}\phi$, leads to:
\begin{equation}
\begin{split}
D\left (x\right ):=&\,-\frac{\delta \Gamma_2^{\text{b.s.}}}{\delta\phi^1\left (x\right )}\Bigg|_{\phi^1=\phi^2=\phi}=-\frac{\delta \Gamma_2^{\text{b.s.}}}{\delta\phi^2\left (x\right )}\Bigg|_{\phi^1=\phi^2=\phi}\\
=&\,-6\ii \lambda^2\int\dd^4y\,\phi\left (y\right )\left [\Theta\left (x^0-y^0\right )\left (\left (\ii G_{\sigma\sigma}^>\left (x,y\right )\right )^3-\left (\ii G_{\sigma\sigma}^<\left (x,y\right )\right )^3\right )\right ]\\
&\, -2\ii \lambda^2\sum_i\int\dd^4y\,\phi\left (y\right )\left [\Theta\left (x^0-y^0\right )\left (\ii G_{\sigma\sigma}^>\left (x,y\right )\left (\ii G_{\pi_i\pi_i}^>\left (x,y\right )\right )^2\right .\right .\\
&\qquad\qquad\qquad\qquad\qquad\qquad\qquad\quad\,\left .\left .-\ii G_{\sigma\sigma}^<\left (x,y\right )\left (\ii G_{\pi_i\pi_i}^<\left (x,y\right )\right )^2\right )\right ]\\
=&\,-6\ii \lambda^2\int_{y^0}^{x^0}\dd y^0\int\dd^3\vec y\,\phi\left (y\right )\left [\mathcal{M}_{\sigma\sigma}\left (x,y\right )+\frac{1}{3}\sum_i\mathcal{M}_{\sigma\pi_i}\left (x,y\right )\right ]\,,
\label{eq:SumDerivPhi}
\end{split}
\end{equation}
where we introduced two kernel functions of the form:
\begin{equation}
\begin{split}
\mathcal{M}_{\sigma\sigma}\left (x,y\right )&:=\left (\ii G_{\sigma\sigma}^>\left (x,y\right )\right )^3-\left (\ii G_{\sigma\sigma}^<\left (x,y\right )\right )^3\,,\\
\mathcal{M}_{\sigma\pi_i}\left (x,y\right )&:=\ii G_{\sigma\sigma}^>\left (x,y\right )\left (\ii G_{\pi_i\pi_i}^>\left (x,y\right )\right )^2-\ii G_{\sigma\sigma}^<\left (x,y\right )\left (\ii G_{\pi_i\pi_i}^<\left (x,y\right )\right )^2\,.
\label{eq:MemFunc}
\end{split}
\end{equation}
By considering the limit $y^0\rightarrow-\infty$ and making use of the
relation \eqref{eq:WignerTransform} we rewrite the term
\eqref{eq:SumDerivPhi} for the interaction between the mean-field and
hard modes:
\begin{equation}
\begin{split}
D\left (x\right )&=-6\ii \lambda^2\int_{-\infty}^{x^0}\dd y^0\int\dd^3\vec y\,\phi\left (y\right )
\mathcal{F}^{-1}_W\left [\mathcal{M}_{\sigma\sigma}\left (x,y\right )+\frac{1}{3}\sum_i\mathcal{M}_{\sigma\pi_i}\left (x,y\right )\right ]\\
&=-6\ii \lambda^2\int_{0}^{\infty}\dd\Delta t\int\frac{\dd^4k}{\left
    (2\pi\right )^4}\,\ee^{-\ii k^0\Delta t}\ee^{\ii \vec k\cdot\vec
  x}\phi\left (t-\Delta t,\vec k\right ) \left [\mathcal{\tilde M}_{\sigma\sigma}\left (X,k\right )+\frac{1}{3}\sum_i\mathcal{\tilde M}_{\sigma\pi_i}\left (X,k\right )\right ]\,,
\label{eq:DampingWigner2}
\end{split}
\end{equation}
where we introduced a more usual notation with $t:=x^0$, $t':=y^0$ and
$\Delta t:=t-t'$ in the last step, showing explicitly the dependence on
the past. Furthermore, we note that a Fourier transformation with
respect to $k^0$ allows us to interpret
$\mathcal{\tilde M}_{\sigma\sigma}$ and
$\mathcal{\tilde M}_{\sigma\pi_i}$ as memory
kernels, which are explicitly given by
Eqs.~\eqref{eq:WignerMemoryKernel1} and \eqref{eq:WignerMemoryKernel2}.

When the relation \eqref{eq:DampingWigner2} is approximated for slowly
changing kernel functions with respect to the averaged coordinate $X$,
the one-particle distribution functions
$f\left (t,\vec x,\vec p\right )$ in
Eqs.~\eqref{eq:WignerMemoryKernel1}, \eqref{eq:WignerMemoryKernel2} can
be evaluated at $X\simeq x$. However, one should keep in mind that such
an approximation will suffer from causality problems as discussed in
Ref.~\cite{PhysRevC.73.034909}. In comparison to the symmetric kernels
of non-zero modes, which are defined in \eqref{eq:b02h1b}, the kernels
$\mathcal{\tilde M}_{\sigma\sigma}$ and
$\mathcal{\tilde M}_{\sigma\pi_i}$ have qualitatively a similar
structure\footnote{This holds true for homogeneous as well as for
  inhomogeneous systems.}, but are antisymmetric with respect to $k^0$.

With relation \eqref{eq:DampingWigner2} the mean-field equation of the
order parameter $\sigma$ becomes (compare with
Eq.~\eqref{eq:MeanFieldEquations})
\begin{equation}
\partial_\mu\partial^\mu\phi+D\left (x\right )+\lambda\left(\phi^2-\nu^2+3G^{11}_{\sigma\sigma}+\sum_\ii G^{11}_{\pi_i\pi_i}\right)\phi-f_\pi m_\pi^2+g\left<\bar\psi\psi\right \rangle=0\,,
\label{eqn:MeanFieldEquation}
\end{equation}
where the terms $G^{11}_{\sigma\sigma}=\ii G^c_{\sigma\sigma}$ and
$G^{11}_{\pi_i\pi_i}=\ii G^c_{\pi_i\pi_i}$ describe local contributions
to the potential, arising from Hartree diagrams. An explicit expression
was given in the previous section along with the calculation of the
scalar density (see Eqs.~\eqref{eq:LoopBosonic} and
\eqref{eq:LoopFermionic}).

For a homogeneous system with initial time $t_0$ the dissipation term
reads
\begin{equation}
\begin{split}
D\left (t\right )&=-6\ii \lambda^2\int_{0}^{t-t_0}\dd\Delta t\,\phi\left (t-\Delta t\right )\int\frac{\dd k^0}{\left (2\pi\right )}\,\ee^{-\ii k^0\Delta t}\left [\mathcal{\tilde M}_{\sigma\sigma}\left (t,k^0\right )+\frac{1}{3}\sum_i\mathcal{\tilde M}_{\sigma\pi_i}\left (t,k^0\right )\right ]\,,
\label{eq:HomDissipationTerm}
\end{split}
\end{equation}

where the kernels
$\mathcal{\tilde M}_{\sigma\sigma}\left (t,k^0\right )$ and
$\mathcal{\tilde M}_{\sigma\pi_i}\left (t,k^0\right )$ follow directly
from the general expressions \eqref{eq:WignerMemoryKernel1},
\eqref{eq:WignerMemoryKernel2} with $\vec k=0$ and $X^0\simeq t$.  Since
the memory kernels are antisymmetric in $k^0$, it is convenient to
integrate \eqref{eq:HomDissipationTerm} by parts with respect to the
time interval $\Delta t$:

\begin{equation}
\begin{split}
D\left (t\right )&=6\lambda^2\phi\left (t-\Delta t\right )\int\frac{\dd k^0}{\left (2\pi\right )}\,\ee^{-\ii k^0\Delta t}\frac{1}{k^0}
\left [\mathcal{\tilde M}_{\sigma\sigma}\left (t,k^0\right )+\frac{1}{3}\sum_i\mathcal{\tilde M}_{\sigma\pi_i}\left (t,k^0\right )\right ]\Bigg |_0^{t-t_0}\\
&\,\quad-6\lambda^2\int_0^{t-t_0}\dd\Delta t\,\phi'\left (t-\Delta t\right )\int\frac{\dd k^0}{\left (2\pi\right )}\,\ee^{-\ii k^0\Delta t}
\frac{1}{k^0}\left [\mathcal{\tilde M}_{\sigma\sigma}\left (t,k^0\right )+\frac{1}{3}\sum_i\mathcal{\tilde M}_{\sigma\pi_i}\left (t,k^0\right )\right ]\\
&=6\lambda^2\left [\Gamma\left (t,\Delta t=t-t_0\right )\phi\left (t_0\right )-\Gamma\left (t,\Delta t=0\right )\phi\left (t\right )\right ]+6\lambda^2\int_{t_0}^{t}\dd t'\dot\phi\left (t'\right )\Gamma\left (t,t-t'\right )\,,
\label{eq:HomDissipationTermPartInt}
\end{split}
\end{equation}
where we introduced a symmetric friction kernel,
\begin{equation}
\Gamma\left (t,\Delta t\right ):=\int\frac{\dd k^0}{\left (2\pi\right )}\,\ee^{-\ii k^0\Delta t}
\frac{1}{k^0}\left [\mathcal{\tilde M}_{\sigma\sigma}\left (t,k^0\right )+\frac{1}{3}\sum_i\mathcal{\tilde M}_{\sigma\pi_i}\left (t,k^0\right )\right ]\,,
\label{eq:FrictionKernel}
\end{equation}
to obtain a Langevin-like form (see for example
\cite{Greiner:1998vd,PhysRevD.55.1026,PhysRevD.62.036012}).

The Fourier transformations for the friction kernel
\eqref{eq:FrictionKernel} and memory kernels \eqref{eq:MemoryKernels}
are numerically evaluated. The pertinent discrete Fourier transform from
the energy to time domain introduces an ultraviolet cut-off with respect
to the energy $k^0$, corresponding to a maximum value of the frequency
for the mean-field. In general a sharp cut-off leads to oscillations as
well as numerical noise for the Fourier transform. The noise can be
significantly reduced by multiplying the memory kernels with a filtering
function $S\left (k^0\right )$ in the energy domain, continiously
suppressing unwanted contributions from high frequencies.  In this paper
the parametrization for the filtering function is of the form of a Fermi
distribution, generating a smooth version of a step function:
\begin{equation}
  S\left (k^0\right ):=\left (\exp\left (\frac{k^0 - \kappa\mu_\phi}{\sigma_\phi}\right )+1\right )^{-1}\,,
\end{equation}
where the product $\kappa\mu_{\phi}$ defines a typical energy scale
beyond which the frequencies are supressed. The third parameter
$\sigma_\phi$ denotes then a transition width of the Fermi-like
function. In numerical calculations of this paper the parameters are
set to $\kappa=4$, $\mu_\phi:=604\,\MeV$ (standing for the vacuum sigma
mass as the relevant estimator of the upper bound for the sigma mass at
temperatures below $T\approx 200\,\MeV$, compare with
Fig.~\ref{fig:PhaseDiagramFull}) and $\sigma_\phi=\mu_\phi/2$.
\label{test}

\begin{figure*}
\centering\includegraphics[width=15.0cm]{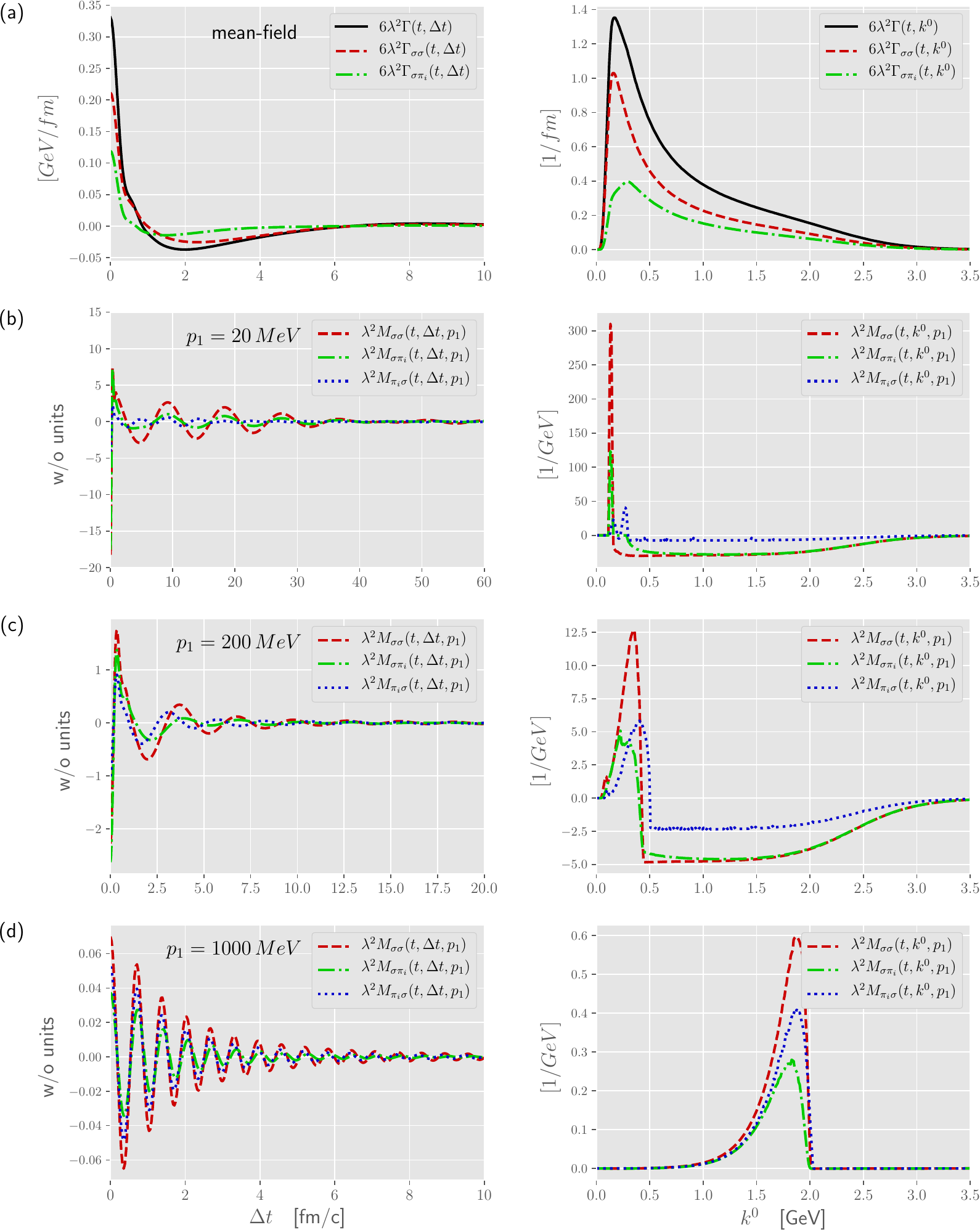}
\caption{Memory kernels near the critical point ($T\approx 105\,\MeV$,
  $\mu=165\,\MeV$) for the mean-field as well as three explicit momentum
  modes in time and energy domains.}
\label{fig:MemoryKernels}
\end{figure*}

Illustratively, Fig.~\ref{fig:MemoryKernels} shows the memory kernels of
the mean field and three explicit momentum modes, thereby the
temperature $T\approx 105\,\MeV$ as well as the chemical potential
$\mu=165\,\MeV$ of the system are chosen to be close to the critical
point as can also be seen from the position of the peak in the energy
domain, moving towards the infrared for decreasing momentum values.

We note, that an effective relaxation time $\tau$ for the mean-field can
be estimated from the memory kernel by considering the temporal average
over the relevant time scale $\xi$ with
$\Gamma(t,\Delta t\ge\xi)\approx 0$, leading to
$\tau \sim \left [\xi\left <\Gamma\left (t,\Delta t\right )\right
  >_{\Delta t <\xi}\right ]^{-1}$.  As seen from
Fig.~\ref{fig:MemoryKernels}, the effective relaxation time becomes
large near the critical point, emphasizing the requirement for a
non-Markovian description in the critical region.

\subsection{Linear harmonic approximation of the dissipation term}

The physics behind the memory kernels \eqref{eq:WignerMemoryKernel1} and
\eqref{eq:WignerMemoryKernel2} becomes more obvious when evaluating the
integral \eqref{eq:DampingWigner2} in the so-called linear harmonic
approximation\footnote{The linear harmonic approximation can be
  understood from considering the differential equation of a simple
  harmonic oscillator in a field theory.},
\begin{equation}
\phi\left (t-\Delta t,\vec k\right )=\phi\left(t,\vec k\right )\cos\left (E_k\Delta t\right )-\partial_t\phi\left (t,\vec k\right )\frac{1}{E_k}\sin\left (E_k\Delta t\right )\,,
\label{eq:lhapprox}
\end{equation}
which contains much more information about the past in comparison to an
instantaneous ansatz for the dynamics of the mean-field as discussed in
Ref.~\cite{PhysRevD.55.1026}.  Inserting the expression
\eqref{eq:lhapprox} into the dissipation term \eqref{eq:DampingWigner2}
requires the evaluation of the following relations:
\begin{equation}
\begin{split}
C\left (k^0,E_k\right ):=&\exp\left (-\ii k^0\Delta t\right )\cos\left (E_k\Delta t\right )\\
=&\frac{1}{2}\left [\exp\left (\ii \left (E_k-k^0\right )\Delta t\right )+\exp\left (-\ii \left (E_k+k^0\right )\Delta t\right )\right ]\,,\\
S\left (k^0,E_k\right ):=&\exp\left (-\ii k^0\Delta t\right )\sin\left (E_k\Delta t\right )\\
=&\frac{1}{2\ii}\left [\exp\left (\ii \left (E_k-k^0\right )\Delta t\right )-\exp\left (-\ii \left (E_k+k^0\right )\Delta t\right )\right ]\,.
\end{split}
\end{equation}
The Fourier transform of a constant function in $\mathds{R}^+$ leads to
\begin{equation}
\int_0^\infty\Delta t\,\ee^{\ii \left (E_k-k^0\right )\Delta t}=\ii\mathcal{P}\frac{1}{E_k-k^0}+\pi\delta\left (E_k-k^0\right )
\label{eq:hftransform}
\end{equation}
with $\mathcal{P}$ denoting the Cauchy principal value. Integrating out the $\Delta t$-dependence on the past, results then in the following expression:
\begin{equation}
\begin{split}
\mathcal{I}_{\Delta t}:=&\int_0^\infty\dd\Delta t\,\ee^{-\ii k^0\Delta t}\phi\left (t-\Delta t,\vec k\right )\\
=&\int_0^\infty\dd\Delta t\left [C\left (k^0,E_k\right )\phi\left (t,\vec k\right )-S\left (k^0,E_k\right )\frac{\partial_t\phi\left (t,\vec k\right )}{E_k}\right ]\\
=&\Bigg\{\frac{1}{2}\phi\left (t,\vec k\right )\left [\ii \mathcal{P}\frac{1}{E_k-k^0}+\pi\delta\left (E_k-k^0\right )-\ii \mathcal{P}\frac{1}{E_k+k^0}+\pi\delta\left (E_k+k^0\right )\right ]\\
&\,\,\,-\frac{1}{2\ii}\frac{\partial_t\phi\left (t,\vec k\right )}{E_k}\left [\ii \mathcal{P}\frac{1}{E_k-k^0}+\pi\delta\left (E_k-k^0\right )+\ii \mathcal{P}\frac{1}{E_k+k^0}-\pi\delta\left (E_k+k^0\right )\right ]\Bigg\}\,.
\label{eq:dkargument}
\end{split}
\end{equation}
With this abbreviation, the general dissipation term
\eqref{eq:DampingWigner2} can be simplified to
\begin{equation}
D(x)\simeq -6\ii \lambda^2\int\frac{\dd^4k}{\left (2\pi\right )^4}\,\ee^{\ii \vec k\cdot\vec x}
\left [\mathcal{\tilde M}_{\sigma\sigma}\left (x,k\right )+\frac{1}{3}\sum_i\mathcal{\tilde M}_{\sigma\pi_i}\left (x,k\right )\right ]\mathcal{I}_{\Delta t}\,.
\label{eq:LinHarmDiss}
\end{equation}
From the antisymmetric property of the memory kernels
\eqref{eq:WignerMemoryKernel1}, \eqref{eq:WignerMemoryKernel2} with
respect to $k^0$,
\begin{equation}
\begin{split}
\mathcal{\tilde M}_{\sigma\sigma}\left (x,k^0,\vec k\right )&=-\mathcal{\tilde M}_{\sigma\sigma}\left (x,-k^0,\vec k\right )\,,\\
\mathcal{\tilde M}_{\sigma\pi_i}\left (x,k^0,\vec k\right )&=-\mathcal{\tilde M}_{\sigma\pi_i}\left (x,-k^0,\vec k\right )\,,
\end{split}
\label{eq:apmkernel}
\end{equation}
it follows that the on-shell contributions from the $\delta$ functions
vanish for the $\phi$-part of Eq.\ \eqref{eq:dkargument}. The same
statement holds for the principal values of the $\partial_t\phi$-part,
and one ends up with
\begin{equation}
\begin{split}
D\left (x\right )&\simeq 6\lambda^2\int\frac{\dd k^4}{\left (2\pi\right )^4}\,\ee^{\ii \vec k\cdot\vec x}\left [\mathcal{\tilde M}_{\sigma\sigma}\left (x,k\right )+\frac{1}{3}\sum_i\mathcal{\tilde M}_{\sigma\pi_i}\left (x,k\right )\right ]
\phi\left (t,\vec k\right )\mathcal{P}\frac{1}{E_k-k^0}\\
&\quad\,+6\lambda^2\int\frac{\dd k^3}{\left (2\pi\right )^3}\,\ee^{\ii \vec k\cdot\vec x}\left [\mathcal{\tilde M}_{\sigma\sigma}\left (x,E_k,\vec k\right )+\frac{1}{3}\sum_i\mathcal{\tilde M}_{\sigma\pi_i}\left (x,E_k,\vec k\right )\right ]
\frac{\partial_t\phi\left (t,\vec k\right )}{2E_k}\,.
\label{eq:dkareduceda}
\end{split}
\end{equation}
For illustration, we explicitly write down the non-Markovian approximation for
a homogeneous system,
\begin{equation}
\begin{split}
D\left (t\right )&\simeq 6\lambda^2\phi\left (t\right )\int\frac{\dd k^0}{\left (2\pi\right )}\left [\mathcal{\tilde M}_{\sigma\sigma}\left (t,k^0\right )+\frac{1}{3}\sum_i\mathcal{\tilde M}_{\sigma\pi_i}\left (t,k^0\right )\right ]
\mathcal{P}\frac{1}{E_k-k^0}\\
&\quad\,+6\lambda^2\dot\phi\left (t\right )\frac{1}{2E_k}\left [\mathcal{\tilde M}_{\sigma\sigma}\left (t,E_k\right )+\frac{1}{3}\sum_i\mathcal{\tilde M}_{\sigma\pi_i}\left (t,E_k\right )\right ]\,.
\label{eq:dkareducedb}
\end{split}
\end{equation}
Thereby, the term with the principal value is an effective mass
contribution for the mean-field equation and would require a proper
renormaliztion scheme.  However, we expect the resulting shift for the
mass spectrum to be small and neglect it in numerical calculations.

\subsection{Final set of evolution equations}
\label{eq:FinalSetEvoEq}

For homogeneous systems the full set of evolution equations consists of
the mean-field equation for the order parameter $\phi:=\sigma$,
\begin{equation}
\begin{split}
&\partial_t^2\phi+D\left (t\right )+J\left (t\right )=0\,,\\
&J\left (t\right ):=\lambda\left(\phi^2-\nu^2+3G^{11}_{\sigma\sigma}+\sum_i G^{11}_{\pi_i\pi_i}\right)\phi-f_\pi m_\pi^2+g\left<\bar\psi\psi\right \rangle
\label{eq:HomMeanFieldEquation}
\end{split}
\end{equation}
with $D$ given by the full non-Markovian Eq.\
(\ref{eq:HomDissipationTermPartInt}) with the divergent
mass-contribution (first term) omitted, and of the following evolution
equations of Boltzmann type for mesons as well as quarks:
\begin{equation}
\begin{split}
\partial_t f^\sigma\left (t,\vec p\right )&=\mathcal{I}_\sigma^{b.}\left (t,\vec p_1\right )+\mathcal{I}_\sigma^{b.s.}\left (t,\vec p_1\right )+\mathcal{I}_\sigma^{f.s.}\left (t,\vec p_1\right )\\
&=\mathcal{C}_{\sigma\sigma\leftrightarrow\sigma\sigma}^{b.}+\sum_i\mathcal{C}_{\sigma\pi_i\leftrightarrow\sigma\pi_i}^{b.}+\sum_i\mathcal{C}_{\sigma\sigma\leftrightarrow\pi_i\pi_i}^{b.}\\
&\quad +\mathcal{C}_{\sigma\phi\leftrightarrow\sigma\sigma}^{b.s.}+\sum_i\mathcal{C}_{\sigma\phi\leftrightarrow\pi_i\pi_i}^{b.s.}+\mathcal{C}_{\sigma\leftrightarrow\psi\bar\psi}^{f.s.}\,,\\
\partial_t f^{\pi_i}\left (t,\vec p\right )&=\mathcal{I}_{\pi_i}^{b.}\left (t,\vec p_1\right )+\mathcal{I}_{\pi_i}^{b.s.}\left (t,\vec p_1\right )+\mathcal{I}_{\pi_i}^{f.s.}\left (t,\vec p_1\right )\\
&=\mathcal{C}_{\pi_i\pi_i\leftrightarrow\pi_i\pi_i}^{b.}+\sum_{j\neq i}\mathcal{C}_{\pi_i\pi_j\leftrightarrow\pi_i\pi_j}^{b.}+\sum_{j\neq i}\mathcal{C}_{\pi_i\pi_i\leftrightarrow\pi_j\pi_j}^{b.}
+\mathcal{C}_{\pi_i\sigma\leftrightarrow\pi_i\sigma}^{b.}+\mathcal{C}_{\pi_i\pi_i\leftrightarrow\sigma\sigma}^{b.}\\
&\quad +\mathcal{C}_{\pi_i\phi\leftrightarrow\pi_i\sigma}^{b.s.}+\mathcal{C}_{\pi_i\leftrightarrow\psi\bar\psi}^{f.s.}\\
\partial_t f^\psi\left (t,\vec p_1\right )&=\mathcal{I}_\psi^{f.s.}\left (t,\vec p_1\right )=\mathcal{C}_{\psi\bar\psi\leftrightarrow\sigma}^{f.s.}+\sum_i\mathcal{C}_{\psi\bar\psi\leftrightarrow\pi_i}^{f.s.}\\
\partial_t f^{\bar\psi}\left (t,\vec p_1\right )&=\mathcal{I}_{\bar\psi}^{f.s.}\left (t,\vec p_1\right )=\mathcal{C}_{\bar\psi\psi\leftrightarrow\sigma}^{f.s.}+\sum_i\mathcal{C}_{\bar\psi\psi\leftrightarrow\pi_i}^{f.s.}\,,
\label{eq:HomBoltzmannEquation}
\end{split}
\end{equation}
where Tab.~\ref{tab:MesonicCollIntegrals} shows the diagramatic
interpretation of the collision integrals. Their explicit expressions
are decomposed with respect to the involved processes in
\ref{chap:app2}, leading to Eqs.\ \eqref{eq:b02h1b},
\eqref{eq:BoltzmannCollSigma},
\eqref{eq:BoltzmannCollSigmaContributions},
\eqref{eq:BoltzmannCollPionContributions}.

The non-Markovian mean-field equation \eqref{eq:HomMeanFieldEquation} is
an ordinary integro-differential equation, which can be solved by
setting a well-defined initial condition $\phi_0$ for the mean-field. In
this case the effective mass terms follow then directly from solving the
self-consistent relations \eqref{eqn:ThermoPotentialMass} for general
one-particle distribution functions, where the loop integrals are
evaluated in accordance with \eqref{eq:AbbreviationLoopInt},
\eqref{eq:LoopBosonic} and \eqref{eq:LoopFermionic}.  This procedure
requires for instance an impoved version of Newton's algorithm for
finding minima (respectively roots) of a multidimensional function.

\begin{table}
\centering
\begin{tabular}{|p{3.0cm}|p{2.0cm}|||p{3.0cm}|p{2.0cm}|}
\hline
collision integral & diagram & collision integral & diagram\\
\hline
\hline
$\mathcal{C}_{\sigma\sigma\leftrightarrow\sigma\sigma}^{b.}$  
&
\begin{minipage}{2cm}
\includegraphics[width=1.5cm]{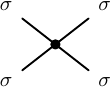}
\end{minipage}
&
$\mathcal{C}_{\pi_i\pi_i\leftrightarrow\pi_i\pi_i}^{b.}$  
&
\begin{minipage}{2cm}
\includegraphics[width=1.5cm]{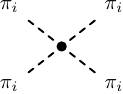}
\end{minipage}\\
\hline
$\mathcal{C}_{\sigma\pi_i\leftrightarrow\sigma\pi_i}^{b.}$ 
& 
\begin{minipage}{2cm}
\includegraphics[width=1.5cm]{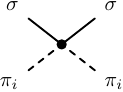}
\end{minipage}
&
$\mathcal{C}_{\pi_i\pi_j\leftrightarrow\pi_i\pi_j}^{b.}$ 
& 
\begin{minipage}{2cm}
\includegraphics[width=1.5cm]{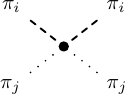}
\end{minipage}\\
\hline
$\mathcal{C}_{\sigma\sigma\leftrightarrow\pi_i\pi_i}^{b.}$ &
\begin{minipage}{2cm}
\includegraphics[width=1.5cm]{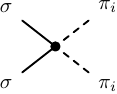}
\end{minipage}
&
$\mathcal{C}_{\pi_i\sigma\leftrightarrow\pi_i\sigma}^{b.}$ 
& 
\begin{minipage}{2cm}
\includegraphics[width=1.5cm]{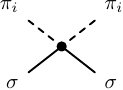}
\end{minipage}\\
\hline
$\mathcal{C}_{\sigma\phi\leftrightarrow\sigma\sigma}^{b.s.}$ & 
\begin{minipage}{2cm}
\includegraphics[width=1.5cm]{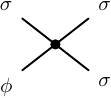}
\end{minipage}
&
$\mathcal{C}_{\pi_i\pi_i\leftrightarrow\pi_j\pi_j}^{b.}$ 
&
\begin{minipage}{2cm}
\includegraphics[width=1.5cm]{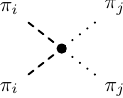}
\end{minipage}\\
\hline
$\mathcal{C}_{\sigma\phi\leftrightarrow\pi_i\pi_i}^{b.s.}$ 
& 
\begin{minipage}{2cm}
\includegraphics[width=1.5cm]{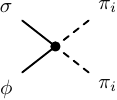}
\end{minipage}
&
$\mathcal{C}_{\pi_i\pi_i\leftrightarrow\sigma\sigma}^{b.}$ & 
\begin{minipage}{2cm}
\includegraphics[width=1.5cm]{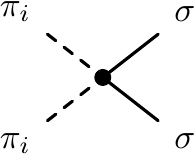}
\end{minipage}\\
\hline
$\mathcal{C}_{\sigma\leftrightarrow\psi\bar\psi}^{f.s.}$ 
& 
\begin{minipage}{2cm}
\includegraphics[width=1.5cm]{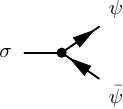}
\end{minipage}
&
$\mathcal{C}_{\pi_i\phi\leftrightarrow\pi_i\sigma}^{b.s.}$ 
& 
\begin{minipage}{2cm}
\includegraphics[width=1.5cm]{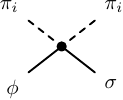}
\end{minipage}\\
\hline
$\mathcal{C}_{\psi\bar\psi\leftrightarrow\sigma}^{f.s.}$ 
& 
\begin{minipage}{2cm}
\includegraphics[width=1.5cm]{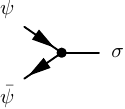}
\end{minipage}
&
$\mathcal{C}_{\pi_i\leftrightarrow\psi\bar\psi}^{f.s.}$ & 
\begin{minipage}{2cm}
\includegraphics[width=1.5cm]{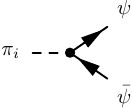}
\end{minipage}\\
\hline
$\mathcal{C}_{\bar\psi\psi\leftrightarrow\sigma}^{f.s.}$ & 
\begin{minipage}{2cm}
\includegraphics[width=1.5cm]{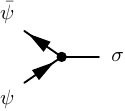}
\end{minipage}
&
$\mathcal{C}_{\psi\bar\psi\leftrightarrow\pi_i}^{f.s.}$ 
& 
\begin{minipage}{2cm}
\includegraphics[width=1.5cm]{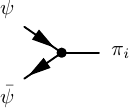}
\end{minipage}\\
\hline
& &
$\mathcal{C}_{\bar\psi\psi\leftrightarrow\pi_i}^{f.s.} $ 
& 
\begin{minipage}{2cm}
\includegraphics[width=1.5cm]{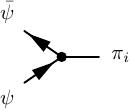}
\end{minipage}\\
\hline
\end{tabular}
\caption{Collision integrals and their diagramatic interpretation for
  the set of evolution equations in \eqref{eq:HomBoltzmannEquation}.}
\label{tab:MesonicCollIntegrals}
\end{table}

For our study in the following, the evolutions equations
\eqref{eq:HomMeanFieldEquation} can be easily rewritten in an isotropic
form by making use of the methods described in
\ref{sec:IsoCollInt1} for bosons and \ref{sec:IsoCollInt2} for the
interaction between bosons and fermions.

\section{The chiral phase transition out of equilibrium}
\label{chap:DynamicalEvol}

The cumulants of conserved quantities are the most promising observables
for finding an experimental evidence for the existence of a critical
point of the chiral phase transition in QCD. As discussed in the
introductory Sec.~\ref{chap:introduction} the cumulant ratios in
equilibrium have been studied using various effective models as well as
within models derived directly from first principles of QCD
\cite{WETTERICH199390,doi:10.1142/S0217751X94000972,BERGES2002223,PhysRevD.66.074507,PhysRevD.95.054504}.
However, the study of the chiral phase transition out of equilibrium
with a full spatial dependence is restricted to simplified models,
assuming local thermal equilibrium and neglecting for example bosonic,
memory and non-Markovian effects
\cite{Herold:2013bi,Nahrgang:2016eou,Herold:2016uvv,Nahrgang:2018afz,Wesp2018,Kitazawa:2020kvc}. Here,
the full set of self-consistent evolution equations of
Sec.~\ref{eq:FinalSetEvoEq}\footnote{Note, that the referred set of
  spatially independent evolution equations can be easily rewritten to a
  form for non-homogeneous systems.} does not assume local thermal
equilibrium and has further advantages by taking also bosonic and memory
effects on the level of one-particle distribution functions into
account, circumventing thus the impact of stochastic effects as well as
the problematic aspect of a finite sample of test particles, which will
be discussed in the Sec.~\ref{sec:InitialConditions}. In its most
general form, the numerical solution for the full set of self-consistent
evolution equations is computationally highly expensive even for a
single simulation run. A study of higher-order cumulants requires many
independent runs, resulting in a very large amount of required memory as
well as computational time.

Thus, in this paper the focus lies on the dynamical evolution of
momentum-dependent cumulants of the initial net-quark density in
homogeneous and isotropic systems, allowing to reduce the complexity of
the evolution equations of Sec.~\ref{eq:FinalSetEvoEq} to a feasible
level without using further approximations. 

Therefore, in the next Sec.~\ref{sec:ExpansionInflation} the derivation
procedure is reformulated for the evolution equations of an expanding homogeneous and
isotropic system by starting from a Friedmann–Lema{\^i}tre–Robertson–Walker
metric in a spatially flat geometry. The result is then applied to the
set of equations of Sec.~\eqref{eq:FinalSetEvoEq}, making it suitable
for the study of the dynamical evolution of the net-quark number in
expanding geometry, which mimics an expanding spherical domain in a
heavy ion collision with a decreasing net-quark density during the
expansion process. We also describe the numerical implementation in
detail and show that in our approach the net-quark number is conserved
to a high level of accuracy within $\sim 0.5-1.5\,\%$ even for rapidly
expanding systems with $v\leq 0.8c$. 

In the subsequent Sec.\ we discuss the initial conditions for the
expanding geometry. Being interested in the cumulants of the net-quark
number, we introduce two types of statistical ensembles for the initial
configurations with a fluctuating net-quark number, which are studied
throughout the last section. Thereby, both approaches are based on
initial fluctuations of Gaussian type for the total net-quark number,
differing in volume and net-quark density fluctuations.

Following the initialization methods, we discuss a crucial advantage of
our approach on the level of one-particle distribution functions in
comparison to test-particle approaches by showing that a
pseudo-stochastic initialization procedure within our framework
significantly outperforms naive Monte Carlo methods, which are generally
used for test particles.

The final Sec.\ then shows a broad study of averaged ensemble quantities
like the averaged quark chemical potential, the temperature, the order
parameter and so forth as a function of the evolving (proper)
time. Moreover, we then discuss a possible signature for the
experimental confirmation of the chiral phase transition in a heavy-ion
collision, by analyzing our results for the cumulant ratios, the
rescaled kurtosis, in various momentum ranges, which depend strongly on
the trajectories of the statistical ensembles for the initial
configuration when passing through the chiral phase transition in the
comover region, or passing close to the critcal point, or passing
through a full first-order phase transition regime.

\subsection{Evolution equations in expanding geometry}
\label{sec:ExpansionInflation}

For deriving evolution equations of an expanding homogeneous and
isotropic system it is convenient to start from the
Friedmann–Lema{\^i}tre–Robertson–Walker (FLRW) metric in Cartesian
coordinates and vanishing curvature (spatially flat geometry) as
described for instance in
\cite{PhysRevLett.116.022301,Bazow_Denicol_Heinz_Martinez_Noronha_2016,book:Bernstein}:
\begin{equation}
\begin{split}
\dd s^2=\dd t^2-a^2\left (t\right )\left (\dd x_1^2+\dd x_2^2+\dd
x_3^2\right )\,\,\,\Rightarrow\,\,\, g_{00}=1,\, \,g_{i0}=g_{0i}=0\,,\,g_{ij}=-a^2\left (t\right )\delta_{ij}\,,
\label{eq:FLRWMetric}
\end{split}
\end{equation}
where $a\left (t\right )$ denotes a time-dependent expansion scale as
known from cosmological models of the Universe.  The Christoffel symbols
of the metric \eqref{eq:FLRWMetric} can be easily calculated as
\begin{equation}
\begin{split}
\Gamma^\mu_{\nu\sigma}&=\frac{1}{2}g^{\mu\lambda}\left (\partial_\nu g_{\sigma\lambda}+\partial_\sigma g_{\nu\lambda}-\partial_\lambda g_{\nu\sigma}\right )
\,\,\,\Rightarrow\,\,\,\Gamma^0_{ij}=a\dot a\delta_{ij}\,,\,\Gamma^i_{0j}=\Gamma^i_{j0}=\frac{\dot a}{a}\delta^i_j\,,\,\Gamma^i_{jk}=0\,.
\label{eq:ChristoffelSymbols}
\end{split}
\end{equation}
One can now rewrite the evolution equations for one-particle distribution
functions \eqref{eq:HomBoltzmannEquation} by taking the expansion via
the metric into account. Therefore, we explicitly calculate the general
expression of the Liouville operator, defined by the left-hand side of
the Boltzmann equation in an arbitrary spacetime geometry. Starting from
the covariant expression for the distribution function
$f=f\left (x^\mu,P^\mu\right )$ and taking its derivative with respect
to an affine parameter $\lambda$, leads to:
\begin{equation}
\begin{split}
\frac{\dd f}{\dd\lambda}&=\frac{\partial f}{\partial x^\mu}\frac{\dd x^\mu}{\dd\lambda}+\frac{\partial f}{\partial P^\mu}\frac{\dd P^\mu}{\dd\lambda}
=P^\mu\frac{\partial f}{\partial x^\mu}-\Gamma^\mu_{\nu\sigma}P^\nu P^\sigma\frac{\partial f}{\partial P^\mu}\,,
\end{split}
\end{equation}
where the second equality follows from the geodesic equation, describing
particle propagation in curved coordinates,
\begin{equation}
\begin{split}
\frac{\dd x^\mu}{\dd\lambda}=P^\mu\,,\quad\frac{\dd^2 x^\mu}{\dd\lambda^2}=\frac{\dd P^\mu}{\dd\lambda}&=-\Gamma^\mu_{\nu\sigma}P^\nu P^{\sigma}\\
\Rightarrow\quad\frac{\dd P^0}{\dd\lambda}&=-\Gamma^0_{ij}P^iP^j=-a\dot a\vec P^2\\
\frac{\dd P^i}{\dd\lambda}&=-2\Gamma^i_{0j}P^0P^j-\Gamma^i_{jk}P^jP^k=-2\frac{\dot a}{a}P^0P^i\,. 
\label{eq:MomEvol}
\end{split}
\end{equation}
Consequently, the general evolution equation becomes
\begin{equation}
\frac{\dd f}{\dd\lambda}=P^0\frac{\partial f}{\partial t}+\vec P\cdot\nabla_{\vec x}f-2\frac{\dot a}{a}P^0P^i\frac{\partial f}{\partial P^i}-a\dot a\vec P^2 \frac{\partial f}{\partial P^0}\,.
\label{eq:GenEvolEq}
\end{equation}
Since we are interested in homogeneous systems, all explicit
dependencies on spatial coordinates will be dropped in the following. In
analogy to the calculation of Sec.~\ref{sec:GenBoltzmannEquation}, it is
convenient to integrate out the energy dependence, leading to the
relativistic dispersion relation,
\begin{equation}
P^2=g_{\mu\nu}P^\mu P^\nu =\left (P^0\right )^2-a^2\vec P^2=m^2\,,
\label{eq:GenEnergyMomentum}
\end{equation}
which then allows to introduce the so-called physical or proper momentum
$p^i:=aP^i\Rightarrow$ $\vec{p}^2=-g_{ij}P^iP^j$. Furthermore, we obtain
a simplified expression by taking into account that the distribution
function depends only \emph{implicitly} on the energy
$p^0:=P^0=\sqrt{\vec p^2+m^2}$ in comparison to \eqref{eq:GenEvolEq}:
\begin{equation}
\begin{split}
\frac{\dd f}{\dd\lambda}&=\frac{\partial f}{\partial x^0}\frac{\dd
  x^0}{\dd\lambda}+\frac{\partial f}{\partial P^i}\frac{\dd
  P^i}{\dd\lambda}
=P^0\frac{\partial f}{\partial t}-2\frac{\dot a}{a}P^0P^i\frac{\partial f}{\partial P^i}\,.\\
\label{eq:ExpansionEquationComoving1}
\end{split}
\end{equation}
Making use of $P^i\left (a,p^i\right )$ being a function of the
time-dependent scale factor $a$ and the proper momentum $p^i$, we can
rewrite the expression \eqref{eq:ExpansionEquationComoving1} in the form
\begin{equation}
  \frac{\dd f}{\dd\lambda}=P^0\frac{\partial f}{\partial
t}+\frac{\partial f}{\partial P^i}\left (\frac{1}{a}\frac{\dd
p^i}{\dd\lambda}-\frac{1}{a^2}\frac{\dd a}{\dd x^0}\frac{\dd
x^0}{\dd\lambda}p^i\right ) =P^0\frac{\partial f}{\partial
t}+\frac{\partial f}{\partial P^i}\left (\frac{1}{a}\frac{\dd
p^i}{\dd\lambda}-\frac{\dot a}{a}P^0P^i\right )
\end{equation}
and thus
\begin{equation}
\frac{\dd p^i}{\dd\lambda}=\frac{\dd a}{\dd
    x^0}\frac{\dd x^0}{\dd\lambda}P^i+a\frac{\dd
    P^i}{\dd\lambda}=-\frac{\dot a}{a}p^0p^i\,.
\label{eq:ExpansionEquationComoving2}
\end{equation}
Consequently, the dynamical evolution of the one-particle distribution
function can be written in terms of the proper momentum
$p^i\left (a,P^i\right )$, resulting in
\begin{equation}
\frac{\dd f}{\dd\lambda}=\frac{\partial f}{\partial x^0}\frac{\dd x^0}{\dd\lambda}+\frac{\partial f}{\partial p^i}\frac{\dd p^i}{\dd\lambda}
=p^0\frac{\partial f}{\partial t}-\frac{\dot a}{a}p^0p^i\frac{\partial f}{\partial p^i}\,.
\label{eq:ExpansionEquationProper}
\end{equation}
After introducing the usual abbreviation $H:=\dot a/a$, known as Hubble
constant, we finally arrive at
\begin{equation}
\left (\frac{\partial }{\partial t}-Hp\frac{\partial}{\partial p}\right
)f=\mathcal{I}\left (t,p\right )\quad \text{with} \quad Hp\frac{\partial}{\partial p}=Hp^i\frac{\partial}{\partial p^i}\,,
\label{eq:BoltzmannExpansion}
\end{equation}
with $p^0$ being absorbed in the collision integral $\mathcal{I}\left (t,p\right )$.

Integrating by parts leads then to the evolution equation of particle
densities
\begin{equation}
\begin{split}
\int\frac{\dd^3\vec p}{\left (2\pi\right )^3}\left (\frac{\partial
    f}{\partial t}-Hp\frac{\partial}{\partial p}f\right
)&=\int\frac{\dd^3\vec p}{\left (2\pi\right )^3}\,\mathcal{I}\left
  (t,p\right ) \; \Rightarrow \; 
\dot n+3Hn=\int\frac{\dd^3\vec p}{\left (2\pi\right )^3}\mathcal{I}\left (t,p\right )\,.
\label{eq:CoolingInflation2}
\end{split}
\end{equation}
The net number of quarks/baryons $N_{\text{q,net}}$ in an expanding
domain with volume $V\left (t\right )$ is a strictly conserved quantity,
as long as the interaction with other domains is not possible.
Therefore, it holds $N_{\text{q,net}}\equiv n_{\text{q,net}}V=\text{const.}$ and one
obtains the following relations between the spherical volume $V$, its radius
$R$ and the internal scale factor $a$:
\begin{equation}
\frac{\dot V}{V}-3H=0\quad\Rightarrow\quad\frac{\dot a}{a}=\frac{\dot R/R_0}{R/R_0}\,.
\label{eq:InflationVolume}
\end{equation}
The general approach with FLRW metric allows to choose between various
parametrizations for the radius as a function of time. Thereby, an
application to heavy-ion collisions is constrained by physical
requirements of a casual evolution for the expansion\footnote{This is
  not required for the expansion of the universe, which can exceed the
  speed of light.}. A possible and rather simple parametrization, used
in the following, is given by a linear growth of the radius with a
constant expansion velocity $v_{e}<c$,
\begin{equation}
R\left (t\right )=R\left (t_0\right )+v_{e}t\,.
\label{eq:RadiusParam}
\end{equation}

The mean-field equation in an expanding homogeneous and isotropic
three-dimensional geometry can be described by transforming the
d'Alembert operator with respect to the FLRW metric (see for instance
\cite{book:Goenner}):

\begin{equation}
\eta^{\mu\nu}\nabla_\mu\nabla_\nu=\left (\partial_t^2-\partial^2_{\vec
    x}\right )\; \longrightarrow \; 
g^{\mu\nu}\nabla_\mu\nabla_\nu=\frac{1}{\sqrt{-g}}\partial_\mu\left (\sqrt{-g}g^{\mu\nu}\partial_\nu\right )
=\partial_t^2+3H\partial_t-a^2\partial^2_{\vec x}=\partial_t^2+3H\partial_t
\label{eq:MeanFieldInflation}
\end{equation}
Here, the damping coefficient $3H$ is directly related to the volume
increase in \eqref{eq:InflationVolume}. Finally, the mean-field equation (\ref{eq:HomMeanFieldEquation})
becomes
\begin{equation}
\partial_t^2\phi+E\left (t\right )+D\left (t\right )+J\left (t\right )=0\quad\text{with}\quad E\left (t\right ):=3H\partial_t\phi
\label{eq:MeanFieldExpansion}
\end{equation}
with $D$ defined by (\ref{eq:dkareducedb}). For expanding geometry the
full set of evolution equations consists of the mean-field equation
\eqref{eq:MeanFieldExpansion} and Eq.~\eqref{eq:BoltzmannExpansion} for
every particle species. Finally, the collision integrals are calculated
as in the case without expansion.

\begin{figure*}
  \centerline{\includegraphics[width=11cm]{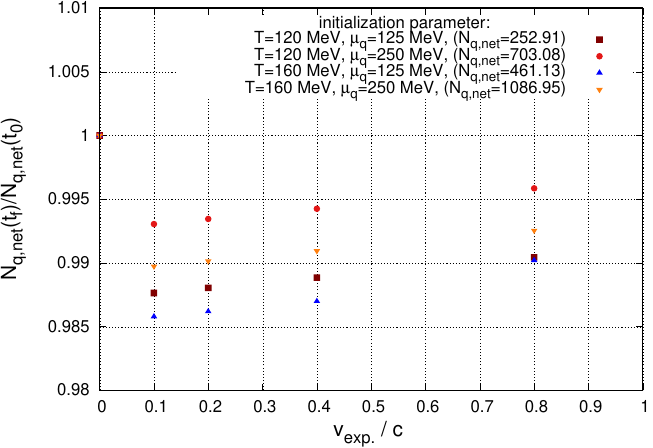}}
     \caption{Ratio of the final over the initial net quark
       number for different initial conditions
       $T, \mu_{\text{q}}\,(N_{\text{q,net}})$ and expansion velocities
       $v_{\text{exp}}$.  The initial and final radii of the system are
       set to $R\left (t_0\right )=5\,\fm$, respectively
       $R\left (t_{\text{f}}\right )=15\,\fm$.}
  \label{fig:ConservationNetQuark}
\end{figure*}

The evolution equation for one-particle distribution functions is solved
numerically by discretizing Eq.~\eqref{eq:BoltzmannExpansion} in
accordance with the linear-implicit finite difference scheme, meaning
that the term including the operator $A:=Hp\frac{\partial}{\partial p}$
is evaluated at the final time $t_{i}+\Delta t$, whereas the collision
term $\mathcal{I}$ is evaluated at the initial time $t_i$,
\begin{equation}
   \begin{split}
     \frac{f_{i+1}-f_i}{\Delta
       t}=Af_{i+1}+\mathcal{I}_i\qquad\Rightarrow\quad f_{i+1}=\left
       (\mathds{1}-\Delta t A\right )^{-1}\left (f_i+\Delta
       t\mathcal{I}_i\right )\,.
     \label{eq:NumCollingInflation}
   \end{split}
\end{equation}
The linear-implicit evaluation of the Integro-PDE improves significantly
the stability region of the finite difference method (see
\cite{book:Lord} and references therein).  Because of the weighted
left-shift operator $A$, generating a particle flow to the infrared
region, the derivative in momentum space is discretized in accordance to
the upwind scheme for the first two grid points of the one-particle
distribution function $f_{i,j}$ with $j\in\{1,...\,,N\}$. This procedure
does not require to specify a boundary condition for the infrared
region. The intermediate region is then discretized by applying a
central differencing scheme, being an approximation of second order with
respect to the momentum interval $\Delta p$: 
\begin{equation}
  A=\frac{H}{\Delta p}\begin{pmatrix} -p^1 & p^1 & 0\\ -p^2 & 0 & p^2\\
    & & \ddots & \\ & & -p^{N-1} & 0 & p^{N-1}\\ & & 0 & -p^{N} &
    p^N \end{pmatrix}\,, \label{eq:ShiftOperator} 
\end{equation}
where the last line of the discrete operator in matrix representation is
used to specify a boundary condition for the highest momentum on the
grid, given here by the backward scheme of the first-order
derivative. An alternative boundary condition is given by a vanishing
first-order derivative with respect to the absolute value of the
momentum $\dd f/\dd p|_{p=p^{N}}= 0$, resulting in a zero line for the
last row of the matrix.  The mean-field equation
\eqref{eq:MeanFieldExpansion} is discretized by replacing exact
derivatives with corresponding second-order finite differences, leading
to the following scheme: 
\begin{equation} 
\begin{split}
    &\frac{\phi_{i+1}-2\phi_i+\phi_{i-1}}{\Delta
      t^2}+3H\frac{\phi_{i+1}-\phi_{i-1}}{2\Delta t}+D_i+J_i=0\\
    &\Rightarrow\phi_{i+1}=\left (1+3H\frac{\Delta t}{2}\right
    )^{-1}\left (2\phi_i-\phi_{i-1}\left (1-3H\frac{\Delta t}{2}\right
      )-\left (D_i+J_i\right )\Delta t^2\right
    ).  
\end{split} 
\label{eq:MeanFieldDiscrete} 
\end{equation} 
In the following we are interested in studying fluctuations of the
net-quark number in various momentum regions, but requiring that the
total quantity has to be conserved. Fig.~\ref{fig:ConservationNetQuark}
explicitly shows that the the total net-quark number is indeed conserved
to a high accuracy of $0.5\%-1.5\%$ even for large running times
$t_{\text{run}}=12.5-100\,\fmc$. Thereby, test simulations were
performed for a wide range of expansion velocities
$v_{\text{exp}}=0.1-0.8\, c$ as well as initial and final radii. Note,
that for the parametrization \eqref{eq:RadiusParam}, the running time of
a simulation is defined via the relation
$t_{\text{run}}=\left (R\left (t_f\right )-R\left (t_0\right )\right
)/v_{\text{exp}}$.

\subsection{Initial conditions for expanding geometry}
\label{sec:InitialConditions}
 
As already discussed in the previous Section, the expansion description
in the following models homogeneous and isotropic systems. Therefore, it
is required to introduce fluctuating initial conditions for the
net-quark number and consider dynamical long-time fluctuations with
respect to \emph{different momentum ranges} instead of widely studied
long-range fluctuations due to spatial inhomogeneities. In detail, we
consider a spherically symmetric bubble/domain with radius $R_0$, which
is modelled to expand according to the evolution equations of
Sec.~\ref{sec:ExpansionInflation}. Thereby, the medium inside the bubble
is initialized in thermal equilibrium $(T_{0},\,\mu_{q,0})$ and the
fluctuations of the initial net quark number are modelled via the
Gaussian distribution,
\begin{equation}
N_{\text{q,net}}\sim\mathcal{N}\left (\left \langle  N_{\text{q,net}}\right \rangle ,\sigma_{\text{q,net}}^2\right )\,,
\label{eq:GaussianNetQuark}
\end{equation}
where $\left \langle N_{\text{q,net}}\right \rangle $ denotes the mean
of the Gaussian distribution and $\sigma_{\text{q,net}}^2$ its variance.
For the thermal choice of $(T_{0},\,\mu_{q,0})$ the mean value of the
Gaussian distribution is defined by calculating the net-quark number
inside of the spherical domain with $R=R_0$,
\begin{equation}
\left \langle  N_{\text{q,net}}\right \rangle :=n_{\text{q,net}}V=\frac{4}{3}\pi R_0^3\int\,\frac{\dd^3\vec p}{\left (2\pi\right )^3}\left (f_{q}-f_{\bar q}\right )\,,
\label{eq:MeanNetQuark}
\end{equation}
being a strictly conserved quantity as already discussed in
Sec.~\ref{sec:ExpansionInflation}.  Numerically, the conservation
property acts as a necessary condition to ensure a stable and precise
result. The standard deviation in our Gaussian description of initial
fluctuations in the total number of net quarks is a parameter and will
be set to
$\sigma_{\text{q,net}}=\left \langle N_{\text{q,net}}\right \rangle /10$
(or also
$\sigma_{\text{q,net}}=\left \langle N_{\text{q,net}}\right \rangle
/5$), being an arbitrary but reasonable choice of relatively small
fluctuations for the net-quark number as one would expect in a
fixed-target experiment or in a well defined centrality class of a
collider. However, as shown in the following, the absolute value of the
standard deviation does not change the qualitative results of the
study. As expected naively, the amplitude of net-quark-number
fluctuations increases with an increasing value of the standard
deviation, without changing significantly the shape of the cumulants and
their ratios.

The cumulants of an arbitrary distribution function can be expressed in
terms of centralized moments\footnote{In contradistinction to the
  moments, the cumulants are pairwise independent quantities of the
  corresponding distribution function.}
$\tilde m_k:=\left \langle (m-\left \langle m\right \rangle )^k \right
\rangle $. In detail, the first six cumulants are given by
\begin{equation}
\begin{split}
\kappa_1&=\left \langle m\right \rangle \,,\\
\kappa_2&=\tilde m_2\equiv\sigma^2\,,\\
\kappa_3&=\tilde m_3\,,\\
\kappa_4&=\tilde m_4-3\tilde m_2^2\,,\\
\kappa_5&=\tilde m_5-10\tilde m_3\tilde m_5\,,\\
\kappa_6&=\tilde m_6-15\tilde m_4\tilde m_2-10\tilde m_3^2+30\tilde m_2^3\,,
\end{split}
\label{eq:CentralMoments}
\end{equation}
where only the first two are non-zero in the case of a Gaussian
distribution function. Note, when later integrating the phase
distribution over \emph{all} momenta, only the first two moments should
be non-zero. On the other hand, when looking in special momentum
classes, this may not be the case and can establish the potential
critical fluctuations close to the phase transition region. The other
momentum regions can act as an effective particle and heat bath,
allowing diffusion into the particular momentum region.

From these moments one can introduce a second class of observables,
given by the following ratios of even and odd cumulants:
\begin{equation}
\begin{split}
R_{4,2}&=\frac{\kappa_4}{\kappa_2}\,,\quad R_{6,2}=\frac{\kappa_6}{\kappa_2}\,,\quad R_{3,1}=\frac{\kappa_3}{\kappa_1}\,,\quad R_{5,1}=\frac{\kappa_5}{\kappa_1}\,,
\end{split}
\label{eq:RatioCumulants}
\end{equation}
where $R_{4,2}$ is known as the rescaled excess kurtosis
$\kappa\sigma_2$ (see also Sec.~\ref{chap:introduction}). Especially,
the last type of observables is sensitive to a critical behavior at the
phase transition, since all of the ratios in \eqref{eq:RatioCumulants}
have a non-trivial dependence on the correlation scale\footnote{In the
  present study the relevant correlation scale is given by the
  correlation time of net-quark-number fluctuations in different
  momentum ranges.} and become zero in case of a Gaussian distribution
function.
 
The most natural way of studying fluctuations is to consider a large set
of stochastically independent, but equally distributed initial
conditions as one also expects for independent heavy ion
collisions. From a numerical point of view the initialization procedure
can be modeled by Monte Carlo methods, which are suited to generate
systems with stochastic initial conditions. Nevertheless, at least
simple MC methods are known for having a very slow convergence rate to
the exact distribution function\footnote{In general, even a small
  deviation from the exact shape of the distribution function means that
  its tails cannot be reproduced at all, leading for higher order
  cumulants to huge uncertainties.}, being the main reason for the usage
of so-called quasi-MC or pseudo-stochastic methods, which are more
suitable for studying higher order cumulants, especially when the
computational time plays a significant role. In our approach, we use a
pseudo-stochastic approach for the initialization of independent
configurations, which significantly improves the convergence rate to the
exact cumulant ratios as discussed in the following.

As already introduced in Eq.~\eqref{eq:GaussianNetQuark}, the initial
probability density of the net quark number is modeled by the Gaussian
distribution function,
\begin{equation}
  \varrho\left (N_{\text{q,net}}\right
  )=\frac{1}{\sigma_{\text{q,net}}\sqrt{2\pi}}\exp\left (-\frac{\left
        (N_{\text{q,net}}-\left \langle  N_{\text{q,net}}\right \rangle
      \right )^2}{2\sigma^2_{\text{q,net}}}\right )\,.
  \label{eq:GaussianDistribution}
\end{equation}
Consequently, a numerical simulation requires an initialization
procedure of initial configurations with respect to this distribution
function. In detail, for a pseudo-stochastic approach it is required to
introduce a grid in a range of allowed net-quark numbers. Thereby, a
physical choice for heavy-ion systems is restricted by a lower boundary
of $N_{\text{q,net}}\geq 0$. Because of symmetry and variance reduction
reasons, it is reasonable to set the upper boundary to
$N_{\text{q,net}}\leq 2\left \langle N_{\text{q,net}}\right \rangle $
for
$\sigma_{\text{q,net}}\ll\left \langle N_{\text{q,net}}\right \rangle $
and distribute the grid points for different initial configurations
within the range of
$\left [ 0,\,2\left \langle N_{2,net}\right \rangle \right ] $.  Even
so, the convergence rate can be improved by using non-uniform grid
points for the initial net-quark number, it is sufficient to discretize
the above range with an equidistant spacing of the form
$\Delta N_{\text{q,net}}:=2\left \langle N_{\text{q,net}}\right \rangle
/M$ for $M+1$ independent initial configurations, leading to the grid
points $N_{\text{q,net},k}=k\Delta N_{\text{q,net}}$ with
$k\in\{0,...,M\}$. This set defines a test sample of initial values for
the net quark number. For each value of the test sample only one
simulation run has to be performed and the final result for an arbitrary
observable $\mathbf{O}$ follows then from calculating the expectation
value of the observable with respect to the test sample,
\begin{equation} \left \langle \mathbf{O}\right \rangle
  =\frac{p_0\mathbf{O}_{0}+p_{M}\mathbf{O}_{M}}{2}+\sum_{k=1}^{M-1} p_k
  \mathbf{O}_{k}\,,\qquad p_k= \varrho\left (N_k\right )/M\,.
  \label{eq:ObservableExpectationValue}
\end{equation}
  
\begin{figure*}
\begin{minipage}{0.45 \linewidth}
\includegraphics[width=\textwidth]{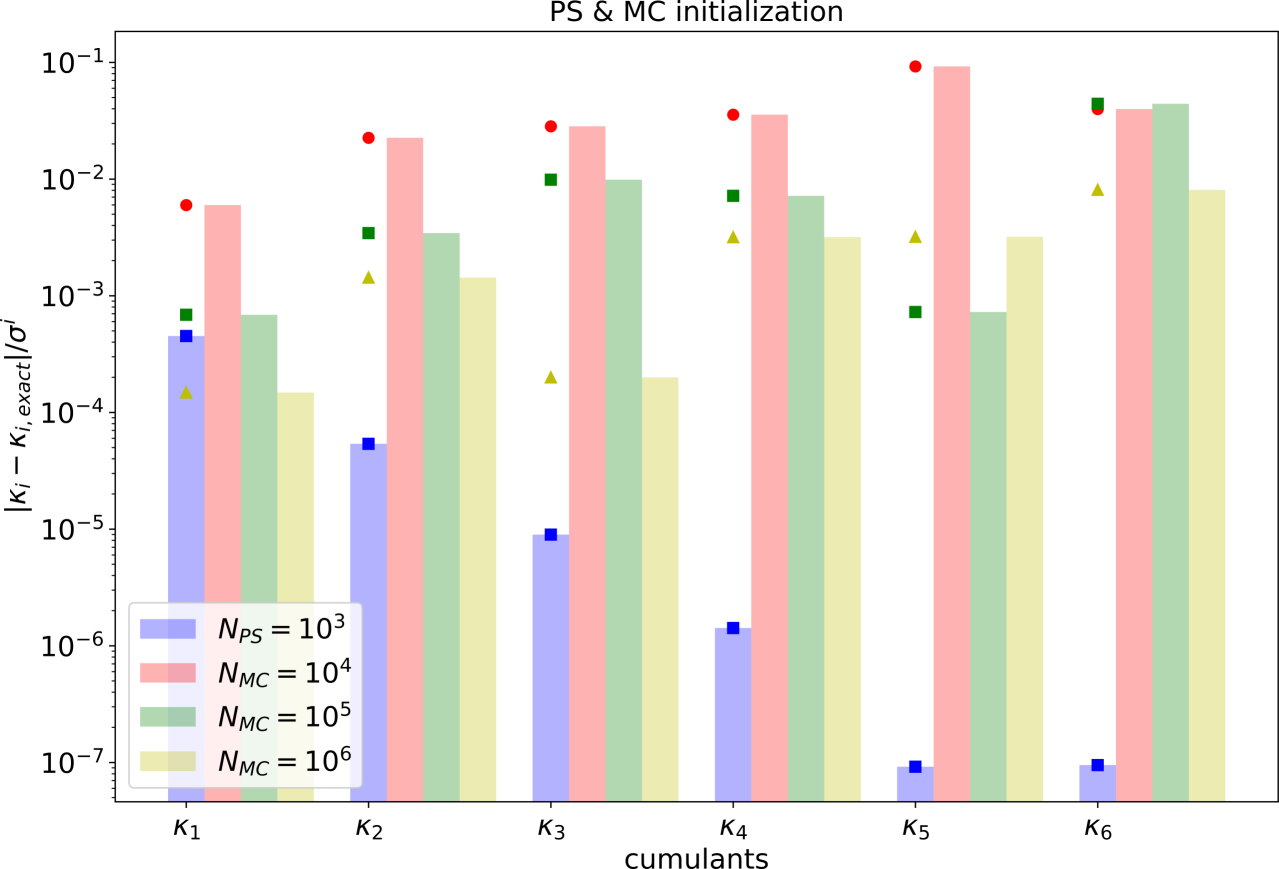}
\end{minipage}\hfill
\begin{minipage}{0.45 \linewidth}
\includegraphics[width=\textwidth]{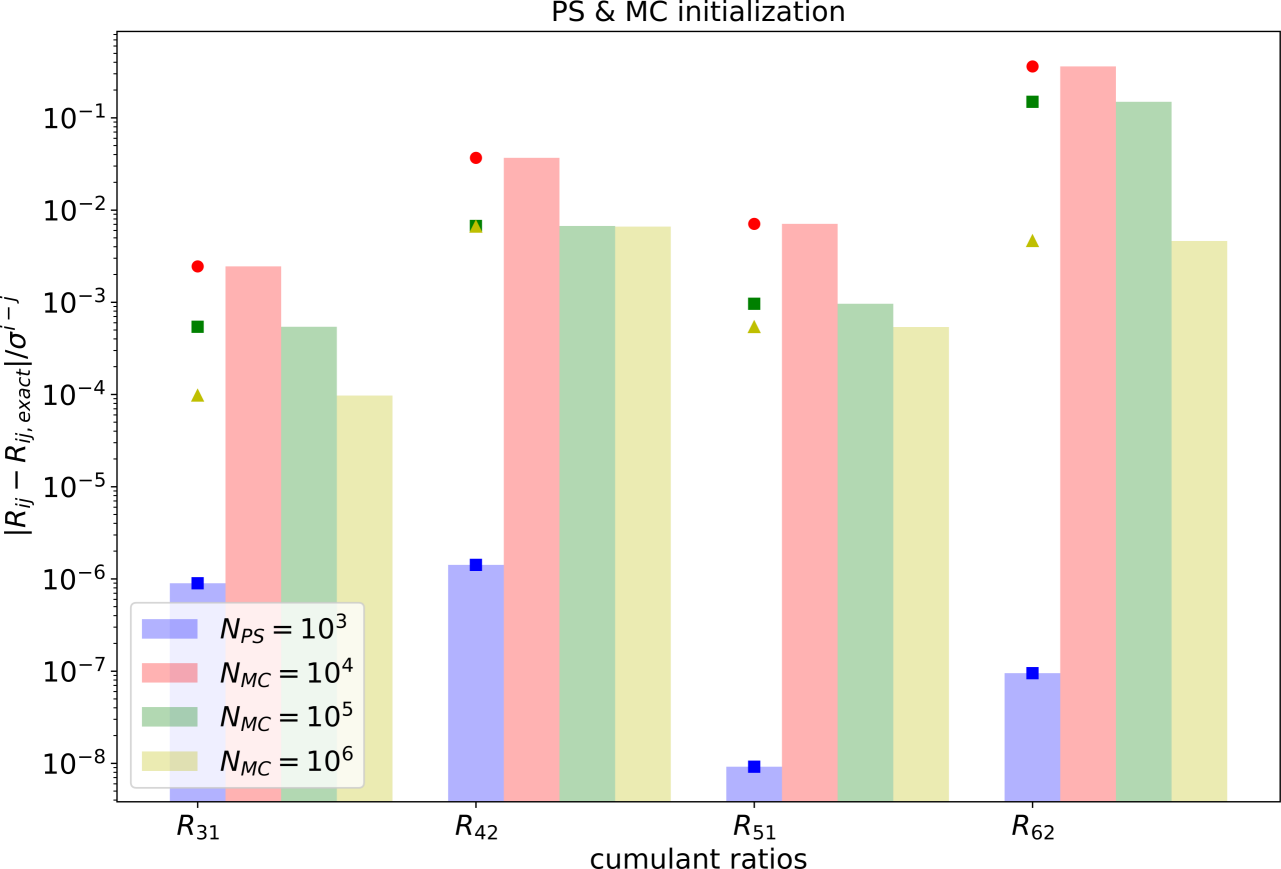}
\end{minipage}
\caption{Left: Dimensionless error for cumulants of the total net-quark
  number for different numbers of independent initial configurations,
  which are generated/sampled with the pseudo-stochastic as well as MC
  methods from the Gaussian distribution function
  \eqref{eq:GaussianDistribution}. Right: The same for the dimensionless
  error of cumulant ratios.}
  \label{fig:CumulantConvergence}
\end{figure*}

Fig.~\ref{fig:CumulantConvergence} shows a numerical comparison between
naive MC sampling for different numbers of initial configurations
$N_{\text{MC}}$ and pseudo-stochastic (PS) initialization method for
$N_{\text{PS}}$ grid points as described above. The numerical results
are compared with exact values for the distribution function
\eqref{eq:GaussianDistribution} by introducing a dimensionless form of
observables from Eqs.\ \eqref{eq:CentralMoments} and
\eqref{eq:RatioCumulants}
\begin{equation}
    \text{err}_\kappa:=\frac{|\kappa_i-\kappa_{i,exact}|}{(\sigma_{\text{q,net}})^i}\,,\quad
    \text{err}_R:=\frac{|R_{ij}-R_{ij,\text{exact}}|}{(\sigma_{\text{q,net}})^{i-j}}\,,
  \label{eq:ErrorCumulantRatio}
\end{equation}
where the cumulants $\kappa_i $ naturally may scale with the standard
deviation $\sigma_{\text{q,net}}$ to the power $i$. Also note that the
cumulants $\kappa_{i,\text{exact}} \equiv 0$ for $i>2 $.

Here, the initialization procedure is also compared to a naive MC method
as shown in Fig.~\ref{fig:CumulantConvergence} for different numbers of
independent initial configurations. As seen from the comparison, even
for very large test samples of MC initializations the pseudo-stochastic
method outperforms the naive MC sampling by up to five orders of
magnitude for the predefined errors, given by
Eq.~\eqref{eq:ErrorCumulantRatio}. Note, that a typical simulation run
with only one initial configuration requires $1-100$ hours of
computational CPU time, depending on the expansion velocity (see
Tab.~\ref{tab:InitialParameters} and
Sec.~\ref{sec:CumulantRatiosExpandingSystems}). Consequently, a simple
MC approach is not applicable, since even $N_{\text{MC}}=10^6$ of
initial configurations result in a rather poor convergence rate. In the
following calculations we restrict the number of independent
pseudo-stochastic initial configurations for a single run to
$N_{\text{PS}}=100-1000$, resulting in a total computational CPU time of
$t_{\text{run}}=10^3-10^4$ hours per run.

At this point, we emphasize that our approach, which is based on
one-particle distribution functions, has a crucial advantage in
comparison to a test particle approach, when studying net-quark-number
fluctuations. The pseudo-stochastic initialization procedure ensures not
only the correct total number of net quarks on average but also the
correct (on average) number of net quarks in an arbitrary sub-range of
the momentum space, which is not realized in a test-particle approach
even for very high statistics. Particularly, in case of test particles
one is restricted to MC methods, making those approaches not suitable
for higher-order-cumulant studies of the net-quark number in an
arbitrary momentum range since test-particle approaches suffer from the
same difficulties as shown in Fig.~\ref{fig:CumulantConvergence} for
every single momentum range.

\begin{figure*}
\centering\includegraphics[width=15cm]{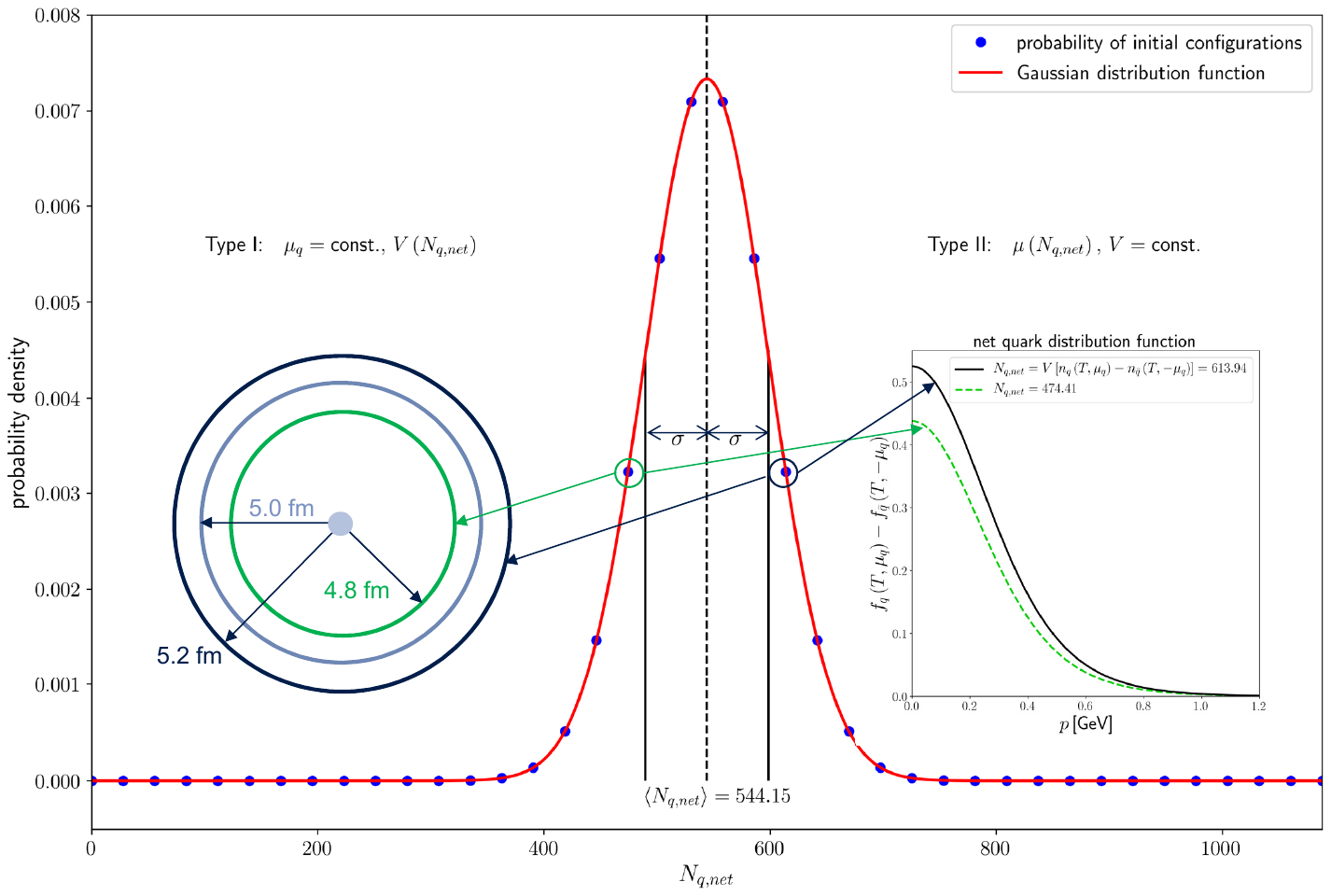}
\caption{Schematic initialization procedure for two types of initial
  conditions with respect to a Gaussian distribution function of net
  quarks. For illustrative reasons, the temperature is set to
  $T=150\,\MeV$ and the quark chemical potential to
  $\mu_{\text{q}}\left (\left \langle N_{\text{q,net}}\right \rangle \right
  )=160\,\MeV$, defining the mean for the Gaussian distribution function
  $\left \langle N_{\text{q,net}}\right \rangle =544.15$. The standard
  deviation is chosen to be
  $\sigma_{\text{q,net}}=\left \langle N_{\text{q,net}}\right \rangle
  /10=54.415$.  All averaged observables from numerical calculations are
  computed with respect to the numerical distribution function of
  initial configurations, given by the blue points. The net-quark
  number from a certain initial configuration is transformed into a
  fluctuation of the radius scale $R\left (N_{\text{q,net}}\right )$
  (type I) or into a fluctuation of the chemical potential
  $\mu_{\text{q}}\left (N_{\text{q,net}}\right )$ (type II) as shown
  schematically. The effective distribution function of the radius scale
  as well as the effective distribution function of the chemical
  potential are shown in Fig.~\ref{fig:StatisticalEnsembles}.  See also
  Tab.~\ref{tab:InitialParameters} and the text for more details.}
\label{fig:InitialConfigurations}
\end{figure*}

\begin{figure*}
\subfloat{\includegraphics[width=15cm]{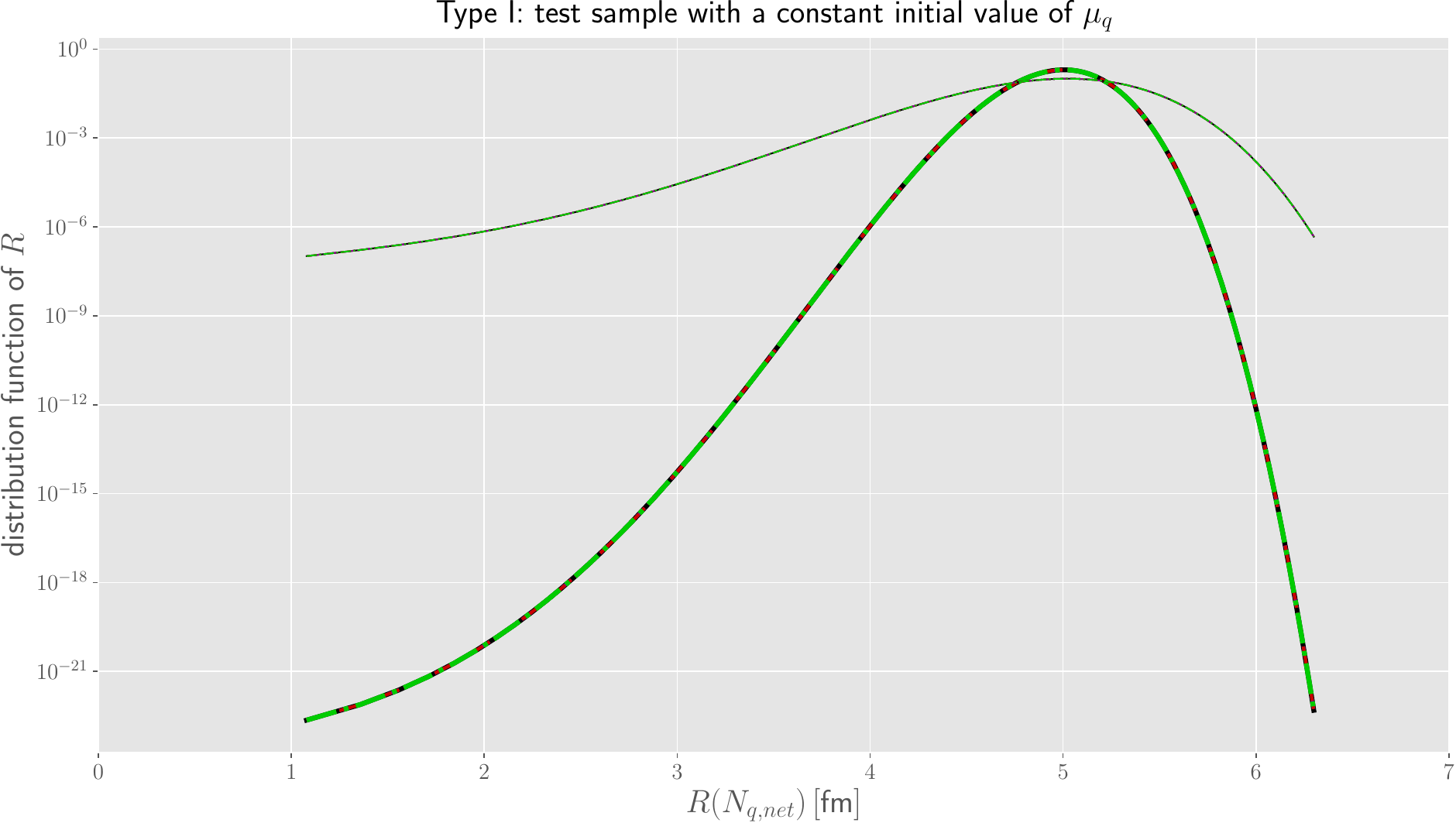}}\\
\subfloat{\includegraphics[width=15cm]{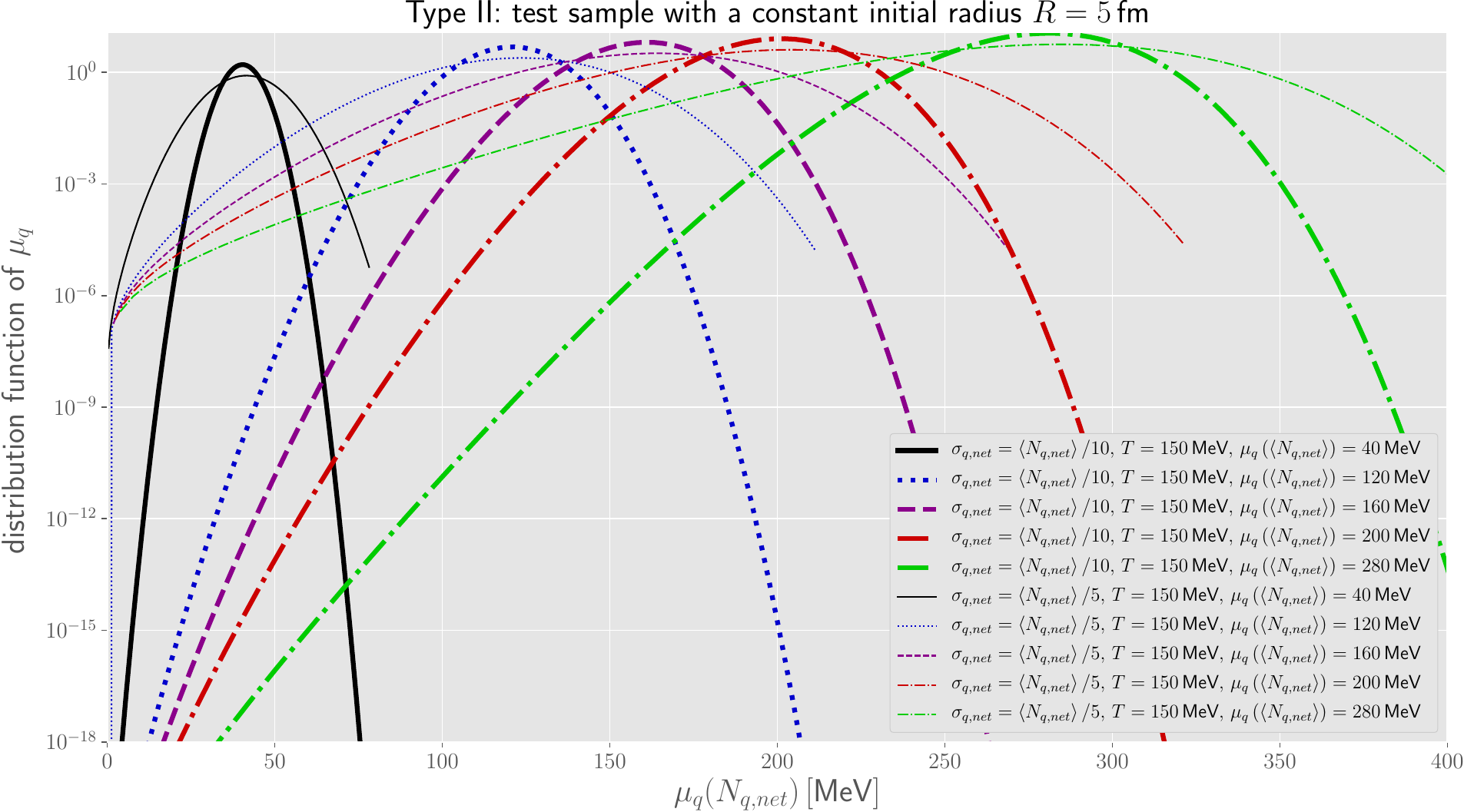}}
\caption{Upper figure: distribution function of the initial radius $R\left (N_{\text{q,net}}\right )$ for the statistical ensemble with initial configuration of type I.\\
  Lower figure: distribution function of the quark chemical potential
  $\mu_{\text{q}}\left (N_{\text{q,net}}\right )$ for the statistical ensemble
  with initial configuration of type II (compare with
  Tab.~\ref{tab:InitialParameters} and
  Fig.~\ref{fig:InitialConfigurations}). Both are shown for
  $T=150\,\MeV$ and various values of the quark chemical potential
  $\mu_{\text{q}}\left (\left \langle N_{\text{q,net}}\right \rangle \right )$ as
  well as two different values of $\sigma_{\text{q,net}}$ (thick lines
  for $\left \langle N_{\text{q,net}}\right \rangle /10$, thin lines for
  $\left \langle N_{\text{q,net}}\right \rangle /5$), denoting the
  standard deviation of the Gaussian distribution
  \eqref{eq:GaussianDistribution}.}
\label{fig:StatisticalEnsembles}
\end{figure*}

Before focusing on expanding systems in the next
Sec.~\ref{sec:CumulantRatiosExpandingSystems}, it is required to specify
statistical ensembles of initial configurations, which are used in the
following. As described above, the initial number of net quarks
$N_{\text{q,net}}$ is initialized according to the Gaussian distribution
function \eqref{eq:GaussianDistribution}, where
$\left \langle N_{\text{q,net}}\right \rangle $ is given by the relation
\eqref{eq:MeanNetQuark} with a certain choice of $T$ and
$\mu_{\text{q}}\left (\left \langle N_{\text{q,net}}\right \rangle
\right )$. Now, we define two statistical ensembles with a fluctuating
net-quark number: The first type is given by holding the physical
parameters $T,\,\mu_{\text{q}}$ constant and changing the radius scale
$R$ for the size of the initial domain, so that a spherical volume with
radius $R\left (N_{\text{q,net}}\right )$ contains the required net
quark number $N_{\text{q,net}}$ from the Gaussian sampling
procedure. Similarly, the second type is given by holding the physical
parameters $R,\,T$ constant and adjusting
$\mu_{\text{q}}\left (N_{\text{q,net}}\right )$ in a self-consistent
way, so that a spherical volume with radius $R=5\,\fm$ contains the
required net quark number $N_{\text{q,net}}$. The type II may be
considered to be more close to the exeprimental setting for various
collision centralities. Both initialization procedures are shown
schematically in Fig.~\ref{fig:InitialConfigurations}. Furthermore, the
general parameters of the corresponding statistical ensembles are
summarized in Tab.~\ref{tab:InitialParameters}, and
Fig.~\ref{fig:StatisticalEnsembles} shows the resulting distribution
functions for $R\left (N_{\text{q,net}}\right )$ in case of type I and
for $\mu_{\text{q}}\left (N_{\text{q,net}}\right )$ in case of type II.

\begin{table}
\centering
\scalebox{0.8}{
\begin{tabular}{|p{0.8cm}|p{1.8cm}|p{2.5cm}|p{2.2cm}|p{1.0cm}|p{1.8cm}|p{1.6cm}|p{1.6cm}|p{2.1cm}|}
  \hline
  type & initial condition & $R\left (\left \langle N_{\text{q,net}}\right \rangle\right )$ $\left [ \fm\right ]$ & $R\left (N_{\text{q,net}}\right )$ $\left [ \fm\right ]$  & $T$ & $\mu_{\text{q}}\left (N_{\text{q,net}}\right )$ & $v$ & $N_{PS}$ & $\sigma_{\text{q,net}}$ \\
  \hline
  \hline
  I & Gaussian & $5.0$ & $\left (\frac{3}{4\pi}\frac{N_{\text{q,net}}}{n_{\text{q,net}}}\right )^{1/3}$ & const. & const. & $0.05-0.8$ & $10^2-10^3$ & $\left \langle N_{\text{q,net}}\right \rangle/10$ \\
  \hline
  II & Gaussian & $5.0$ & $5.0$ & const. & variable & $0.05-0.8$ &
                                                                   $10^2-10^3$ & $\left \langle N_{\text{q,net}}\right \rangle/10$ \\
  \hline
\end{tabular}
}
\caption{Initial parameters for two different types of initial conditions:\\ I. The first type 
stands for a fixed combination of the temperature $T$ and chemical potential $\mu_{\text{q}}\left (\left \langle N_{\text{q,net}}\right \rangle\right )$. Here, the radius of the system is adjusted to 
match the net quark number from the Gaussian sampling procedure.\\ 
II. The second type stands for a fixed combination of the temperature $T$ and initial radius $R_0=5\,\fm$. 
Here, the chemical potential of quarks $\mu_{\text{q}}\left (N_{\text{q,net}}\right )$ is adjusted to match the net quark number from the Gaussian sampling procedure, requiring a self-consistent initialization method.}
\label{tab:InitialParameters}
\end{table}

\begin{figure*}
	\centering\includegraphics[width=15cm]{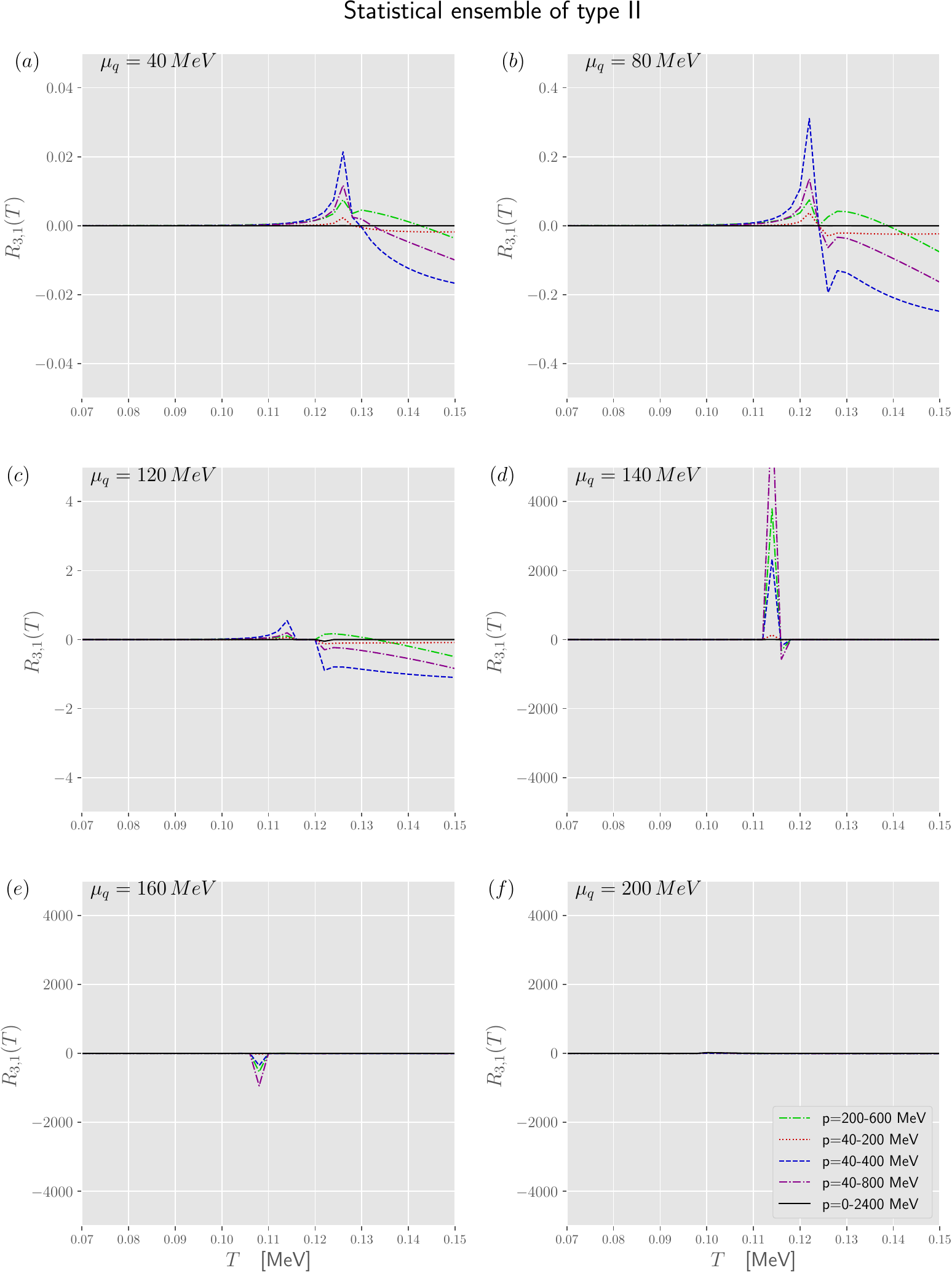}
	\caption{Cumulant ratio $R_{3,1}$ as a function of the temperature $T$
		with respect to $5$ momentum intervals and $6$ different values of the
		mean quark chemical potential
		$\mu_{\text{q}}\left (\left \langle N_{\text{q,net}}\right \rangle \right )$ in
		case of the statistical ensemble of type II.}
	\label{fig:ThermalCumulantRatioR31}
\end{figure*}

\begin{figure*}
	\centering\includegraphics[width=15cm]{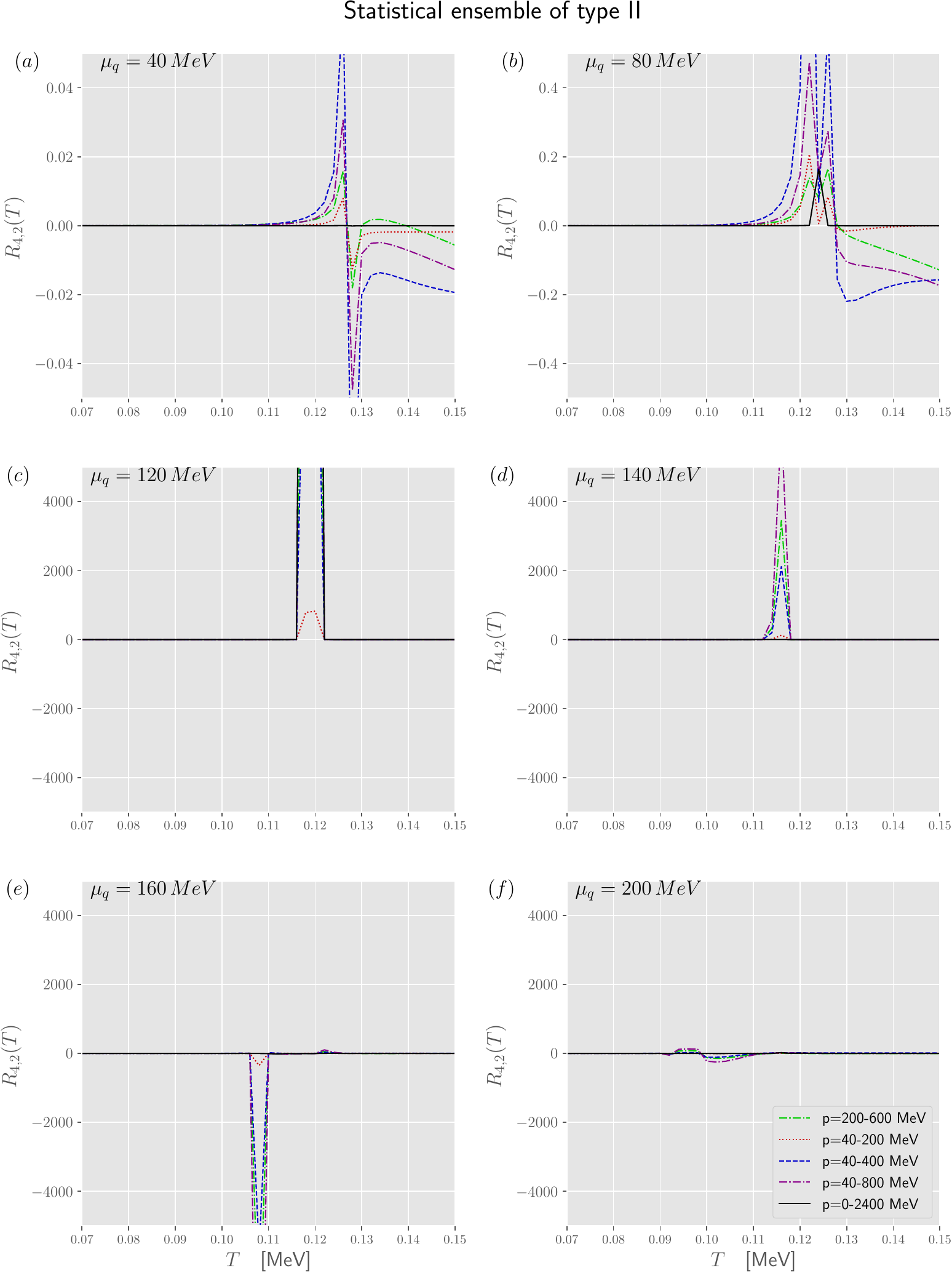}
	\caption{Same as Fig.~\ref{fig:ThermalCumulantRatioR31} but for the cumulant ratio $R_{4,2}$.}
	\label{fig:ThermalCumulantRatioR42}
\end{figure*}

\begin{figure*}
\centering\includegraphics[width=14cm]{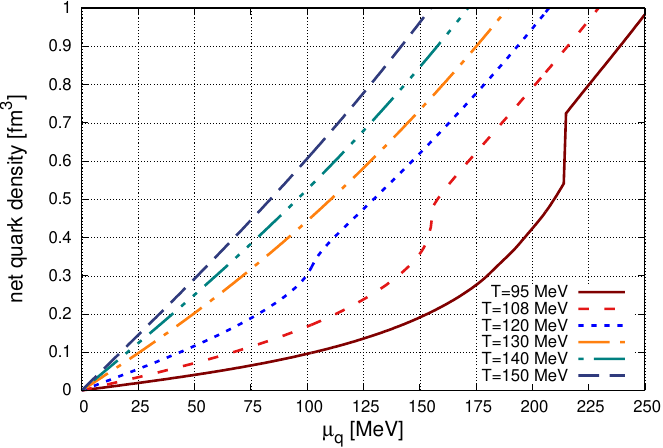}
\caption{Net-quark density in thermal and chemical equilibrium as a
  function of the quark chemical potential $\mu_{\text{q}}$, shown for several
  values of the temperature $T$.}
\label{fig:NetQuarkDensityThermal}
\end{figure*}

We close this Section with illustrating and demonstrating the purpose of
the two types of fluctuations as just described and calculate the
cumulant ratios $R_{3,1}$ and $R_{4,2}$ in thermal equilibrium for the
quark-meson $\sigma $-model as described in Sec.~\ref{chap:EvolEq}: For
a test sample of type I (see Tab.~\ref{tab:InitialParameters}),
fluctuations in momentum space can be expected only for systems out of
equilibrium, since the one-particle distribution function of quarks is
equal for all independent configurations of a certain choice of $T$ and
$\mu_{\text{q}}\left (\left \langle N_{\text{q,net}}\right \rangle
\right )$. Hence, the higher-order cumulants have to vanish also in each
different and specified momentum window. Therefore, we only show the
calculation of cumulant ratios in equilibrium to initial conditions of
type II (see Tab.~\ref{tab:InitialParameters}), where the net-quark
number $N_{\text{q,net}}$ from the sampling procedure defines the
chemical potential $\mu_{\text{q}}\left (N_{\text{q,net}}\right )$ of
quarks for a fixed value of $T$. For each combination of $T$ and
$\mu_{\text{q}}\left (\left \langle N_{\text{q,net}}\right \rangle
\right )\Leftrightarrow\left \langle N_{\text{q,net}}\right \rangle $,
we initialize $N_{\text{PS}}=1000$ of independent initial configurations
according to the Gaussian distribution function
\eqref{eq:GaussianDistribution} as described above. Following that, we
compute the fluctuations of the net quark number by cumulants and their
ratios (see Eqs. \eqref{eq:CentralMoments} and
\eqref{eq:RatioCumulants}) for different momentum ranges by assuming
thermal as well as chemical equilibrium for the quark distribution
functions of the quark-meson $\sigma $-model. The final result for the
cumulant ratios $R_{3,1}$ and $R_{4,2}$ is depicted in
Figs.~\ref{fig:ThermalCumulantRatioR31} and
\ref{fig:ThermalCumulantRatioR42} with respect to $T$,
$\mu_{\text{q}}\left (\left \langle N_{\text{q,net}}\right \rangle
\right )$ as well as five momentum intervals. In both cases the ratios
show a critical and even divergent behavior near the phase transition
(see the phase diagram in Sec.~\ref{chap:EvolEq} and also compare to
Fig.~\ref{fig:NetQuarkDensityThermal}, where the net-quark density is
given for various isotherms as function of $\mu_q $). Especially, in the
vicinity of the critical point for $\mu_{\text{q}}=160\,\MeV$ and
$T\approx 108\,\MeV$ there is a remarkable sign change in the cumulant
ratios. This observation underlines the fact that the cumulant ratios of
the net-quark number in momentum space are suitable observables for
studying the chiral phase transition within our approach. A careful
discussion for these findings of the cumulant ratios we postpone to the
next section, when we consider that the phyiscal system expands and
cools through the phase diagram.

\subsection{Dynamical evolution of the cumulant ratios}
\label{sec:CumulantRatiosExpandingSystems}

\begin{figure*}
\centering\includegraphics[width=14cm]{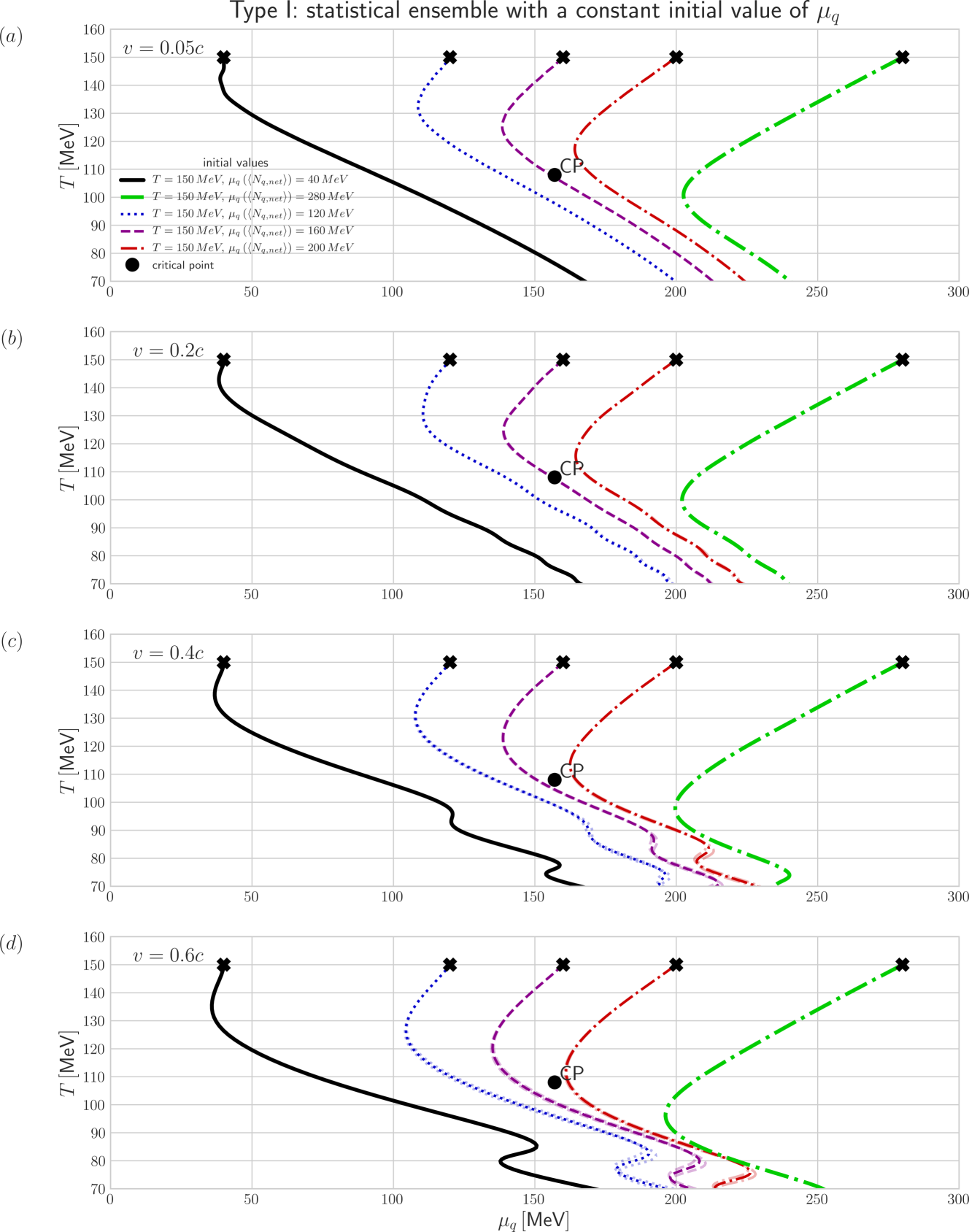}
\caption{Thick lines are averaged trajectories of initialized
  statistical ensembles in $T$-$\mu_{\text{q}}$ space. Transparent lines
  are boundary trajectories of the one-$\sigma$ width around the mean
  trajectory with
  $N_{\text{q,net}}=\left \langle N_{\text{q,net}}\right \rangle $ .}
\label{fig:PhaseDiagramTmuDynamicTypeI}
\end{figure*}

\begin{figure*}
\centering\includegraphics[width=14cm]{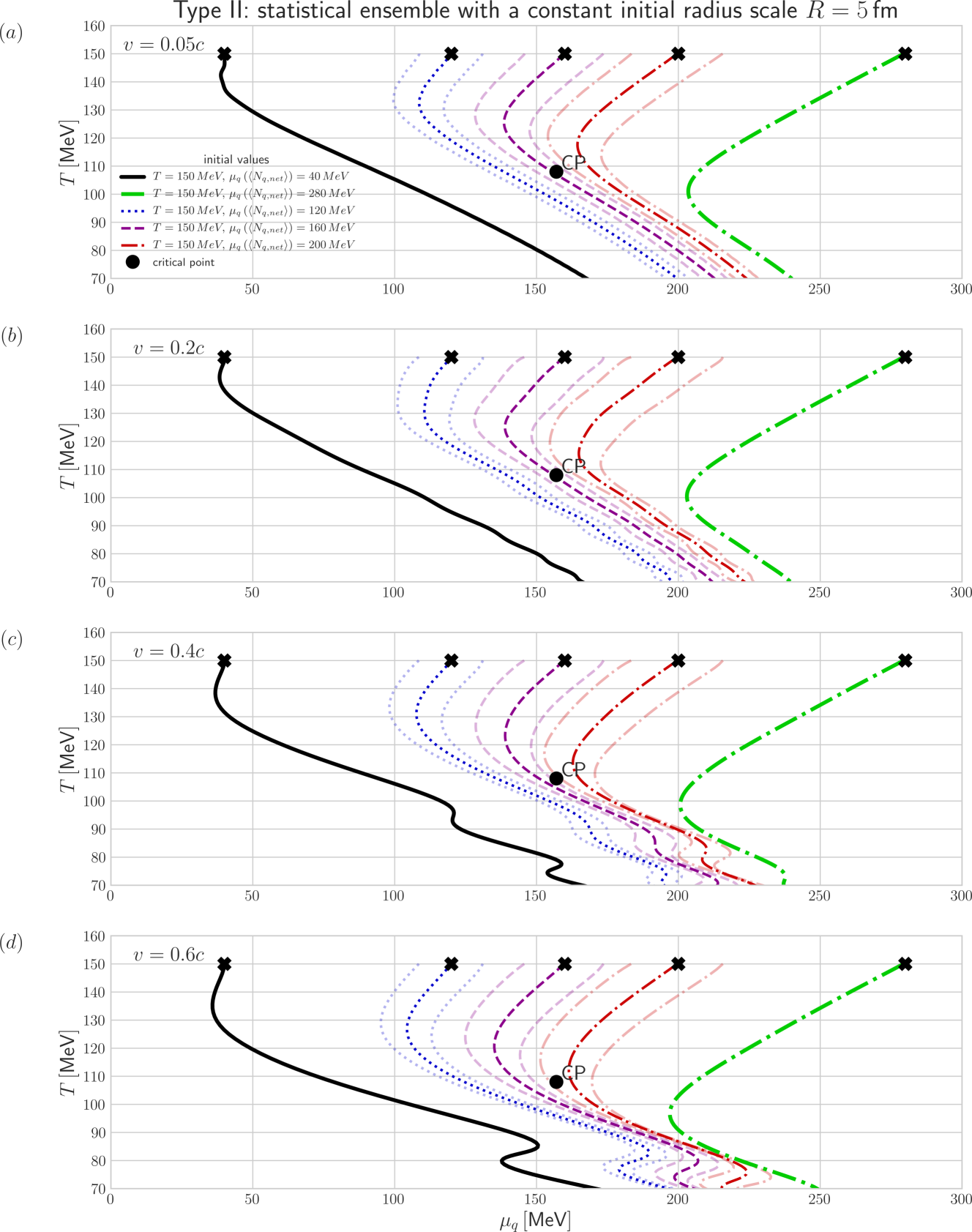}
\caption{Same as Fig.~\ref{fig:PhaseDiagramTmuDynamicTypeI} but for
  initial condition of type II.}
\label{fig:PhaseDiagramTmuDynamicTypeII}
\end{figure*}

After discussing the numerical initialization procedures for the
statistical ensembles of types I and II in the previous Section, we now
focus on the dynamical evolution of systems with such initial
conditions. We start with the dynamical phase diagram in an effective
$T$-$\mu_{\text{q}}$ plane, which is obtained from a fit of the
numerical quark distribution function $f_{\text{q}}\left (t,p\right )$
to the Fermi distribution function, allowing to extract an effective
value for the temperature $T$ as well as the chemical potential
$\mu_{\text{q}}$. Thereby, the fitting procedure is applied to the
(lower) momentum interval $p=100-300\,\MeV$. The results for
$T$-$\mu_{\text{q}}$ trajectories of ensemble averages are shown in
Figs.~\ref{fig:PhaseDiagramTmuDynamicTypeI} and
\ref{fig:PhaseDiagramTmuDynamicTypeII} for
$\sigma_{\text{q,net}}=\left \langle N_{\text{q,net}}\right \rangle
/10$. On the level of averaged ensemble trajectories there are no
significant differences between initial conditions of types I and II.
However, on the level of single ensemble trajectories the differences
are more pronounced as seen from the boundaries of the one-$\sigma$
width trajectories (transparent lines with the same color), which show a
significant deviation from the averaged trajectories in case of a type
II initialization. For type I small deviations can only be observed at
high expansion velocities, underlining the fact that the phase
transition as well as the thermalization process are determined by the
particle density of the system, being constant in case of type I, where
the ensemble trajectories differs only with respect to the initial
radius scale $R\left (t_0\right )$.

\begin{figure*}
\centering\includegraphics[width=14cm]{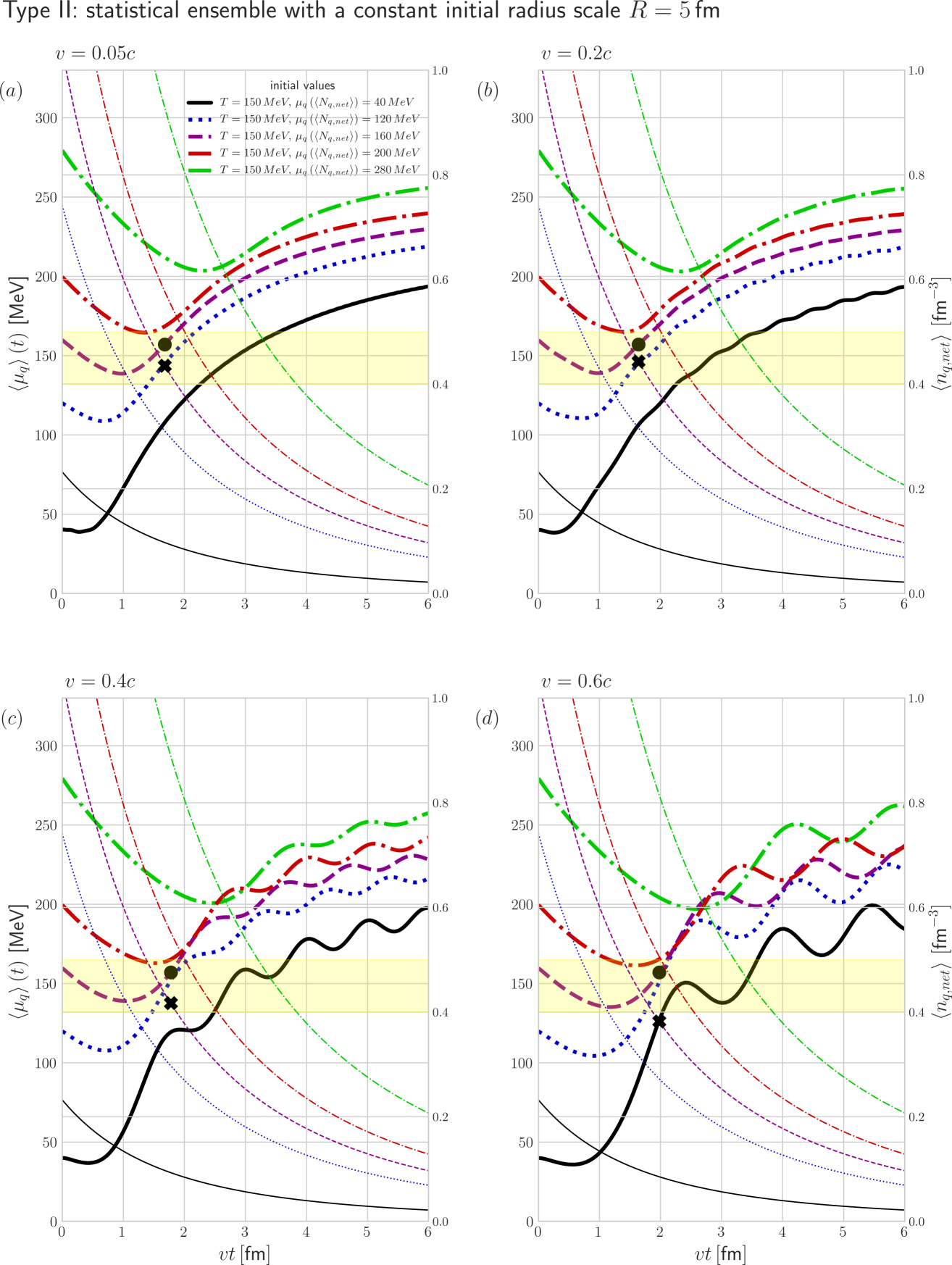}
\caption{Ensemble average of the effective chemical potential of quarks
  (thick lines), extracted from a fit to the Fermi distribution
  function. The averaged net-quark density (thin lines) is calculated
  directly from numerical realizations of the one-particle distribution
  functions.}
\label{fig:EffChemPotNetQuarkDensityMu}
\end{figure*}

\begin{figure*}
\centering\includegraphics[width=14cm]{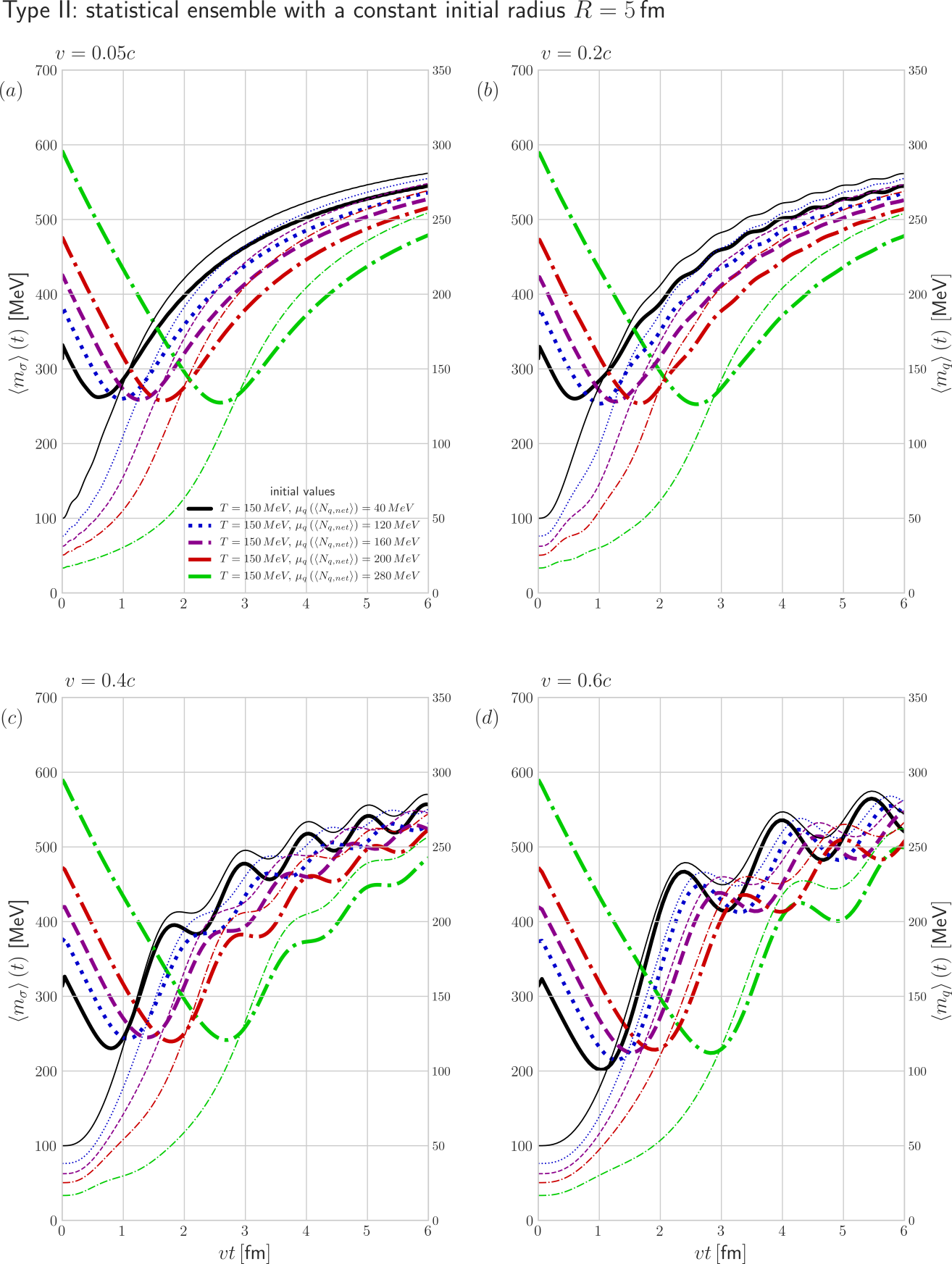}
\caption{Ensemble average of $\sigma$-meson (thick lines) and quark
  (thin lines) masses. The axis range for quark masses is chosen to be
  exactly one-half of the axis range for sigma masses. The intersection
  point between corresponding lines (same color and style) defines a
  time scale $\tau_{\sigma\rightarrow q\bar q}$ for the
  $q\bar q$-production from sigma decay.}
\label{fig:PhiMass}
\end{figure*}

Since the averaged trajectories are very similar for both initialization
methods, we thus focus on type II for the moment.
Fig.~\ref{fig:EffChemPotNetQuarkDensityMu} shows the dynamical evolution
of the averaged net-quark density
$\left \langle n_{\text{q}}\right \rangle \left (t\right )$ (thin lines)
as well as the averaged effective quark chemical potential
$\left \langle \mu_{\text{q}}\right \rangle \left (t\right )$ (thick
lines).  The yellow band stands for the critical region of the net-quark
density in equilibrium as calculated near the critical point for
$T\approx 108\,\MeV$ and $\mu_{\text{q}}\approx 157\,\MeV$ (compare with
Fig.~\ref{fig:NetQuarkDensityThermal}). A black cross marks the exact
value of the averaged net-quark density, for which the statistical
ensemble has an effective averaged quark chemical potential of
$\left \langle \mu_{\text{q}}\right \rangle \approx 157\,\MeV$, being
represented by a black circle with the equilibrium net-quark density. A
large separation between those two points indicates that the averaged
quantities of the statistical ensemble deviate significantly from the
equilibrium values of the critical point. Indeed, for fast expanding
systems with $v\gtrsim 0.4$ one observes strong oscillations of the
averaged quantities, and the net-quark density
$\left \langle n_{\text{q}}\right \rangle \left (t\right )$ is even
outside of the yellow region, meaning that the system dilutes much
faster than it is the case for $v\lesssim 0.2$, circumventing the system
from passing through the critical point. During the expansion process a
similar situation is also observed for the mass evolution of the
$\sigma$-meson $\left \langle m_\sigma\right \rangle \left (t\right )$
and of quarks
$\left \langle m_{\text{q}}\right \rangle \left (t\right )$ as shown in
Fig.~\ref{fig:PhiMass}. Here, the axis range for quark masses is chosen
to be exactly one-half of the axis range for sigma masses. Consequently,
a sigma decay to a quark-antiquark pair is only allowed, when the thick
curve stays above the corresponding (same color and style) thin curve,
defining an important time scale, denoted with
$\tau_{\sigma\rightarrow q\bar q}$ in the following. Notable is also the
minimum of the dynamical sigma mass for the ensemble average, arising at
the time scale $\tau_{m_{\sigma,\,\mathrm{min.}}}$ and staying in all
cases above $200\,\MeV$, which makes it difficult to distinguish between
different orders of the phase transition from averaged quantities.

\begin{figure*}
\centering\includegraphics[width=14cm]{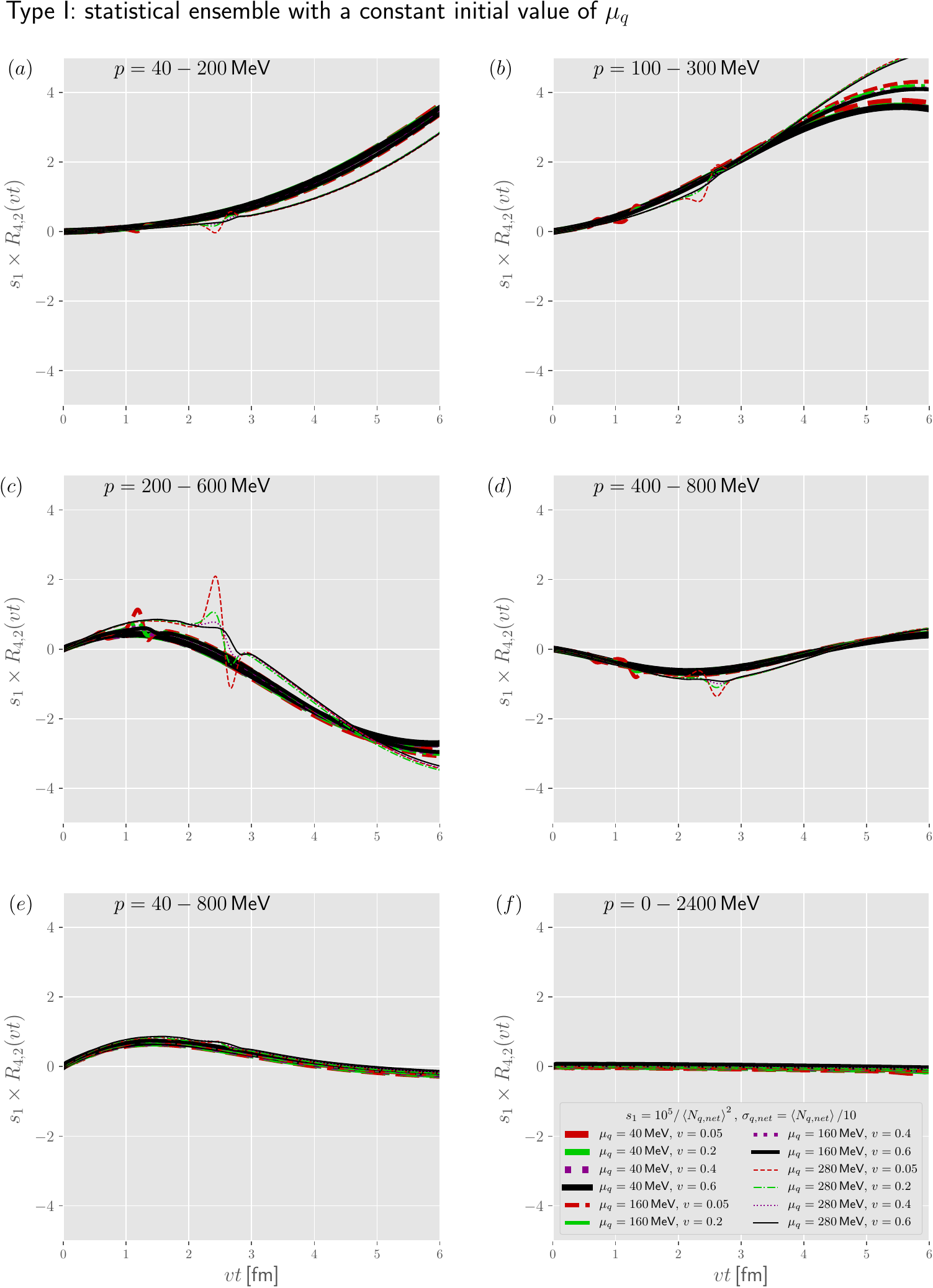}
\caption{Dynamical evolution of the rescaled cumulant ratio $R_{4,2}$ in
  different momentum ranges as a function of the growing radius scale
  $vt=:\Delta R\left (t\right )$ in case of the statistical ensemble of
  type I.  Different expansion velocities are shown with different
  colors and the thickness of curves refer to different values of the
  quark chemical potential
  $\mu_{\text{q}}\left (\left \langle N_{\text{q,net}}\right \rangle
  \right )$.}
\label{fig:R42_RFull}
\end{figure*}

We now return to the discussion of statistical ensembles of type I with
$\sigma_{\text{q,net}}=\left \langle N_{\text{q,net}}\right \rangle /10$
and different initial values of the quark chemical potential
$\mu_{\text{q}}=\mu_{\text{q}}\left (\left \langle
    N_{\text{q,net}}\right \rangle \right )$. We choose three different
initial chemical potentials: (1) $\mu_q(t_0=0)=40 \, \MeV$ and thus
$\left \langle N_{\text{q,net}}\right \rangle = 122.7$, (2)
$\mu_q(t_0=0)=160 \, \MeV$ and thus
$\left \langle N_{\text{q,net}}\right \rangle =544.15$, (3)
$\mu_q(t_0=0)=280\, \MeV$ and thus
$\left \langle N_{\text{q,net}}\right \rangle = 1158.4$.
Fig.~\ref{fig:R42_RFull} shows the rescaled cumulant ratio
$s_1\times R_{4,2}$ for various momentum ranges, where the color-marked
curves refer to different expansion velocities. Furthermore, the curve
thickness refers to different choices of $\mu_{\text{q}}$. Since the
results for $R_{4,2}$ vary in absolute values, they are rescaled by the
factor $s_1=10^5/\left \langle N_{\text{q,net}}\right \rangle ^2$,
accounting for the increasing net-quark number
$\left \langle N_{\text{q,net}}\right \rangle $ or the increasing
variance $\sigma_{\text{q,net}}$ alongside with an increasing chemical
potential $\mu_{\text{q}}$, which leads to the same range of the y-axis
for all shown simulation runs. We note, that only in case of the highest
net-quark density one observes critical behavior in the range of
$vt=2-3\,\fm$. However, this time scale is significantly below the
typical time scale for nearly vanishing interaction rates (``thermal
freezeout''), meaning that this signature of the critical point is not
observable at the end of the simulation.

From the thermal calculations of Figs.~\ref{fig:ThermalCumulantRatioR31}
and \ref{fig:ThermalCumulantRatioR42} and the discussion in this regard,
we know, that in thermal and chemical equilibrium the cumulant ratios
for our initial distribution function of net quarks are zero for type I
fluctuations, and converge to zero for decreasing temperatures for type
II fluctuations. In our dynamical simulations without chemical
equilibration we now observe significant deviations from zero, which
could be used as a weak indication for the phase transition in
experiments, when also the cumulant ratios of the baryon number
fluctuations deviate significantly from the crossover behavior with a
non-trivial dependence on the center-of-momentum energy as well as
centrality classes, being indeed observable in experiments (compare our
discussion in Sec.~\ref{chap:introduction}). However, generally for
type I initial conditions the cumulant ratios show only a weak
dependence on the expansion velocity, leading to significant
oscillations only in the vicinity of the phase transition and at highest
net-quark densities. Thereby, the oscillations are stronger pronounced
for smaller expansion velocities. Following that, a critical behavior at
the phase transition due to simple volume fluctuations of the initial
domain is hardly observable at the end of the simulation.

\begin{figure*}
\centering\includegraphics[width=14cm]{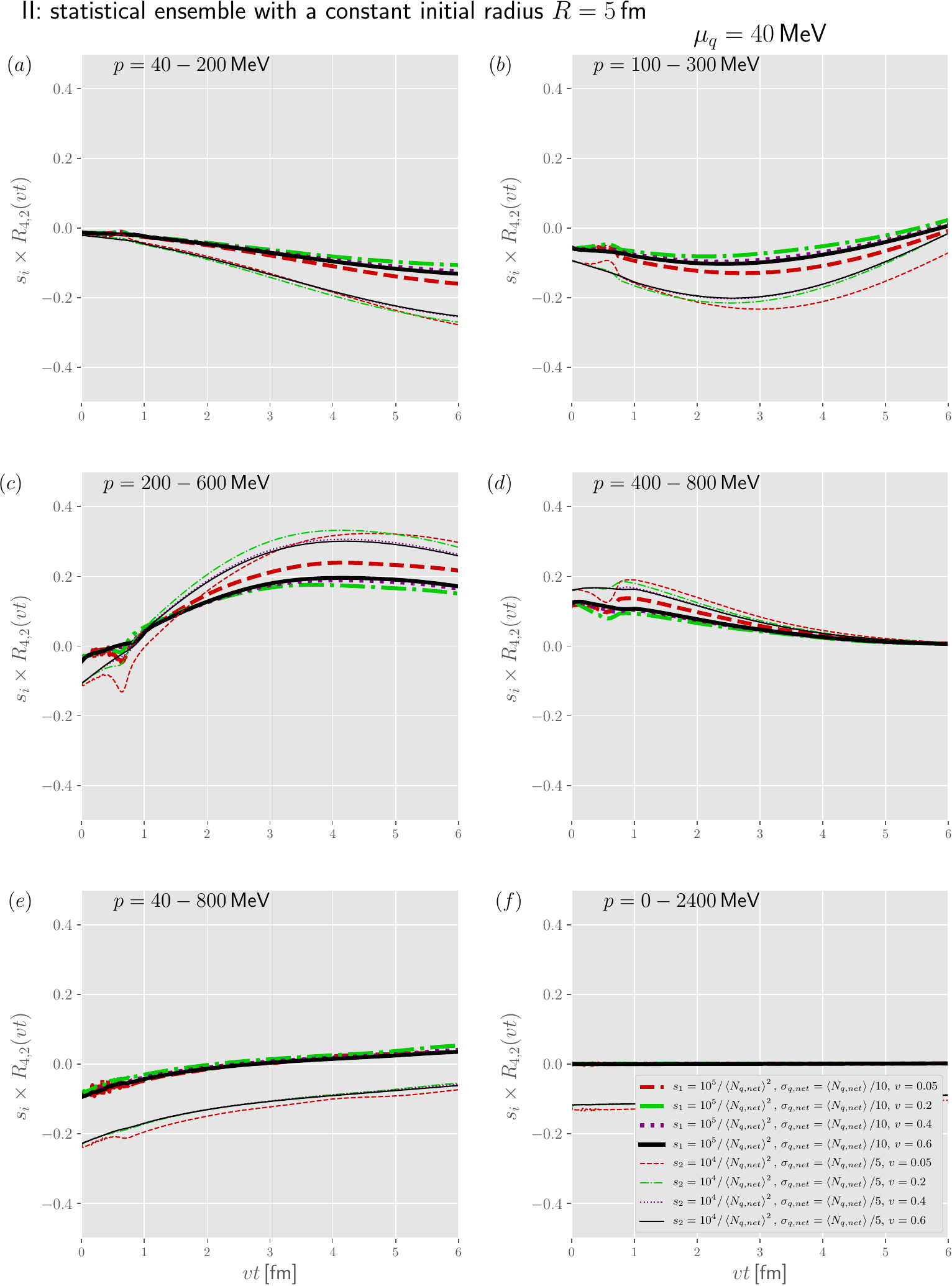}
\caption{Dynamical evolution of the rescaled cumulant ratio $R_{4,2}$ in
  different momentum ranges as a function of the growing radius scale
  $vt=:\Delta R\left (t\right )$ in case of the statistical ensemble of
  type II. Different expansion velocities are shown with different
  colors and the quark chemical potential of the mean trajectory is set
  to
  $\mu_{\text{q}}\left (\left \langle N_{\text{q,net}}\right \rangle
  \right )=40\,\MeV$.}
\label{fig:R42Mu40}
\end{figure*}

\begin{figure*}
\centering\includegraphics[width=14cm]{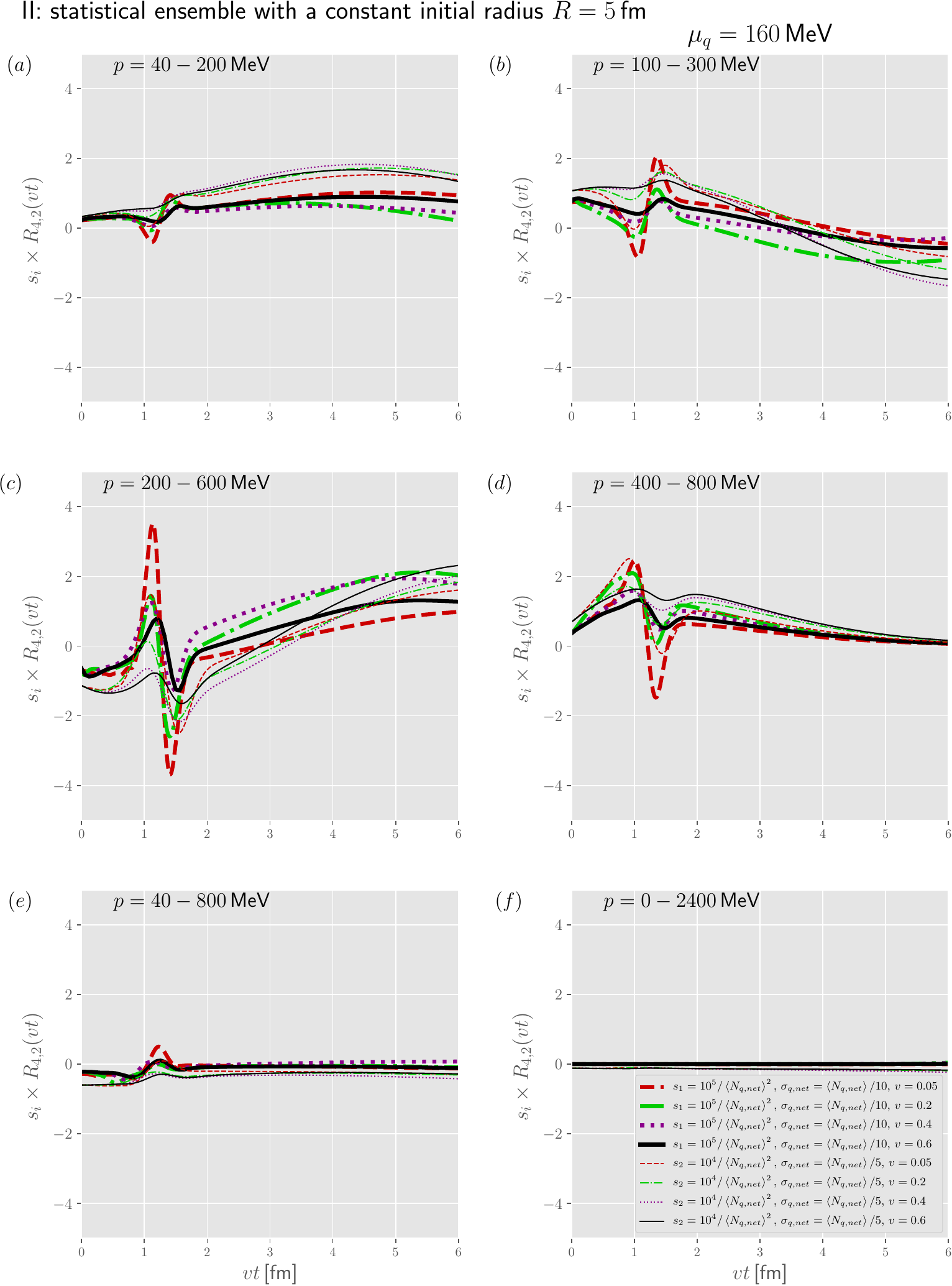}
\caption{Same as Fig.~\ref{fig:R42Mu40} but for
  $\mu_{\text{q}}\left (\left \langle N_{\text{q,net}}\right \rangle
  \right )=160\,\MeV$ as the quark chemical potential of the mean
  trajectory.}
\label{fig:R42Mu160}
\end{figure*}

\begin{figure*}
\centering\includegraphics[width=14cm]{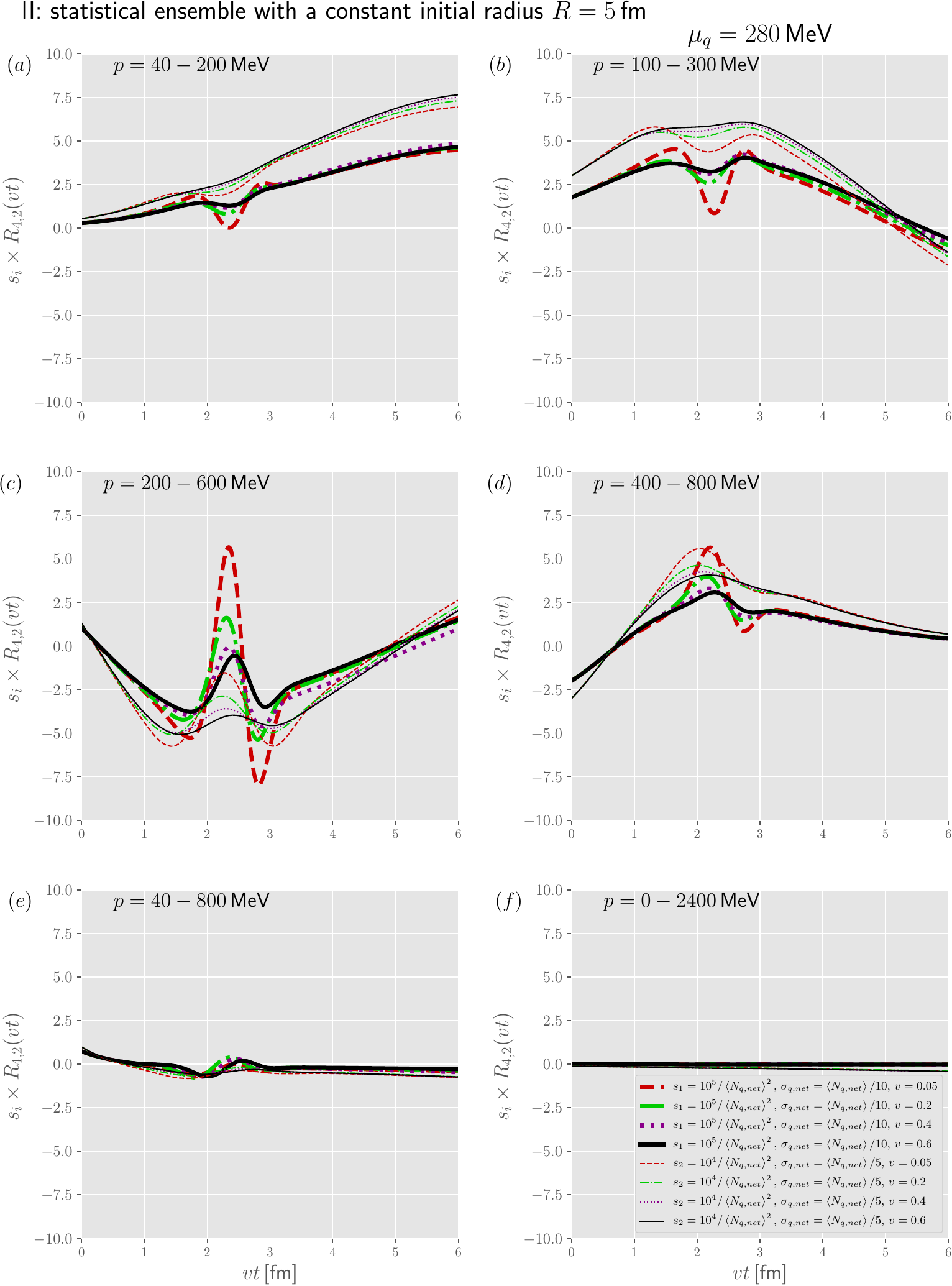}
\caption{Same as Fig.~\ref{fig:R42Mu40} but for
  $\mu_{\text{q}}\left (\left \langle N_{\text{q,net}}\right \rangle
  \right )=280\,\MeV$ as the quark chemical potential of the mean
  trajectory.}
\label{fig:R42Mu280}
\end{figure*}

Finally, we consider the rescaled cumulant ratio $s_i R_{4,2}$ for
initial configurations of type II, where the rescaling factor $s_i$ with
$i\in\{1,\,2\}$ accounts for the increasing net-quark number
$N_{\text{q,net}}$ as well as for different choices of the standard
deviation $\sigma_{\text{q,net}}$. The results are shown in
Figs.~\ref{fig:R42Mu40}, \ref{fig:R42Mu160} and \ref{fig:R42Mu280} for
different values of
$\mu_{\text{q}}\left (\left \langle N_{\text{q,net}}\right \rangle
\right )$, which can be referred to the crossover region for
$\mu_{\text{q}}=40\,\MeV$, the critical region of second order for
$\mu_{\text{q}}=160\,\MeV$ and the region of a first order phase
transition for $\mu_{\text{q}}=280\,\MeV$ (compare also with
Fig.~\ref{fig:PhaseDiagramTmuDynamicTypeII}). All calculations are
performed for two values of the standard deviation
$\sigma_{\text{q,net}}$ for Gaussian-type fluctuations of the net-quark
number. Thereby, the cumulant ratios are by an order of magnitude larger
in case of
$\sigma_{\text{q,net}}=\left \langle N_{\text{q,net}}\right \rangle /5$
(thin lines) compared to
$\sigma_{\text{q,net}}=\left \langle N_{\text{q,net}}\right \rangle /10$
(thick lines). Note also, that for the higher value of
$\sigma_{\text{q,net}}$ the cumulant ratio $R_{4,2}$ of the total
net-quark number (Subfig.~(f)) deviates from a vanishing one as expected
for a Gaussian distribution function, even though the full range
$\left [0,\,2\left \langle N_{\text{q,net}}\right \rangle \right ]$ of
the numerical distribution function \eqref{eq:GaussianDistribution} is
given by $10\sigma_{\text{q,net}}$, which again underlines the
requirement of an extremely high accuracy for the study of cumulants.
However, the qualitative behavior of our results is independent of
$\sigma_{\text{q,net}}$ and we should rather focus on details of the
dynamical evolution, depending on the expansion velocity.

For all simulations of type II the intermediate time scale of the most
pronounced fluctuations is highly correlated to the critical time scales
$\tau_{m_{\sigma,\,\mathrm{min.}}}$ (dynamical minimum of the
$\sigma$-meson mass) and $\tau_{\sigma\rightarrow q\bar q}$
($q\bar q$-production from sigma decay), as discussed above in
the context of Fig.~\ref{fig:PhiMass}. Furthermore, as expected
intuitively, we observe that the most pronounced fluctuations are formed
for the smallest expansion velocity $v=0.05 \, c$ (and one may regard as
a quasi-adiabatic expansion), since the system stays longer in the
critical region, which is also supported by the analysis of the relevant
time scales $\tau_{m_{\sigma,\,\mathrm{min.}}}$ as well as
$\tau_{\sigma\rightarrow q\bar q}$. Following that, our results confirm
that the dynamical evolution of the $\sigma$-meson mass has a large
impact on the fluctuating behavior of the net-quark number in different
momentum ranges. However, referring to real experiments the critical
effects from intermediate time scales cannot be directly observed and
one is restricted to the products of a heavy ion collision, meaning in
our case that the final realization at $vt\geq 6$ has to be considered,
resulting by construction in a final radius scale of $R\geq 11\,\fm$.
  
All our simulations start with $T=150\,\MeV$, leading to a time shift of
the phase transition to higher values of $vt$ with an increasing initial
chemical potential $\mu_{\text{q}}$ of the ensemble. Even so, the final
fluctuations from the simulation run with $\mu_{\text{q}}=160\,\MeV$,
which is expected to pass through the critical region of second order,
show the most interesting behavior. Here, the most pronounced deviations
from zero in the \emph{final state} are obtained for intermediate
expansion velocities $v=0.2-0.4$, whereas in case of a first order phase
transition the final results depend only weakly on the expansion
velocity for the most interesting momentum range of $p=200-600\,\MeV$,
allowing to distinguish, at least in numerical simulations, between
different orders of the phase transition. This indicates non-equilibrium
behavior in the critical region of second order due to an increasing
relaxation time, which is supported by the fact, that the strongest
fluctuations from intermediate time scales $vt=1-2\,\fm$ in case of the
smallest expansion velocity result in the most suppressed deviations
from zero in the final state. In detail, the system requires a longer
time to equilibrate and one observes the remnants of the critical
behavior at the phase transition, which now depends on the expansion
velocity in contrast to type I initial conditions (compare with
Fig.~\ref{fig:R42_RFull}). Following that, a successful observation of
strong fluctuations in cumulant ratios is an interplay between the
formation time of fluctuations, which is needed to be as long as
possible and the time scale, which is needed to wash out those
fluctuations.

We finalize our studies of the cumulant ratios by pointing out, that the
absolute value of the cumulant ratio $R_{4,2}$ increases continuously
with an increasing net-quark density in systems of type I and II. This
effect is significantly reduced by taking the rescaling factors
$s_1,\,s_2\sim 1/\left \langle N_{\text{q,net}}\right \rangle ^2$ into
account as discussed before. Nevertheless, an increase of absolute
values for $s_iR_{4,2}$ is still observable, which results from
non-linear scaling of the net-quark density with respect to the
increasing chemical potential $\mu_{\text{q}}$ as seen from
Fig.~\ref{fig:NetQuarkDensityThermal}. Although the fluctuations in an
expanding system out of equilibrium are less pronounced compared to the
expectations from equilibrium systems (see
Figs.~\ref{fig:ThermalCumulantRatioR31} and
\ref{fig:ThermalCumulantRatioR42}), we are able to confirm that a
significant deviation from the crossover behavior is observable, when
considering higher-order cumulant ratios of the net-quark number in
different momentum ranges, being thus a promising candidate for an
experimental signature of the chiral phase transition.

\pagebreak

\section{Conclusions and Outlook}

In this paper the evolution of net-baryon-number fluctuations in an
expanding fireball of strongly interacting matter similar as produced in
ultrarelativistic heavy-ion collisions at the Large-Hadron Collider
(CERN) or at the Relativistic Heavy Ion Collider (BNL) is
investigated. In thermal equilibrium the ``grand-canonical
fluctuations'' of the net-baryon number in different momentum and
rapidity ranges of produced hadrons are the most promising observables,
indicating phase transitions of the strongly interacting medium. From
lattice-QCD calculations it is known that at vanishing net-baryon number
($\mu_{\text{B}}=0$) both the deconfinement and the chiral transition
are smooth cross-over transitions occuring at a ``pseudo-critical
temperature'' of about $155 \; \text{MeV}$. From effective models at
$\mu_{\text{B}} \neq 0$ one expects a first-order transition line in the
$T$-$\mu_{\text{B}}$ phase diagram, ending in a critical point with a
second-order phase transition. In thermal equilibrium and in the
thermodynamic limit the susceptibilities of conserved quantum numbers,
like the net-baryon number, diverge at the critical point and thus are
the most promising candidates to observe these different types of phase
transitions and maybe constrain the location of the critical point in
the phase diagram.

However, the fireball of strongly interacting matter created in
heavy-ion collisions is rapidly expanding and cooling, which implies
that a realistic theoretical description has to take this
off-equilibrium situation as well as the finite volume and lifetime of
the fireball into account.

Thus, in this work the restoration of explicitly and spontaneously
broken chiral symmetry is studied with respect to the $\sigma$ field as
the order parameter of the chiral symmetry within a linear quark-meson
$\sigma$ model, describing pions, $\sigma$-mesons and constituent quarks
and anti-quarks. For this system, a set of coupled and self-consistent
evolution equations is derived in Sec.~\ref{chap:EvolEq} by considering
a truncation scheme of the underlying two-particle irreducible (2PI)
quantum effective action, which still contains the most relevant
diagrams, including also dissipation with and without memory effects,
arising from the interaction between the slowly changing chiral
mean-field and mesonic excitations. Applying a gradient expansion to the
exact evolution equations for one- and two-point functions of the
quark-meson model, one obtains generalized transport equations, allowing
to interpret the full dynamics of two-point functions in terms of
mesonic interactions. Thereby, the interaction between soft and hard
modes is encoded in a dissipation kernel, leading to memory effects,
which become particularly important near the critical point.

In this work a simplified homogeneous and isotropic ``Hubble-like'' expanding
fireball is used to solve the derived set of semiclassical kinetic and
mean-field equations, including memory effects in terms of a
non-Markovian evolution of the $\sigma$ field, i.e., the order parameter
of chiral symmetry coupled to a set of Boltzmann-Uehling-Uhlenbeck-type
transport equations for the phase-space-distribution functions of the
$\sigma$-mesons, pions, quarks and anti-quarks. 

To quantify the fluctuations it is necessary to evaluate the dynamical
evolution of the higher-order cumulants of the net-baryon-number
distribution and their ratios, which is numerically challenging within
the usually employed test-particle Monte Carlo approach. Therefore a
numerical integration method for the coupled set of integro-differential
equations for two statistical ensembles with initial conditions of
Gaussian-type fluctuations of the net-quark number, differing in volume
and particle-density fluctuations has been developed.

Simulating the expanding system with different initial conditions and
net-baryon densities leads to a dynamical evolution of the medium along
different effective trajectories in the phase diagram of the strongly
interacting medium, which can stay within the cross-over region as well
as run through a first-order phase transition or close to the
second-order phase transition at the critical point. Depending on the
effective fireball-expansion rate and for trajectories across the
first-order-transition line or close to the critical point, larger
fluctuations, as quantified by the higher cumulants and cumulant ratios,
can build up, but are also damped out again in the later stages. The
simulations show that the dynamical evolution of the $\sigma$ mass
thereby has a large impact on the behavior of the net-quark-number
functuations in different momentum ranges, since these fluctuations are
highly correlated to the critical time scale to reach the dynamical
minimum of the $\sigma$ mass during the expansion process. At low
expansion rates, i.e., under quite ``adiabatic conditions'', the
fluctuations of the net-baryon number can be observed in the final
cumulant ratios in the range of small and intermediate momenta up to
$\sim 600 \,\MeV$.

From the point of view of observations in heavy-ion collisions the
measurable cumulant ratios may be used to distinguish between a
dynamical evolution of the fireball with the medium undergoing different
types of transition (i.e., cross-over or first- and second-order phase
transitions), provided the fireball expansion rate is in an intermediate
range, such that on one hand off-equilibrium effects in the vicinity of
the first-order-transition line or the critical point become strong but
on the other hand, due to the enhanced relaxation times towards the
range of low $\sigma$ masses characterizing this region of the phase
diagram, the system stays for a sufficient time close to the transition
line or the critical point. The simulations indeed show that the
cumulant ratios in the final state, observable in heavy-ion collisions,
are quite sensitive to the fireball-expansion velocity which makes a
proper interperation and quantitative determination of the location of
the critical point in the phase diagram, based on a measurement of the
cumulants and cumulant ratios of the net-baryon number, challenging.

For a more realistic description of the fireball evolution in heavy-ion
collisions an extension of the numerical solution of the set of kinetic
equations for anistropic distribution functions in a (2+1)-dimensional
Bjorken-like scenario would be desirable, which however makes a further
increase in computational performance necessary.

\section*{Acknowledgment}

The authors are grateful to the LOEWE Center for Scientific Computing
(LOEWE-CSC) at Frankfurt for providing computing resources. We
acknowledge support by the Deutsche Forschungsgemeinschaft (DFG, German
Research Foundation) through the CRC-TR 211 `Strong-interaction matter
under extreme conditions' -- project number 315477589 -- TRR 211.

\appendix

\section{Symmetry and multiplicity of diagrams}
\label{chap:app} The multiplicity of a diagram in QFT is directly
connected to the number of all possible contractions of two fields,
diveded by the number of identical contractions. Different diagrams
follow from the generating functional by applying derivatives with
respect to external sources. However, depending on the theory it becomes
a tedious procedure since the multiplicity of diagrams grows extremely
fast with the number of vertices. To avoid such calculations one can use
simple combinatorical arguments, leading to a formula which simplifies
enormously the calculation of higher order diagrams and is applicable to
a large class of theories. The basic expression of this formula for
$\mathrm{O}\left (4\right )$ theories\footnote{As seen in the following, the
formula \eqref{eq:a01} is also applicable to the Yukawa part of the
Lagrangian \eqref{eq:LinSigmaModell}.} has the following structure
\cite{PhysRevE.62.1537,book:16401}:
\begin{equation}
\begin{split}
M&=\frac{\left(cM_v\right)^{n}n!n_e!}{\left(2!\right)^{s+d}\left(3!\right)^t\left(4!\right)^fN_v}\,.
 \label{eq:a01}
\end{split}
\end{equation}
\begin{table}[h!]  \centering
\begin{tabular}{|p{2.0cm}|p{10.0cm}|} \hline parameter & description \\
\hline \hline $c$ & prefactor of the vertex from the expansion of the
Lag\-rangian \\ \hline $M_v$ & multiplicity of the vertex (product of
permutations for equal fields) \\ \hline $n$ & number of vertices \\
\hline $n_e$ & number of external lines \\ \hline $s,d,t,f$ & number of
self-, double-, triple- and four-field connections, leaving the diagram
unchanged \\ \hline $N_v$ & number of vertex permutations, leading to
the same diagram (vertices are connected to the same lines)\\ \hline
\end{tabular}
\label{tab:601}
\end{table}
\begin{figure} \captionsetup[subfloat]{labelformat=empty}
\text{without $\left <\sigma \right >$:}\qquad 
\qquad\subfloat[$1$]{\includegraphics[width=2.3cm]{vertex4s}}\qquad\quad
\subfloat[$2$]{\includegraphics[width=2.3cm]{vertex2s2p}}\qquad\quad
\subfloat[$1$]{\includegraphics[width=2.3cm]{vertex4p}}\qquad\quad
\subfloat[$2$]{\includegraphics[width=2.3cm]{vertex2p2p}}\\
\subfloat[$4$]{\includegraphics[width=3.1cm]{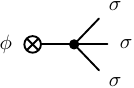}}\qquad
\subfloat[$4$]{\includegraphics[width=3.1cm]{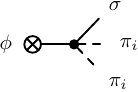}}\qquad
\subfloat[$1$]{\includegraphics[width=2.3cm]{vertex1s2q}}\qquad\quad
\subfloat[$1$]{\includegraphics[width=2.3cm]{vertex1p2q}}
\caption{Relevant vertices for the quantum effective action
\eqref{eq:2PIfull}.}
\label{fig:app101}
\end{figure}

Usually, one defines the symmetry factor $S_{n}$ of a diagram as a product of
the diagramatic multiplicity with a weight factor $w_{n}$, denoting the
prefactor from the power series of the generating functional for
connected diagrams (as discussed for instance in
Sec.~\ref{sec:1PIAction} and Sec.~\ref{sec:2PIAction}) with respect to
the coupling constant $\alpha$ of the theory: 
\begin{equation}
W=\sum_{n}\frac{1}{n!}\left (V_\alpha\right )^{n}W^{\left (n\right )}=:\sum_n\left (-i\alpha\right )^nw_nW^{\left (n\right )}\,.
\label{eq:a02a} 
\end{equation} 
In case of the Lagrangian \eqref{eq:LinSigmaModell},
there are two expressions of the form $V_\alpha$. The first one is given
by the Yukawa interaction between bosonic $\sigma,\vec\pi$- and
fermionic $\psi_i,\bar\psi_i$-fields, whereas the second treats the pure
$\phi^4$ interaction of $\sigma,\pi$-fields:
\begin{equation}
\begin{split} V_g&=-\ii g\,,\quad
w_n=\frac{1}{n!}\qquad\,\quad\Rightarrow\qquad S_{n}=\frac{1}{n!}M\,,\\
V_\lambda&=\frac{-\ii \lambda}{4}\,,\quad
w_{n}=\frac{1}{4^{n}n!}\qquad\Rightarrow\qquad S_n=\frac{1}{4^nn!}M\,.
\label{eq:a03a}
\end{split} \end{equation}
Now, we apply the formula \eqref{eq:a01} to the truncation of $\Gamma_2$
as shown in Fig.~\ref{fig:2PIpart} and derive the multiplicity factors
for all relevant diagrams.  The results are summerized in
Tab.~\ref{tab:app01} and \ref{tab:app02}, thereby the different vertices
for the $\mathrm{O}\left (N\right )$ part of the Lagrangian
\eqref{eq:LinSigmaModell} can be extracted from the expansion of the
interaction part: 
\begin{equation}
  V\left (\sigma,\vec\pi\right )=\frac{\lambda}{4}\left (\sigma^2+\vec\pi^2-\nu^2\right )^2=\frac{\lambda}{4}\left
    (\sigma^4+2\sigma^2\sum_{i}\pi_i^2-2\nu^2\sigma^2+\sum_{i}\pi_i^4 +
    2\sum_{i,j>i}\pi_i^2\pi_j^2-2\nu^2\sum_{i}\pi_i^2+\nu^4\right )\,.
\label{eq:InteractionPartExpansion}
\end{equation} 
From considering individual terms with respect to the transformation
$\sigma\rightarrow\sigma+\phi$, where $\phi$ denotes a finite value for
the mean-field part of the $\sigma$-field\footnote{Analogously, also
  finite mean-field values for pionic fields can be introduced, but they
  are not considered in this paper.}, one derives all possible bosonic
vertices. Furthermore, there are $2$ additional vertices, coming from
the interaction between bosonic and fermionic fields.  Consequently, one
obtains for the 2PI quantum effective action $8$ relevant vertices (see
Sec.~\ref{sec:2PIAction} for more details) with their prefactors given
directly by the expansion \eqref{eq:InteractionPartExpansion}, as shown
in Fig.~\ref{fig:app101}.
\begin{sidewaystable} \centering
\begin{tabular}{|p{3.1cm}|p{11.0cm}|p{2.25cm}|p{2.25cm}|p{1.2cm}|}
\hline type & vertex/diagram/expression & $\left (c, M_v, n, n_e\right )$ &
$\left (s, d, t, f, N_v\right )$ & M, S \\ \hline \hline Hartree:\\
$4\sigma$-vertex &
\begin{minipage}{6cm}
\includegraphics[width=1.8cm]{vertex4s}
\includegraphics[width=1.8cm]{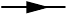}
\includegraphics[width=1.8cm]{hartree4s}
\end{minipage}
\begin{minipage}{3.5cm} \footnotesize{
\begin{equation*} \sim -\frac{3}{4}\lambda\int_{\mathcal
C}\,G_{\sigma\sigma}^2
\end{equation*}}
\end{minipage} & $\left (1, 4!, 1, 0\right )$ & $\left (2, 1, 0, 0, 1\right )$ & $3,
3/4$\\ \hline Hartree:\\ $2\sigma-2\pi_i$-vertex &
\begin{minipage}{6cm}
\includegraphics[width=1.8cm]{vertex2s2p}
\includegraphics[width=1.8cm]{arrowright}
\includegraphics[width=1.8cm]{hartree2s2p}
\end{minipage}
\begin{minipage}{3.5cm} \footnotesize{
\begin{equation*} \sim -\frac{1}{2}\lambda\sum_i\int_{\mathcal
C}\,G_{\sigma\sigma}G_{\pi_i\pi_i}
\end{equation*}}
\end{minipage} & $\left (2, 4\,, 1, 0\right )$ & $\left (2, 0, 0, 0, 1\right )$ & $2,
1/2$\\ \hline Hartree: \\ $4\pi_i$-vertex &
\begin{minipage}{6cm}
\includegraphics[width=1.8cm]{vertex4p}
\includegraphics[width=1.8cm]{arrowright}
\includegraphics[width=1.8cm]{hartree4p}
\end{minipage}
\begin{minipage}{3.5cm} \footnotesize{
\begin{equation*} \sim -\frac{3}{4}\lambda\sum_i\int_{\mathcal
C}\,G_{\pi_i\pi_i}^2
\end{equation*}}
\end{minipage} & $\left (1, 4!\,, 1, 0\right )$ & $\left (2, 1, 0, 0, 1\right )$ & $3,
3/4$\\ \hline Hartree: $i\neq j$\\ $2\pi_i-2\pi_j$-vertex &
\begin{minipage}{6cm}
\includegraphics[width=1.8cm]{vertex2p2p}
\includegraphics[width=1.8cm]{arrowright}
\includegraphics[width=1.8cm]{hartree2p2p}
\end{minipage}
\begin{minipage}{4.5cm} \footnotesize{
\begin{equation*} \sim -\frac{1}{2}\lambda\sum_{i,j>i}\int_{\mathcal
C}\,G_{\pi_i\pi_i}G_{\pi_j\pi_j}
\end{equation*}}
\end{minipage} & $\left (2, 4\,, 1, 0\right )$ & $\left (2, 0, 0, 0, 1\right )$ & $2,
1/2$\\ \hline Sunset: \\ $\phi\sigma-2\sigma$-vertex &
\begin{minipage}{6cm}
\includegraphics[width=1.8cm]{vertex1m3s}
\includegraphics[width=1.8cm]{arrowright}
\includegraphics[width=1.8cm]{sunset1m3s}
\end{minipage}
\begin{minipage}{4.5cm} \footnotesize{
\begin{equation*} \sim 3\ii \lambda^2\int_{\mathcal C}\,\int_{\mathcal
C}\,\phi\,G_{\sigma\sigma}^3\,\phi
\end{equation*}}
\end{minipage} & $\left (4, 3!\,, 2, 0\right )$ & $\left (0, 0, 1, 0, 2\right )$ & $96,
3$\\ \hline Sunset: \\ $\phi\sigma-2\pi_i$-vertex &
\begin{minipage}{6cm}
\includegraphics[width=1.8cm]{vertex1m1s2p}
\includegraphics[width=1.8cm]{arrowright}
\includegraphics[width=1.8cm]{sunset1m1s2p}
\end{minipage}
\begin{minipage}{4.5cm} \footnotesize{
\begin{equation*} \sim \ii \lambda^2\sum_{i}\int_{\mathcal
C}\,\int_{\mathcal C}\,\phi\,G_{\sigma\sigma}G_{\pi_i\pi_i}^2\,\phi
\end{equation*}}
\end{minipage} & $\left (4, 2\,, 2, 0\right )$ & $\left (0, 1, 0, 0, 2\right )$ & $32,
1$\\ \hline
\end{tabular}
\caption{Multiplicities of Hartree and bosonic sunset diagrams.}
\label{tab:app01}
\end{sidewaystable}

\begin{sidewaystable} \centering
\begin{tabular}{|p{3.1cm}|p{11cm}|p{2.25cm}|p{2.25cm}|p{1.2cm}|} \hline
type & vertex/diagram/expression & $\left (c, M_v, n, n_e\right )$ & $\left (s, d, t,
f, N_v\right )$ & M, S \\ \hline \hline Basketball:\\ $4\sigma$-vertex &
\begin{minipage}{6cm}
\includegraphics[width=1.8cm]{vertex4s}
\includegraphics[width=1.8cm]{arrowright}
\includegraphics[width=1.8cm]{basketball4s}
\end{minipage}
\begin{minipage}{3.5cm} \footnotesize{
\begin{equation*} \sim \frac{3}{4}\ii \lambda^2\int_{\mathcal
C}\,\int_{\mathcal C}\,G_{\sigma\sigma}^4
\end{equation*}}
\end{minipage} & $\left (1, 4!, 2, 0\right )$ & $\left (0, 0, 0, 1, 2\right )$ & $24,
3/4$\\ \hline Basketball:\\ $2\sigma-2\pi_i$-vertex &
\begin{minipage}{6cm}
\includegraphics[width=1.8cm]{vertex2s2p}
\includegraphics[width=1.8cm]{arrowright}
\includegraphics[width=1.8cm]{basketball2s2p}
\end{minipage}
\begin{minipage}{3.5cm} \footnotesize{
\begin{equation*} \sim \frac{1}{2}\ii \lambda^2\sum_i\int_{\mathcal
C}\,\int_{\mathcal C}\,G_{\sigma\sigma}^2G_{\pi_i\pi_i}^2
\end{equation*}}
\end{minipage} & $\left (2, 4\,, 2, 0\right )$ & $\left (0, 0, 0, 1, 2\right )$ & $16,
1/2$\\ \hline Basketball: \\ $4\pi_i$-vertex &
\begin{minipage}{6cm}
\includegraphics[width=1.8cm]{vertex4p}
\includegraphics[width=1.8cm]{arrowright}
\includegraphics[width=1.8cm]{basketball4p}
\end{minipage}
\begin{minipage}{3.5cm} \footnotesize{
\begin{equation*} \sim \frac{3}{4}\ii \lambda^2\sum_i\int_{\mathcal
C}\,\int_{\mathcal C}\,G_{\pi_i\pi_i}^4
\end{equation*}}
\end{minipage} & $\left (1, 4!\,, 2, 0\right )$ & $\left (0, 0, 0, 1, 2\right )$ & $24,
3/4$\\ \hline Basketball: $i\neq j$\\ $2\pi_i-2\pi_j$-vertex &
\begin{minipage}{6cm}
\includegraphics[width=1.8cm]{vertex2p2p}
\includegraphics[width=1.8cm]{arrowright}
\includegraphics[width=1.8cm]{basketball2p2p}
\end{minipage}
\begin{minipage}{4.5cm} \footnotesize{
\begin{equation*} \sim \frac{1}{2}\ii \lambda^2\sum_{i,j>i}\int_{\mathcal
C}\,\int_{\mathcal C}\,G_{\pi_i\pi_i}^2G_{\pi_j\pi_j}^2
\end{equation*}}
\end{minipage} & $\left (2, 4\,, 2, 0\right )$ & $\left (0, 0, 0, 1, 2\right )$ & $16,
1/2$\\ \hline Sunset: \\ $\sigma-q\bar q$-vertex &
\begin{minipage}{6cm}
\includegraphics[width=1.8cm]{vertex1s2q}
\includegraphics[width=1.8cm]{arrowright}
\includegraphics[width=1.8cm]{sunset1s2q}
\end{minipage}
\begin{minipage}{4.5cm} \scriptsize{
\begin{equation*} \sim -\frac{1}{2}ig^2\int_{\mathcal C}\,\int_{\mathcal
C}\,\text{Tr}\left [D_{xy}D_{yx}\right ]\,G_{\sigma\sigma}
\end{equation*}}
\end{minipage} & $\left (1, 1\,, 2, 0\right )$ & $\left (0, 0, 0, 0, 2\right )$ & $1,
1/2$\\ \hline Sunset: \\ $\pi_i-q\bar q$-vertex &
\begin{minipage}{6cm}
\includegraphics[width=1.8cm]{vertex1p2q}
\includegraphics[width=1.8cm]{arrowright}
\includegraphics[width=1.8cm]{sunset1p2q}
\end{minipage}
\begin{minipage}{4.5cm} \scriptsize{
\begin{equation*} \sim -\frac{1}{2}ig^2\sum_i\int_{\mathcal
C}\,\int_{\mathcal C}\,\text{Tr}\left [D_{xy}D_{yx}\right ]\,G_{\pi_i\pi_i}
\end{equation*}}
\end{minipage} & $\left (1, 1\,, 2, 0\right )$ & $\left (0, 0, 0, 0, 2\right )$ & $1,
1/2$\\ \hline
\end{tabular}
\caption{Multiplicities of basketball and fermionic sunset diagrams.}
\label{tab:app02}
\end{sidewaystable}

\subsection{Self-energies and collision integrals}

\label{chap:app2} In this chapter we calculate all relevant
selfenergies, which arise from the analysis of the diagrams given in
Tab. \ref{tab:app01} and \ref{tab:app02}.

\textbf{1. Hartree diagrams (h.)}
\begin{equation}
  \begin{split} \Gamma_2^{\text{h.}}=\,&-\frac{3}{4}\lambda\int_{\mathcal
      C}\dd^4x\,G_{\sigma\sigma}^2\left (x,x\right )-\frac{1}{2}\lambda\sum_i\int_{\mathcal
      C}\dd^4x\,G_{\sigma\sigma}\left (x,x\right )G_{\pi_i\pi_i}\left (x,x\right )\\\
    &-\frac{3}{4}\lambda\sum_i\int_{\mathcal
      C}\dd^4x\,G_{\pi_i\pi_i}^2\left (x,x\right )-\frac{1}{2}\lambda\sum_{i,j>i}\int_{\mathcal
      C}\dd^4x\,G_{\pi_i\pi_i}\left (x,x\right )G_{\pi_j\pi_j}\left (x,x\right )\,.
\label{eq:GammaHartree}
\end{split} 
\end{equation} 
Using the relations
\eqref{eq:ProperSelfEnergy} one obtains for the self-energies: 
\begin{equation}
  \begin{split} \Pi_{\sigma\sigma}\left (x,x\right )=&\,-3\ii\lambda
    G_{\sigma\sigma}\left (x,x\right )-\ii \lambda \sum_i G_{\pi_i\pi_i}\left (x,x\right )\\
    \Pi_{\pi_i\pi_i}\left (x,x\right )=&\,-3\ii\lambda
    G_{\pi_i\pi_i}\left (x,x\right )-\ii \lambda\sum_{j\neq
      i}G_{\pi_j\pi_j}\left (x,x\right )-\ii \lambda G_{\sigma\sigma}\left (x,x\right )\,.
\label{eq:SelfEnergyHartree}
\end{split} 
\end{equation} 
Obviously, the self-energies from Hartree diagrams are
local contributions to the full expression of the proper self-energy and
can be interpreted as an effective mass (see Sec.~\ref{sec:EffMass}).

\textbf{2. Bosonic sunset diagrams (b.s.)} 
\begin{equation}
\begin{split} \Gamma_2^{\text{b.s.}}=\,&3\ii\lambda^2\int_{\mathcal
C}\dd^4x\int_{\mathcal
C}\dd^4y\,\phi\left (x\right )\,G_{\sigma\sigma}^3\left (x,y\right )\,\phi\left (y\right )\\ &
+\ii \lambda^2\sum_{i}\int_{\mathcal C}\dd^4x\int_{\mathcal
C}\dd^4y\,\phi\left (x\right )\,G_{\sigma\sigma}\left (x,y\right )G_{\pi_i\pi_i}^2\left (x,y\right )\,\phi\left (y\right )\,.
\label{eq:b01a}
\end{split} 
\end{equation} 
In analogy to \eqref{eq:SelfEnergyHartree} one obtains
for the self-energies: 
\begin{equation}
\begin{split}
  i\Pi_{\sigma\sigma}^{<,>}\left (x,y\right )&=18\lambda^2\phi\left (x\right )\left (iG_{\sigma\sigma}^{<,>}\left (x,y\right )\right )^2\phi\left (y\right )+2\lambda^2\sum_i\phi\left (x\right )\left (iG_{\pi_i\pi_i}^{<,>}\left (x,y\right )\right )^2\phi\left (y\right )\,,\\
  i\Pi_{\pi_i\pi_i}^{<,>}\left (x,y\right )&=4\lambda^2\phi\left
    (x\right )\left (iG_{\sigma\sigma}^{<,>}\left (x,y\right )\right
  )\left (iG_{\pi_i\pi_i}^{<,>}\left (x,y\right )\right )\phi\left
    (y\right )\,.
\label{eq:SelfEnergyBosonicSunsetDiagram}
\end{split} 
\end{equation} 
After introducing auxiliary functions over an inner loop,
\begin{equation}
\begin{split}
\tilde{\mathcal{I}}_{G_{\sigma\sigma}^2}^{<,>}\left (X,k\right ):=&\int\dd^4\Delta
x\,\ee^{\ii k^\mu\Delta x_\mu}\left (iG_{\sigma\sigma}^{<,>}\left (x,y\right )\right )^2\\
=&\int\frac{\dd^4
p_3}{\left (2\pi\right )^4}\ii \tilde{G}_{\sigma\sigma}^{<,>}\left (X,p_3\right )\ii \tilde{G}_{\sigma\sigma}^{<,>}\left (X,k-p_3\right )\,,\\
\tilde{\mathcal{I}}_{G_{\pi_i\pi_i}^2}^{<,>}\left (X,k\right ):=&\int\dd^4\Delta
x\,\ee^{\ii k^\mu\Delta x_\mu}\left (iG_{\pi_i\pi_i}^{<,>}\left (x,y\right )\right )^2\\
=&\int\frac{\dd^4
p_3}{\left (2\pi\right )^4}\ii \tilde{G}_{\pi_i\pi_i}^{<,>}\left (X,p_3\right )\ii \tilde{G}_{\pi_i\pi_i}^{<,>}\left (X,k-p_3\right )\,,\\
\tilde{\mathcal{I}}_{G_{\sigma\pi_i}^2}^{<,>}\left (X,k\right ):=&\int\dd^4\Delta
x\,\ee^{\ii k^\mu\Delta
x_\mu}\left (iG_{\sigma\sigma}^{<,>}\left (x,y\right )\right )\left (iG_{\pi_i\pi_i}^{<,>}\left (x,y\right )\right )\\
=&\int\frac{\dd^4
p_3}{\left (2\pi\right )^4}\ii \tilde{G}_{\sigma\sigma}^{<,>}\left (X,p_3\right )\ii \tilde{G}_{\pi_i\pi_i}^{<,>}\left (X,k-p_3\right )\,,
\label{eq:AuxFunc}
\end{split} \end{equation}

the Wigner transforms of the self-energies become:
\begin{equation}
\begin{split}
  \ii \tilde\Pi_{\sigma\sigma}^{<,>}\left (X,p_1\right
  )&=18\lambda^2\int\dd^4\Delta x\,\ee^{\ii p_1^\mu\Delta
    x_\mu}\phi\left (x\right )\left (iG_{\sigma\sigma}^{<,>}\left (x,y\right )\right )^2\phi\left (y\right )\\
  &\quad+2\lambda^2\sum_i\int\dd^4\Delta x\,\ee^{\ii p_1^\mu\Delta
    x_\mu}\phi\left (x\right )\left (iG_{\pi_i\pi_i}^{<,>}\left (x,y\right )\right )^2\phi\left (y\right )\\
  &=18\lambda^2\int\dd^4\Delta x\,\ee^{\ii p_1^\mu\Delta x_\mu}\phi\left
    (x\right )\phi\left (y\right )\left [\int\frac{\dd^4k}{\left
        (2\pi\right )^4}\,\ee^{-\ii k^\mu\Delta
      x_\mu}\,\tilde{\mathcal{I}}_{G_{\sigma\sigma}^2}^{<,>}\left (X,k\right )\right ]\\
  &\quad+2\lambda^2\sum_i\int\dd^4\Delta x\,\ee^{\ii p_1^\mu\Delta
    x_\mu}\phi\left (x\right )\phi\left (y\right )\left
    [\int\frac{\dd^4k}{\left (2\pi\right )^4}\,\ee^{-\ii k^\mu\Delta
      x_\mu}\,\tilde{\mathcal{I}}_{G_{\pi_i\pi_i}^2}^{<,>}\left (X,k\right )\right ]\\
  &=18\lambda^2\int\dd^4\Delta x\,\phi\left (x\right )\phi\left
    (x-\Delta x\right )\left [\int\frac{\dd^4k}{\left (2\pi\right
      )^4}\,\ee^{-\ii \left (k-p_1\right )^\mu\Delta
      x_\mu}\,\tilde{\mathcal{I}}_{G_{\sigma\sigma}^2}^{<,>}\left (X,k\right )\right ]\\
  &\quad+2\lambda^2\sum_i\int\dd^4\Delta x\,\phi\left (x\right
  )\phi\left (x-\Delta x\right )\left [\int\frac{\dd^4k}{\left
        (2\pi\right )^4}\,\ee^{-\ii \left (k-p_1\right )^\mu\Delta
      x_\mu}\,\tilde{\mathcal{I}}_{G_{\pi_i\pi_i}^2}^{<,>}\left
      (X,k\right )\right ]\,,
\label{eq:b02b}
\end{split} \end{equation}
\begin{equation}
\begin{split}
  \ii \tilde\Pi_{\pi_i\pi_i}^{<,>}\left (X,p_1\right
  )&=4\lambda^2\int\dd^4\Delta x\,\ee^{\ii p_1^\mu\Delta
    x_\mu}\phi\left (x\right )\left (iG_{\sigma\sigma}^{<,>}\left (x,y\right )\right )\left (iG_{\pi_i\pi_i}^{<,>}\left (x,y\right )\right )\phi\left (y\right )\\
  &=4\lambda^2\int\dd^4\Delta x\,\phi\left (x\right )\phi\left (x-\Delta
    x\right )\left [\int\frac{\dd^4k}{\left (2\pi\right )^4}\,\ee^{-\ii \left
        (k-p_1\right )^\mu\Delta
      x_\mu}\,\tilde{\mathcal{I}}_{G_{\sigma\pi_i}^2}^{<,>}\left
      (X,k\right )\right ]\,.
\label{eq:b02c}
\end{split} \end{equation}

Now, we explicitly calculate the expressions in \eqref{eq:AuxFunc} by
using the on-shell ansatz for the propagators in Wigner space
\eqref{eq:WignerPropagatorBoson}
\begin{equation}
\begin{split} 
\tilde{\mathcal{I}}_{G_{\sigma\sigma}^2}^{<}\left (X,k^0,\vec
k\right )\simeq &\,\,\frac{\pi}{2}\int\frac{\dd^3\vec
p_3}{\left (2\pi\right )^3}\int\dd^3\vec p_4\frac{\delta^{\left (3\right )}\left (\vec k -\vec
p_3-\vec p_4\right )}{E_3^\sigma E_4^\sigma}\\ &\times\Big\{\left [f_3^\sigma
f_4^\sigma\right ]\delta\left (k^0-E_3^\sigma-E_4^\sigma\right )\\ &\qquad
+\left [\left (1+f_3^\sigma\right )f_4^\sigma\right ]\delta\left (k^0+E_3^\sigma-E_4^\sigma\right )\\
&\qquad
+\left [f_3^\sigma\left (1+f_4^\sigma\right )\right ]\delta\left (k^0-E_3^\sigma+E_4^\sigma\right )\\
&\qquad
+\left [\left (1+f_3^\sigma\right )\left (1+f_4^\sigma\right )\right ]\delta\left (k^0+E_3^\sigma+E_4^\sigma\right )\Big\}\,,
\label{eq:b05a}
\end{split} 
\end{equation} 
\begin{equation}
\begin{split} \tilde{\mathcal{I}}_{G_{\sigma\sigma}^2}^{>}\left (X,k^0,\vec
k\right )\simeq &\,\,\frac{\pi}{2}\int\frac{\dd^3\vec
p_3}{\left (2\pi\right )^3}\int\dd^3\vec p_4\frac{\delta^{\left (3\right )}\left (\vec k -\vec
p_3-\vec p_4\right )}{E_3^\sigma E_4^\sigma}\\
&\times\Big\{\left [\left (1+f_3^\sigma\right )\left (1+f_4^\sigma\right )\right ]\delta\left (k^0-E_3^\sigma-E_4^\sigma\right )\\
&\qquad
+\left [f_3^\sigma\left (1+f_4^\sigma\right )\right ]\delta\left (k^0+E_3^\sigma-E_4^\sigma\right )\\
&\qquad
+\left [\left (1+f_3^\sigma\right )f_4^\sigma\right ]\delta\left (k^0-E_3^\sigma+E_4^\sigma\right )\\
&\qquad +\left [f_3^\sigma
f_4^\sigma\right ]\delta\left (k^0+E_3^\sigma+E_4^\sigma\right )\Big\}\,.\,\quad\qquad\qquad
\label{eq:b05b}
\end{split} 
\end{equation}

In analogy one derives similar relations for
$\tilde{\mathcal{I}}_{G_{\pi_i\pi_i}^2}^{<}$ and
$\tilde{\mathcal{I}}_{G_{\pi_i\pi_i}^2}^{>}$ by replacing
$E_3^\sigma,f_3^\sigma,E_4^\sigma, f_4^\sigma$ with
$E_3^{\pi_i},f_3^{\pi_i},E_4^{\pi_i},f_4^{\pi_i}$.  Consequently, for
the terms $\tilde{\mathcal{I}}_{G_{\sigma\pi_i}^2}^{<}$ and
$\tilde{\mathcal{I}}_{G_{\sigma\pi_i}^2}^{>}$ one replaces
$E_4^\sigma,f_4^\sigma$ with $E_4^{\pi_i},f_4^{\pi_i}$.

In the following we consider a homogeneous system with mean-fields
evaluated at $t:=\max\left (x^0,y^0\right )$ and $t':=\min\left (x^0,y^0\right )$.  Taking
into account both combinations for $\Delta t:=t-t'$ one obtains for the
self-energies \eqref{eq:b02b}, \eqref{eq:b02c}
\begin{equation}
\begin{split} \ii \tilde\Pi_{\sigma\sigma}^{<,>}\left (X^0,p_1\right )
&=18\lambda^2\phi\left (t\right )\int_0^\infty\dd\Delta t\,\phi\left (t-\Delta
t\right )\left [\int\frac{\dd k^0}{\left (2\pi\right )}\,\ee^{-\ii \left (k^0-p_1^0\right )\Delta
t}\,\tilde{\mathcal{I}}_{G_{\sigma\sigma}^2}^{<,>}\left (X^0,k^0,\vec
p_1\right )\right ]\\ &\quad+2\lambda^2\phi\left (t\right )\sum_i\int_0^\infty\dd\Delta
t\,\phi\left (t-\Delta t\right )\left [\int\frac{\dd
k^0}{\left (2\pi\right )}\,\ee^{-\ii \left (k^0-p_1^0\right )\Delta
t}\,\tilde{\mathcal{I}}_{G_{\pi_i\pi_i}^2}^{<,>}\left (X^0,k^0,\vec
p_1\right )\right ]\\ &\quad+18\lambda^2\phi\left (t\right )\int^{\infty}_0\dd\Delta
t\,\phi\left (t-\Delta t\right )\left [\int\frac{\dd
k^0}{\left (2\pi\right )}\,\ee^{-\ii \left (p_1^0-k^0\right )\Delta
t}\,\tilde{\mathcal{I}}_{G_{\sigma\sigma}^2}^{<,>}\left (X^0,k^0,\vec
p_1\right )\right ]\\ &\quad+2\lambda^2\phi\left (t\right )\sum_i\int^{\infty}_0\dd\Delta
t\,\phi\left (t-\Delta t\right )\left [\int\frac{\dd
k^0}{\left (2\pi\right )}\,\ee^{-\ii \left (p_1^0-k^0\right )\Delta
t}\,\tilde{\mathcal{I}}_{G_{\pi_i\pi_i}^2}^{<,>}\left (X^0,k^0,\vec
p_1\right )\right ]\,,
\label{eq:b02d}
\end{split} 
\end{equation} 
\begin{equation}
\begin{split} \ii \tilde\Pi_{\pi_i\pi_i}^{<,>}\left (X^0,p_1\right )
&=4\lambda^2\phi\left (t\right )\int_0^\infty\dd\Delta t\,\phi\left (t-\Delta
t\right )\left [\int\frac{\dd k^0}{\left (2\pi\right )}\,\ee^{-\ii \left (k^0-p_1^0\right )\Delta
t}\,\tilde{\mathcal{I}}_{G_{\sigma\pi_i}^2}^{<,>}\left (X^0,k^0,\vec
p_1\right )\right ]\\ &\quad+4\lambda^2\phi\left (t\right )\int^{\infty}_0\dd\Delta
t\,\phi\left (t-\Delta t\right )\left [\int\frac{\dd
k^0}{\left (2\pi\right )}\,\ee^{-\ii \left (p_1^0-k^0\right )\Delta
t}\,\tilde{\mathcal{I}}_{G_{\sigma\pi_i}^2}^{<,>}\left (X^0,k^0,\vec
p_1\right )\right ]\,.
\label{eq:b02e}
\end{split} 
\end{equation}
Inserting the on-shell approximations \eqref{eq:b05a},
\eqref{eq:b05b} in \eqref{eq:b02d}, \eqref{eq:b02e} and evaluating the
integral over $k^0$ at time $X^0\simeq t$ for slowly changing
one-particle distribution functions as well as substituting the
integration variable $k^0\rightarrow p_1^0\pm k^0$, leads to
\begin{equation}
\begin{split} 
  \ii \tilde\Pi_{\sigma\sigma}^{<}\left (t,p_1\right )\simeq
  9\pi\lambda^2\phi\left (t\right ) &\int_0^\infty\dd\Delta t\,\phi\left
    (t-\Delta
    t\right )\int\frac{\dd k^0}{\left (2\pi\right )}\,\ee^{-\ii k^0\Delta t}\\
  &\times\Bigg[\int\frac{\dd^3\vec p_3}{\left (2\pi\right
    )^3}\int\dd^3\vec p_4\frac{\delta^{\left (3\right )}\left (\vec p_1
      -\vec p_3-\vec p_4\right )}{E_3^\sigma E_4^\sigma}\\
  &\qquad\times\Bigg\{\left [f_3^\sigma
    f_4^\sigma\right ]\delta\left (k^0+E_1^\sigma-E_3^\sigma-E_4^\sigma\right )\\
  &\qquad\qquad
  +\left [\left (1+f_3^\sigma\right )f_4^\sigma\right ]\delta\left (k^0+E_1^\sigma+E_3^\sigma-E_4^\sigma\right )\\
  &\qquad\qquad
  +\left [f_3^\sigma\left (1+f_4^\sigma\right )\right ]\delta\left (k^0+E_1^\sigma-E_3^\sigma+E_4^\sigma\right )\\
  &\qquad\qquad
  +\left [\left (1+f_3^\sigma\right )\left (1+f_4^\sigma\right )\right ]\delta\left (k^0+E_1^\sigma+E_3^\sigma+E_4^\sigma\right )\\
  &\qquad\qquad +\left [f_3^\sigma
    f_4^\sigma\right ]\delta\left (k^0-E_1^\sigma+E_3^\sigma+E_4^\sigma\right )\\
  &\qquad\qquad
  +\left [\left (1+f_3^\sigma\right )f_4^\sigma\right ]\delta\left (k^0-E_1^\sigma-E_3^\sigma+E_4^\sigma\right )\\
  &\qquad\qquad
  +\left [f_3^\sigma\left (1+f_4^\sigma\right )\right ]\delta\left (k^0-E_1^\sigma+E_3^\sigma-E_4^\sigma\right )\\
  &\qquad\qquad +\left [\left (1+f_3^\sigma\right )\left
      (1+f_4^\sigma\right )\right ]\delta\left
    (k^0-E_1^\sigma-E_3^\sigma-E_4^\sigma\right ) \Bigg\}\Bigg]\\
  +\pi\lambda^2&\phi\left (t\right )\sum_i\int_0^\infty\dd\Delta
  t\,\phi\left (t-\Delta t\right )\int\frac{\dd k^0}{\left (2\pi\right
    )}\,\ee^{-\ii k^0\Delta t}\\ &\,\times\Bigg[\int\frac{\dd^3\vec
    p_3}{\left (2\pi\right )^3}\int\dd^3\vec p_4\frac{\delta^{\left
        (3\right )}\left (\vec p_1 -\vec p_3-\vec p_4\right
    )}{E_3^{\pi_i} E_4^{\pi_i}}\\ &\qquad\times\Bigg\{\left [f_3^{\pi_i}
    f_4^{\pi_i}\right ]\delta\left (k^0+E_1^\sigma-E_3^{\pi_i}-E_4^{\pi_i}\right )\\
  &\qquad\qquad
  +\left [\left (1+f_3^{\pi_i}\right )f_4^{\pi_i}\right ]\delta\left (k^0+E_1^\sigma+E_3^{\pi_i}-E_4^{\pi_i}\right )\\
  &\qquad\qquad
  +\left [f_3^{\pi_i}\left (1+f_4^{\pi_i}\right )\right ]\delta\left (k^0+E_1^\sigma-E_3^{\pi_i}+E_4^{\pi_i}\right )\\
  &\qquad\qquad
  +\left [\left (1+f_3^{\pi_i}\right )\left (1+f_4^{\pi_i}\right )\right ]\delta\left (k^0+E_1^\sigma+E_3^{\pi_i}+E_4^{\pi_i}\right )\\
  &\qquad\qquad +\left [f_3^{\pi_i}
    f_4^{\pi_i}\right ]\delta\left (k^0-E_1^\sigma+E_3^{\pi_i}+E_4^{\pi_i}\right )\\
  &\qquad\qquad
  +\left [\left (1+f_3^{\pi_i}\right )f_4^{\pi_i}\right ]\delta\left (k^0-E_1^\sigma-E_3^{\pi_i}+E_4^{\pi_i}\right )\\
  &\qquad\qquad
  +\left [f_3^{\pi_i}\left (1+f_4^{\pi_i}\right )\right ]\delta\left (k^0-E_1^\sigma+E_3^{\pi_i}-E_4^{\pi_i}\right )\\
  &\qquad\qquad +\left [\left (1+f_3^{\pi_i}\right )\left
      (1+f_4^{\pi_i}\right )\right ]\delta\left
    (k^0-E_1^\sigma-E_3^{\pi_i}-E_4^{\pi_i}\right )\Bigg\}\Bigg]\,,
\label{eq:b02f}
\end{split} 
\end{equation}
\begin{equation}
  \begin{split} \ii \tilde\Pi_{\sigma\sigma}^{>}\left (t,p_1\right )\simeq
    9\pi\lambda^2\phi\left (t\right ) &\int_0^\infty\dd\Delta t\,\phi\left (t-\Delta
      t\right )\int\frac{\dd k^0}{\left (2\pi\right )}\,\ee^{-\ii k^0\Delta t}\\
    &\times\Bigg[\int\frac{\dd^3\vec p_3}{\left (2\pi\right )^3}\int\dd^3\vec
    p_4\frac{\delta^{\left (3\right )}\left (\vec p_1 -\vec p_3-\vec p_4\right )}{E_3^\sigma
      E_4^\sigma}\\
    &\qquad\times\Bigg\{\left [\left (1+f_3^\sigma\right )\left (1+f_4^\sigma\right )\right ]\delta\left (k^0+E_1^\sigma-E_3^\sigma-E_4^\sigma\right )\\
    &\qquad\qquad
    +\left [f_3^\sigma\left (1+f_4^\sigma\right )\right ]\delta\left (k^0+E_1^\sigma+E_3^\sigma-E_4^\sigma\right )\\
    &\qquad\qquad
    +\left [\left (1+f_3^\sigma\right )f_4^\sigma\right ]\delta\left (k^0+E_1^\sigma-E_3^\sigma+E_4^\sigma\right )\\
    &\qquad\qquad +\left [f_3^\sigma
      f_4^\sigma\right ]\delta\left (k^0+E_1^\sigma+E_3^\sigma+E_4^\sigma\right )\\
    &\qquad\qquad
    +\left [\left (1+f_3^\sigma\right )\left (1+f_4^\sigma\right )\right ]\delta\left (k^0-E_1^\sigma+E_3^\sigma+E_4^\sigma\right )\\
    &\qquad\qquad
    +\left [f_3^\sigma\left (1+f_4^\sigma\right )\right ]\delta\left (k^0-E_1^\sigma-E_3^\sigma+E_4^\sigma\right )\\
    &\qquad\qquad
    +\left [\left (1+f_3^\sigma\right )f_4^\sigma\right ]\delta\left (k^0-E_1^\sigma+E_3^\sigma-E_4^\sigma\right )\\
    &\qquad\qquad +\left [f_3^\sigma
      f_4^\sigma\right ]\delta\left (k^0-E_1^\sigma-E_3^\sigma-E_4^\sigma\right )\Bigg\}\\
    +\pi\lambda^2&\phi\left (t\right ) \sum_i\int_0^\infty\dd\Delta
    t\,\phi\left (t-\Delta t\right )\int\frac{\dd k^0}{\left (2\pi\right )}\,\ee^{-\ii k^0\Delta
      t}\\ &\times\Bigg[\int\frac{\dd^3\vec p_3}{\left (2\pi\right )^3}\int\dd^3\vec
    p_4\frac{\delta^{\left (3\right )}\left (\vec p_1 -\vec p_3-\vec p_4\right )}{E_3^{\pi_i}
      E_4^{\pi_i}}\\
    &\qquad\times\Bigg\{\left [\left (1+f_3^{\pi_i}\right )\left
        (1+f_4^{\pi_i}\right )\right ]\delta\left
      (k^0+E_1^\sigma-E_3^{\pi_i}-E_4^{\pi_i}\right )\\ 
    &\qquad\qquad
    +\left [f_3^{\pi_i}\left (1+f_4^{\pi_i}\right )\right ]\delta\left (k^0+E_1^\sigma+E_3^{\pi_i}-E_4^{\pi_i}\right )\\
    &\qquad\qquad
    +\left [\left (1+f_3^{\pi_i}\right )f_4^{\pi_i}\right ]\delta\left (k^0+E_1^\sigma-E_3^{\pi_i}+E_4^{\pi_i}\right )\\
    &\qquad\qquad +\left [f_3^{\pi_i}
      f_4^{\pi_i}\right ]\delta\left (k^0+E_1^\sigma+E_3^{\pi_i}+E_4^{\pi_i}\right )\\
    &\qquad\qquad
    +\left [\left (1+f_3^{\pi_i}\right )\left (1+f_4^{\pi_i}\right )\right ]\delta\left (k^0-E_1^\sigma+E_3^{\pi_i}+E_4^{\pi_i}\right )\\
    &\qquad\qquad
    +\left [f_3^{\pi_i}\left (1+f_4^{\pi_i}\right )\right ]\delta\left (k^0-E_1^\sigma-E_3^{\pi_i}+E_4^{\pi_i}\right )\\
    &\qquad\qquad
    +\left [\left (1+f_3^{\pi_i}\right )f_4^{\pi_i}\right ]\delta\left (k^0-E_1^\sigma+E_3^{\pi_i}-E_4^{\pi_i}\right )\\
    &\qquad\qquad +\left [f_3^{\pi_i}
      f_4^{\pi_i}\right ]\delta\left (k^0-E_1^\sigma-E_3^{\pi_i}-E_4^{\pi_i}\right )\Bigg\}\Bigg]\,,
\label{eq:b02g}
\end{split} 
\end{equation} 
In analogy to \eqref{eq:b02f} and \eqref{eq:b02g} one
derives the self-energies $\ii \tilde\Pi_{\pi_i\pi_i}^{<,>}$ by replacing
$\tilde{\mathcal{I}}_{G_{\sigma\sigma}^2}^{<,>}$ and
$\tilde{\mathcal{I}}_{G_{\pi_i\pi_i}^2}^{<,>}$ with the terms
$\tilde{\mathcal{I}}_{G_{\sigma\pi_i}^2}^{<,>}$. The self-energies
define the scattering-in and -out rates for the momentum mode $p_1$. By
combining in- as well as out-rates and including the external
propagator, one obtains the contributions for the collision integral due
to time-dependent interactions with the mean-field mode
\begin{equation}
\begin{split} \mathcal{I}^{b.s.}_{\sigma}\left (t,\vec
p_1\right )&=\frac{1}{2E_1^\sigma}\left (\left (1+f_1^\sigma\right )\ii \tilde\Pi_{\sigma\sigma}^{<}\left (t,p_1\right )-f_1^\sigma
\ii \tilde\Pi_{\sigma\sigma}^{>}\left (t,p_1\right )\right )\\
&=\frac{9}{2}\pi\lambda^2\phi\left (t\right ) \int_0^\infty\dd\Delta
t\,\phi\left (t-\Delta t\right )\int\frac{\dd k^0}{\left (2\pi\right )}\,\ee^{-\ii k^0\Delta
t}\\ &\quad\times\Bigg[\int\frac{\dd^3\vec
p_3}{\left (2\pi\right )^3}\int\dd^3\vec p_4\frac{\delta^{\left (3\right )}\left (\vec p_1
-\vec p_3-\vec p_4\right )}{E_1^\sigma E_3^\sigma E_4^\sigma}\\
&\qquad\times\Bigg\{\left [\left (1+f_1^\sigma\right )f_3^\sigma
f_4^\sigma-f_1^\sigma
\left (1+f_3^\sigma\right )\left (1+f_4^\sigma\right )\right ]\delta\left (k^0+E_1^\sigma-E_3^\sigma-E_4^\sigma\right )\\
&\qquad\qquad
+\left [\left (1+f_1^\sigma\right )\left (1+f_3^\sigma\right )f_4^\sigma-f_1^\sigma
f_3^\sigma\left (1+f_4^\sigma\right )\right ]\delta\left (k^0+E_1^\sigma+E_3^\sigma-E_4^\sigma\right )\\
&\qquad\qquad
+\left [\left (1+f_1^\sigma\right )f_3^\sigma\left (1+f_4^\sigma\right )-f_1^\sigma\left (1+f_3^\sigma\right )f_4^\sigma\right ]\delta\left (k^0+E_1^\sigma-E_3^\sigma+E_4^\sigma\right )\\
&\qquad\qquad
+\left [\left (1+f_1^\sigma\right )\left (1+f_3^\sigma\right )\left (1+f_4^\sigma\right )-f_1^\sigma
f_3^\sigma
f_4^\sigma\right ]\delta\left (k^0+E_1^\sigma+E_3^\sigma+E_4^\sigma\right )\\
&\qquad\qquad +\left [\left (1+f_1^\sigma\right )f_3^\sigma
f_4^\sigma-f_1^\sigma\left (1+f_3^\sigma\right )\left (1+f_4^\sigma\right )\right ]\delta\left (k^0-E_1^\sigma+E_3^\sigma+E_4^\sigma\right )\\
&\qquad\qquad
+\left [\left (1+f_1^\sigma\right )\left (1+f_3^\sigma\right )f_4^\sigma-f_1^\sigma
f_3^\sigma\left (1+f_4^\sigma\right )\right ]\delta\left (k^0-E_1^\sigma-E_3^\sigma+E_4^\sigma\right )\\
&\qquad\qquad
+\left [\left (1+f_1^\sigma\right )f_3^\sigma\left (1+f_4^\sigma\right )-f_1^\sigma\left (1+f_3^\sigma\right )f_4^\sigma\right ]\delta\left (k^0-E_1^\sigma+E_3^\sigma-E_4^\sigma\right )\\
&\qquad\qquad
+\left [\left (1+f_1^\sigma\right )\left (1+f_3^\sigma\right )\left (1+f_4^\sigma\right )-f_1^\sigma
f_3^\sigma f_4^\sigma\right ]\delta\left (k^0-E_1^\sigma-E_3^\sigma-E_4^\sigma\right )
\Bigg\}\Bigg]\\ &\quad
+\frac{1}{2}\pi\lambda^2\phi\left (t\right )\sum_i\int_0^\infty\dd\Delta
t\,\phi\left (t-\Delta t\right )\int\frac{\dd k^0}{\left (2\pi\right )}\,\ee^{-\ii k^0\Delta
t}\\ &\quad\times\Bigg[\int\frac{\dd^3\vec
p_3}{\left (2\pi\right )^3}\int\dd^3\vec p_4\frac{\delta^{\left (3\right )}\left (\vec p_1
-\vec p_3-\vec p_4\right )}{E_1^\sigma E_3^{\pi_i} E_4^{\pi_i}}\\
&\qquad\times\Bigg\{\left [\left (1+f_1^\sigma\right )f_3^{\pi_i}
f_4^{\pi_i}-f_1^\sigma\left (1+f_3^{\pi_i}\right )\left (1+f_4^{\pi_i}\right )\right ]\delta\left (k^0+E_1^\sigma-E_3^{\pi_i}-E_4^{\pi_i}\right )\\
&\qquad\qquad
+\left [\left (1+f_1^\sigma\right )\left (1+f_3^{\pi_i}\right )f_4^{\pi_i}-f_1^\sigma
f_3^{\pi_i}\left (1+f_4^{\pi_i}\right )\right ]\delta\left (k^0+E_1^\sigma+E_3^{\pi_i}-E_4^{\pi_i}\right )\\
&\qquad\qquad
+\left [\left (1+f_1^\sigma\right )f_3^{\pi_i}\left (1+f_4^{\pi_i}\right )-f_1^\sigma\left (1+f_3^{\pi_i}\right )f_4^{\pi_i}\right ]\delta\left (k^0+E_1^\sigma-E_3^{\pi_i}+E_4^{\pi_i}\right )\\
&\qquad\qquad
+\left [\left (1+f_1^\sigma\right )\left (1+f_3^{\pi_i}\right )\left (1+f_4^{\pi_i}\right )-f_1^\sigma
f_3^{\pi_i}
f_4^{\pi_i}\right ]\delta\left (k^0+E_1^\sigma+E_3^{\pi_i}+E_4^{\pi_i}\right )\\
&\qquad\qquad +\left [\left (1+f_1^\sigma\right )f_3^{\pi_i}
f_4^{\pi_i}-f_1^\sigma\left (1+f_3^\pi\right )\left (1+f_4^\pi\right )\right ]\delta\left (k^0-E_1^\sigma+E_3^{\pi_i}+E_4^{\pi_i}\right )\\
&\qquad\qquad
+\left [\left (1+f_1^\sigma\right )\left (1+f_3^{\pi_i}\right )f_4^{\pi_i}-f_1^\sigma
f_3^{\pi_i}\left (1+f_4^{\pi_i}\right )\right ]\delta\left (k^0-E_1^\sigma-E_3^{\pi_i}+E_4^{\pi_i}\right )\\
&\qquad\qquad
+\left [\left (1+f_1^\sigma\right )f_3^{\pi_i}\left (1+f_4^{\pi_i}\right )-f_1^\sigma\left (1+f_3^{\pi_i}\right )f_4^{\pi_i}\right ]\delta\left (k^0-E_1^\sigma+E_3^{\pi_i}-E_4^{\pi_i}\right )\\
&\qquad\qquad
+\left [\left (1+f_1^\sigma\right )\left (1+f_3^{\pi_i}\right )\left (1+f_4^{\pi_i}\right )-f_1^\sigma
f_3^{\pi_i}
f_4^{\pi_i}\right ]\delta\left (k^0-E_1^\sigma-E_3^{\pi_i}-E_4^{\pi_i}\right )\Bigg\}\Bigg]\\
\label{eq:b02h1a}
\end{split} 
\end{equation} 
\begin{equation}
\begin{split} \mathcal{I}^{b.s.}_{\sigma}\left (t,\vec
p_1\right )&:=\mathcal{C}_{\sigma\phi\leftrightarrow\sigma\sigma}^{b.s.}+\sum_i\mathcal{C}_{\sigma\phi\leftrightarrow\pi_i\pi_i}^{b.s.}\\
&:=\frac{9}{2}\pi\lambda^2\phi\left (t\right )\int_0^\infty\dd\Delta
t\,\phi\left (t-\Delta t\right )\int\frac{\dd k^0}{\left (2\pi\right )}\,\ee^{-\ii k^0\Delta
t}\,\mathcal{M}_{\sigma\sigma}\left (t,k^0,p_1\right )\\ &\quad
+\frac{1}{2}\pi\lambda^2\phi\left (t\right )\sum_i\int_0^\infty\dd\Delta
t\,\phi\left (t-\Delta t\right )\int\frac{\dd k^0}{\left (2\pi\right )}\,\ee^{-\ii k^0\Delta
t}\,\mathcal{M}_{\sigma\pi_i}\left (t,k^0,p_1\right )\,,\\
\mathcal{I}^{b.s.}_{\pi_i}\left (t,\vec
p_1\right )&:=\mathcal{C}_{\pi_i\phi\leftrightarrow\pi_i\sigma}^{b.s.}\\
&:=\pi\lambda^2\phi\left (t\right )\int_0^\infty\dd\Delta t\,\phi\left (t-\Delta
t\right )\int\frac{\dd k^0}{\left (2\pi\right )}\,\ee^{-\ii k^0\Delta
t}\,\mathcal{M}_{\pi_i\sigma}\left (t,k^0,p_1\right )\,,
\label{eq:b02h1b}
\end{split} 
\end{equation} 
where the calculation for pions is analougs to that of
$\sigma$ modes.  $\mathcal{M}_{\sigma\sigma}\left (t,k^0,p_1\right )$,
$\mathcal{M}_{\sigma\pi_i}\left (t,k^0,p_1\right )$ and \linebreak
$\mathcal{M}_{\pi_i\sigma}\left (t,k^0,p_1\right )$ define memory kernels with
respect to $\Delta t$ via the Fourier transform: 
\begin{equation}
\begin{split} \mathcal{M}_{\sigma\sigma}\left (t,\Delta
t,p_1\right )&:=\int\frac{\dd k^0}{\left (2\pi\right )}\,\ee^{-\ii k^0\Delta
t}\mathcal{M}_{\sigma\sigma}\left (t,k^0,p_1\right )\,,\\
\mathcal{M}_{\sigma\pi_i}\left (t,\Delta t,p_1\right )&:=\int\frac{\dd
k^0}{\left (2\pi\right )}\,\ee^{-\ii k^0\Delta
t}\mathcal{M}_{\sigma\pi_i}\left (t,k^0,p_1\right )\,,\\
\mathcal{M}_{\pi_i\sigma}\left (t,\Delta t,p_1\right )&:=\int\frac{\dd
k^0}{\left (2\pi\right )}\,\ee^{-\ii k^0\Delta
t}\mathcal{M}_{\pi_i\sigma}\left (t,k^0,p_1\right )\,.\\
\label{eq:MemoryKernels}
\end{split} 
\end{equation} 
Obviously, we recover contributions in \eqref{eq:b02h1a} which do not
conserve the particle number and therefore should be essential for the
chemical equilibration. However, an inclusion of such inelastic
processes requires a proper renormalization, since they contain a
divergent part and we decide not to consider those contributions in our
first study.

In numerical simulations one has different possibilities for computing
the collision integrals \eqref{eq:b02h1b}, either by explicitly
evaluating the kernel functions
$\mathcal{M}_{\sigma\sigma}\left (t,k^0,p_1\right )$,
$\mathcal{M}_{\sigma\pi_i}\left (t,k^0,p_1\right )$,
$\mathcal{M}_{\pi_i\sigma}\left (t,k^0,p_1\right )$ and taking the Fourier
transform or directly by making use of the symmetry with
respect to $k^0$ and integrating out the energy $\delta$-function.  By
doing the latter, the integration over the antisymmetric and complex
part vanishes, leading to a time integration which contains a
cosine. Such behavior is known from oscillatory systems and a pure time
integration with cosine as argument converges to a delta like
function\footnote{In a strict sense a sequence of sinc functions does
not converge to a delta function because of negative
contributions. Nevertheless, an integral over such a sequence with an
arbitrary test function has the same properties as known for a delta
function.} for large time differences: 
\begin{equation}
\lim_{t-t_0\rightarrow\infty}\int^{t-t_0}_0\dd\Delta t\,\cos\left (k^0\Delta
t\right )=\lim_{t-t_0\rightarrow\infty}\frac{1}{k^0}\sin\left (k^0\left (t-t_0\right )\right )\simeq\pi\delta\left (k^0\right )\,.
\label{eq:b02j} 
\end{equation} 
Therefore, we note that a local time approximation
with $\phi\left (t-\Delta t\right )\simeq\phi\left (t\right )$ leads to a simple on-shell
gradient expression for the sunset diagram and cannot account for
important dissipation phenomena from the mean-field to hard modes of the
same particle species. 

Due to detailed balance also the mean-field
equation contains memory kernels, which are defined in
Sec.~\ref{sec:DissipationTerm}. By making use of the decomposition
\eqref{eq:WignerPropagatorBoson} for the Green's functions, applying the
on-shell ansatz \eqref{eq:SpectralFunctions} for the spectral function
in Wigner space and combining that with relations \eqref{eq:b05a},
\eqref{eq:b05b} results in the following form for the kernels of the
mean-field equation: 
\begin{equation}
\begin{split} \mathcal{\tilde
M}_{\sigma\sigma}\left (X,k\right )&=\int\dd^4\Delta x\,\ee^{\ii k^\mu\Delta
x_\mu}\left [\left (iG_{\sigma\sigma}^{>}\left (x,y\right )\right )^3-\left (iG_{\sigma\sigma}^{<}\left (x,y\right )\right )^3\right ]\\
&=:\left (\mathcal{\tilde I}^{>}_{G_{\sigma\sigma}}*\mathcal{\tilde
I}^{>}_{G_{\sigma\sigma}^2}\right )\left (X,k\right )-\left (\mathcal{\tilde
I}^{<}_{G_{\sigma\sigma}}*\mathcal{\tilde
I}^{<}_{G_{\sigma\sigma}^2}\right )\left (X,k\right )\\
&=\int\frac{\dd^4p_1}{\left (2\pi\right )^4}\left [\ii \tilde{G}^{>}\left (X,p_1\right )\mathcal{\tilde
I}^{>}_{G_{\sigma\sigma}^2}\left (X,k-p_1\right )-\ii \tilde{G}^{<}\left (X,p_1\right )\mathcal{\tilde
I}^{<}_{G_{\sigma\sigma}^2}\left (X,k-p_1\right )\right ]\\
&\simeq\frac{\pi}{4}\int\frac{\dd\vec
p_1}{\left (2\pi\right )^3}\int\frac{\dd\vec p_3}{\left (2\pi\right )^3}\int\dd\vec
p_4\frac{\delta^{\left (3\right )}\left (\vec k+\vec p_1-\vec p_3-\vec
p_4\right )}{E_1^\sigma E_3^\sigma E_4^\sigma}\\
&\quad\times\Bigg\{\left [f_1^\sigma
\left (1+f_3^\sigma\right )\left (1+f_4^\sigma\right )-\left (1+f_1^\sigma\right )f_3^\sigma
f_4^\sigma\right ]\delta\left (k^0+E_1^\sigma-E_3^\sigma-E_4^\sigma\right )\\
&\qquad\qquad +\left [f_1^\sigma
f_3^\sigma\left (1+f_4^\sigma\right )-\left (1+f_1^\sigma\right )\left (1+f_3^\sigma\right )f_4^\sigma\right ]\delta\left (k^0+E_1^\sigma+E_3^\sigma-E_4^\sigma\right )\\
&\qquad\qquad
+\left [f_1^\sigma\left (1+f_3^\sigma\right )f_4^\sigma-\left (1+f_1^\sigma\right )f_3^\sigma\left (1+f_4^\sigma\right )\right ]\delta\left (k^0+E_1^\sigma-E_3^\sigma+E_4^\sigma\right )\\
&\qquad\qquad +\left [f_1^\sigma f_3^\sigma
f_4^\sigma-\left (1+f_1^\sigma\right )\left (1+f_3^\sigma\right )\left (1+f_4^\sigma\right )\right ]\delta\left (k^0+E_1^\sigma+E_3^\sigma+E_4^\sigma\right )\\
&\qquad\qquad +\left [\left (1+f_1^\sigma\right )f_3^\sigma
f_4^\sigma-f_1^\sigma\left (1+f_3^\sigma\right )\left (1+f_4^\sigma\right )\right ]\delta\left (k^0-E_1^\sigma+E_3^\sigma+E_4^\sigma\right )\\
&\qquad\qquad
+\left [\left (1+f_1^\sigma\right )\left (1+f_3^\sigma\right )f_4^\sigma-f_1^\sigma
f_3^\sigma\left (1+f_4^\sigma\right )\right ]\delta\left (k^0-E_1^\sigma-E_3^\sigma+E_4^\sigma\right )\\
&\qquad\qquad
+\left [\left (1+f_1^\sigma\right )f_3^\sigma\left (1+f_4^\sigma\right )-f_1^\sigma\left (1+f_3^\sigma\right )f_4^\sigma\right ]\delta\left (k^0-E_1^\sigma+E_3^\sigma-E_4^\sigma\right )\\
&\qquad\qquad
+\left [\left (1+f_1^\sigma\right )\left (1+f_3^\sigma\right )\left (1+f_4^\sigma\right )-f_1^\sigma
f_3^\sigma f_4^\sigma\right ]\delta\left (k^0-E_1^\sigma-E_3^\sigma-E_4^\sigma\right )
\Bigg\}\,
\label{eq:WignerMemoryKernel1}
\end{split} \end{equation}
\begin{equation}
\begin{split} \mathcal{\tilde M}_{\sigma\pi_i}\left (X,k\right )
&\simeq\frac{\pi}{4}\int\frac{\dd\vec
p_1}{\left (2\pi\right )^3}\int\frac{\dd\vec p_3}{\left (2\pi\right )^3}\int\dd\vec
p_4\frac{\delta^{\left (3\right )}\left (\vec k+\vec p_1-\vec p_3-\vec
p_4\right )}{E_1^\sigma E_3^{\pi_i} E_4^{\pi_i}}\\
&\qquad\times\Bigg\{\left [f_1^\sigma\left (1+f_3^{\pi_i}\right )\left (1+f_4^{\pi_i}\right )-\left (1+f_1^\sigma\right )f_3^{\pi_i}
f_4^{\pi_i}\right ]\delta\left (k^0+E_1^\sigma-E_3^{\pi_i}-E_4^{\pi_i}\right )\\
&\qquad\qquad +\left [f_1^\sigma
f_3^{\pi_i}\left (1+f_4^{\pi_i}\right )-\left (1+f_1^\sigma\right )\left (1+f_3^{\pi_i}\right )f_4^{\pi_i}\right ]\delta\left (k^0+E_1^\sigma+E_3^{\pi_i}-E_4^{\pi_i}\right )\\
&\qquad\qquad
+\left [f_1^\sigma\left (1+f_3^{\pi_i}\right )f_4^{\pi_i}-\left (1+f_1^\sigma\right )f_3^{\pi_i}\left (1+f_4^{\pi_i}\right )\right ]\delta\left (k^0+E_1^\sigma-E_3^{\pi_i}+E_4^{\pi_i}\right )\\
&\qquad\qquad +\left [f_1^\sigma f_3^{\pi_i}
f_4^{\pi_i}-\left (1+f_1^\sigma\right )\left (1+f_3^{\pi_i}\right )\left (1+f_4^{\pi_i}\right )\right ]\delta\left (k^0+E_1^\sigma+E_3^{\pi_i}+E_4^{\pi_i}\right )\\
&\qquad\qquad +\left [\left (1+f_1^\sigma\right )f_3^{\pi_i}
f_4^{\pi_i}-f_1^\sigma\left (1+f_3^{\pi_i}\right )\left (1+f_4^{\pi_i}\right )\right ]\delta\left (k^0-E_1^\sigma+E_3^{\pi_i}+E_4^{\pi_i}\right )\\
&\qquad\qquad
+\left [\left (1+f_1^\sigma\right )\left (1+f_3^{\pi_i}\right )f_4^{\pi_i}-f_1^\sigma
f_3^{\pi_i}\left (1+f_4^{\pi_i}\right )\right ]\delta\left (k^0-E_1^\sigma-E_3^{\pi_i}+E_4^{\pi_i}\right )\\
&\qquad\qquad
+\left [\left (1+f_1^\sigma\right )f_3^{\pi_i}\left (1+f_4^{\pi_i}\right )-f_1^\sigma\left (1+f_3^{\pi_i}\right )f_4^{\pi_i}\right ]\delta\left (k^0-E_1^\sigma+E_3^{\pi_i}-E_4^{\pi_i}\right )\\
&\qquad\qquad
+\left [\left (1+f_1^\sigma\right )\left (1+f_3^{\pi_i}\right )\left (1+f_4^{\pi_i}\right )-f_1^\sigma
f_3^{\pi_i}
f_4^{\pi_i}\right ]\delta\left (k^0-E_1^\sigma-E_3^{\pi_i}-E_4^{\pi_i}\right )\Bigg\}\,.
\label{eq:WignerMemoryKernel2}
\end{split} \end{equation}

\textbf{3. Basketball diagrams (b.)} 
\begin{equation}
\begin{split}
\Gamma_2^{\text{b.}}=\,&\frac{3}{4}\ii \lambda^2\int_{\mathcal
C}\dd^4x\int_{\mathcal C}\dd^4y\,G_{\sigma\sigma}^4\left (x,y\right )
+\frac{3}{4}\ii \lambda^2\sum_i\int_{\mathcal C}\dd^4x\int_{\mathcal
C}\dd^4y\,G_{\pi_i\pi_i}^4\left (x,y\right )\\
&+\frac{1}{2}\ii \lambda^2\sum_i\int_{\mathcal C}\dd^4x\int_{\mathcal
C}\dd^4y\,G_{\sigma\sigma}^2\left (x,y\right )G_{\pi_i\pi_i}^2\left (x,y\right )\\
&+\frac{1}{2}\ii \lambda^2\sum_{i,j>i}\int_{\mathcal C}\dd^4x\int_{\mathcal
C}\dd^4y\,G_{\pi_i\pi_i}^2\left (x,y\right )G_{\pi_j\pi_j}^2\left (x,y\right )\,.
\label{eq:b03}
\end{split} 
\end{equation} 
In analogy to \eqref{eq:SelfEnergyBosonicSunsetDiagram}
one obtains for the self-energies: 
\begin{equation}
\begin{split}
i\Pi_{\sigma\sigma}^{<,>}\left (x,y\right )=&\,\,6\lambda^2\left (iG_{\sigma\sigma}^{<,>}\left (x,y\right )\right )^3+2\lambda^2\sum_i\left (iG_{\sigma\sigma}^{<,>}\left (x,y\right )\right )\left (iG_{\pi_i\pi_i}^{<,>}\left (x,y\right )\right )^2\\
i\Pi_{\pi_i\pi_i}^{<,>}\left (x,y\right )=&\,\,6\lambda^2\left (iG_{\pi_i\pi_i}^{<,>}\left (x,y\right )\right )^3\\
&+2\lambda^2\left (iG_{\pi_i\pi_i}^{<,>}\left (x,y\right )\right )\left [\left (iG_{\sigma\sigma}^{<,>}\left (x,y\right )\right )^2+\sum_{j\neq
i}\left (iG_{\pi_j\pi_j}^{<,>}\left (x,y\right )\right )^2\right ]\,.
\label{eq:b04a}
\end{split} \end{equation} 
With the auxiliary functions \eqref{eq:AuxFunc}, it
follows for the Wigner transforms: 
\begin{equation}
\begin{split}
\ii \tilde{\Pi}_{\sigma\sigma}^{<,>}\left (X,p_1\right )=&\,\,6\lambda^2\left (\ii \tilde{G}_{\sigma\sigma}^{<,>}\ast\tilde{\mathcal{I}}_{G_{\sigma\sigma}^2}^{<,>}\right )\left (X,p_1\right )
+2\lambda^2\sum_i\left (\ii \tilde{G}_{\sigma\sigma}^{<,>}\ast\tilde{\mathcal{I}}_{G_{\pi_i\pi_i}^2}^{<,>}\right )\left (X,p_1\right )\,,\\
\ii \tilde{\Pi}_{\pi_i\pi_i}^{<,>}\left (X,p_1\right )=&\,\,6\lambda^2\left (\ii \tilde{G}_{\pi_i\pi_i}^{<,>}\ast\tilde{\mathcal{I}}_{G_{\pi_i\pi_i}^2}^{<,>}\right )\left (X,p_1\right )\\
&+2\lambda^2\left (\ii \tilde{G}_{\pi_i\pi_i}^{<,>}\ast\left [\tilde{\mathcal{I}}_{G_{\sigma\sigma}^2}^{<,>}+\sum_{j\neq
i}\tilde{\mathcal{I}}_{G_{\pi_j\pi_j}^2}^{<,>}\right ]\right )\left (X,p_1\right )\,.
\label{eq:b04b}
\end{split} 
\end{equation} 
Inserting the on-shell ansatz for the propagators in
Wigner space \eqref{eq:WignerPropagatorBoson} leads then to the
following gain and loss contributions (compare also with
\eqref{eq:b05a}): \begin{equation}
\begin{split}
\left (\ii \tilde{G}_{\sigma\sigma}^{<}\ast\tilde{\mathcal{I}}_{G_{\sigma\sigma}^2}^{<}\right )\left (X,p_1\right )=&\,\,
\frac{\pi}{4}\int\frac{\dd^3\vec p_2}{\left (2\pi\right )^3}\int\frac{\dd^3\vec
p_3}{\left (2\pi\right )^3}\int\dd^3\vec p_4\frac{\delta^{\left (3\right )}\left (\vec
p_1-\vec p_2-\vec p_3+\vec p_4\right )}{E_2^\sigma E_3^\sigma E_4^\sigma}\\
&\times\Big\{\left [\left (1+f_2^\sigma\right ) f_3^\sigma
f_4^\sigma\right ]\delta\left (E_1+E_2-E_3-E_4\right )\\ &\qquad
+\left [f_2^\sigma\left (1+f_3^\sigma\right )f_4^\sigma\right ]\delta\left (E_1-E_2+E_3-E_4\right )\\
&\qquad +\left [f_2^\sigma
f_3^\sigma\left (1+f_4^\sigma\right )\right ]\delta\left (E_1-E_2-E_3+E_4\right )\Big\}\,,
\label{eq:ConvolutionInt1}
\end{split} 
\end{equation} 
\begin{equation}
\begin{split}
\left (\ii \tilde{G}_{\sigma\sigma}^{>}\ast\tilde{\mathcal{I}}_{G_{\sigma\sigma}^2}^{>}\right )\left (X,p_1\right )=&\,\,
\frac{\pi}{4}\int\frac{\dd^3\vec p_2}{\left (2\pi\right )^3}\int\frac{\dd^3\vec
p_3}{\left (2\pi\right )^3}\int\dd^3\vec p_4\frac{\delta^{\left (3\right )}\left (\vec
p_1-\vec p_2-\vec p_3+\vec p_4\right )}{E_2^\sigma E_3^\sigma E_4^\sigma}\\
&\times\Big\{\left [f_2^\sigma\left (1+f_3^\sigma\right )\left (1+f_4^\sigma\right )\right ]\delta\left (E_1+E_2-E_3-E_4\right )\\
&\qquad
+\left [\left (1+f_2^\sigma\right )f_3^\sigma\left (1+f_4^\sigma\right )\right ]\delta\left (E_1-E_2+E_3-E_4\right )\\
&\qquad +\left [\left (1+f_2^\sigma\right )\left (1+f_3^\sigma\right )
f_4^\sigma\right ]\delta\left (E_1-E_2-E_3+E_4\right )\Big\}\,,
\label{eq:ConvolutionInt2}
\end{split} 
\end{equation} 
where we neglected all energy conserving
$\delta$-functions, which cannot be fullfilled on-shell.

In analogy one derives similar expressions for
$\left (\ii \tilde{G}_{\sigma\sigma}^{<}\ast\tilde{\mathcal{I}}_{G_{\pi_i\pi_i}^2}^{<}\right )$
and
$\left (\ii \tilde{G}_{\sigma\sigma}^{>}\ast\tilde{\mathcal{I}}_{G_{\pi_i\pi_i}^2}^{>}\right )$
by replacing $E_3^\sigma,f_3^\sigma,E_4^\sigma, f_4^\sigma$ with
$E_3^{\pi_i},f_3^{\pi_i},E_4^{\pi_i},f_4^{\pi_i}$.  Consequently, for
the terms
$\left (\ii \tilde{G}_{\pi_i\pi_i}^{<}\ast\tilde{\mathcal{I}}_{G_{\sigma\sigma}^2}^{<}\right )$
and
$\left (\ii \tilde{G}_{\pi_i\pi_i}^{>}\ast\tilde{\mathcal{I}}_{G_{\sigma\sigma}^2}^{>}\right )$
one replaces $E_2^\sigma,f_2^\sigma$ with $E_2^{\pi_i},f_2^{\pi_i}$.
Finally, terms of the form $\left (\ii \tilde{G}_{\pi_i\pi_i}^{<}\ast\right .$
$\left ( \tilde{\mathcal{I}}_{G_{\pi_i\pi_i}^2}^{<}\right )$,
$\left (\ii \tilde{G}_{\pi_i\pi_i}^{>}\ast\tilde{\mathcal{I}}_{G_{\pi_i\pi_i}^2}^{>}\right )$,
$\left (\ii \tilde{G}_{\pi_i\pi_i}^{<}\ast\tilde{\mathcal{I}}_{G_{\pi_j\pi_j}^2}^{<}\right )$
as well as
$\left (\ii \tilde{G}_{\pi_i\pi_i}^{>}\ast\tilde{\mathcal{I}}_{G_{\pi_j\pi_j}^2}^{>}\right )$
contain only $\pi_i$ and $\pi_j$ as upper indices, but have the same
structure like equations \eqref{eq:ConvolutionInt1},
\eqref{eq:ConvolutionInt2}. 

Combining in- as well as out-rates and summing up identical
contributions leads to the collision integrals of elastic scatterings
between mesons: 
\begin{equation}
  \begin{split} \mathcal{I}^{b.}_{\sigma}\left (t,\vec
      p_1\right )&=\frac{1}{2E_1^\sigma}\left (\left (1+f_1^\sigma\right )\ii \tilde\Pi_{\sigma\sigma}^{<}\left (t,p_1\right )-f_1^\sigma
      \ii \tilde\Pi_{\sigma\sigma}^{>}\left (t,p_1\right )\right )\\
    &=:\mathcal{C}_{\sigma\sigma\leftrightarrow\sigma\sigma}^{b.} +
    \sum_i\mathcal{C}_{\sigma\pi_i\leftrightarrow\sigma\pi_i}^{b.} +\sum_i\mathcal{C}_{\sigma\sigma\leftrightarrow\pi_i\pi_i}^{b.}\,,
    \\
    \mathcal{I}^{b.}_{\pi_i}\left (t,\vec
      p_1\right )&=\frac{1}{2E_1^{\pi_i}}\left (\left (1+f_1^{\pi_i}\right )\ii \tilde\Pi_{\pi_i\pi_i}^{<}\left (t,p_1\right )-f_1^{\pi_i}
      \ii \tilde\Pi_{\pi_i\pi_i}^{>}\left (t,p_1\right )\right )\\
    &=:\mathcal{C}_{\pi_i\pi_i\leftrightarrow\pi_i\pi_i}^{b.}+\sum_{j\neq
      i}\mathcal{C}_{\pi_i\pi_j\leftrightarrow\pi_i\pi_j}^{b.}+\sum_{j\neq
      i}\mathcal{C}_{\pi_i\pi_i\leftrightarrow\pi_j\pi_j}^{b.}
    +\mathcal{C}_{\pi_i\sigma\leftrightarrow\pi_i\sigma}^{b.}+\mathcal{C}_{\pi_i\pi_i\leftrightarrow\sigma\sigma}^{b.}\,,\\
\label{eq:BoltzmannCollSigma}
\end{split} 
\end{equation} 
where the following decomposition with respect to different processes is
applied:
\begin{equation*}
\begin{split}
\mathcal{C}_{\sigma\sigma\leftrightarrow\sigma\sigma}^{b.}:=&\frac{9}{4}\pi\lambda^2\int\frac{\dd^3\vec
p_2}{\left(2\pi\right)^3} \int\frac{\dd^3\vec
p_3}{\left (2\pi\right )^3}\int\dd^3\vec p_4\frac{\delta^{\left (3\right )}\left (\vec p_1 +
\vec p_2-\vec p_3-\vec p_4\right )}{E_1^\sigma E_2^\sigma E_3^\sigma
E_4^\sigma}\\
&\qquad\times\Big\{\left [\left (1+f_1^\sigma\right )\left (1+f_2^\sigma\right )f_3^\sigma
f_4^\sigma\right .\\ &\qquad\qquad\left .-f_1^\sigma
f_2^\sigma\left (1+f_3^\sigma\right )\left (1+f_4^\sigma\right )\right ]\delta\left (E_1^\sigma +
E_2^\sigma-E_3^\sigma-E_4^\sigma\right )\Big\}\,,
\end{split}
\end{equation*}
\begin{equation*}
\begin{split}
\mathcal{C}_{\sigma\pi_i\leftrightarrow\sigma\pi_i}^{b.}:=&
\frac{2}{4}\pi\lambda^2\int\frac{\dd^3\vec p_2}{\left(2\pi\right)^3}
\int\frac{\dd^3\vec p_3}{\left (2\pi\right )^3}\int\dd^3\vec
p_4\frac{\delta^{\left (3\right )}\left (\vec p_1 + \vec p_2-\vec p_3-\vec
p_4\right )}{E_1^\sigma E_2^{\pi_i} E_3^\sigma E_4^{\pi_i}}\\
&\qquad\times\Big\{\left [\left (1+f_1^\sigma\right )\left (1+f_2^{\pi_i}\right )f_3^\sigma
f_4^{\pi_i}\right .\\ &\qquad\qquad\left .-f_1^\sigma
f_2^{\pi_i}\left (1+f_3^\sigma\right )\left (1+f_4^{\pi_i}\right )\right ]\delta\left (E_1^\sigma+E_2^{\pi_i}-E_3^\sigma-E_4^{\pi_i}\right )\Big\}\,,
\end{split}
\end{equation*}
\begin{equation}
\begin{split}
\mathcal{C}_{\sigma\sigma\leftrightarrow\pi_i\pi_i}^{b.}:=&
\frac{1}{4}\pi\lambda^2\int\frac{\dd^3\vec p_2}{\left(2\pi\right)^3}
\int\frac{\dd^3\vec p_3}{\left (2\pi\right )^3}\int\dd^3\vec
p_4\frac{\delta^{\left (3\right )}\left (\vec p_1 + \vec p_2-\vec p_3-\vec
p_4\right )}{E_1^\sigma E_2^\sigma E_3^{\pi_i} E_4^{\pi_i}}\\
&\qquad\times\Big\{\left [\left (1+f_1^\sigma\right )\left (1+f_2^\sigma\right )f_3^{\pi_i}
f_4^{\pi_i}\right .\\ &\qquad\qquad\left .-f_1^\sigma
f_2^\sigma\left (1+f_3^{\pi_i}\right )\left (1+f_4^{\pi_i}\right )\right ]\delta\left (E_1^\sigma+E_2^\sigma-E_3^{\pi_i}-E_4^{\pi_i}\right )\Big\}\,,
\label{eq:BoltzmannCollSigmaContributions}
\end{split}
\end{equation}
\begin{equation*}
\begin{split}
\mathcal{C}_{\pi_i\pi_i\leftrightarrow\pi_i\pi_i}^{b.}:=&\frac{9}{4}\pi\lambda^2\int\frac{\dd^3\vec
p_2}{\left(2\pi\right)^3} \int\frac{\dd^3\vec
p_3}{\left (2\pi\right )^3}\int\dd^3\vec p_4\frac{\delta^{\left (3\right )}\left (\vec p_1 +
\vec p_2-\vec p_3-\vec p_4\right )}{E_1^{\pi_i} E_2^{\pi_i} E_3^{\pi_i}
E_4^{\pi_i}}\\
&\qquad\times\Big\{\left [\left (1+f_1^{\pi_i}\right )\left (1+f_2^{\pi_i}\right )f_3^{\pi_i}
f_4^{\pi_i}\right .\\ &\qquad\qquad\left .-f_1^{\pi_i}
f_2^{\pi_i}\left (1+f_3^{\pi_i}\right )\left (1+f_4^{\pi_i}\right )\right ]\delta\left (E_1^{\pi_i}
+ E_2^{\pi_i}-E_3^{\pi_i}-E_4^{\pi_i}\right )\Big\}\\
\end{split}
\end{equation*}
\begin{equation*}
\begin{split} \mathcal{C}_{\pi_i\pi_j\leftrightarrow\pi_i\pi_j}^{b.}:=&
\frac{2}{4}\pi\lambda^2\int\frac{\dd^3\vec p_2}{\left(2\pi\right)^3}
\int\frac{\dd^3\vec p_3}{\left (2\pi\right )^3}\int\dd^3\vec
p_4\frac{\delta^{\left (3\right )}\left (\vec p_1 + \vec p_2-\vec p_3-\vec
p_4\right )}{E_1^{\pi_i} E_2^{\pi_j} E_3^{\pi_i} E_4^{\pi_j}}\\
&\qquad\times\Big\{\left [\left (1+f_1^{\pi_i}\right )\left (1+f_2^{\pi_j}\right )f_3^{\pi_i}
f_4^{\pi_j}\right .\\ &\qquad\qquad\left .-f_1^{\pi_i}
f_2^{\pi_j}\left (1+f_3^{\pi_i}\right )\left (1+f_4^{\pi_j}\right )\right ]\delta\left (E_1^{\pi_i}+E_2^{\pi_j}-E_3^{\pi_i}-E_4^{\pi_j}\right )\Big\}
\end{split}
\end{equation*}
\begin{equation*}
\begin{split} \mathcal{C}_{\pi_i\pi_i\leftrightarrow\pi_j\pi_j}^{b.}:=&
\frac{1}{4}\pi\lambda^2\int\frac{\dd^3\vec p_2}{\left(2\pi\right)^3}
\int\frac{\dd^3\vec p_3}{\left (2\pi\right )^3}\int\dd^3\vec
p_4\frac{\delta^{\left (3\right )}\left (\vec p_1 + \vec p_2-\vec p_3-\vec
p_4\right )}{E_1^{\pi_i} E_2^{\pi_i} E_3^{\pi_j} E_4^{\pi_j}}\\
&\qquad\times\Big\{\left [\left (1+f_1^{\pi_i}\right )\left (1+f_2^{\pi_i}\right )f_3^{\pi_j}
f_4^{\pi_j}\right .\\ &\qquad\qquad\left .-f_1^{\pi_i}
f_2^{\pi_i}\left (1+f_3^{\pi_j}\right )\left (1+f_4^{\pi_j}\right )\right ]\delta\left (E_1^{\pi_i}+E_2^{\pi_i}-E_3^{\pi_j}-E_4^{\pi_j}\right )\Big\}\,,\\
\end{split}
\end{equation*}
\begin{equation*}
\begin{split}
\mathcal{C}_{\pi_i\sigma\leftrightarrow\pi_i\sigma}^{b.}:=&\frac{2}{4}\pi\lambda^2\int\frac{\dd^3\vec
p_2}{\left(2\pi\right)^3} \int\frac{\dd^3\vec
p_3}{\left (2\pi\right )^3}\int\dd^3\vec p_4\frac{\delta^{\left (3\right )}\left (\vec p_1 +
\vec p_2-\vec p_3-\vec p_4\right )}{E_1^{\pi_i} E_2^\sigma E_3^{\pi_i}
E_4^\sigma}\\
&\qquad\times\Big\{\left [\left (1+f_1^{\pi_i}\right )\left (1+f_2^\sigma\right )f_3^{\pi_i}
f_4^\sigma\right .\\ &\qquad\qquad\left .-f_1^{\pi_i}
f_2^\sigma\left (1+f_3^{\pi_i}\right )\left (1+f_4^\sigma\right )\right ]\delta\left (E_1^{\pi_i}+E_2^\sigma-E_3^{\pi_i}-E_4^\sigma\right )\Big\}\,,\\
\end{split}
\end{equation*}
\begin{equation}
\begin{split}
\mathcal{C}_{\pi_i\pi_i\leftrightarrow\sigma\sigma}^{b.}:=&\frac{1}{4}\pi\lambda^2\int\frac{\dd^3\vec
p_2}{\left(2\pi\right)^3} \int\frac{\dd^3\vec
p_3}{\left (2\pi\right )^3}\int\dd^3\vec p_4\frac{\delta^{\left (3\right )}\left (\vec p_1 +
\vec p_2-\vec p_3-\vec p_4\right )}{E_1^{\pi_i} E_2^{\pi_i} E_3^\sigma
E_4^\sigma}\\
&\qquad\times\Big\{\left [\left (1+f_1^{\pi_i}\right )\left (1+f_2^{\pi_i}\right )f_3^\sigma
f_4^\sigma\right .\\ &\qquad\qquad\left .-f_1^{\pi_i}
f_2^{\pi_i}\left (1+f_3^\sigma\right )\left (1+f_4^\sigma\right )\right ]\delta\left (E_1^{\pi_i}+E_2^{\pi_i}-E_3^\sigma-E_4^\sigma\right )\Big\}\,.
\label{eq:BoltzmannCollPionContributions}
\end{split}
\end{equation} 
Note: since the collision integrals
\eqref{eq:BoltzmannCollSigmaContributions},
\eqref{eq:BoltzmannCollPionContributions} for elastic scatterings have
locally (in space) the same form for homogeneous and inhomogeneous
systems, we skip the possible dependence on the position vector $\vec
x$.

\textbf{4. Fermionic sunset diagrams (f.s.)}

\begin{equation}
\begin{split} \Gamma_2^{\text{f.s.}}=\,&-\frac{1}{2}ig^2\int_{\mathcal
C}\dd^4x\int_{\mathcal
C}\dd^4y\,\text{Tr}\left [D\left (x,y\right )D\left (y,x\right )\right ]\,G_{\sigma\sigma}\left (x,y\right )\\
\quad &-\frac{1}{2}ig^2\sum_i\int_{\mathcal C}\dd^4x\int_{\mathcal
C}\dd^4y\,\text{Tr}\left [D\left (x,y\right )D\left (y,x\right )\right ]\,G_{\pi_i\pi_i}\left (x,y\right )\,.
\label{eq:b01b}
\end{split} 
\end{equation} 
In analogy to \eqref{eq:SelfEnergyBosonicSunsetDiagram} one obtains for
the self-energies
\begin{equation}
\begin{split}
i\Pi_{\sigma\sigma}^{<,>}\left (x,y\right )=&\,-g^2\text{Tr}\left [iD^{<,>}\left (x,y\right )iD^{>,<}\left (y,x\right )\right ]\,,\\
i\Pi_{\pi_i\pi_i}^{<,>}\left (x,y\right )=&\,-g^2\text{Tr}\left [iD^{<,>}\left (x,y\right )iD^{>,<}\left (y,x\right )\right ]\,.
\label{eq:b04c}
\end{split} 
\end{equation} 
Consequently, for the Wigner transforms it follows
\begin{equation}
\begin{split}
\ii \tilde{\Pi}_{\sigma\sigma}^{<,>}\left (X,p_1\right )=&\,-g^2\Tr \left (\ii \tilde
D^{<,>}\ast \ii \tilde D^{>,<}\right )\left (X,p_1\right )\,,\\
\ii \tilde{\Pi}_{\pi_i\pi_i}^{<,>}\left (X,p_1\right )=&\,-g^2\Tr \left (\ii \tilde
D^{<,>}\ast \ii \tilde D^{>,<}\right )\left (X,p_1\right )\,.
\label{eq:b04d}
\end{split} \end{equation} 
Since the expressions for sigma and pions have identical
structure, it is suitable to skip the mesonic index in the following.

Inserting the on-shell ansatz for the propagators in Wigner space (see
Eqs.~\eqref{eq:WignerPropagatorFermion}, \eqref{eq:SpectralFunctions})
results in (analogous calculation for $\ii \tilde\Pi^<\left (X,p_1\right
)$: 
\begin{equation}
\begin{split}
\ii \tilde\Pi^>\left (X,p_1\right )&=\pi^2g^2\int\frac{\dd^4p_2}{\left (2\pi\right )^4}\frac{\Tr \left [\left (\gamma^\mu\left (p_{1,\mu}
+p_{2,\mu}\right )+M_\psi\right )\left (\gamma^\nu
p_{2,\nu}+M_\psi\right )\right ]}{E^\psi\left (X,p_1+p_2\right )E^\psi\left (X,p_2\right )}\\
&\qquad\quad\times\left [\delta\left (p_1^0+p_2^0-E^\psi\left (X,p_1+p_2\right )\right )-\delta\left (p_1^0+p_2^0+E^\psi\left (X,p_1+p_2\right )\right )\right ]\\
&\qquad\quad\times\left [\delta\left (p_2^0-E^\psi\left (X,p_2\right )\right )-\delta\left (p_2^0+E^\psi\left (X,p_2\right )\right )\right ]\\
&\qquad\quad\times N_\psi^>\left (X,p_1+p_2\right )N_\psi^<\left (X,p_2\right )\,.
\label{eq:SelfEnegySigmaQuark1}
\end{split} 
\end{equation} 
After evaluating the trace over Dirac and flavor indices,
\begin{equation}
\begin{split} &\Tr \left [\left (\gamma^\mu\left (p_{1,\mu}
+p_{2,\mu}\right )+M_\psi\right )\left (\gamma^\nu p_{2,\nu}+M_\psi\right )\right ]\\
&\,=\Tr \left [\gamma^\mu\gamma^\nu\left (p_{1,\mu}+p_{2,\nu}\right )p_{2,\nu}+M_\psi^2\right ]=4d_\psi\left (\eta^{\mu\nu}\left (p_{1,\mu}+p_{2,\mu}\right )p_{2,\nu}+M_\psi^2\right )\\
&\,=4d_\psi\left (\left (p_1^\nu+p_2^\nu\right )p_{2,\nu}+M_\psi^2\right )=4d_\psi\left (\left (p_1^0+p_2^0\right )p_{2,0}-\left (\vec
p_1+\vec p_2\right )\cdot\vec p_2+M_\psi^2\right )\,,\\
\label{eq:TraceDirac}
\end{split} 
\end{equation} 
the relevant self-energies with $p_1^0\geqq0$ become:
\begin{equation}
\begin{split} \ii \tilde\Pi^>\left (X,p_1\right )&=2\pi g^2d_\psi\int\frac{\dd^3\vec
p_2}{\left (2\pi\right )^3} \frac{p_1^0E^{\bar\psi}_{\vec p_2}-\vec p_1\cdot\vec
p_2-2M_\psi^2}{E^\psi_{\vec p_1-\vec p_2}E^{\bar\psi}_{\vec p_2}}\\
&\qquad\qquad\,\times\delta\left (p_1^0-E^\psi_{\vec p_1-\vec
p_2}-E^{\bar\psi}_{\vec p_2}\right )\Big(1-f^\psi\left (\vec p_1-\vec
p_2\right )\Big)\left (1-f^{\bar\psi}\left (\vec p_2\right )\right )\,,\\
\ii \tilde\Pi^<\left (X,p_1\right )&=2\pi g^2d_\psi\int\frac{\dd^3p_2}{\left (2\pi\right )^3}
\frac{p_1^0E^{\bar\psi}_{\vec p_2}-\vec p_1\cdot\vec
p_2-2M_\psi^2}{E^\psi_{\vec p_1 -\vec p_2}E^{\bar\psi}_{\vec p_2}}\\
&\qquad\qquad\,\times\delta\left (p_1^0-E^\psi_{\vec p_1-\vec
p_2}-E^{\bar\psi}_{\vec p_2}\right )f^\psi\left (\vec p_1-\vec
p_2\right )f^{\bar\psi}\left (\vec p_2\right )\,,
\label{eq:SelfEnegySigmaQuark2}
\end{split} 
\end{equation} 
where all energy conserving $\delta$-functions were neglected, which
cannot be fullfilled on-shell. 

Combining in- as well as out-rates and including the external
propagatoras, leads to the collision integrals of elastic scatterings
between mesons and quarks:
\begin{equation}
\begin{split}
\mathcal{C}_{\sigma\leftrightarrow\psi\bar\psi}^{f.s.}:=&\,\pi
g^2d_\psi\int\frac{\dd^3\vec p_2}{\left(2\pi\right)^3} \int\dd^3\vec
p_3\frac{\delta^{\left (3\right )}\left (\vec p_1 - \vec p_2-\vec
p_3\right )}{E_1^{\sigma}}\left (1-\frac{\vec p_2\cdot\vec p_3}{E_2^{\bar\psi}
E_3^{\psi}}-\frac{M_\psi^2}{E_2^{\bar\psi} E_3^{\psi}}\right )\\
&\quad\times\Big\{\left [\left (1+f_1^{\sigma}\right )f_2^{\bar\psi}f_3^{\psi}-f_1^{\sigma}\left (1-f_2^{\bar\psi}\right )\left (1-f_3^{\psi}\right )\right ]\delta\left (E_1^{\sigma}-E_2^{\bar\psi}-E_3^{\psi}\right )\Big\}\,,\\
\mathcal{C}_{\pi_i\leftrightarrow\psi\bar\psi}^{f.s.}:=&\,\pi
g^2d_\psi\int\frac{\dd^3\vec p_2}{\left(2\pi\right)^3} \int\dd^3\vec
p_3\frac{\delta^{\left (3\right )}\left (\vec p_1 - \vec p_2-\vec
p_3\right )}{E_1^{\pi_i}}\left (1-\frac{\vec p_2\cdot\vec p_3}{E_2^{\bar\psi}
E_3^{\psi}}-\frac{M_\psi^2}{E_2^{\bar\psi} E_3^{\psi}}\right )\\
&\quad\times\Big\{\left [\left (1+f_1^{\pi_i}\right )f_2^{\bar\psi}f_3^{\psi}-f_1^{\pi_i}\left (1-f_2^{\bar\psi}\right )\left (1-f_3^{\psi}\right )\right ]\delta\left (E_1^{\pi_i}-E_2^{\bar\psi}-E_3^{\psi}\right )\Big\}\,.
\label{eq:BosonFermionEquation}
\end{split} 
\end{equation} 
Instead of an explicit and tedious calculation for the collision
integrals of quarks, we make use of detailed balance, leading directly
to the following expressions: 
\begin{equation}
\begin{split} \mathcal{I}^{f.s.}_{\psi}\left (t,\vec
p_1\right )&:=\mathcal{C}_{\psi\bar\psi\leftrightarrow\sigma}^{f.s.}+\sum_i\mathcal{C}_{\psi\bar\psi\leftrightarrow\pi_i}^{f.s.}\,,\\
\mathcal{C}_{\psi\bar\psi\leftrightarrow\sigma}^{f.s.}:=&\,\pi
g^2\int\frac{\dd^3\vec p_2}{\left(2\pi\right)^3} \int\dd^3\vec
p_3\frac{\delta^{\left (3\right )}\left (\vec p_1 + \vec p_2-\vec
p_3\right )}{E_3^{\sigma}}\left (1-\frac{\vec p_1\cdot\vec p_2}{E_1^\psi
E_2^{\bar\psi}}-\frac{M_\psi^2}{E_1^\psi E_2^{\bar\psi}}\right )\\
&\quad\times\Big\{\left [\left (1-f_1^{\psi}\right )\left (1-f_2^{\bar\psi}\right )f_3^{\sigma}-f_1^{\psi}f_2^{\bar\psi}\left (1+f_3^{\sigma}\right )\right ]\delta\left (E_1^{\psi}+E_2^{\bar\psi}-E_3^{\sigma}\right )\Big\}\,,\\
\mathcal{C}_{\psi\bar\psi\leftrightarrow\pi_i}^{f.s.}:=&\,\pi
g^2\int\frac{\dd^3\vec p_2}{\left(2\pi\right)^3} \int\dd^3\vec
p_3\frac{\delta^{\left (3\right )}\left (\vec p_1 + \vec p_2-\vec
p_3\right )}{E_1^{\pi_i}}\left (1-\frac{\vec p_1\cdot\vec p_2}{E_1^\psi
E_2^{\bar\psi}}-\frac{M_\psi^2}{E_1^\psi E_2^{\bar\psi}}\right )\\
&\quad\times\Big\{\left [\left (1-f_1^{\psi}\right )\left (1-f_2^{\bar\psi}\right )f_3^{\pi_i}-f_1^{\psi}f_2^{\bar\psi}\left (1+f_3^{\pi_i}\right )\right ]\delta\left (E_1^{\psi}+E_2^{\bar\psi}-E_3^{\pi_i}\right )\Big\}\,,\\
\label{eq:FermionBosonEquation}
\end{split} 
\end{equation} 
Consequently, the collision integral for anti-quarks
$\mathcal{I}^{f.s.}_{\bar\psi}\left (t,\vec p_1\right )$ follows by
exchanging fermionic field indices $\psi \leftrightarrow\bar\psi$.

\section{Thermodynamic calculations within the imaginary time formalism}
\label{sec:ThermoPotentialImaginary}

\subsection{Bosonic loop integrals}
\label{sec:ThermoBosonicLoopIntegrals}

As already discussed in Sec.~\ref{sec:EffMass} the propagators of a
system in equilibrium depend only on the relative space-time difference,
allowing to perform all relevant calculations in momentum
space. Additionally, by focusing only on momentum independent mass terms
from local part of the self-energy (see propagator relations
\ref{eq:HartreePropagatorMomentum}), one obtains
\begin{equation}
\begin{split}
  \frac{\ii}{2}\Tr \ln\left [G^{-1}_{\varphi_a\varphi_a}\right
  ]&=\frac{\ii}{2}\int_x\int_k\ln\left [G^{-1}_{\varphi_a\varphi_a}\left
      (k\right )\right ] =\frac{\ii}{2}V\sum_n\int\frac{\dd^3\vec
    k}{\left (2\pi\right )^3}\ln\left [-i\left (k_n^2-M_{\varphi_a}^2\right )\right ]\\
  =\frac{\ii}{2}V\sum_n\int\frac{\dd^3\vec
    p}{\left (2\pi\right )^3}\ln\left (k_n^2-M_{\varphi_a}^2\right )\\
\label{eq:LogarithmImaginaryTimeFormalismBosons}
\end{split} 
\end{equation} 
The sum over bosonic Matsubara frequencies can be evaluated by applying
following steps (see also Ref.~\cite{PhysRevD.9.3320}):
\begin{equation}
\begin{split}
  f\left (E_{\varphi_a}\right ):=&\,\sum_n\ln\left [i\left (-k_n^2+M_{\varphi_a}^2\right )\right ]=\sum_n\ln\left [i\left (\frac{4\pi^2n^2}{\beta^2}+E_{\varphi_a}^2\right )\right ]\,,\\
  \frac{\partial f}{\partial
    E_{\varphi_a}}=&\,\sum_n\frac{2E_{\varphi_a}}{\frac{4\pi^2n^2}{\beta^2}+E_{\varphi_a}^2}
  =\beta\left (2\sum_{n=1}^\infty\frac{\left (\frac{\beta
          E_{\varphi_a}}{2}\right )}{\pi^2n^2+\left (\frac{\beta
          E_{\varphi_a}}{2}\right )^2}+\frac{1}{\left (\frac{\beta E_{\varphi_a}}{2}\right )}\right )\\
  =&\,\beta\coth\left (\frac{\beta E_{\varphi_a}}{2}\right )=2\beta\left (\frac{1}{2}+\frac{1}{\ee^{\beta E_{\varphi_a}}-1}\right )\\
  &\Rightarrow f\left (E_{\varphi_a}\right )=2\beta\left
    (\frac{E_{\varphi_a}}{2}+\frac{1}{\beta}\ln\left [1-\ee^{-\beta
        E_{\varphi_a}}\right ]\right )+\text{const.}
\label{eq:SumEvalBosons}
\end{split}
\end{equation}

Neglecting energy independent terms results for
Eq.~\eqref{eq:LogarithmImaginaryTimeFormalismBosons}
in: 
\begin{equation} \frac{\ii}{2}\Tr \ln\left
    [G^{-1}_{\varphi_a\varphi_a}\right ]=\frac{\ii}{2}\beta
  V\int\frac{\dd^3\vec k}{\left (2\pi\right )^3}E_{\varphi_a}
  +iV\int\frac{\dd^3\vec k}{\left (2\pi\right )^3}\ln\left [1-\ee^{-\beta
      E_{\varphi_a}}\right ]\,,
\label{eq:LogarithmImaginaryTimeFormalismFinala} 
\end{equation} 
where the first term requires a proper renormalization. 

Within the Hartree approximation and imaginary time formalism, the
simple loop integral becomes:
\begin{equation}
\begin{split} G_{\varphi_a\varphi_a}&=\int_k
G_{\varphi_a\varphi_a}\left (k\right )=\frac{1}{-\ii \beta}\sum_n\int\frac{\dd^3\vec
k}{\left (2\pi\right )^3}\frac{\ii}{\left (k_n^2-M_{\varphi_a}^2\right )}\\
&=\frac{1}{\beta}\sum_n\int\frac{\dd^3\vec
k}{\left (2\pi\right )^3}\frac{1}{\left (\frac{4\pi^2 n^2}{\beta^2}+E^2\right )}
=\frac{\beta}{4}\sum_n\int\frac{\dd^3\vec
k}{\left (2\pi\right )^3}\frac{1}{\left (\frac{\beta
E_{\varphi_a}}{2}\right )^2+\pi^2n^2}\\ &=\frac{\beta}{4}\int\frac{\dd^3\vec
k}{\left (2\pi\right )^3}\frac{1}{\left (\frac{\beta
E_{\varphi_a}}{2}\right )}\coth\left (\frac{\beta E_{\varphi_a}}{2}\right )\\
&=\frac{1}{2}\int\frac{\dd^3\vec
k}{\left (2\pi\right )^3}\frac{1}{E_{\varphi_a}}+\int\frac{\dd^3\vec
k}{\left (2\pi\right )^3}\frac{1}{E_{\varphi_a}}\frac{1}{\ee^{\beta
E_{\varphi_a}}-1}\,,
\label{eq:LoopImaginaryTimeFormalisma}
\end{split} 
\end{equation} 
leading obviously to the same result as already known
from Eq.~\eqref{eq:LoopBosonic}, when the one-particle distribution
function is replaced by the Bose distribution.

\subsection{Fermionic loop integrals}
\label{sec:ThermoFermionicLoopIntegrals}

The logarithmic term for fermionic propagators has the following form:
\begin{equation}
\begin{split}
  i\Tr \ln\left [D^{-1}_{\psi_i\psi_i}\right ]&=i\int_x\int_k\ln\left [-i\det_{Dirac}\left (k\!\!\!/-M_{\psi_i}\right )\right ]\\
  &=id_{\psi_i} V\sum_n\int\frac{\dd^3\vec k}{\left (2\pi\right
    )^3}\ln\left [i\left (-k_n^2+M_{\psi_i}^2\right )\right ]\\
  &=id_{\psi_i} V\sum_n\int\frac{\dd^3\vec
    k}{\left (2\pi\right )^3}\ln\left [i\left (\frac{\left (2n+1\right )^2\pi^2}{\beta^2}+E_{\psi_i}^2\right )\right ]\,,\\
  =\frac{\ii}{2}V\sum_n\int\frac{\dd^3\vec p}{\left (2\pi\right
    )^3}\ln\left (k_n^2-M_{\varphi_a}^2\right )
\label{eq:LogarithmImaginaryTimeFormalismFermions}
\end{split} 
\end{equation} 
where $d_{\psi_i}=N_cN_s=6$ denotes the degeneracy
factor of a fermionic flavor $\psi_i$ with $N_c=3$ colors and $N_s=2$
spin states.

In analogy to \eqref{eq:SumEvalBosons} the sum over fermionic Matsubara
frequencies can be evaluated by applying following steps (see also
Ref.~\cite{PhysRevD.9.3320}): 
\begin{equation}
\begin{split}
f\left (E_{\psi_i}\right ):=&\,\sum_n\ln\left [\ii \left (\frac{\left (2n+1\right )^2\pi^2}{\beta^2}+E_{\psi_i}^2\right )\right ]\,,\\
\frac{\partial f}{\partial
E_{\psi_i}}=&\,\sum_n\frac{2E_{\psi_i}}{\frac{\left (2n+1\right )^2\pi^2}{\beta^2}+E_{\psi_i}^2}
=2\beta\sum_n\frac{\beta E_{\psi_i}}{\left (2n+1\right )^2\pi^2+\left (\beta
E_{\psi_i}\right )^2}\\ =&\,2\beta\left (\frac{1}{2}+\frac{1}{\ee^{\beta
E_{\psi_i}}+1}\right )\\ &\Rightarrow
f\left (E_{\psi_i}\right )=2\beta\left (\frac{E_{\psi_i}}{2}+\frac{1}{\beta}\ln\left [1+\ee^{-\beta
E_{\psi_i}}\right ]\right )+\text{const.}
\label{eq:SumEvalFermions}
\end{split} 
\end{equation} 
By introducing energy shifts due to the quark chemical potential
$\mu_\psi$ and neglecting energy independent terms, leads for
Eq.~\eqref{eq:LogarithmImaginaryTimeFormalismFermions} to the following
result: 
\begin{equation}
\begin{split} 
\ii\Tr \ln\left [D^{-1}_{\psi_i\psi_i}\right ]&=\ii\beta
d_{\psi_i} V\int\frac{\dd^3\vec k}{\left (2\pi\right )^3}E_{\psi_i}\\ &\quad
+\ii d_{\psi_i} V\int\frac{\dd^3\vec
k}{\left (2\pi\right )^3}\left (\ln\left [1+\ee^{-\beta\left (E_{\psi_i}-\mu_{\psi_i}\right )}\right ]
+\ln \left [1+\ee^{-\beta\left (E_{\psi_i}+\mu_{\psi_i}\right )}\right ]\right )\,.
\label{eq:LogarithmImaginaryTimeFormalismFinalb}
\end{split} 
\end{equation} 
Within the Hartree approximation and imaginary time
formalism, the simple loop integral becomes 
\begin{equation}
\begin{split} D_{\psi_a\varphi_a}&=\int_k
G_{\varphi_a\varphi_a}\left (k\right )=\frac{1}{-\ii \beta}\sum_n\int\frac{\dd^3\vec
k}{\left (2\pi\right )^3}\frac{\ii}{\left (k_n^2-M_{\varphi_a}^2\right )}\\
&=\frac{1}{\beta}\sum_n\int\frac{\dd^3\vec
k}{\left (2\pi\right )^3}\frac{1}{\left (\frac{4\pi^2 n^2}{\beta^2}+E^2\right )}
=\frac{\beta}{4}\sum_n\int\frac{\dd^3\vec
k}{\left (2\pi\right )^3}\frac{1}{\left (\frac{\beta
E_{\varphi_a}}{2}\right )^2+\pi^2n^2}\\ &=\frac{\beta}{4}\int\frac{\dd^3\vec
k}{\left (2\pi\right )^3}\frac{1}{\left (\frac{\beta
E_{\varphi_a}}{2}\right )}\coth\left (\frac{\beta E_{\varphi_a}}{2}\right )\\
&=\frac{1}{2}\int\frac{\dd^3\vec
k}{\left (2\pi\right )^3}\frac{1}{E_{\varphi_a}}+\int\frac{\dd^3\vec
k}{\left (2\pi\right )^3}\frac{1}{E_{\varphi_a}}\frac{1}{\ee^y{\beta
E_{\varphi_a}}-1}\,,
\label{eq:LoopImaginaryTimeFormalismb}
\end{split} 
\end{equation} 
leading obviously to the same result as already known
from Eq.~\eqref{eq:LoopFermionic}, when the one-particle distribution
function is replaced by the Bose distribution.

\section{Isotropic Boltzmann equations}
\label{chap:IsoBoltzmann}
\subsection{Isotropic collisions between scalar bosons}
\label{sec:IsoCollInt1}

In the case of a $\phi^4$-interaction with a constant matrix element one
can integrate out the angular dependence of a general
Boltzmann-Uehling-Uhlenbeck equation,
\begin{equation}
\begin{split}
  \partial_t f\left(t, \vec{p}_1\right) =
  &\frac{1}{2E_1}\int\frac{\dd^3\vec
    p_2}{\left(2\pi\right)^32E_2}\int\frac{\dd^3\vec{p}_3}{\left(2\pi\right)^32E_3}
  \int\frac{\dd^3\vec{p}_4}{\left(2\pi\right)^32E_4} \\
  &\times\left(2\pi\right)^4\delta^{(4)}\left(P_1+P_2-P_3-P_4\right)\frac{\left
      |\mathcal{M}_{12\rightarrow 34} \right|^2}{\nu} \\ &\times
  [\left(1+f_1\right)\left(1+f_2\right)f_3f_4
  -f_1f_2\left(1+f_3\right)\left(1+f_4\right) ]\,,
\label{eq:200}
\end{split}
\end{equation}
by simply using the Fourier transform of the momentum conserving
$\delta$-function
\begin{equation}
\begin{split} \left (2\pi\right )^3\delta^{(3)}\left (\vec{p}\right
)&=\int_{\R^3}\dd^3q\exp\left (-\ii\,\vec{p}\cdot\vec{q}\right )\\
&=\int_0^{\infty} \,\dd q \, q^2 \int_0^{2 \pi}\, \dd \varphi
\int_0^{\pi} \dd \vartheta_q\sin\vartheta_q\exp\left
[-\ii\,pq\cos\vartheta_q\right ]\\ &=\int_0^{\infty} \dd q \, q^2\left
[\frac{4\pi}{pq}\sin\left (pq\right )\right ]\,,
\label{eq:a02b}
\end{split}
\end{equation} 
leading to the isotropic form
\begin{equation}
\begin{split} 
\partial_t f_1\sim &\int_0^{\infty}\,\dd p_2\,\int_0^{\infty}\,\dd p_3\,\int_0^{\infty}\,\dd
p_4\,\int_0^{\infty}\,\dd q \frac{p_2p_3p_4}{E_{1}E_{2}E_{3}E_{4}}\delta\left
(E_1+E_2-E_3-E_4\right )\\ &\times\frac{\sin\left
(p_1q\right )\sin\left (p_2q\right )\sin\left (p_3q\right )\sin\left
(p_4q\right )}{p_1q^2} [\left (1+f_1\right )\left (1+f_2\right )f_3f_4-f_1f_2\left (1+f_3\right )\left (1+f_4\right ) ]\,.
\label{eq:a03b}
\end{split}
\end{equation} 
Defining an auxiliary function which depends on the involved momenta
\begin{equation} \mathcal{F}_{1,2\leftrightarrow
3,4}^{p_1}=\int_0^{\infty}\,\dd q \frac{\sin\left (p_1q\right )\sin\left
(p_2q\right )\sin\left (p_3q\right )\sin\left (p_4q\right )}{p_1q^2} \,
,
\label{eq:a04}
\end{equation} 
and integrating out the $q$-dependence leads to
\begin{align}
\begin{split} \mathcal{F}_{1,2\leftrightarrow
3,4}^{p_1=0}=\frac{\pi}{8}\Bigg[&\sign\left(p_2+p_3-p_4\right)-\sign\left(p_2-p_3-p_4\right)
\\
&+\sign\left(p_2-p_3+p_4\right)-\sign\left(p_2+p_3+p_4\right)\Bigg]\,,
\label{eq:a05}
\end{split}\\
\begin{split} \mathcal{F}_{1,2\leftrightarrow
3,4}^{p_1>0}=\frac{\pi}{16p_1} \Bigg[ &\left |p_1-p_2-p_3-p_4\right |
-\left |p_1+p_2-p_3-p_4\right | \\ &-\left |p_1-p_2+p_3-p_4\right |
+\left |p_1+p_2+p_3-p_4\right | \\ &-\left |p_1-p_2-p_3+p_4\right |
+\left |p_1+p_2-p_3+p_4\right | \\ &+\left |p_1-p_2+p_3+p_4\right |
-\left |p_1+p_2+p_3+p_4\right | \Bigg]\,,
\label{eq:a06}
\end{split}
\end{align} 
where the form of $\mathcal{F}_{1,2\leftrightarrow 3,4}^{p_1}$ depends
especially on the incoming momentum $p_1$.  The remaining energy
conserving $\delta$-function is used to reduce the dimension of the
integral by writing
\begin{equation} 
\delta\left (E_1+E_2-E_3-E_4\right
)=\frac{E_4}{p_4}\delta\left (p_4-\tilde p_4\right )\,.
\label{eq:a07}
\end{equation} 
Here $\tilde p_4$ is given via energy-momentum conservation with the
relativistic dispersion relation
\begin{equation} 
\tilde p_4=\sqrt{\left (E_1+E_2-E_3\right )^2-m_4^2}\,.
\label{eq:a08}
\end{equation} 
Finally, one ends up with the following expression for the isotropic
Boltzmann equation
\begin{equation}
\begin{split} \partial_t f_1\sim & \int_0^{\infty}\,\dd
p_2\int_0^{\infty}\,\dd p_3\frac{p_2p_3}{E_1E_2E_3}\theta\left (\tilde
p_4^2\right )\mathcal{F}_{1,2\leftrightarrow 3,4}^{p_1} \\ &\times
[\left (1+f_1\right )\left (1+f_2\right )f_3f_4 -f_1f_2\left
(1+f_3\right )\left (1+f_4\right ) ]\,,
\label{eq:a09}
\end{split}
\end{equation} 
where the $\theta$-function ensures that $\tilde{p}_4^2
\geq 0$.

In analogy to \eqref{eq:a04} one can define the auxiliary functions
\begin{equation}
\begin{split} \mathcal{F}_{1,2\leftrightarrow
3,4}^{p_2=0}&=\int_0^{\infty}\,\dd q \frac{\sin\left (p_1q\right
)\sin\left (p_3q\right )\sin\left (p_4q\right )}{p_1q}, \\
\mathcal{F}_{1,2\leftrightarrow 3,4}^{p_3=0}&=\int_0^{\infty}\,\dd q
\frac{\sin\left (p_1q\right )\sin\left (p_2q\right )\sin\left
(p_4q\right )}{p_1q}, \\ \mathcal{F}_{1,2\leftrightarrow
3,4}^{p_4=0}&=\int_0^{\infty}\,\dd q \frac{\sin\left (p_1q\right
)\sin\left (p_2q\right )\sin\left (p_3q\right )}{p_1q}\,,
\label{eq:a10}
\end{split}
\end{equation} 
allowing for the description of condensation and mean-field dynamics in
isotropic systems, when one of the internal modes has zero momentum.

\subsection{Isotropic collisions between scalar bosons and fermions}
\label{sec:IsoCollInt2}

Compared to relations of Sec.~\ref{sec:IsoCollInt1}, the isotropic
integrations for fermions are carried out along similar lines. The main
difference arises from the scalar product of fermionic momenta (compare
with Eqs.~\eqref{eq:BosonFermionEquation} and
\eqref{eq:FermionBosonEquation}). Writing the scalar product between the
fermions in spherical coordinates leads to
\begin{equation}
\begin{split} \vec p_2\cdot\vec
p_3&=p_2p_3\Big[\sin\left (\vartheta_2\right )\sin\left (\vartheta_3\right )\cos\left (\varphi_2\right )\cos\left (\varphi_3\right )\\
&\qquad\qquad
+\sin\left (\vartheta_2\right )\sin\left (\vartheta_3\right )\sin\left (\varphi_2\right )\sin\left (\varphi_3\right )+\cos\left (\vartheta_2\right )\cos\left (\vartheta_3\right )\Big]\,.
\label{eq:SphericalScalarProduct}
\end{split} 
\end{equation} 
Without loss of generality, the integration over the
inner momenta $\vec p_2$ and $\vec p_3$ can be performed by choosing the
$z$-axis in direction of the auxiliary vector $\vec q$, which is used for
the Fourier transform of the momentum conserving $\delta$-function (see
Eq.~\eqref{eq:a02b}).  For the energy conserving $\delta$-function it
follows
\begin{equation}
\begin{split}
\delta\left (E_1^{\varphi_a}-E_2^{\bar\psi}-E_3^{\psi}\right )=&\,\frac{E_2^{\bar\psi}}{p_2}\delta\left (p_2-\tilde
p_2\right )\,,\qquad \tilde
p_2:=\sqrt{\left (E_1^{\varphi_a}-E_3^\psi\right )^2-M^2_\psi}\,.
%\label{eq:}
\end{split} 
\end{equation} 
In case of fermions, it is convenient to introduce some additional
auxiliary functions and analytically integrate out the dependence on the
vector $\vec q$:
\begin{equation*}
\begin{split} \mathcal{F}_{1\leftrightarrow 2,3}^{1,p_1} &=
\int_0^\infty\,\dd
q\frac{\sin\left (p_1q\right )\sin\left (p_2q\right )\sin\left (p_3q\right )}{p_1q}\\
&=\frac{\pi}{8p_1}\Big[\sign\left (p_1+p_2-p_3\right )-\sign\left (p_1-p_2-p_3\right )\\
&\qquad\qquad +\sign\left (p_1-p_2+p_3\right )-\sign\left (p_1+p_2+p_3\right )\Big]\,,\\
\mathcal{F}_{1\leftrightarrow 2,3}^{2,p_1} &= \int_0^\infty\,\dd
q\frac{\sin\left (p_1q\right )\sin\left (p_2q\right )\sin\left (p_3q\right )}{p_1q^3}\\
&=\frac{\pi}{16p_1}\Big[\left (p_1+p_2+p_3\right )^2\sign\left (p_1+p_2+p_3\right )\\
&\qquad\qquad +\left (p_2+p_3-p_1\right )^2\sign\left (p_1-p_2-p_3\right )\\ &\qquad\qquad
-\left (p_1+p_2-p_3\right )^2\sign\left (p_1+p_2-p_3\right )\\ &\qquad\qquad
-\left (p_1-p_2+p_3\right )^2\sign\left (p_1-p_2+p_3\right )\Big]\,,\\
\end{split}
\end{equation*} 
\begin{equation}
\begin{split} \mathcal{F}_{1\leftrightarrow 2,3}^{3,p_1} &=
\int_0^\infty\,\dd
q\frac{\sin\left (p_1q\right )\sin\left (p_2q\right )\cos\left (p_3q\right )}{p_1q^2}\\
&=\frac{\pi}{8p_1}\Big[p_1+p_2+p_3+|p_1+p_2-p_3|\\
&\qquad\qquad-|p_1-p_2+p_3|-|p_2+p_3-p_1|\Big]\,,\\
\mathcal{F}_{1\leftrightarrow 2,3}^{4,p_1} &= \int_0^\infty\,\dd
q\frac{\sin\left (p_1q\right )\cos\left (p_2q\right )\sin\left (p_3q\right )}{p_1q^2}\\
&=\frac{\pi}{8p_1}\Big[p_1+p_2+p_3+|p_1-p_2+p_3|\\ &\qquad\qquad
-|p_1+p_2-p_3|-|p_2+p_3-p_1|\Big]\,,\\ \mathcal{F}_{1\leftrightarrow
2,3}^{5,p_1} &= \int_0^\infty\,\dd
q\frac{\sin\left (p_1q\right )\cos\left (p_2q\right )\cos\left (p_3q\right )}{p_1q}\\
&=\frac{\pi}{8p_1}\Big[\sign\left (p_1-p_2-p_3\right )+\sign\left (p_1+p_2-p_3\right )\\
&\qquad\qquad +\sign\left (p_1-p_2+p_3\right )+\sign\left (p_1+p_2+p_3\right )\Big]\,.\\
\label{eq:AuxiliaryFunctionsFermion}
\end{split} 
\end{equation} 
Finally, one arrives at the following form for the collision
integrals: 
\begin{equation}
\begin{split} \partial_t f^{\varphi_a}_1 &\sim\int\dd
p_3\Bigg[\left (\frac{p_3E_2^{\bar\psi}}{E_1^{\varphi_a}}-\frac{p_3
M^2_\psi}{E_1^{\varphi_a}E_3^\psi}\right )\mathcal{F}_{1\leftrightarrow
2,3}^{1,p_1}+\frac{p_3}{E_1^{\varphi_a}E_3^\psi}\mathcal{F}_{1\leftrightarrow2,3}^{2,p_1}\\
&\qquad\qquad\quad-\frac{p_3^2}{E_1^{\varphi_a}E_3^\psi}\mathcal{F}_{1\leftrightarrow
2,3}^{3,p_1}-\frac{p_2p_3}{E_1^{\varphi_a}E_3^\psi}\mathcal{F}_{1\leftrightarrow
2,3}^{4,p_1}+\frac{p_2p_3^2}{E_1^{\varphi_a}E_3^{\psi}}\mathcal{F}_{1\leftrightarrow
2,3}^{5,p_1}\Bigg]\theta\left (\tilde p_2^2\right )\\
&\qquad\qquad\times\left [\left (1+f_1^{\varphi_a}\right
  )f_2^{\bar\psi}f_3^{\psi}-f_1^{\varphi_a}\left (1-f_2^{\bar\psi}\right
  )\left (1-f_3^{\psi}\right )\right ]\,, 
\label{eq:IsotropicBoltzmannEquationFermion1a}
\end{split} 
\end{equation} 
\begin{equation}
  \begin{split} \partial_t f^{\psi}_1 &\sim\int\dd
    p_3\Bigg[\left (\frac{p_3^2E_2^{\bar\psi}}{p_1E_3^{\varphi_a}}-\frac{p_3^2
        M^2_\psi}{p_1E_1^{\psi}E_3^{\varphi_a}}\right )\mathcal{F}_{1\leftrightarrow
      2,3}^{1,p_3}+\frac{p_3^2}{p_1E_1^{\psi}E_3^{\varphi_a}}\mathcal{F}_{1\leftrightarrow2,3}^{2,p_3}\\
    &\qquad\qquad\quad-\frac{p_3^2}{E_1^{\psi}E_3^{\varphi_a}}\mathcal{F}_{1\leftrightarrow
      2,3}^{3,p_3}-\frac{p_2p_3^2}{p_1E_1^{\psi}E_3^{\varphi_a}}\mathcal{F}_{1\leftrightarrow
      2,3}^{4,p_3}+\frac{p_2p_3^2}{E_1^{\psi}E_3^{\varphi_a}}\mathcal{F}_{1\leftrightarrow
      2,3}^{5,p_3}\Bigg]\theta\left (\tilde p_2^2)\right )\\
    &\qquad\qquad\times\left [\left (1-f_2^{\psi}\right )\left
        (1-f_2^{\bar\psi}\right
      )f_3^{\varphi_a}-f_1^{\psi}f_2^{\bar\psi}\left
        (1+f_3^{\varphi_a}\right )\right ]\,,
\label{eq:IsotropicBoltzmannEquationFermion1b}
\end{split} 
\end{equation} 
where the auxiliary functions with the index $p_3$ are obtained by
exchanging and renaming the momentum $p_1$ of the boson with $p_3$ of
the fermion in \eqref{eq:AuxiliaryFunctionsFermion}.  The expression for
$\partial_t f^{\bar\psi}_1$ follows analogously to
Eq.~\eqref{eq:IsotropicBoltzmannEquationFermion1b}.

\pagebreak

\bibliographystyle{els-hvh}

\begin{thebibliography}{10}
\providecommand{\url}[1]{\texttt{#1}}
\providecommand{\urlprefix}{URL }
\providecommand{\eprint}[2][]{\url{#2}}

\bibitem{ARSENE20051}
I.~A. et~al., Nuclear Physics A 757 (2005) 1 , first Three Years of Operation
  of RHIC, \urlprefix\url{https://doi.org/10.1016/j.nuclphysa.2005.02.130}.

\bibitem{FODOR200287}
Z.~Fodor and S.~Katz, Physics Letters B 534 (2002) 87 ,
  \urlprefix\url{https://doi.org/10.1016/S0370-2693(02)01583-6}.

\bibitem{Aoki:2006we}
Y.~Aoki, G.~Endrodi, Z.~Fodor, S.~D. Katz, and K.~K. Szabo, Nature 443 (2006)
  675, \urlprefix\url{https://dx.doi.org/10.1038/nature05120}.

\bibitem{doi:10.1142/9789814663717_0001}
H.-T. Ding, F.~Karsch, and S.~Mukherjee, \emph{{Thermodynamics of
  Strong-Interaction Matter from Lattice QCD}} (World Scientific, 2016), 1--65,
  \urlprefix\url{https://dx.doi.org/10.1142/9789814663717_0001}.

\bibitem{PhysRev.122.345}
Y.~Nambu and G.~Jona-Lasinio, Phys. Rev. 122 (1961) 345,
  \urlprefix\url{https://dx.doi.org/10.1103/PhysRev.122.345}.

\bibitem{PhysRev.124.246}
Y.~Nambu and G.~Jona-Lasinio, Phys. Rev. 124 (1961) 246,
  \urlprefix\url{https://dx.doi.org/10.1103/PhysRev.124.246}.

\bibitem{Jungnickel:1995fp}
D.~U. Jungnickel and C.~Wetterich, Phys. Rev. D 53 (1996) 5142.

\bibitem{doi:10.1142/S0217751X03014034}
J.~Berges, D.-U. Jungnickel, and C.~Wetterich, International Journal of Modern
  Physics A 18 (2003) 3189,
  \urlprefix\url{https://dx.doi.org/10.1142/S0217751X03014034}.

\bibitem{PhysRevD.73.014019}
C.~Ratti, M.~A. Thaler, and W.~Weise, Phys. Rev. D 73 (2006) 014019,
  \urlprefix\url{https://dx.doi.org/10.1103/PhysRevD.73.014019}.

\bibitem{PhysRevD.75.034007}
S.~R\"o\ss{}ner, C.~Ratti, and W.~Weise, Phys. Rev. D 75 (2007) 034007,
  \urlprefix\url{https://dx.doi.org/10.1103/PhysRevD.75.034007}.

\bibitem{PhysRevD.76.074023}
B.-J. Schaefer, J.~M. Pawlowski, and J.~Wambach, Phys. Rev. D 76 (2007) 074023,
  \urlprefix\url{https://dx.doi.org/10.1103/PhysRevD.76.074023}.

\bibitem{HERBST201158}
T.~K. Herbst, J.~M. Pawlowski, and B.-J. Schaefer, Physics Letters B 696 (2011)
  58 , \urlprefix\url{https://doi.org/10.1016/j.physletb.2010.12.003}.

\bibitem{MASAYUKI1989668}
A.~Masayuki and Y.~Koichi, Nuclear Physics A 504 (1989) 668 ,
  \urlprefix\url{https://doi.org/10.1016/0375-9474(89)90002-X}.

\bibitem{PhysRevD.58.096007}
M.~A. Halasz, A.~D. Jackson, R.~E. Shrock, M.~A. Stephanov, and J.~J.~M.
  Verbaarschot, Phys. Rev. D 58 (1998) 096007,
  \urlprefix\url{https://dx.doi.org/10.1103/PhysRevD.58.096007}.

\bibitem{doi:10.1142/S0217751X92001757}
F.~Wilczek, International Journal of Modern Physics A 07 (1992) 3911,
  \urlprefix\url{https://dx.doi.org/10.1142/S0217751X92001757}.

\bibitem{PhysRevD.75.085015}
B.-J. Schaefer and J.~Wambach, Phys. Rev. D 75 (2007) 085015,
  \urlprefix\url{https://dx.doi.org/10.1103/PhysRevD.75.085015}.

\bibitem{Asakawa:2019kek}
M.~Asakawa, M.~Kitazawa, and B.~Müller, Phys.\ Rev.\ C 101 (2020) 034913,
  \urlprefix\url{https://dx.doi.org/10.1103/PhysRevC.101.034913}.

\bibitem{PhysRevD.60.114028}
M.~Stephanov, K.~Rajagopal, and E.~Shuryak, Phys. Rev. D 60 (1999) 114028,
  \urlprefix\url{https://dx.doi.org/10.1103/PhysRevD.60.114028}.

\bibitem{PhysRevLett.85.2072}
M.~Asakawa, U.~Heinz, and B.~M\"uller, Phys. Rev. Lett. 85 (2000) 2072,
  \urlprefix\url{https://dx.doi.org/10.1103/PhysRevLett.85.2072}.

\bibitem{PhysRevLett.102.032301}
M.~A. Stephanov, Phys. Rev. Lett. 102 (2009) 032301,
  \urlprefix\url{https://dx.doi.org/10.1103/PhysRevLett.102.032301}.

\bibitem{PhysRevLett.103.262301}
M.~Asakawa, S.~Ejiri, and M.~Kitazawa, Phys. Rev. Lett. 103 (2009) 262301,
  \urlprefix\url{https://dx.doi.org/10.1103/PhysRevLett.103.262301}.

\bibitem{PhysRevD.82.074008}
C.~Athanasiou, K.~Rajagopal, and M.~Stephanov, Phys. Rev. D 82 (2010) 074008,
  \urlprefix\url{https://dx.doi.org/10.1103/PhysRevD.82.074008}.

\bibitem{WETTERICH199390}
C.~Wetterich, Physics Letters B 301 (1993) 90 ,
  \urlprefix\url{https://doi.org/10.1016/0370-2693(93)90726-X}.

\bibitem{doi:10.1142/S0217751X94000972}
T.~R. Morris, International Journal of Modern Physics A 09 (1994) 2411,
  \urlprefix\url{https://dx.doi.org/10.1142/S0217751X94000972}.

\bibitem{BERGES2002223}
J.~Berges, N.~Tetradis, and C.~Wetterich, Physics Reports 363 (2002) 223 ,
  renormalization group theory in the new millennium. IV,
  \urlprefix\url{https://doi.org/10.1016/S0370-1573(01)00098-9}.

\bibitem{PhysRevC.83.054904}
V.~Skokov, B.~Friman, and K.~Redlich, Phys. Rev. C 83 (2011) 054904,
  \urlprefix\url{https://dx.doi.org/10.1103/PhysRevC.83.054904}.

\bibitem{PhysRevD.66.074507}
C.~R. Allton, S.~Ejiri, S.~J. Hands, O.~Kaczmarek, F.~Karsch, E.~Laermann,
  C.~Schmidt, and L.~Scorzato, Phys. Rev. D 66 (2002) 074507,
  \urlprefix\url{https://dx.doi.org/10.1103/PhysRevD.66.074507}.

\bibitem{PhysRevD.95.054504}
A.~Bazavov, H.-T. Ding, P.~Hegde, O.~Kaczmarek, F.~Karsch, E.~Laermann,
  Y.~Maezawa, S.~Mukherjee, H.~Ohno, P.~Petreczky, H.~Sandmeyer,
  P.~Steinbrecher, C.~Schmidt, S.~Sharma, W.~Soeldner, and M.~Wagner, Phys.
  Rev. D 95 (2017) 054504,
  \urlprefix\url{https://dx.doi.org/10.1103/PhysRevD.95.054504}.

\bibitem{PhysRevLett.105.022302}
M.~M. e.~a. Aggarwal (STAR Collaboration), Phys. Rev. Lett. 105 (2010) 022302,
  \urlprefix\url{https://dx.doi.org/10.1103/PhysRevLett.105.022302}.

\bibitem{LUO201675}
X.~Luo, Nuclear Physics A 956 (2016) 75 ,
  \urlprefix\url{https://doi.org/10.1016/j.nuclphysa.2016.03.025}.

\bibitem{HE2016296}
S.~He, X.~Luo, Y.~Nara, S.~Esumi, and N.~Xu, Physics Letters B 762 (2016) 296 ,
  \urlprefix\url{https://doi.org/10.1016/j.physletb.2016.09.053}.

\bibitem{PhysRevLett.107.052301}
M.~A. Stephanov, Phys. Rev. Lett. 107 (2011) 052301,
  \urlprefix\url{https://dx.doi.org/10.1103/PhysRevLett.107.052301}.

\bibitem{Meistrenko:2013yya}
A.~Meistrenko, C.~Wesp, H.~van Hees, and C.~Greiner, J. Phys. Conf. Ser. 503
  (2014) 012003,
  \urlprefix\url{https://dx.doi.org/10.1088/1742-6596/503/1/012003}.

\bibitem{Wesp2018}
C.~Wesp, H.~van Hees, A.~Meistrenko, and C.~Greiner, Eur. Phys. J. A 54 (2018)
  24, \urlprefix\url{https://dx.doi.org/10.1140/epja/i2018-12464-y}.

\bibitem{Shen:2020jya}
L.~Shen, J.~Berges, J.~M. Pawlowski, and A.~Rothkopf, Phys. Rev. D 102 (2020)
  016012, \urlprefix\url{https://doi.org/10.1103/PhysRevD.102.016012}.

\bibitem{dgh92}
J.~F. Donoghue, E.~Golowich, and B.~R. Holstein, \emph{{{Dynamics} of the
  {Standard} {Model}}} (Cambridge University press, 1992).

\bibitem{doi:10.1142/S0218301397000147}
V.~Koch, International Journal of Modern Physics E 06 (1997) 203,
  \urlprefix\url{https://dx.doi.org/10.1142/S0218301397000147}.

\bibitem{Patrignani:2016xqp}
C.~Patrignani et~al. (Particle Data Group), Chin. Phys. C 40 (2016) 100001,
  \urlprefix\url{https://dx.doi.org/10.1088/1674-1137/40/10/100001}.

\bibitem{PhysRev.175.2195}
M.~Gell-Mann, R.~J. Oakes, and B.~Renner, Phys. Rev. 175 (1968) 2195,
  \urlprefix\url{https://dx.doi.org/10.1103/PhysRev.175.2195}.

\bibitem{PhysRev.127.965}
J.~Goldstone, A.~Salam, and S.~Weinberg, Phys. Rev. 127 (1962) 965,
  \urlprefix\url{https://dx.doi.org/10.1103/PhysRev.127.965}.

\bibitem{PhysRevC.64.045202}
O.~Scavenius, A.~M\'ocsy, I.~N. Mishustin, and D.~H. Rischke, Phys. Rev. C 64
  (2001) 045202, \urlprefix\url{https://dx.doi.org/10.1103/PhysRevC.64.045202}.

\bibitem{vanHees:2013qla}
H.~van Hees, C.~Wesp, A.~Meistrenko, and C.~Greiner, Acta Phys. Polon. Supp. 7
  (2014) 59, \urlprefix\url{https://dx.doi.org/10.5506/APhysPolBSupp.7.59}.

\bibitem{Wesp:2014xpa}
C.~Wesp, H.~van Hees, A.~Meistrenko, and C.~Greiner, Phys. Rev. E 91 (2015)
  043302, \urlprefix\url{https://dx.doi.org/10.1103/PhysRevE.91.043302}.

\bibitem{Greiner:2015tra}
C.~Greiner, C.~Wesp, H.~van Hees, and A.~Meistrenko, J. Phys. Conf. Ser. 636
  (2015) 012007,
  \urlprefix\url{https://dx.doi.org/10.1088/1742-6596/636/1/012007}.

\bibitem{Wesp2015}
C.~Wesp, \emph{{Dynamical simulation of a linear sigma model near the chiral
  phase transition}}, Ph.D. thesis, Goethe-Universit{\"a}t Frankfurt (2015).

\bibitem{doi:10.1063/1.1703727}
J.~Schwinger, Journal of Mathematical Physics 2 (1961) 407,
  \urlprefix\url{https://dx.doi.org/10.1063/1.1703727}.

\bibitem{Keldysh:1964ud}
L.~V. Keldysh, Zh. Eksp. Teor. Fiz. 47 (1964) 1515,
  \urlprefix\url{http://www.jetp.ac.ru/cgi-bin/e/index/e/20/4/p1018?a=list}.

\bibitem{book:16449}
D.~V.~S. Michael E.~Peskin, \emph{{An introduction to quantum field theory}},
  Frontiers in Physics (Addison-Wesley Pub. Co, 1995).

\bibitem{doi:10.1143/PTP.14.351}
T.~Matsubara, Progress of Theoretical Physics 14 (1955) 351,
  \urlprefix\url{https://dx.doi.org/10.1143/PTP.14.351}.

\bibitem{PhysRevD.9.3320}
L.~Dolan and R.~Jackiw, Phys. Rev. D 9 (1974) 3320,
  \urlprefix\url{https://dx.doi.org/10.1103/PhysRevD.9.3320}.

\bibitem{doi:10.1063/1.1843591}
J.~Berges, AIP Conference Proceedings 739 (2004) 3,
  \urlprefix\url{https://dx.doi.org/10.1063/1.1843591}.

\bibitem{Gelis:1994dp}
F.~Gelis, Z. Phys. C 70 (1996) 321,
  \urlprefix\url{https://dx.doi.org/10.1007/s002880050109}.

\bibitem{Gelis:1999nx}
F.~Gelis, Phys. Lett. B 455 (1999) 205, \eprint{hep-ph/9901263},
  \urlprefix\url{https://dx.doi.org/10.1016/S0370-2693(99)00460-8}.

\bibitem{PhysRevD.9.1686}
R.~Jackiw, Phys. Rev. D 9 (1974) 1686,
  \urlprefix\url{https://dx.doi.org/10.1103/PhysRevD.9.1686}.

\bibitem{phd:HendrikVanHees}
H.~van Hees, \emph{{Renormierung selbstkonsistenter Näherungen in der
  Quantenfeldtheorie bei endlichen Temperaturen}}, Ph.D. thesis, Technische
  Universi{\"a}t Darmstadt (2000).

\bibitem{PhysRevD.10.2428}
J.~M. Cornwall, R.~Jackiw, and E.~Tomboulis, Phys. Rev. D 10 (1974) 2428,
  \urlprefix\url{https://dx.doi.org/10.1103/PhysRevD.10.2428}.

\bibitem{PhysRevD.47.2356}
G.~Amelino-Camelia and S.-Y. Pi, Phys. Rev. D 47 (1993) 2356,
  \urlprefix\url{https://dx.doi.org/10.1103/PhysRevD.47.2356}.

\bibitem{PhysRevD.65.025010}
H.~van Hees and J.~Knoll, Phys. Rev. D 65 (2001) 025010,
  \urlprefix\url{https://dx.doi.org/10.1103/PhysRevD.65.025010}.

\bibitem{PhysRevD.65.105005}
H.~van Hees and J.~Knoll, Phys. Rev. D 65 (2002) 105005,
  \urlprefix\url{https://dx.doi.org/10.1103/PhysRevD.65.105005}.

\bibitem{PhysRev.127.1391}
G.~Baym, Phys. Rev. 127 (1962) 1391,
  \urlprefix\url{https://dx.doi.org/10.1103/PhysRev.127.1391}.

\bibitem{IVANOV1999413}
Y.~Ivanov, J.~Knoll, and D.~Voskresensky, Nuclear Physics A 657 (1999) 413 ,
  \urlprefix\url{https://dx.doi.org/10.1016/S0375-9474(99)00313-9}.

\bibitem{IVANOV2000313}
Y.~Ivanov, J.~Knoll, and D.~Voskresensky, Nuclear Physics A 672 (2000) 313 ,
  \urlprefix\url{https://dx.doi.org/10.1016/S0375-9474(99)00559-X}.

\bibitem{KNOLL2001126}
J.~Knoll, Y.~Ivanov, and D.~Voskresensky, Annals of Physics 293 (2001) 126 ,
  \urlprefix\url{http://dx.doi.org/10.1006/aphy.2001.6185}.

\bibitem{Ivanov2003}
Y.~B. Ivanov, J.~Knoll, and D.~N. Voskresensky, Physics of Atomic Nuclei 66
  (2003) 1902, \urlprefix\url{https://dx.doi.org/10.1134/1.1619502}.

\bibitem{PhysRevD.15.2897}
G.~Baym and G.~Grinstein, Phys. Rev. D 15 (1977) 2897,
  \urlprefix\url{https://dx.doi.org/10.1103/PhysRevD.15.2897}.

\bibitem{PhysRevD.66.025028}
H.~van Hees and J.~Knoll, Phys. Rev. D 66 (2002) 025028,
  \urlprefix\url{https://dx.doi.org/10.1103/PhysRevD.66.025028}.

\bibitem{PILAFTSIS2013594}
A.~Pilaftsis and D.~Teresi, Nuclear Physics B 874 (2013) 594 ,
  \urlprefix\url{https://doi.org/10.1016/j.nuclphysb.2013.06.004}.

\bibitem{book:1046942}
J.~Maciejko, \emph{{An Introduction to Non-equilibrium Many-Body Theory}}
  (Springer, Lecture Notes, 2007).

\bibitem{book:304428}
J.~Rammer, \emph{{Quantum Field Theory of Non-equilibrium States}} (Cambridge
  University Press, 2007).

\bibitem{Juchem:2004cs}
S.~Juchem, W.~Cassing, and C.~Greiner, Nucl. Phys. A 743 (2004) 92,
  \urlprefix\url{https://dx.doi.org/10.1016/j.nuclphysa.2004.07.010}.

\bibitem{Juchem:2003bi}
S.~Juchem, W.~Cassing, and C.~Greiner, Phys. Rev. D 69 (2004) 025006,
  \urlprefix\url{https://dx.doi.org/10.1103/PhysRevD.69.025006}.

\bibitem{Rischke:1998qy}
D.~H. Rischke, Phys. Rev. C 58 (1998) 2331,
  \urlprefix\url{https://dx.doi.org/10.1103/PhysRevC.58.2331}.

\bibitem{0954-3899-26-4-309}
J.~T. Lenaghan and D.~H. Rischke, J. Phys. G 26 (2000) 431,
  \urlprefix\url{https://dx.doi.org/10.1088/0954-3899/26/4/309}.

\bibitem{Blaizot:2003br}
J.-P. Blaizot, E.~Iancu, and U.~Reinosa, Phys. Lett. B 568 (2003) 160,
  \urlprefix\url{https://doi.org/10.1016/j.physletb.2003.06.008}.

\bibitem{PhysRevC.73.034909}
B.~Schenke and C.~Greiner, Phys. Rev. C 73 (2006) 034909,
  \urlprefix\url{https://dx.doi.org/10.1103/PhysRevC.73.034909}.

\bibitem{Greiner:1998vd}
C.~Greiner and S.~Leupold, Ann. Phys. 270 (1998) 328,
  \urlprefix\url{https://dx.doi.org/10.1006/aphy.1998.5849}.

\bibitem{PhysRevD.55.1026}
C.~Greiner and B.~M\"uller, Phys. Rev. D 55 (1997) 1026,
  \urlprefix\url{https://dx.doi.org/10.1103/PhysRevD.55.1026}.

\bibitem{PhysRevD.62.036012}
Z.~Xu and C.~Greiner, Phys. Rev. D 62 (2000) 036012,
  \urlprefix\url{https://dx.doi.org/10.1103/PhysRevD.62.036012}.

\bibitem{Herold:2013bi}
C.~Herold, M.~Nahrgang, I.~Mishustin, and M.~Bleicher, Phys. Rev. C 87 (2013)
  014907, \urlprefix\url{https://dx.doi.org/10.1103/PhysRevC.87.014907}.

\bibitem{Nahrgang:2016eou}
M.~Nahrgang and C.~Herold, Eur. Phys. J. A 52 (2016) 240,
  \urlprefix\url{https://dx.doi.org/10.1140/epja/i2016-16240-9}.

\bibitem{Herold:2016uvv}
C.~Herold, M.~Nahrgang, Y.~Yan, and C.~Kobdaj, Phys. Rev. C 93 (2016) 021902,
  \urlprefix\url{https://dx.doi.org/10.1103/PhysRevC.93.021902}.

\bibitem{Nahrgang:2018afz}
M.~Nahrgang, M.~Bluhm, T.~Schaefer, and S.~A. Bass, Phys. Rev. D 99 (2019)
  116015, \urlprefix\url{https://dx.doi.org/10.1103/PhysRevD.99.116015}.

\bibitem{Kitazawa:2020kvc}
M.~Kitazawa, G.~Pihan, N.~Touroux, M.~Bluhm, and M.~Nahrgang, Nucl. Phys. A
  1005 (2021) 121797,
  \urlprefix\url{https://doi.org/10.1016/j.nuclphysa.2020.121797}.

\bibitem{PhysRevLett.116.022301}
D.~Bazow, G.~S. Denicol, U.~Heinz, M.~Martinez, and J.~Noronha, Phys. Rev.
  Lett. 116 (2016) 022301,
  \urlprefix\url{https://dx.doi.org/10.1103/PhysRevLett.116.022301}.

\bibitem{Bazow_Denicol_Heinz_Martinez_Noronha_2016}
D.~Bazow, G.~S. Denicol, U.~Heinz, M.~Martinez, and J.~Noronha, Physical Review
  D 94 (2016), \urlprefix\url{https://dx.doi.org/10.1103/PhysRevD.94.125006}.

\bibitem{book:Bernstein}
J.~Bernstein, \emph{{Kinetic Theory in the Expanding Universe (Cambridge
  Monographs on Mathematical Physics)}} (Cambridge University Press, 1988).

\bibitem{book:Goenner}
H.~Goenner, \emph{{Einf{\"u}hrung in die Kosmologie}} (Spektrum Akamedischer
  Verlag, 1994).

\bibitem{book:Lord}
G.~J.~L. et~al., \emph{{An Introduction to Computational Stochastic PDEs}},
  Cambridge Texts in Applied Mathematics (Cambridge University Press, 2014), 1
  ed.

\bibitem{PhysRevE.62.1537}
H.~Kleinert, A.~Pelster, B.~Kastening, and M.~Bachmann, Phys. Rev. E 62 (2000)
  1537, \urlprefix\url{https://dx.doi.org/10.1103/PhysRevE.62.1537}.

\bibitem{book:16401}
H.~Kleinert, \emph{{Particles and quantum fields}} (World Scientific Pub Co
  Inc, 2016).

\end{thebibliography}

\end{document}